\documentclass[
11pt, 
english, 
singlespacing, 
headsepline, 
]{MastersDoctoralThesis} 

\usepackage[utf8]{inputenc} 
\usepackage[T1]{fontenc} 
\usepackage{mathpazo} 
\usepackage{amsmath}
\usepackage{graphicx}
\usepackage{lscape}
\usepackage{caption}
\usepackage{makecell}
\usepackage{placeins}
\usepackage[flushleft]{threeparttable}
\usepackage{ragged2e}
\usepackage[backend=biber,style=authoryear,natbib=true]{biblatex} 
\usepackage{tabularx}
\usepackage[autostyle=true]{csquotes} 
\usepackage{titlesec}
\usepackage{bookmark}
\usepackage{xcolor,hyperref}
\usepackage{tikz}
\usetikzlibrary{calc}
\usetikzlibrary{decorations.pathmorphing}
\usepackage{lipsum}

\titleformat{\paragraph}
{\normalfont\normalsize\bfseries}{\theparagraph}{1em}{}
\titlespacing*{\paragraph}
{0pt}{3.25ex plus 1ex minus .2ex}{1.5ex plus .2ex}

\thesistitle{Exoplanets Prediction in Multi-Planetary Systems and Determining the Correlation Between the Parameters of Planets and Host Stars Using Artificial Intelligence} 
\supervisor{Prof. Davood \textsc{M. Jassur}\\Dr. Ghassem \textsc{Gozaliasl}} 
\examiner{} 
\degree{Doctor of Philosophy} 
\author{Mahdiyar \textsc{Mousavi-Sadr}} 
\addresses{} 

\keywords{} 
\university{University of Tabriz} 
\department{Department of Theoretical Physics and Astrophysics} 
\faculty{Faculty of Physics} 

\AtBeginDocument{
\hypersetup{pdftitle=\ttitle} 
\hypersetup{pdfauthor=\authorname} 
\hypersetup{pdfkeywords=\keywordnames} 
\hypersetup{citecolor=blue}
\hypersetup{linkcolor=mdtRed}
\hypersetup{urlcolor=mdtRed}
}

\begin{document}

\frontmatter 
\pagestyle{plain} 

\defcitealias{2013MNRAS.435.1126B}{BL13}
\defcitealias{2015MNRAS.448.3608B}{BL15}
\defcitealias{2017A&A...604A..83B}{B17}
\defcitealias{2019A&A...630A.135U}{Ulmer19}

\begin{titlepage}
\begin{center}
\vspace*{-55pt}
\includegraphics[scale=0.06]{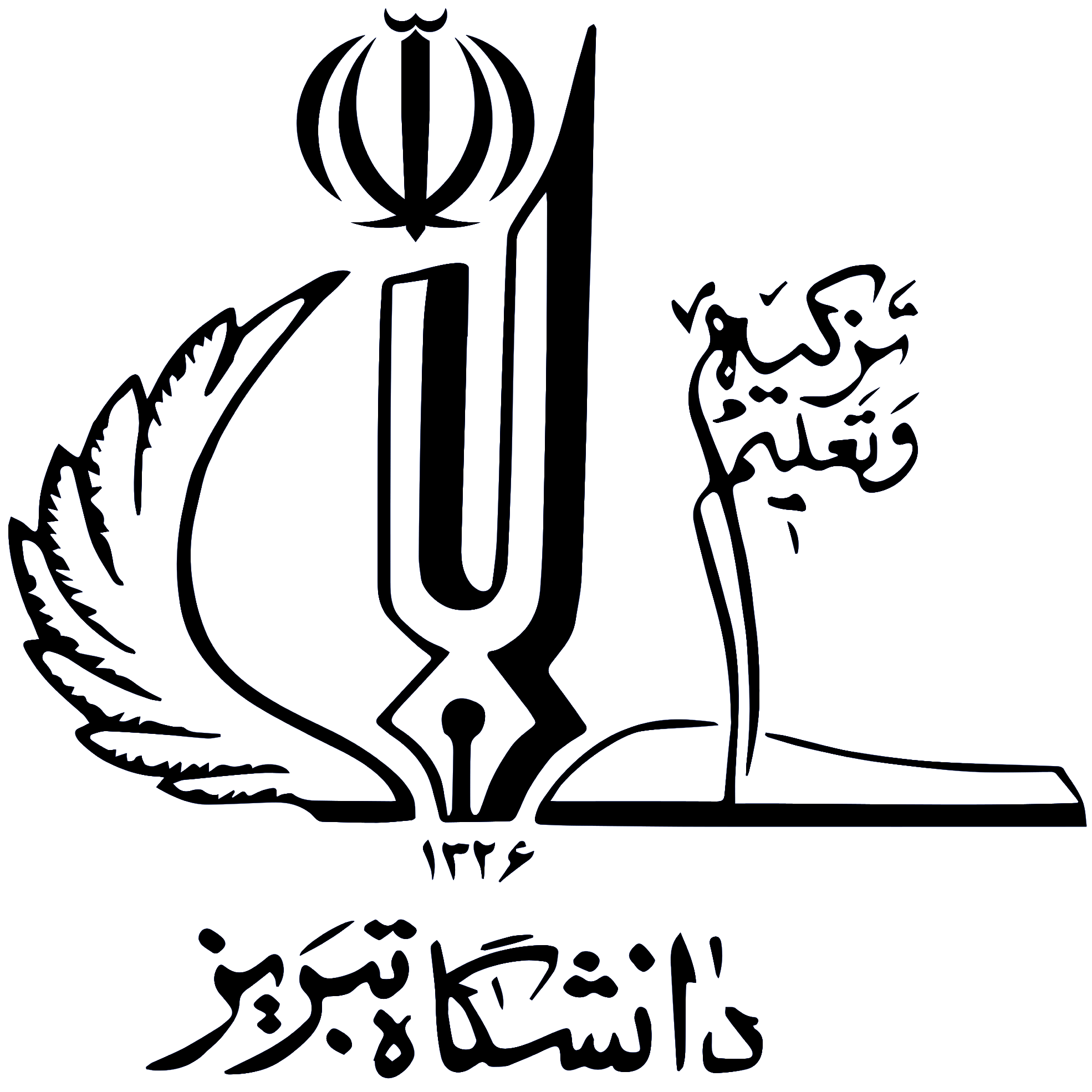} 
\vspace{-0.015\textheight}

{\scshape\LARGE \univname\\ \Large\facname\\\deptname\par}\vspace{0.75cm} 
\textsc{\Large Doctoral Dissertation}\\[0.20cm] 

\HRule \\
{\huge \bfseries \ttitle\par}
\HRule \\[0.20cm] 
 
\begin{minipage}[t]{0.4\textwidth}
\begin{flushleft} \large
\emph{Author:}\\
{\href{https://scholar.google.com/citations?user=xo3DPZAAAAAJ&hl=en&oi=ao}\authorname} 
\end{flushleft}
\end{minipage}
\begin{minipage}[t]{0.4\textwidth}
\begin{flushright} \large
\emph{Supervisors:} \\
{\supname} 
\end{flushright}
\end{minipage}\\[1cm]

\large \textit{A thesis submitted in fulfillment of the requirements\\ for the degree of \degreename}\\[0.3cm] 
\textit{in the}\\[0.3cm]
\textit{Astronomy and Astrophysics}\\[1.75cm]
{\large February 2024}\\[4cm] 

\vfill
\end{center}
\end{titlepage}




 

\cleardoublepage


\vspace*{0.01\textheight}
\noindent\enquote{\itshape Our posturings, our imagined self-importance, the delusion that we have some privileged position in the Universe, are challenged by this point of pale light. Our planet is a lonely speck in the great enveloping cosmic dark.}\bigbreak

\hfill Carl Sagan
\bigbreak
\bigbreak

\begin{center}
\includegraphics[scale=0.082]{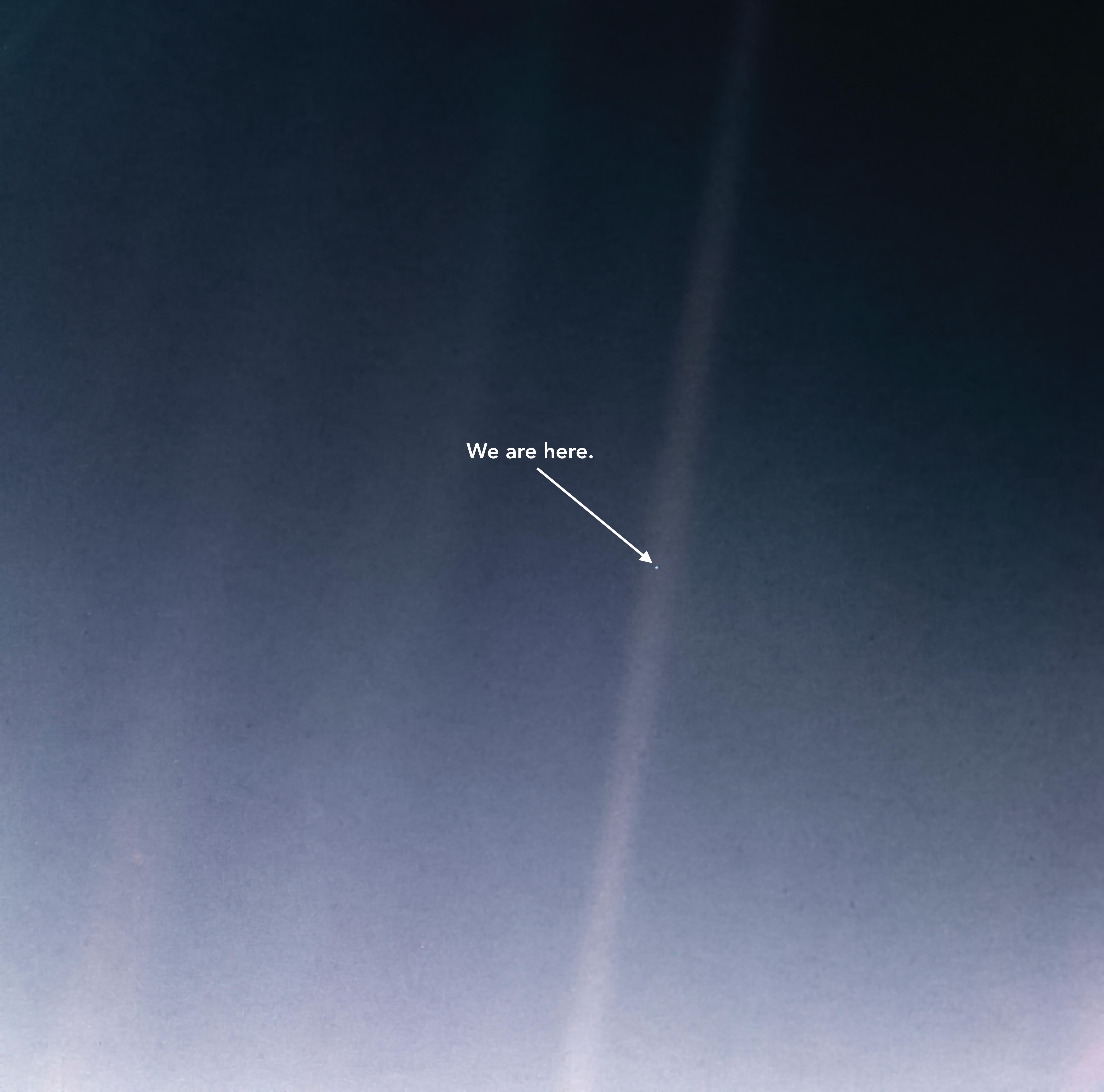}

\small The Pale Blue Dot: This is a photograph of Earth taken in 1990 by the Voyager 1 at a distance of 6 billion kilometers from the Sun.
\end{center}
\vspace*{0.1\textheight}


\dedicatory{To my family and all my teachers.}


\begin{acknowledgements}
\addchaptertocentry{\acknowledgementname} 
Words cannot express my gratitude to the people without whom I could not have completed my Ph.D. I would like to express my deepest appreciation to my supervisors, Prof. Davood M. Jassur and Dr. Ghassem Gozaliasl, for their enthusiasm for this project and their support, encouragement, and patience. It is with great acknowledgment that I appreciate my caring and kind mother for always selflessly helping me achieve my goals. I am extremely grateful to my father, who made me interested in astronomy when I was an elementary student through a book entitled \emph{I Wonder Why Stars Twinkle and Other Questions About Space} by Carole Stott. Many thanks to my loving sister, who has always been by my side like a friend. Lastly, thanks should go to my uncle, Mr. Mir Vali Mousavi Sadr, who has always kept my spirit and motivation high during this process.
\end{acknowledgements}

\begin{abstract}
\addchaptertocentry{\abstractname} 
\begin{tikzpicture}[overlay, remember picture]
\draw [line width=0.25mm, decorate]
($ (current page.north west) + (2.78cm, -2.5cm) $)rectangle($ (current page.south east) + (-2.25cm, 2.25cm) $);\end{tikzpicture}\textbf{Abstract:} The number of extrasolar planets discovered is increasing, so that more than five thousand exoplanets have been confirmed to date, and there are also thousands of planetary candidates whose existence requires more detailed observations and studies. The Kepler space telescope and TESS play a prominent role in detecting planets and increasing our knowledge about them. The growing number of exoplanet discoveries and advances in machine learning (ML) techniques have opened new avenues for exploring and understanding the characteristics of worlds beyond our Solar System. Now we have an opportunity to test the validity of the laws governing planetary systems and take steps to discover the relationships between the physical parameters of planets and stars.

Firstly, we present the results of a search for additional exoplanets in all multi-planetary systems discovered to date, employing a logarithmic spacing between planets in our Solar System known as the Titius-Bode (TB) relation. We use the Markov Chain Monte Carlo (MCMC) method and separately analyze 229  multi-planetary systems that house at least three or more confirmed planets. We find that the planets in $\sim53\%$ of these systems adhere to a logarithmic spacing relation remarkably better than the Solar System planets. Using the TB relation, we predict the presence of 426 additional exoplanets in 229 multi-planetary systems, of which 197 candidates are discovered by interpolation and 229 by extrapolation. Altogether, 47 predicted planets are located within the habitable zone (HZ) of their host stars, and five of the 47 planets have a maximum mass limit of 0.1-2$M_{\oplus}$ and a maximum radius lower than 1.25$R_{\oplus}$. Our results and prediction of additional planets agree with previous studies' predictions; however, we improve the uncertainties in the orbital period measurement for the predicted planets significantly.
\begin{tikzpicture}[overlay, remember picture]
\draw [line width=0.25mm, decorate]
($ (current page.north west) + (2.25cm, -2.5cm) $)rectangle($ (current page.south east) + (-2.65cm, 7.5cm) $);\end{tikzpicture}

Secondly, we employ efficient machine learning approaches to analyze a dataset comprising 762 confirmed exoplanets and eight Solar System planets, aiming to characterize their fundamental quantities. By applying different unsupervised clustering algorithms, we classify the data into two main classes: “small” and “giant” planets, with cut-off values at $R_{p}=8.13R_{\oplus}$ and $M_{p}=52.48M_{\oplus}$. This classification reveals an intriguing distinction: giant planets have lower densities, suggesting higher H-He mass fractions, while small planets are denser, composed mainly of heavier elements. We apply various regression models to uncover correlations between physical parameters and their predictive power for exoplanet radius. Our analysis highlights that planetary mass, orbital period, and stellar mass play crucial roles in predicting exoplanet radius. Among the models evaluated, the Support Vector Regression consistently outperforms others, demonstrating its promise for obtaining accurate planetary radius estimates. Furthermore, we derive parametric equations using the M5P and MCMC methods. Notably, our study reveals a noteworthy result: small planets exhibit a positive linear mass-radius relation, aligning with previous findings. Conversely, for giant planets, we observe a strong correlation between planetary radius and the mass of their host stars, which might provide intriguing insights into the relationship between giant planet formation and stellar characteristics.
\end{abstract}


\tableofcontents 

\listoffigures 

\listoftables 


\begin{abbreviations}{ll} 

\textbf{AU} & \textbf{A}stronomical \textbf{U}nit\\
\textbf{FS} & \textbf{F}eature \textbf{S}election \\
\textbf{HZ} & \textbf{H}abitable \textbf{Z}one\\
\textbf{JWST} & \textbf{J}ames \textbf{W}ebb \textbf{S}pace \textbf{T}elescope\\
\textbf{LOF} & \textbf{L}ocal \textbf{O}utlier \textbf{F}actor \\
\textbf{MAE} & \textbf{M}ean \textbf{A}bsolute \textbf{E}rror \\
\textbf{MCMC} & \textbf{M}arkov \textbf{C}hain \textbf{M}onte \textbf{C}arlo\\
\textbf{ML} & \textbf{M}achine \textbf{L}earning\\
\textbf{NMAD} & \textbf{N}ormalized \textbf{M}edian \textbf{A}bsolute \textbf{D}eviation\\
\textbf{RMSE} & \textbf{R}oot \textbf{M}eans \textbf{S}quare \textbf{E}rror\\
\textbf{SED} & \textbf{S}pectral \textbf{E}nergy \textbf{D}istribution \\
\textbf{SNR} & \textbf{S}ignal-to-\textbf{N}oise \textbf{R}atio \\
\textbf{SSD} & \textbf{S}um (of) \textbf{S}quared \textbf{D}istances \\
\textbf{SVR} & \textbf{S}upport \textbf{V}ector \textbf{R}egression \\
\textbf{TB} & \textbf{T}itius \textbf{B}ode\\
\textbf{TESS} & \textbf{T}ransiting \textbf{E}xoplanet \textbf{S}urvey \textbf{S}atellite\\
\textbf{YSO} & \textbf{Y}ong \textbf{S}tellar \textbf{O}bject\\
\textbf{ZAMS} & \textbf{Z}ero \textbf{A}ge \textbf{M}ain-sequence \textbf{S}tar\\

\end{abbreviations}







\begin{symbols}{lll} 

$M_{\odot}$ & mass of Sun & \\
$R_{\odot}$ & radius of Sun & \\
$M_{\oplus}$ & mass of Earth & \\
$R_{\oplus}$ & radius of Earth & \\
$d$ & days on Earth \\
$M_{J}$ & mass of Jupiter & \\
$R_{J}$ & radius of Jupiter & \\
$M_{s}$ & stellar mass & \\
$R_{s}$ & stellar radius & \\
$M_{p}$ & planetary mass & \\
$R_{p}$ & planetary radius & \\
$T_{\text{equ}}$ & planet's equilibrium temperature \\
$P$ & orbital period \\
$a$ & orbital semi-major axis \\
$e$ & orbital eccentricity & \\
$T_{\text{eff}}$ & star's effective temperature \\
$L$ & star's luminosity \\
Fe/H & star's metallicity & \\
$\Delta$ & dynamical spacing criterion \\
$\rho^{2}$ & coefficient of determination \\

\addlinespace 
\end{symbols}

\begin{introduction}
\addchaptertocentry{\introductionname}
The number of detected new worlds beyond the Solar System, known as exoplanets, is growing rapidly, so that over five thousand exoplanets have been detected and confirmed to date. There are also thousands of other candidate exoplanets that require further follow-up observation. Our comprehension of exoplanets, their population, and diversity come largely from the latest generation of modern satellites. The Kepler space mission, the Transiting Exoplanet Survey Satellite, the James Webb Space Telescope, and many ground-based observatories make important contributions to detecting and characterizing exoplanets \citep{2007PASP..119..923P,2010Sci...327..977B,2014PASP..126.1134B}. The data generated by these state-of-the-art instruments are now available to everyone. Researchers skilled in data science, data analytics, or machine learning (ML) and neural network techniques study and analyze these data to predict, identify, characterize, and classify the exoplanets \citep{2019A&A...626A..21A,2019MNRAS.487.5062M,2020EPSC...14..833B,2020AJ....159...41T,2021PASA...38...15M,2021A&A...656A..73S,2021MNRAS.507.2154V,2023A&A...670A..68M,2023NatAs...7....8M,2023MNRAS.525.3469M}. In addition, the observational data are not only used to study exoplanets but they are also applied to peruse entire planetary and stellar science. As many planets are found around other stars, they have provided us with an opportunity to understand the main ways of planet formation and evolution and to put our Solar System in a broader context \citep{2018MNRAS.473..784K,2020apfs.book.....A,2020AJ....159..281G,2021NewA...8901625M,mishra2023framework}.

Photometry with the transit method is the most successful exoplanet discovery way. However, this method has its own difficulties. For example, transits are detectable only when the planet's orbit happens to be almost exactly aligned with the observer's line-of-sight. This covers only a small fraction of exoplanets. Furthermore, the planet's transit lasts for a small fraction of its total orbital period. As a result, it is not very likely to detect planets' transits, especially those with long orbital periods. The first part of this study sets out to predict the existence of additional undetected planets in multiple exoplanet systems. We test the adherence of 229 multiple exoplanet systems with at least three detected planets to the TB relation, and compare their adherence rate with the Solar System’s. Next, we aim to predict the undetected exoplanets and estimate their physical properties, either maximum mass or maximum radius. Moreover, we highlight exoplanets located within the HZ of their host stars. Using a sample of seven multiple exoplanet systems with detected planets after predictions made by \citetalias{2015MNRAS.448.3608B}, we also aim to determine if the TB relation is a reliable method of identifying undetected member planets. Finally, we compare our predictions with those from \citetalias{2015MNRAS.448.3608B}.

The planetary mass and radius are two fundamental physical properties to understand a planet’s composition. However, not all discovered exoplanets have both the measured mass and radius \citep{2010exop.book.....S,2018haex.bookE.117D}. As the number of discovered exoplanets rapidly increases, ML techniques can be used to investigate correlations between planets and their host stars. In the second part of this study, we implement various ML regression algorithms to find the potential relationships between physical parameters in exoplanet systems. The Markov Chain Monte Carlo (MCMC) \citep[see][]{goodman10,emcee3} as a precise method for regression, is used to quantify the uncertainties of the best-fit parameters. In addition, different ML clustering algorithms are used to group the exoplanets and study their properties.

This dissertation is organized as follows: chapter~\ref{chapter1} describes some basic concepts and reviews previous studies. In chapter~ \ref{chapter2}, the data samples and methods are introduced. Chapter~\ref{chapter3} presents the results, discussion, and future work.
\end{introduction}


\mainmatter 

\pagestyle{thesis} 



\chapter{Literature Review} 

\label{chapter1} 



\newpage
\section{The origin of planets}
\subsection{History}
The origin of the solar system has been a mystery for astronomers for a couple of reasons. Firstly, it is challenging to observe the planet-formation regions because the protoplanetary disks around young stars are mostly covered by gas and dust. Secondly, the processes of planet formation are immensely shorter than the lifetime of a star. It takes tens to hundreds of millions of years for a planet to form, which is a tiny fraction of a typical star’s main-sequence lifetime \citep{johnson2015you}.

The history of scientific thought about the formation and evolution of the solar system includes various hypotheses, e.g., Laplace’s nebular hypothesis, the tidal hypothesis, the Chamberlin-Moulton model, the Lyttleton’s scenario, the capture hypothesis, etc. Laplace stated four facts about the solar system that must be explained by any planet-formation theory. First, the orbits of planets are all almost in the same plane. Second, all the planets revolve around the Sun in the same direction, counter-clockwise when viewed from above the Sun’s north pole. Third, the orbits of all the planets are roughly circular. Fourth, every planet except for Venus and Uranus rotates counter-clockwise as seen from the planet’s north pole, which is the same direction that they revolve around the Sun \citep{1975orpl.book.....W,1989ossc.book.....D}.

Additionally, other complicated phenomena in the solar system need to be considered. For example, even though 99.9 percent of the solar system’s mass is concentrated in its center, the Sun has only one percent of the total angular momentum of the system. Moreover, the distances of the planets from the Sun follow a simple arithmetic progression known as the Titius-Bode law. Furthermore, the difference between the formation and evolution of terrestrial planets, which have solid cores, and gas giant planets like Jupiter, or the origin of planetary satellites like the Moon, should be addressed by any hypothesis.

\subsection{Modern nebular theory}
The most widely accepted model of planetary formation is known as the modern nebular theory. This model posits that the solar system originated in a gravitationally unstable giant molecular cloud. The contraction of this gas and dust cloud under its own gravitational pull and, hence the small net rotation of the cloud fragment created a disk around the central condensation. Eventually, the central condensation formed the Sun, while small condensations in the disk formed the planets.

\subsubsection{Young stellar objects}
\begin{figure}[t!]
\centering
\captionsetup{width=0.70\paperwidth}
	\includegraphics[width=0.70\paperwidth]{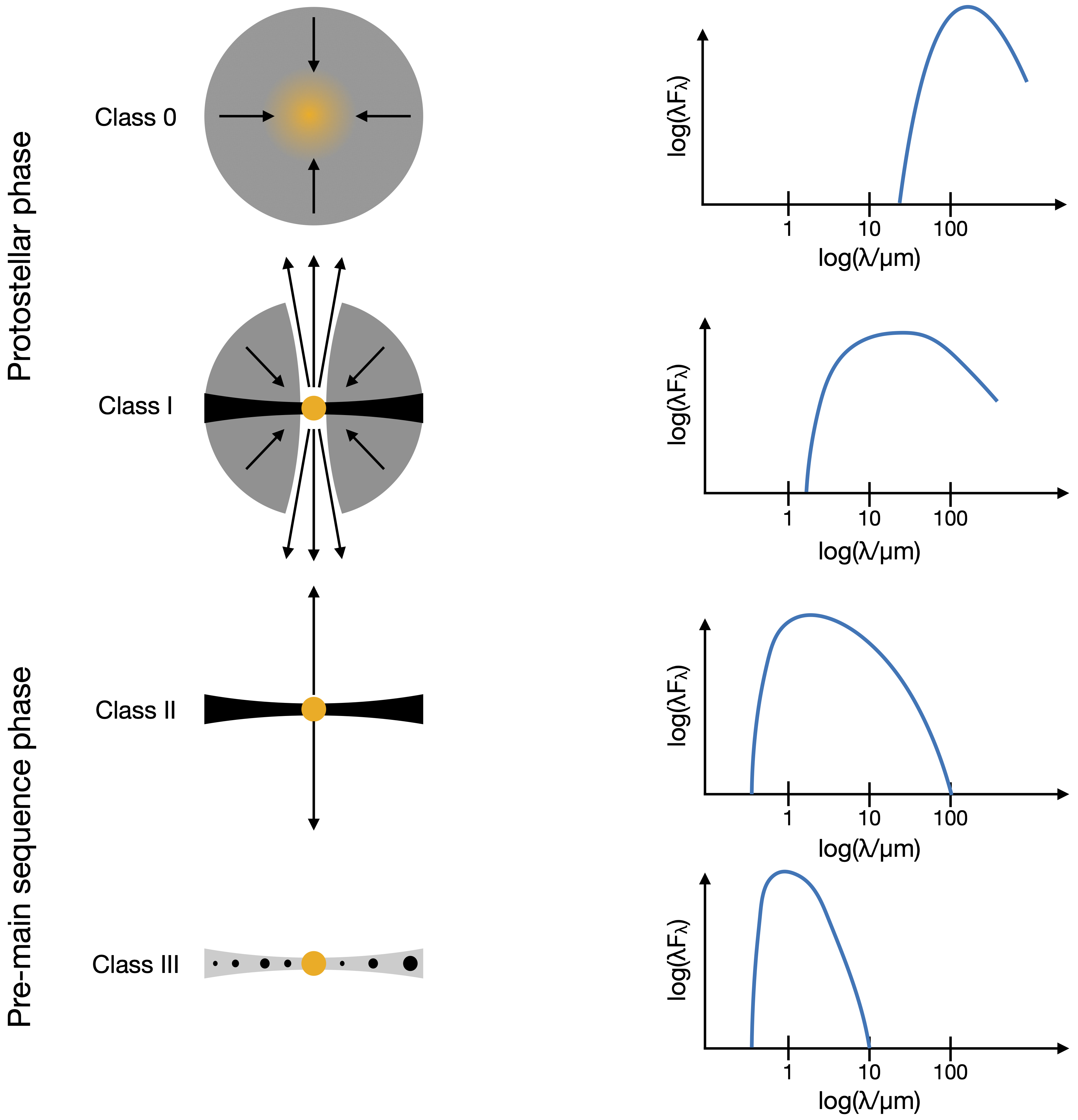}
    \caption[Classification of YSOs based on the slope of their SED]{Classification of young stellar objects (YSOs) based on the slope of their spectral energy distribution (SED). \textit{[The image has been reproduced from \citet{2021PhDT........30S}.]}}
    \label{figure1.1}
\end{figure}

The evolution of a young stellar object (YSO) is classified into four main stages (see figure~\ref{figure1.1}) based on the slope of its spectral energy distribution (SED), i.e., values of intervals of spectral index ($\alpha$), which is calculated between wavelength of about 2 and 25 $\mu m$ (near- and mid-infrared region) \citep{1984ApJ...287..610L,1987IAUS..115....1L}:

\begin{equation}
\alpha=\frac{d\log \lambda F_{\lambda}}{d\log \lambda}.
\label{eq2.1}
\end{equation}

\noindent Here, $\lambda$ is wavelength, and $F_{\lambda}$ is flux density.

\begin{figure}[h!]
\centering
\captionsetup{width=0.70\paperwidth}
	\includegraphics[width=0.70\paperwidth]{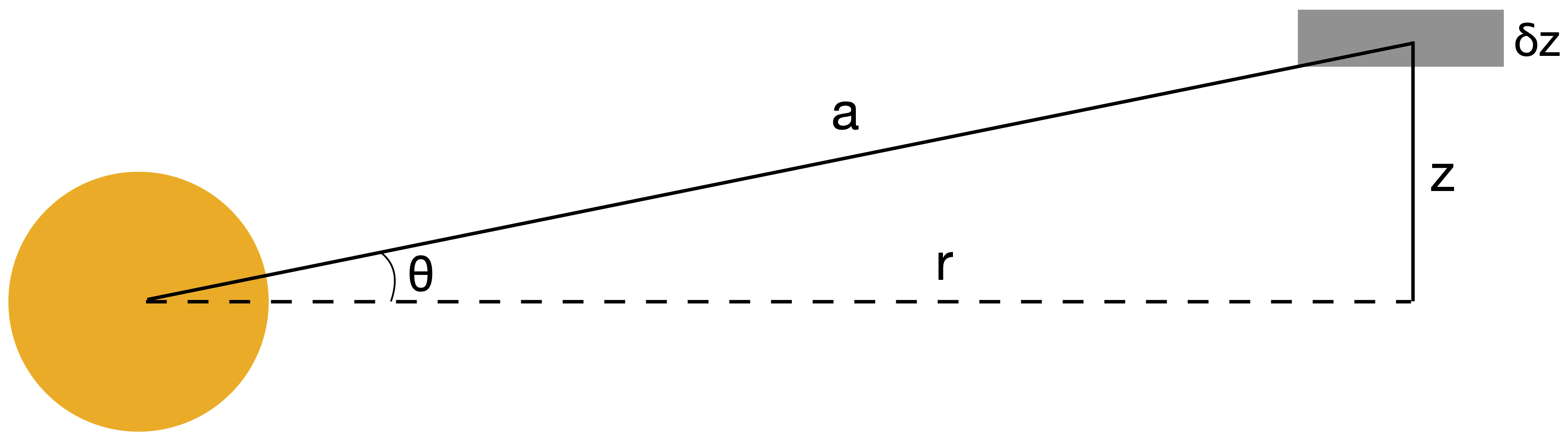}
    \caption[A small element located in the disk around the star]{A small element is located in the disk at a distance $a$ from the star. The distance of the element along the disk is $r$, and its distance from the mid-plane of the disk is represented by $z$. \textit{[The image has been reproduced from \citet{2009pps..book.....E}.]}}
    \label{figure1.2}
\end{figure}

In the beginning, the molecular cloud collapses into a protostar fully embedded in a surrounding envelope of gas and dust. So, this warm envelope can only be observed in far-infrared or mm wavelengths. The SED of class 0 objects is undetectable at $\lambda<20\mu m$. In the following, owing to the angular momentum conservation, the falling material forms a disk buried in an envelope. The SED of class I objects has a positive slope in the near- and mid-infrared ($\alpha>0.3$). The disk continues to accrete onto the protostar, and some material is lost through the outflow to preserve the angular momentum. In the third stage, due to the accretion and molecular outflows, the envelope has been scattered, and the disk is clear. In the center of the disk, a classical T Tauri star, which is an optically visible pre-main-sequence star heated mostly by the gravitational energy of its contraction, exists. The SED of class II objects has either a flat or negative slope ($-1.6<\alpha<-0.3$). Class III objects have lost their accretion disks and correspond approximately to weak-line T Tauri stars. At this stage, the pre-main-sequence star is surrounded by a disk of debris, which is a gas-poor disk dominated by dust. The SED of class III objects ($\alpha<-1.6$) can be fit with a reddened blackbody function. In the end, the pre-main-sequence star moves onto the main-sequence with continuous hydrogen burning, known as a zero-age main-sequence star \citep{2011ARA&A..49...67W,2021PhDT........30S}.

\subsubsection{Planets formation}
In the collapsing molecular cloud, due to the conversion of gravitational energy into heat, the pressure increases, and the contraction stops. At this time, no solid objects have been formed out of the protoplanetary disk. We assume that the star has already formed in the center of the protoplanetary disk. In figure~\ref{figure1.2}, a small element is located in the disk at a distance $a$ from the star. The distance of the element along the disk is $r$, and its distance from the mid-plane of the disk is represented by $z$. The vertical component of the gravitational force on the element is given by the equation~\ref{eq2.2}, where $M_{s}$ is the mass of star, $\rho$ is the density of gas the element contains, $A$ is the cross-section area of the element, and $\delta z$ is its thickness. If the angle $\theta$ is very small, $a=r$ and $sin\theta\approx z/a\approx z/r$. So, the vertical component of the gravitational force can be obtained by the equation~\ref{eq2.3} \citep{2009pps..book.....E}.

\begin{equation}
F_{g}=\frac{GM_{s}\rho A\delta z\sin \theta}{a^{2}}
\label{eq2.2}
\end{equation}

\begin{equation}
F_{g}=\frac{GM_{s}\rho Az\delta z}{r^{3}}
\label{eq2.3}
\end{equation}

Using the principle of hydrostatic equilibrium, there is a balance between gravitational force and pressure for any element of the disk. Therefore, $F_{g}$ is balanced by the force due to the pressure difference between the lower and upper face of the element or $A\delta P$. The perfect gas law is presented by equation~\ref{eq2.4}, where $<\mu_{A}>$ is the mean molecular weight, $m_{amu}$ is the atomic mass unit, $T$ is the temperature, and $k$ is Boltzmann’s constant. Using this equation and equating the gravitational force and the pressure, equation~\ref{eq2.5} is obtained, in which the negative sign indicates the reverse link between the pressure and height. By integrating the equation~\ref{eq2.5} and introducing the scale height $H$, which represents a height at which the pressure has fallen by a factor of $e^{-1}$ (see equation~\ref{eq2.6}), the relationship between height and pressure when the disk is balanced can be obtained by equation~\ref{eq2.7}. In this equation, $P_{0}$ is the pressure at the mid-plane of the disk or $z=0$. These two equations demonstrate that with increasing distance from the star, the thickness of the disk increases \citep{2009pps..book.....E,2020apfs.book.....A}.

\begin{equation}
P<\mu_{A}>m_{\text{amu}}=\rho KT
\label{eq2.4}
\end{equation}

\begin{equation}
\frac{\delta P}{P}=-\frac{<\mu_{A}>m_{\text{amu}}GM_{s}z\delta z}{kTr^{3}}
\label{eq2.5}
\end{equation}

\begin{equation}
H=\sqrt{\frac{2r^3 kT}{<\mu_{A}>m_{\text{amu}}GM_{s}}}
\label{eq2.6}
\end{equation}

\begin{equation}
P=P_{0}e^{-z^{2}/H^{2}}
\label{eq2.7}
\end{equation}

The composition of planets in the solar system is diverse even though they are all formed from the same nebula. On the one hand, the four inner planets are rocky, small, and very dense, with solid surfaces and thin atmospheres of heavier gases such as carbon dioxide, nitrogen, and oxygen. On the other hand, the five outer planets are gas giants, much larger with thick atmospheres of lighter gases such as hydrogen, helium, ammonia, methane, and other hydrocarbons, but much less dense. This diversity is rooted in the condensation of matter in protoplanetary disks. In a protoplanetary disk, different elements condense sequentially based on the temperature of the nebula. Metals like calcium, aluminum, and nickel that have high boiling points condense first. The molecular cloud continues to cool, and compounds with lower boiling points condense. This can happen to even very volatile compounds like water, methane, and ammonia \citep{2009pps..book.....E}.

\begin{figure}
\centering
\captionsetup{width=0.60\paperwidth}
	\includegraphics[width=0.60\paperwidth]{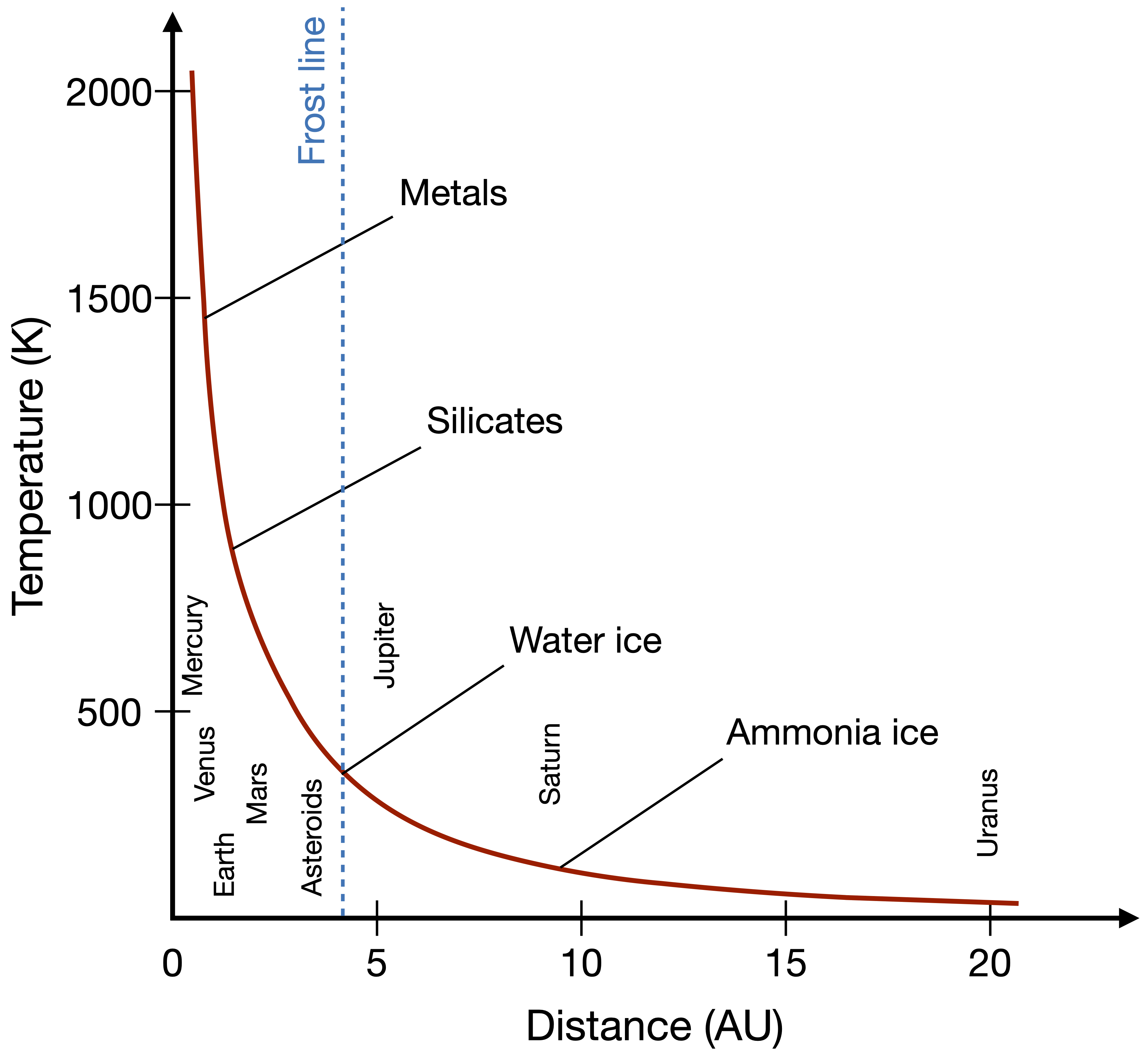}
    \caption[The condensation sequence]{The condensation sequence is the order in which chemical compounds go from gas to solid in a protoplanetary disk based on the condensation temperature of each compound. It explains why the solar system has two groups of planets: large, low-density Jovian planets in the outer parts and small, high-density terrestrial planets in the inner parts.}
    \label{figure1.3}
\end{figure}

Figure~\ref{figure1.3} presents the condensation sequence in which the different distances from the Sun (or disparate temperatures) allowed different elements to be used in the creation of planets. The terrestrial planets, which are closer to the Sun, formed at much higher temperatures than the Jovian planets. At such a high temperature, the only materials that could condense out were the metals and silicates. The outer planets, however, formed from the condensation of water, ammonia, and methane ice in a much colder environment. In figure~\ref{figure1.3}, the blue dashed line (at around 4 AU from the Sun) represents the frost line, also known as the snow line or ice line. It is a boundary zone in the solar system, where it becomes cold enough for volatile compounds like water, ammonia, and methane to condense into solid grains. Inside, the frost line is so warm that only rocks and metals can condense. Outside, the frost line is cold enough for hydrogen compounds (water, ammonia, and methane) to turn into ice.

\begin{figure}
\centering
\captionsetup{width=0.60\paperwidth}
	\includegraphics[width=0.60\paperwidth]{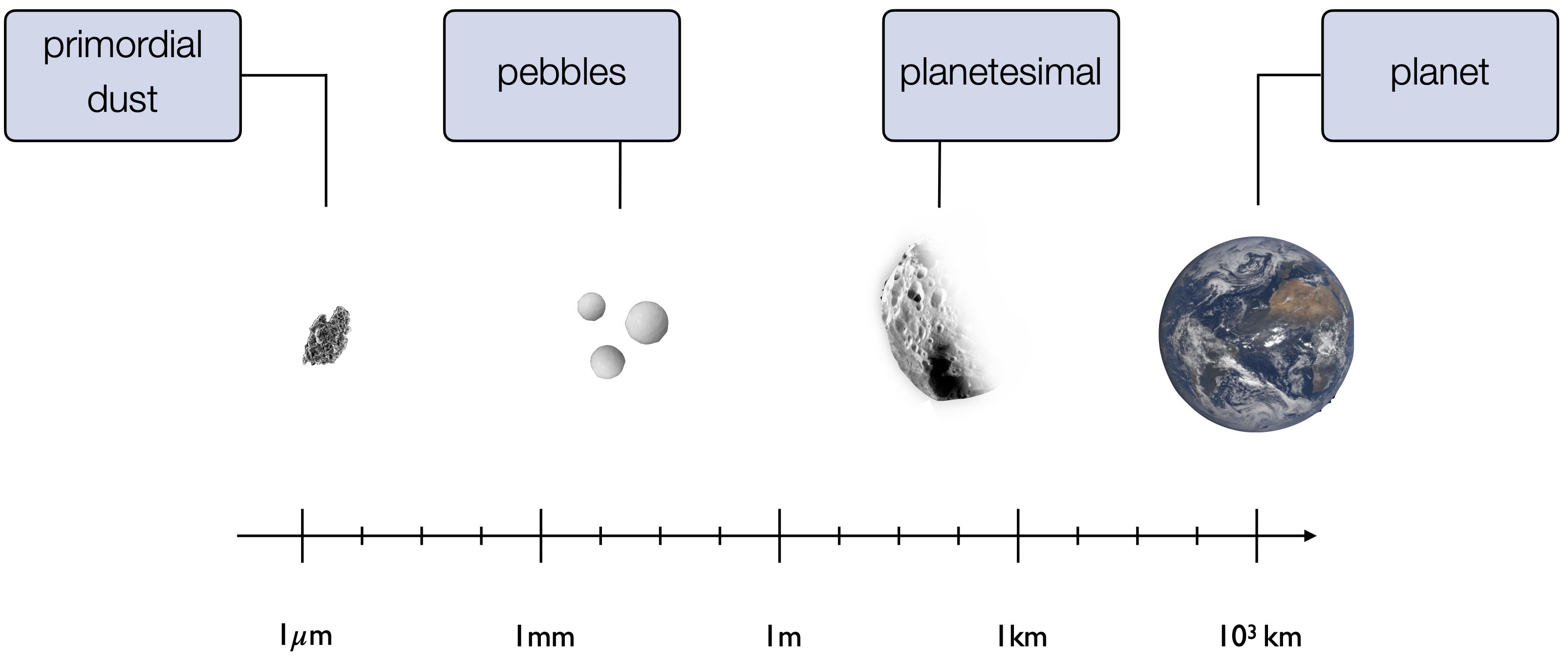}
    \caption[An overview of the planet formation process]{Overview of the planet formation process. The bottom axis shows the size of objects involved in the process. \textit{[Image source: \citet{2021PhDT........30S}.]}}
    \label{figure1.4}
\end{figure}

\begin{figure}
\centering
\captionsetup{width=0.60\paperwidth}
	\includegraphics[width=0.60\paperwidth]{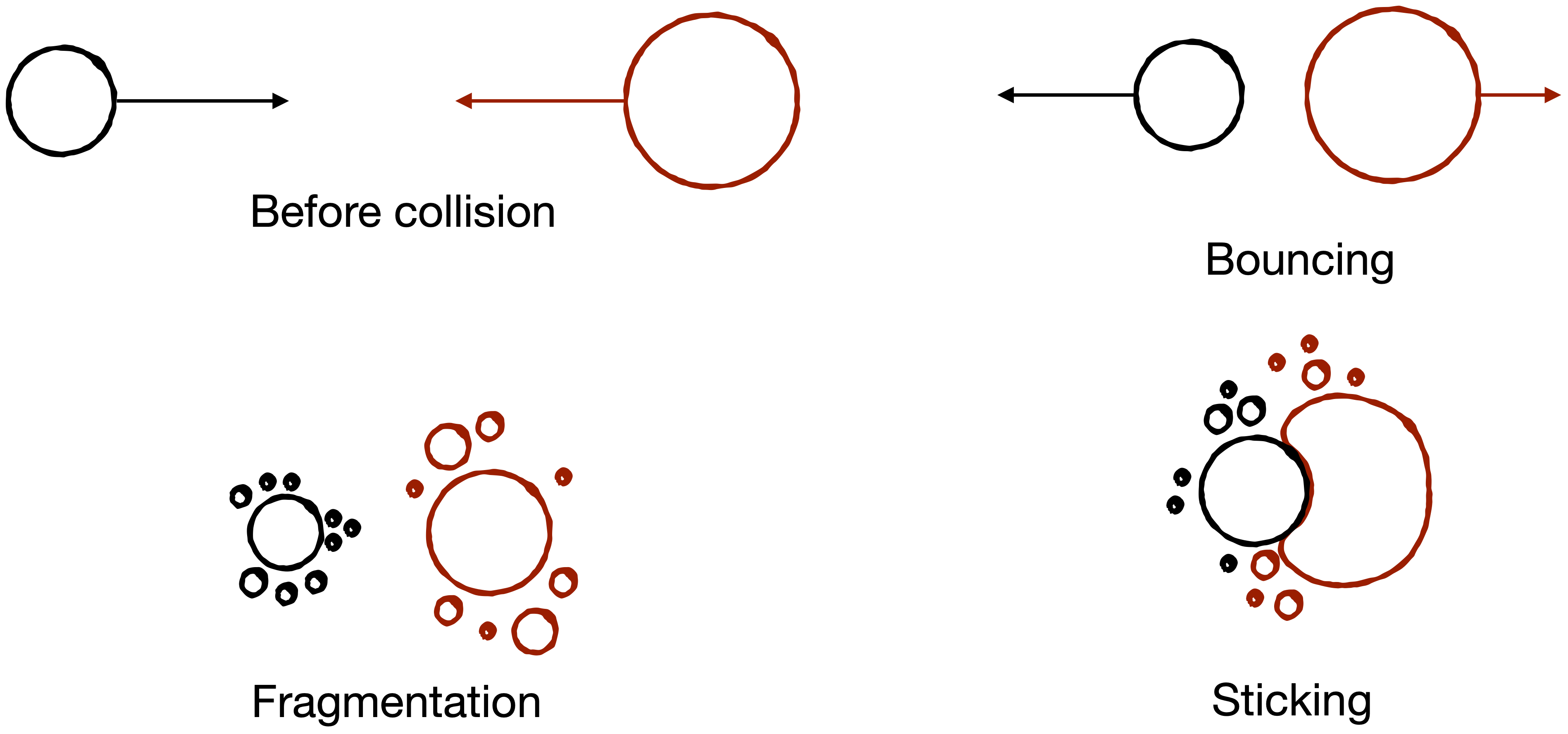}
    \caption[Three types of collisions between planetesimals]{Three types of collisions between planetesimals. The two planetesimals can bounce off each other, break into pieces, or stick together.}
    \label{figure1.5}
\end{figure}

Once the condensation process in a planetary nebula is completed, there are plenty of micron-sized grains. Through the accumulation of these ice and dust particles, centimeter-size pebbles, and subsequently, kilometer-sized planetesimals are formed (see figure~\ref{figure1.4}). Nevertheless, this part of the story is the most uncertain, and some fundamental problems need to be tackled.

The story becomes simpler when planetesimals dominate the protoplanetary disk. At this stage, the influence of the gas on the nebula is negligible. Gravity is the only significant force, which leads to three types of collisions between planetesimals (see figure~\ref{figure1.5}): the two planetesimals bounce off each other, one or both of the planetesimals are fragmented, and the two planetesimals stick together. The collision between a small planetesimal and a much larger planetesimal is almost certainly a coalescence.

We can use the rate of collisions between atoms in the kinetic theory of gases to estimate the planetesimal’s rate of mass increase (RMI). Providing that the relative velocity between the planetesimals is $v$, the mass of material swept up by a small planetesimal each second is given by equation~\ref{eq2.8}, in which $\rho_{n}$ is the density of the solid material in the nebula and $R$ is the radius of the planetesimal.

\begin{equation}
\frac{\text{swept-up mass}}{\text{1 second}}=\rho_{n}\pi R^{2}v
\label{eq2.8}
\end{equation}

However, this is only true if we ignore the effect of gravity. A large planetesimal collides with more objects due to its greater gravitational field. Therefore, the true rate of increase of the planetesimal’s mass can be given by adding a new term to this equation (see equation~\ref{eq2.9}, in which $v_{e}$ represents the escape velocity of the planetesimal).

\begin{equation}
RMI=\rho_{n}\pi R^{2}v\biggl(1+\frac{v_{e}^{2}}{v^{2}}\biggr)
\label{eq2.9}
\end{equation}

If the planetesimal is so massive that the first term in equation~\ref{eq2.9} is ignorable, the RMI is obtained by equation~\ref{eq2.10}. Using the escape velocity formula ($v_{e}^{2}=2GM/R$) and replacing the planetesimal’s mass by its average density ($\rho_{p}$) times its volume, the rate of increase in the mass of a planetesimal is given by equation~\ref{eq2.11}. Comparing this equation with equation~\ref{eq2.8} indicates that the larger planetesimals ($\propto R^{4}$) grow faster than smaller ones ($\propto R^{2}$) \citep{2009pps..book.....E}.

\begin{equation}
RMI\simeq\rho_{n}\pi R^{2}\frac{v_{e}^{2}}{v}
\label{eq2.10}
\end{equation}

\begin{equation}
RMI=\frac{8\pi^{2}\rho_{n}\rho_{p}GR^{4}}{3v}
\label{eq2.11}
\end{equation}

As a result, there is one planetesimal that becomes a planetary embryo in any region of the protoplanetary disk, which has a typical diameter greater than 1,000 kilometers. It is probable that the planetary embryos located in the inner portion of the protoplanetary disk run out of solid materials, cease to grow, and form planets. The computer simulations suggest that the number of early planetary embryos was more than the number of planets in the solar system today. Due to the gravitational forces between planetary embryos, their orbits were altering remarkably, with some being ejected from the solar system entirely and others merging to create the terrestrial planets we know today.

\begin{figure}[b!]
\centering
\captionsetup{width=0.50\paperwidth}
	\includegraphics[width=0.50\paperwidth]{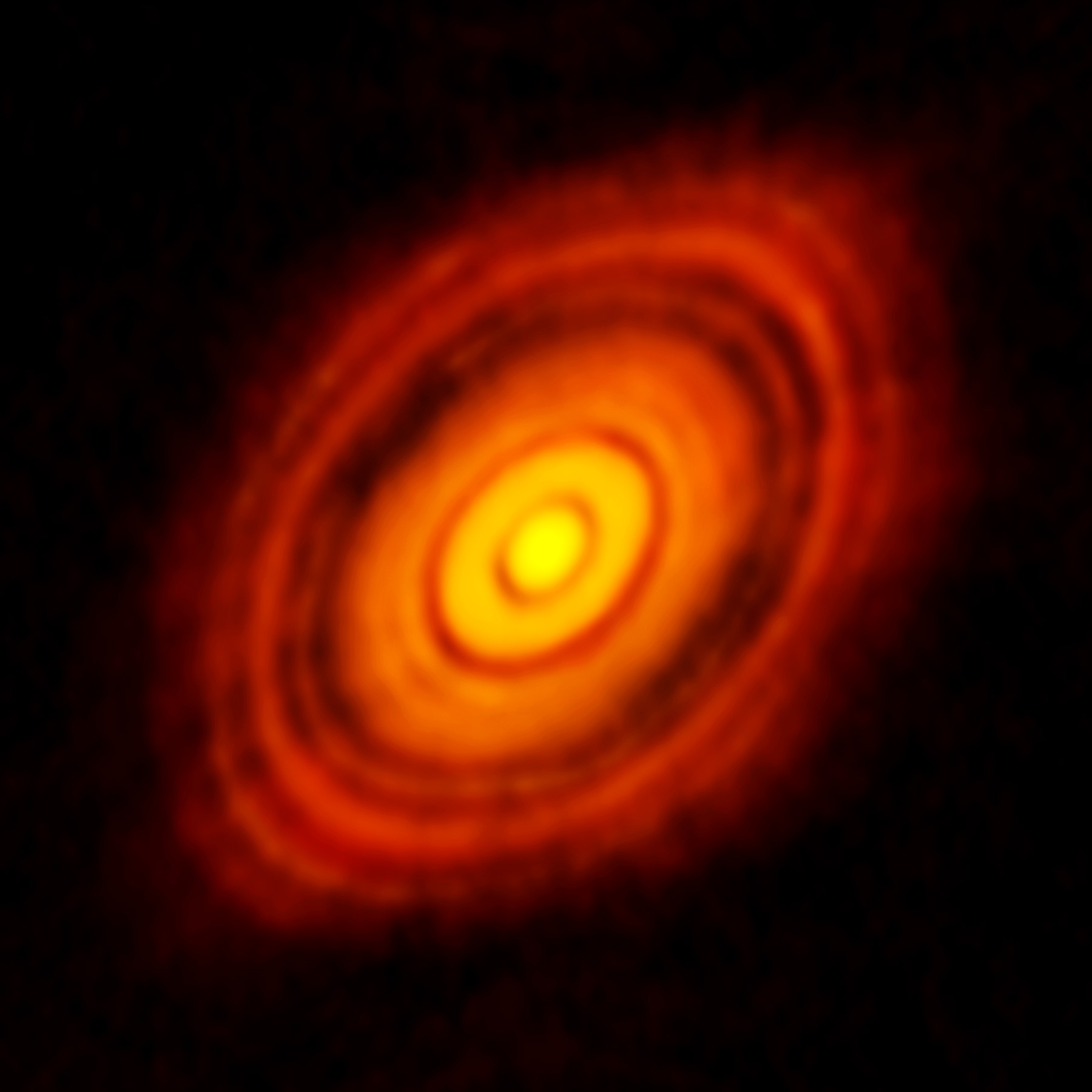}
    \caption[A protoplanetary disk surrounding the HL Tauri                                                                                                                                                                                                                                                                                                                                                                                                            ]{A protoplanetary disk surrounding the HL Tauri, which is a young newly formed star. \textit{[Image source: ALMA (ESO/NAOJ/NRAO)]}.}
    \label{figure1.6}
\end{figure}

Figure~\ref{figure1.6} shows the image of a protoplanetary disk captured by the Atacama Large Millimeter/sub-millimeter Array (ALMA). It has been suggested that the dark gaps in the protoplanetary disk must be caused by planets that formed in the disk and carved out material from their surroundings. Therefore, these gaps can indicate the presence of exoplanets.

\subsubsection{Gas giants}
The method by which the jovian planets of the solar system were formed is possibly different. There are two theories to explain this big remaining question in planetary science: the core-accretion theory and the gravitational-collapse theory. The first one claims that a giant planet consists of a solid core surrounded by a large amount of gas. The rocky and icy core of a giant planet forms like the terrestrial planets. In the protoplanetary nebula, the planetary embryos accrete gas besides the solid material. The rate of gas accretion is determined by the speed at which the gas will cool down after being captured: the faster the cooling rate, the more gas accretion rate. The detailed models indicate that when the planetary embryo is small, the gas does not accrete to the planetary embryo very fast. Once the planetary embryo reaches a mass of 10 to 20 times the mass of the Earth, the gas cools rapidly, and then the planetary embryo absorbs a large amount of it. According to this theory, the large gaseous envelopes of outer planets in the solar system formed when their rocky cores reached this critical mass.

The second theory is that the outer planets may look like mini-solar systems. It maintains that the gas giants are formed by the instant gravitational collapse of the massive parts of the protoplanetary disk, much like how stars themselves form \citep{2009pps..book.....E,2020apfs.book.....A}.

For each theory, there is some evidence that is not conclusive. The elemental abundances of the giant planets and the solar nebula should be very similar if the planets were formed by gravitational collapse. The spectroscopic observations show that gas giants contain larger amounts of heavy elements in their atmospheres, those with atomic weights greater than helium. Nonetheless, this is uncertain because we can only observe the upper layers of the atmosphere, which comet impacts may have contaminated. On the other hand, the giant planets have an extensive system of moons, which can be a piece of evidence for the gravitational collapse theory. Their numerous moons were probably formed by the same process that formed the planets. Moreover, if the core-accretion theory is correct, the giant planets must have a rocky core, but the evidence for this is not strong \citep{2009pps..book.....E,2021PhDT........30S}.

There are many hot Jupiters, giant planets that are very close to their stars, discovered around other stars. The planetary migration theory explains this notable difference between the solar system and extrasolar systems. In the outermost part of the solar system, there is an icy shell known as the Oort Cloud that contains billions of planetesimals ejected from the solar system by the gravitational interactions with other planets. The law of conservation of energy requires that the planet itself ought to move to a lower position in the Sun’s gravitational field if a planetesimal is ejected from the inner solar system. Although this change in position is tiny, the planet could move a considerable distance from its original location after millions of encounters.

According to the planetary migration theory, hot Jupiters were formed far away from their host stars but migrated inwards because of the gravitational interactions with the remaining planetesimals. A plausible reason why there is no hot Jupiter in the solar system is its lower mass compared to the mass of disks around other stars. In a low-mass planetary disk, there is a smaller number of planetesimals and thus less chance for gravitational interactions \citep{2009pps..book.....E}.


\section{Exoplanet detection methods}
So far, more than five thousand exoplanets have been discovered. Four methods are mainly used to detect planets that orbit stars other than the Sun: radial velocity, transit, gravitational microlensing, and direct imaging. The vast majority of all known exoplanets have been detected using the transit and radial velocity methods, which are most efficient in discovering exoplanets close to their stars, such as hot Jupiters. Besides, gravitational microlensing and direct imaging techniques are implemented to find exoplanets that are widely separated from their host stars. Therefore, by using these exoplanet detection methods, we can have a complete view of the orbital architecture of exoplanetary systems resulting from different mechanisms of planet formation \citep{beky2014development,johnson2015you}. Other techniques also contribute to the discovery of exoplanets. They are timing variation, orbital brightness modulation, astrometry, and polarimetry.

\subsection{Radial velocity}\label{rv}
\begin{figure}[t!]
\centering
\captionsetup{width=0.65\paperwidth}
\includegraphics[width=0.65\paperwidth]{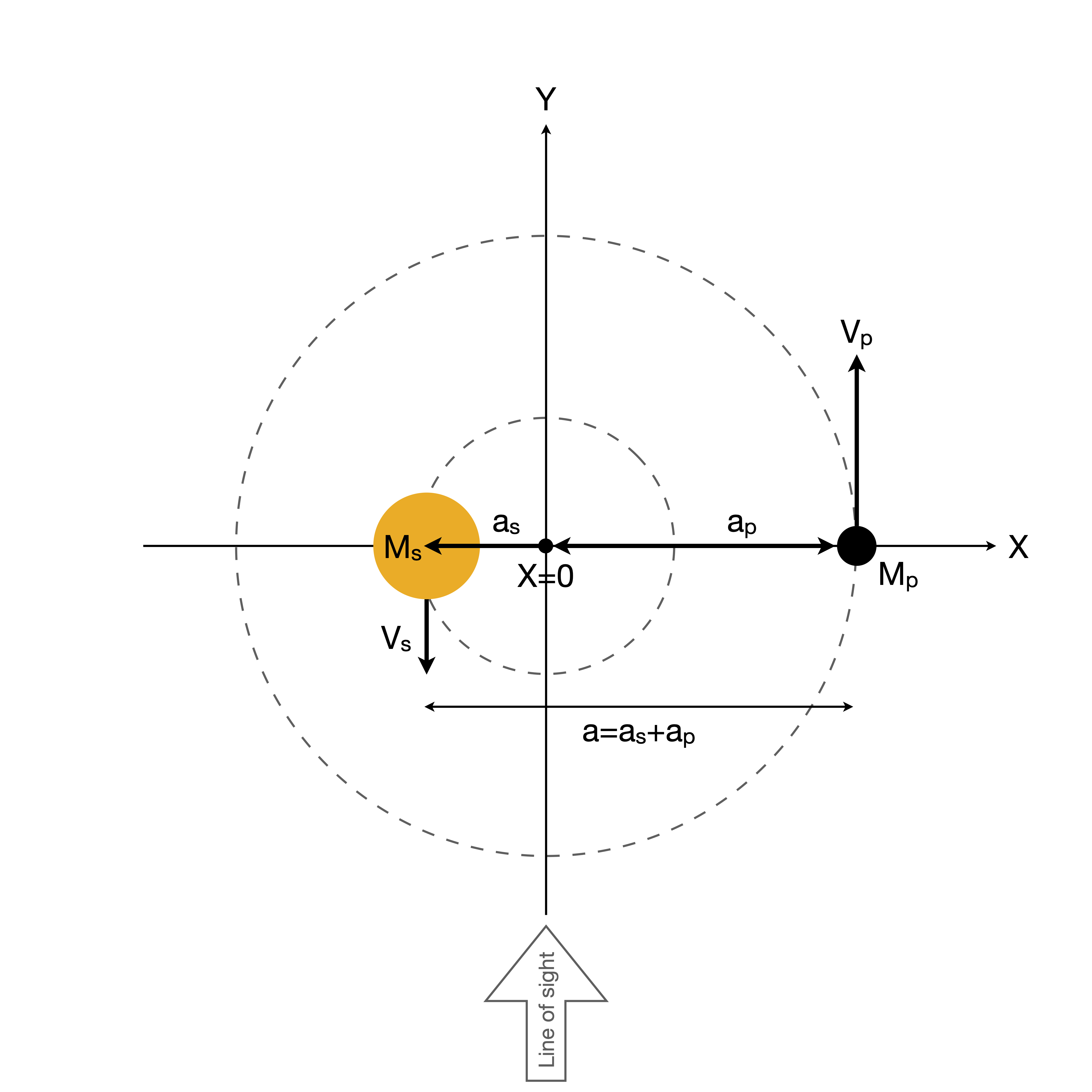}
\caption[Pole-on perspective of a planetary system]{Pole-on perspective of a planetary system. The star (yellow circle) and planet (black circle) revolve around the barycenter, corresponding to $x=0$. The star's distance from the barycenter, semi-major axis, is shown by $a_{s}$, while that of the planet is $a_{p}$. The mean semi-major axis is $a=a_{p}+a_{s}$. The stellar and planetary mass are denoted by $M_{s}$ and $M_{p}$, respectively. Moreover, $v_{s}$ is the speed of star and $v_{p}$ is that of the planet. \textit{[The image has been reproduced from \citet{johnson2015you}.]}}
    \label{figure1.7}
\end{figure}

The radial velocity method looks for tiny backward and forward motion of the host star due to the changing direction of the gravitational tug from an orbiting exoplanet. Since the wavelength of absorption lines in the stellar spectrum is affected by the star's radial motion via the Doppler effect, we can measure the radial velocity of the host star via spectroscopic observations \citep{marcy1992precision,2010exop.book.....S}. Figure \ref{figure1.7} shows the pole-on view of a typical planetary system where the planet's orbit is considered to be circular (zero eccentricity, $e=0$). If the barycenter, system's center of mass, is assigned to $x=0$, equation \ref{eq1.1} will be obtained. In this equation, the planet's distance from the barycenter (semi-major axis) is shown by $a_{p}$, while that of the star is $a_{s}$. The masses of the planet and star are $M_{p}$ and $M_{s}$, respectively. Furthermore, the speeds of the planet and star are $v_{p}$ and $v_{s}$, respectively. Solving this equation for the star's semi-major axis gives equation \ref{eq1.2}.

\begin{equation}
\bar{x}=\frac{-M_{s}a_{s}+M_{p}a_{p}}{M_{s}+M_{p}}=0
\label{eq1.1}
\end{equation}

\begin{equation}
a_{s}=\frac{M_{p}}{M_{s}}a_{p}
\label{eq1.2}
\end{equation}

\begin{figure}[b!]
\centering
\captionsetup{width=0.60\paperwidth}
\includegraphics[width=0.60\paperwidth]{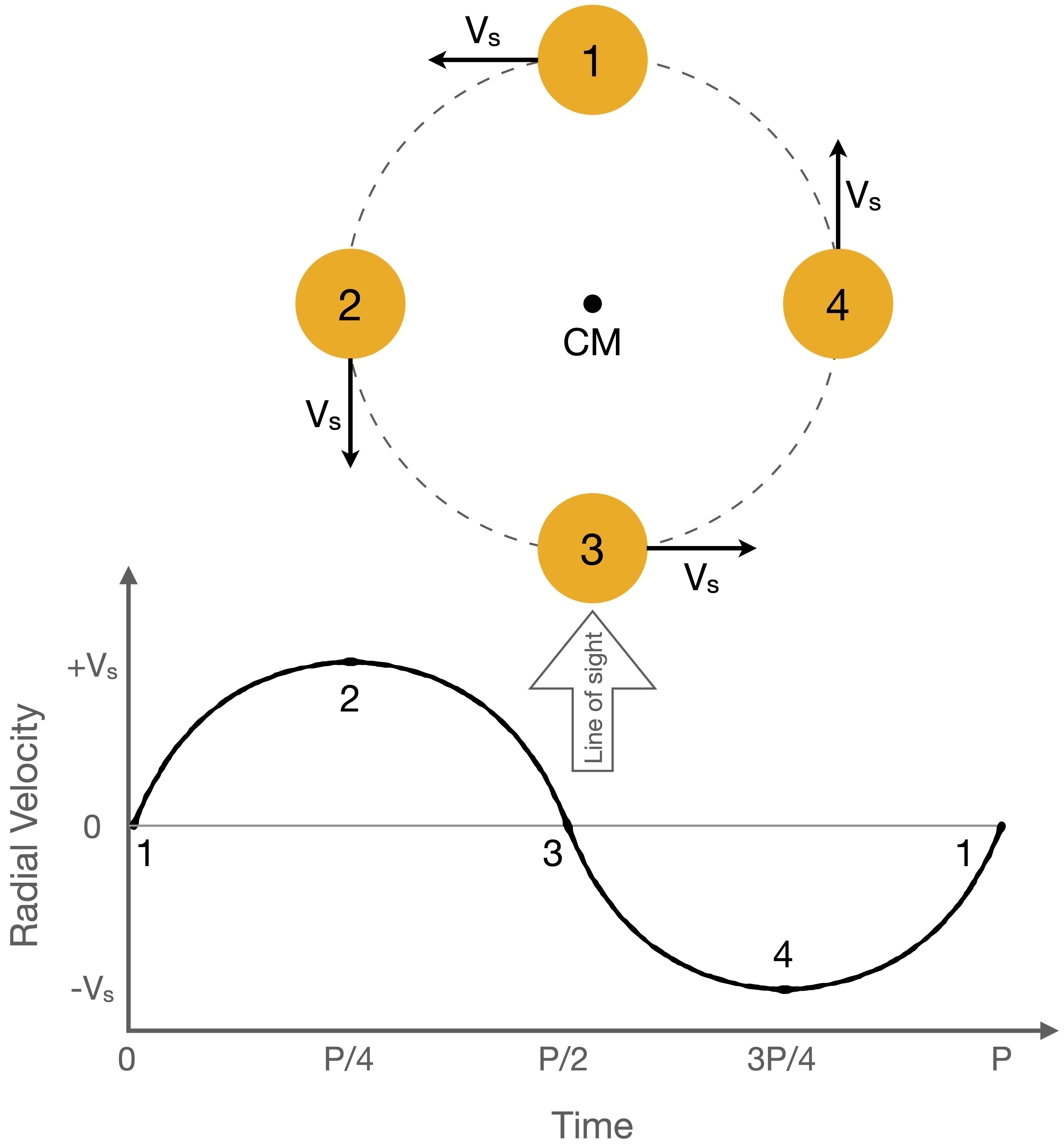}
\caption[A typical star orbiting the system's center of mass]{Illustration of a typical star (yellow circle) orbiting the system's center of mass with velocity $V_{s}$. The star's radial velocity values are shown by the black curve. Periodic shifts are detected using Doppler spectroscopy. When the star moves away from the observer, its spectrum is red-shifted, while it is blue-shifted when it moves towards the observer. \textit{[The image has been reproduced from \citet{johnson2015you}.]}}
    \label{figure1.8}
\end{figure}

Figure~\ref{figure1.8} illustrates a star (yellow circle) revolving around the system's center of mass with velocity $V_{s}$. The star's radial velocity measurements (black curve) show periodic behavior, detected using Doppler spectroscopy. When the star moves away from the observer, its spectrum is red-shifted, while it is blue-shifted when it moves towards the observer. So far, more than nine hundred exoplanets have been detected via radial velocity measurements.

The planet and star go around their orbits, $2\pi a_{p}$ and $2\pi a_{s}$, respectively, once every orbital period $P$. So, the star's speed is obtained by equation \ref{eq1.3}. Because the planet's mass is much less than the mass of the star, $M_{p}\ll M_{s}$, the speed and semi-major axis of the star will be far less than those of the planet.

\begin{equation}
v_{s}=\frac{2\pi a_{s}}{P}=\frac{M_{p}}{M_{s}}v_{p}
\label{eq1.3}
\end{equation}

Kepler's third law demonstrates that the square of the orbital period of a planet is proportional to the cube of its semi-major axis, which can be seen in equation \ref{eq1.4}. In mechanics, the virial theorem relates the total kinetic energy, $T$, of a gravitationally stable system to the gravitational potential energy, $U$, of the system. Equation\ref{eq1.5} represents the virial theorem, where $a_{p}$ has been replaced with $a$, which is the mean semi-major axis, $a=a_{p}+a_{s}$. $G$ is the gravitational constant. Solving this equation for the speed of the planet gives equation \ref{eq1.6}.

\begin{equation}
P^{2}=\frac{4\pi^{2}}{G}\biggl(\frac{a^{3}}{M_{s}+M_{p}}\biggr)
\label{eq1.4}
\end{equation}

\begin{gather}
\nonumber2T+U=0\\
\frac{1}{2}M_{p}v_{p}^{2}=-\frac{1}{2}\biggl(-\frac{GM_{p}M_{s}}{a}\biggr)
\label{eq1.5}
\end{gather}

\begin{equation}
v_{p}=\sqrt{\frac{GM_{s}}{a}}
\label{eq1.6}
\end{equation}

By substituting $v_{p}=v_{s}(M_{s}/M_{p})$ in equation \ref{eq1.6}, and replacing $a$ with the orbital period of the planet, $P$, and the star's mass, $M_{s}$, using Kepler's third law results in equation \ref{eq1.7} for the speed of star. As an example, in a one-year orbit around a solar-mass star, a Jupiter-mass planet will cause its star to move at a speed of $28.4\;m/s$. As a typical star like the Sun revolves around the center of the Milky Way galaxy at a speed of 220 $km/s$, the velocity changes caused by planets are, therefore, a small part of the movement of a star \citep{johnson2015you}.

\begin{equation}
v_{s}=(2\pi G)^{1/3}\frac{M_{p}}{(M_{s}+M_{p})^{2/3}}P^{-1/3}
\label{eq1.7}
\end{equation}

Throughout this derivation, the line-of-sight component of the star's velocity, its radial velocity $v_{\text{rad}}$, is at its largest value because we have assumed that we are looking at the planet's orbit edge-on. However, this particular condition is not true in most cases, and we can see only the projection of the star's speed. Although the radial velocity method is important for determining the mass of planets, in the absence of other methods, radial velocity measurements cannot resolve the degeneracy between orbital inclination ($i$) and planetary mass ($M_{p}$), and only yield the product $M_{p}\sin i$. Despite the fact that all the planets of the solar system, except Mercury, orbit around the Sun in almost circular paths with $e<0.1$, the orbital eccentricity of exoplanets can vary from almost zero up to a comet-like orbit such as HD 20782b ($e=0.97$) and HD 80606b ($e=0.93$) \citep{naef2001hd,o2009selection}. Consequently, if we do not neglect the effects of eccentricity and express the planet mass, orbital period, and stellar mass in terms of the Jupiter mass ($M_{J}$), days on Earth ($d$), and mass of the Sun ($M_{\odot}$), respectively, the radial velocity of the star is given by equation \ref{eq1.8} \citep{1999ApJ...526..890C}.

\begin{equation}
\frac{v_{\text{rad}}}{\text{ms}^{-1}}=203\biggl(\frac{P}{\text{d}}\biggr)^{-1/3}\frac{(M_{p}/M_{J})\sin i}{((M_{s}/M_{\odot})+9.548\times10^{-4}(M_{p}/M_{J}))^{2/3}}(1-e^{2})^{-1/2}
\label{eq1.8}
\end{equation}

There are three biases associated with this method: first, close-in massive planets have a larger gravitational influence on their host star than low-mass planets with long orbital periods. Hence, the radial velocity method is more interested in planets with short orbital periods and those with high masses. Second, very young stars and massive main sequence A members are examples of stars that rotate rapidly. These kinds of stars are normally omitted from radial velocity observing programs because they have wide spectral features that inhibit accurate radial velocity measurements. Third, since the starlight needs to be dispersed into a spectrum with high resolution, the radial velocity observations are biased toward brighter stars, stars with a V-band magnitude of about $V=8$. In other words, spectroscopic reference lines are densest at optical wavelengths. So, radial velocity surveys are biased against stars with cooler temperatures, such as M dwarfs, as most of their emission is in the infrared band.

\subsection{Transit}\label{transit}
At the time of writing this thesis, more than 3800 alien worlds have been detected by observing the slight decrease in brightness of the host star caused by the transit of a planet in front of it, which is known as the transit method. Transit is a popular and relatively economical method, so a transiting survey allows us to monitor thousands of stars simultaneously. It can reveal the stellar density, as well as the planet's radius relative to that of the star \citep{seager2003unique,winn2011chapter}. Nevertheless, the transit method can only detect exoplanets that have the needed geometry to eclipse; that is, at a given semi-major axis $a$, only a limited range of inclinations, $i$, can carry the exoplanet in front of its host star \citep{beky2014development,johnson2015you}.

\begin{figure}[b!]
\centering
\captionsetup{width=0.60\paperwidth}
\includegraphics[width=0.60\paperwidth]{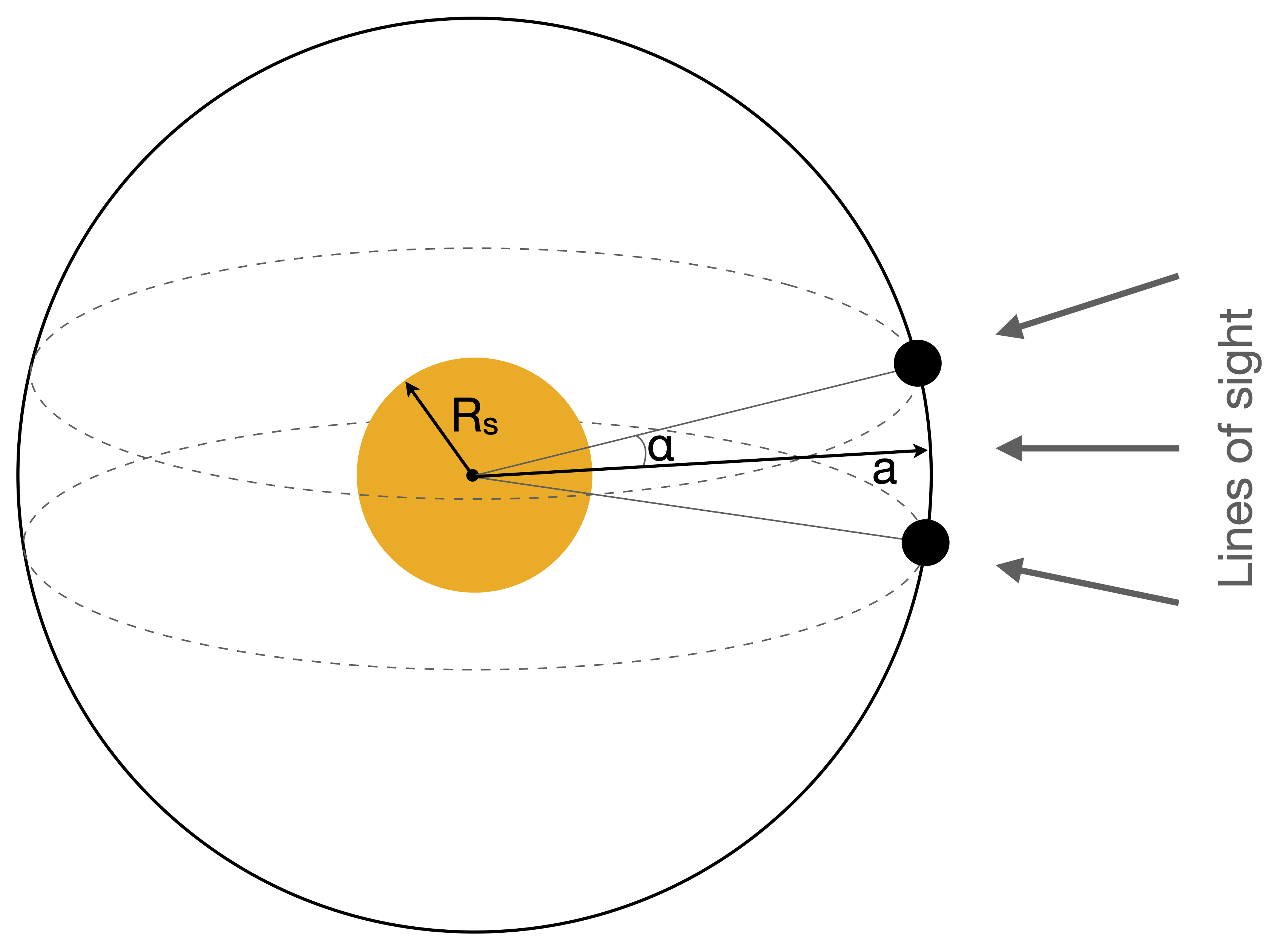}
\caption[The geometry required to observe a transiting exoplanet]{The geometry required to observe a transiting exoplanet. The orbital semi-major axis is $a$. The angel $\alpha$ is measured from the horizontal direction. Dashed lines represent the strip of solid angles in which the planet (black circle), with a radius $R_{p}$, transit its star (yellow circle), with a radius $R_{s}$, as viewed from the observer on Earth. The height of strip is $\sin^{-1}[(R_{s}+R{p})/a]$. \textit{[The image has been reproduced from \citet{johnson2015you}.]}}
    \label{figure1.9}
\end{figure}

Figure~\ref{figure1.9} illustrates the strip of solid angles (dashed lines) in which the planet, with a radius $R_{p}$, transit its star, with a radius $R_{s}$, as viewed from the observer on Earth. The angle $\alpha$ is measured from the horizontal direction, and the strip's height is $\sin^{-1}[(R_{s}+R{p})/a]$. By integrating the height $2\pi$ around the star, the strip's solid angle is calculated. The transit probability of an exoplanet is given by the strip's solid angle divided by total solid angle, $4\pi$ steradians. Hence, the transit probability is given by equation \ref{eq1.9}, where $\alpha_{\text{min}}=-\sin^{-1}[(R_{s}+R{p})/a]$, $\alpha_{\text{max}}=\sin^{-1}[(R_{s}+R{p})/a]$, $\theta_{\text{min}}=0$, and $\theta_{\text{max}}=2\pi$. Solving this integral results in equation \ref{eq1.10}. For instance, an observer located beyond the solar system would calculate a transit probability of about 0.5\% for planet Earth.

\begin{equation}
P_{\text{transit}}=\frac{\int_{\alpha_{\text{min}}}^{\alpha_{\text{max}}}\int_{0}^{2\pi}\cos{(\alpha)}d\alpha d\theta}{4\pi}
\label{eq1.9}
\end{equation}

\begin{equation}
P_{\text{transit}}=\frac{R_{s}+R_{p}}{a}
\label{eq1.10}
\end{equation}

\begin{figure}[hb!]
\centering
\captionsetup{width=0.60\paperwidth}
\includegraphics[width=0.60\paperwidth]{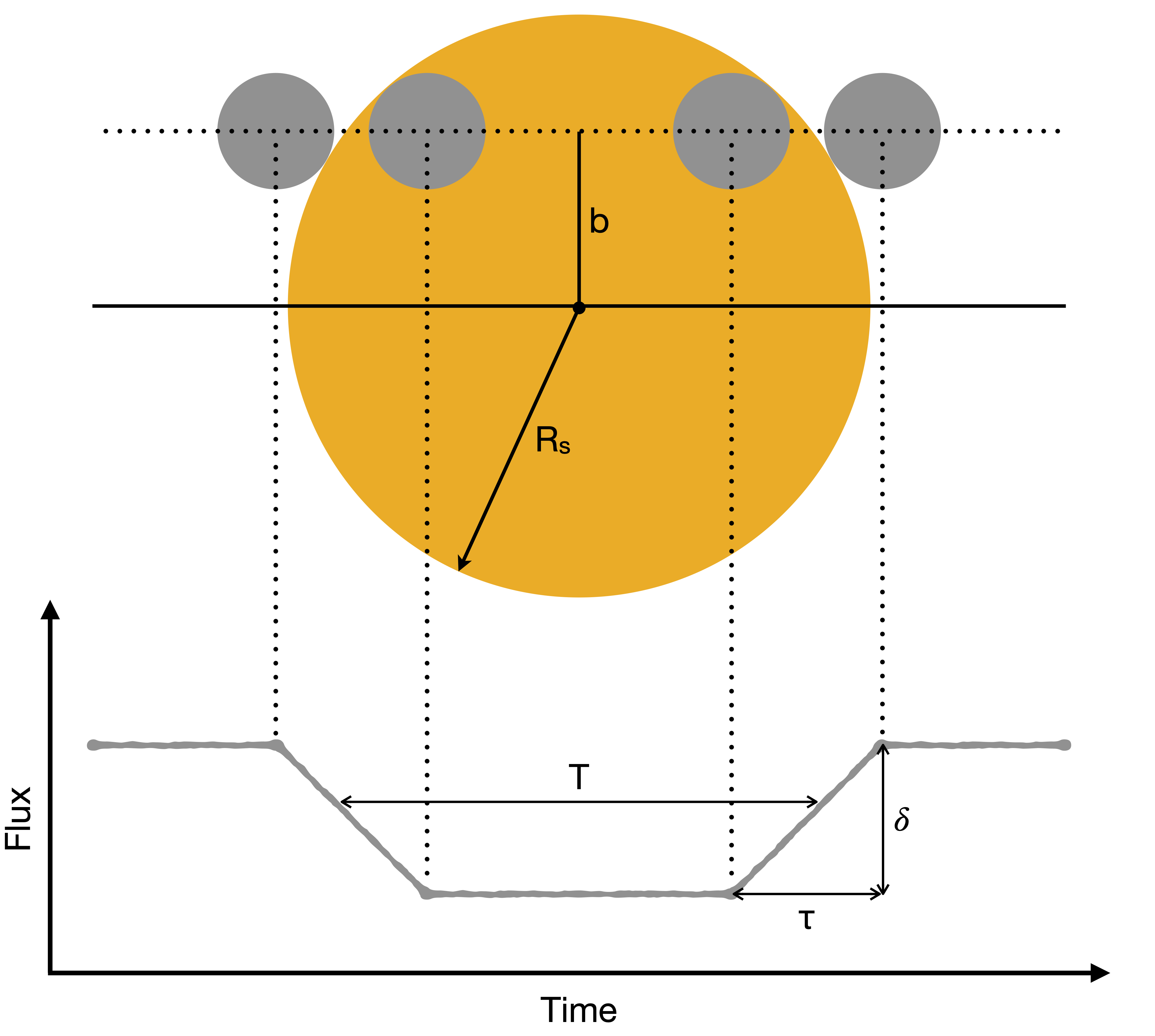}
\caption[The light curve of an exoplanet]{Illustration of an exoplanet (gray circle) transiting in front of its host star (yellow circle) and corresponding light curve (gray line). The impact parameter $b$ represents the perpendicular distance between the planet's transit path and the center of the star. The ingress/egress duration is $\tau$, while the total duration is $T$. The transit depth is $\delta=(R_{p}/R_{s})^{2}$, where $R_{p}$ and $R_{s}$ are the radius of the exoplanet and star, respectively. \textit{[The image has been reproduced from \citet{johnson2015you}.]}}
    \label{figure1.10}
\end{figure}

Transiting of an exoplanet (gray circle) in front of its host star (yellow circle) along with the corresponding light curve (gray line) is illustrated in figure~\ref{figure1.10}. The impact parameter $b$ represents the perpendicular distance between the planet's transit path and the center of the star, which is equal to $(a/R_{s})\cos{i}$. For $b=0$, the exoplanet transits along the stellar equator, while for $b>0$, it transits on a shorter path along the stellar surface. The transit depth is $\delta=(R_{p}/R_{s})^{2}$, which is equal to the amount of light blocked by the exoplanet, that is the ratio of areas. The ingress/egress duration is $\tau$, while the total duration of the transit is $T$, which is approximately the full width at half of the maximum depth.

Assuming a circular orbit ($e=0$), the speed of the planet, $v_{p}$, is constant throughout its orbit and is equal to $2\pi a/P$, where $P$ is the orbital period. If the impact parameter is $b=0$, the ingress/egress duration is obtained by equation \ref{eq1.11}. Furthermore, the transit time, which is the amount of time it takes for a planet to travel around a star, is calculated using equation \ref{eq1.12}.

\begin{equation}
\tau=\frac{2R_{p}}{v_{p}}=\frac{PR_{p}}{\pi a}
\label{eq1.11}
\end{equation}

\begin{equation}
T=\frac{2R_{s}}{v_{p}}=\frac{PR_{s}}{\pi a}
\label{eq1.12}
\end{equation}

In a more general case where the planet traverses its star along a chord rather than the stellar equator ($b>0$), the ingress/egress and total duration are given by equations \ref{eq1.13} and \ref{eq1.14}, respectively. As a result, transits with nonzero impact parameter have longer $\tau$ and shorter $T$ than those with $b=0$. In other words, the higher the $b$, the more V-shaped the light curve is.

\begin{equation}
\tau=\frac{PR_{p}}{\pi a}\frac{1}{\sqrt{1-b^{2}}}
\label{eq1.13}
\end{equation}

\begin{equation}
T=\frac{PR_{s}}{\pi a}\sqrt{1-b^{2}}
\label{eq1.14}
\end{equation}

Transit parameters $T$, $\tau$, and $\delta$ are all measurable from the light curve and can be related to the physical features of the system using equations \ref{eq1.15} to \ref{eq1.17}. We can assume that $P$ is a constant because the orbital periods of transiting exoplanets are typically measured with an accuracy of 1 second or better. If we express the semi-major axis in terms of the stellar mass ($M_{s}$) and orbital period ($P$) using Newton's version of Kepler's third law, the stellar density is obtained by equation \ref{eq1.18}.

\begin{equation}
\frac{R_{p}}{R_{s}}=\sqrt{\delta}
\label{eq1.15}
\end{equation}

\begin{equation}
b^{2}=1-\frac{\delta^{1/2}T}{\tau}
\label{eq1.16}
\end{equation}

\begin{equation}
\frac{a}{R_{s}}=\frac{P\delta^{1/4}}{2\pi}\biggl(\frac{4}{T\tau}\biggr)^{1/2}
\label{eq1.17}
\end{equation}

\begin{equation}
\frac{a}{R_{s}}=\biggl(\frac{GP^{2}}{4\pi^{2}}\biggr)^{1/3}\biggl(\frac{M_{s}^{1/3}}{R_{s}}\biggr)\sim\rho_{s}^{1/3}
\label{eq1.18}
\end{equation}

For a non-circular orbit, there is another factor named $g(e,\omega)$. As presented in equation \ref{eq1.19}, beside the orbital eccentricity ($e$), this factor depends on the orbital orientation ($\omega$), where different values of $\omega$ rotate the orbit relative to the observer. The density function for systems with $e>0$ is defined by equation \ref{eq1.20}, where $\rho_{\text{circ}}$ is the density of a circular system and is obtained by equation \ref{eq1.21}.

\begin{equation}
g(e,\omega)=\frac{1+e\sin{\omega}}{\sqrt{1-e^{2}}}
\label{eq1.19}
\end{equation}

\begin{equation}
\rho_{s}=g(e,\omega)^{3}\rho_{\text{circ}}
\label{eq1.20}
\end{equation}

\begin{equation}
\rho_{\text{circ}}=\frac{3P}{G\pi^{2}}\biggl(\frac{\delta^{1/4}}{\sqrt{T\tau}}\biggr)^{3}
\label{eq1.21}
\end{equation}

Radial velocity observations have shown that only 1\% of stars host a hot Jupiter with a mass more than one-tenth of the mass of Jupiter that revolves in an orbit with a period of less than ten days \citep{wright2012frequency}. If stars with masses and radii similar to the Sun's are considered, the typical transit probability of a hot Jupiter is about 10\%. So, for a single transit, one would need to search for $1/(0.1\times0.01)=1000$ stars. As wide-field transit surveys need to observe many stars to detect a single transit, they are biased toward fainter stars.

It will be beneficial if one studies a planetary system using both transit and radial velocity observations. The transit method gives us the planet's size, and the radial velocity method limits the planet's mass (see section \ref{rv}). Consequently, we can determine the bulk density of exoplanets and learn about their compositions.

\subsection{Gravitational microlensing}
The microlensing method uses the effects of the planet's gravitational field on the passing light of a distant background star to discover an exoplanet. So far, there are more than 130 exoplanets that have been detected using the gravitational microlensing method. During a gravitational lensing event, the gravitational field of a massive foreground object (the lens) bends some light paths emitted from a background object (the source). As a consequence, the observer sees distorted, magnified, and brightened images of the source object \citep{1964MNRAS.128..295R}. If the observer's line-of-sight is exactly toward the lens and source stars, the Einstein ring, the most elegant manifestation of the lensing phenomenon, will be produced. If not, the observer might see multiple images of the source star \citep{2010exop.book...79G}. A microlensing event can make the effective size of a telescope bigger and let us see distant objects, like stars on the other side of the Galaxy, that are usually too faint to study spectroscopically.

\begin{figure}[t!]
\centering
\captionsetup{width=0.60\paperwidth}
	\includegraphics[width=0.60\paperwidth]{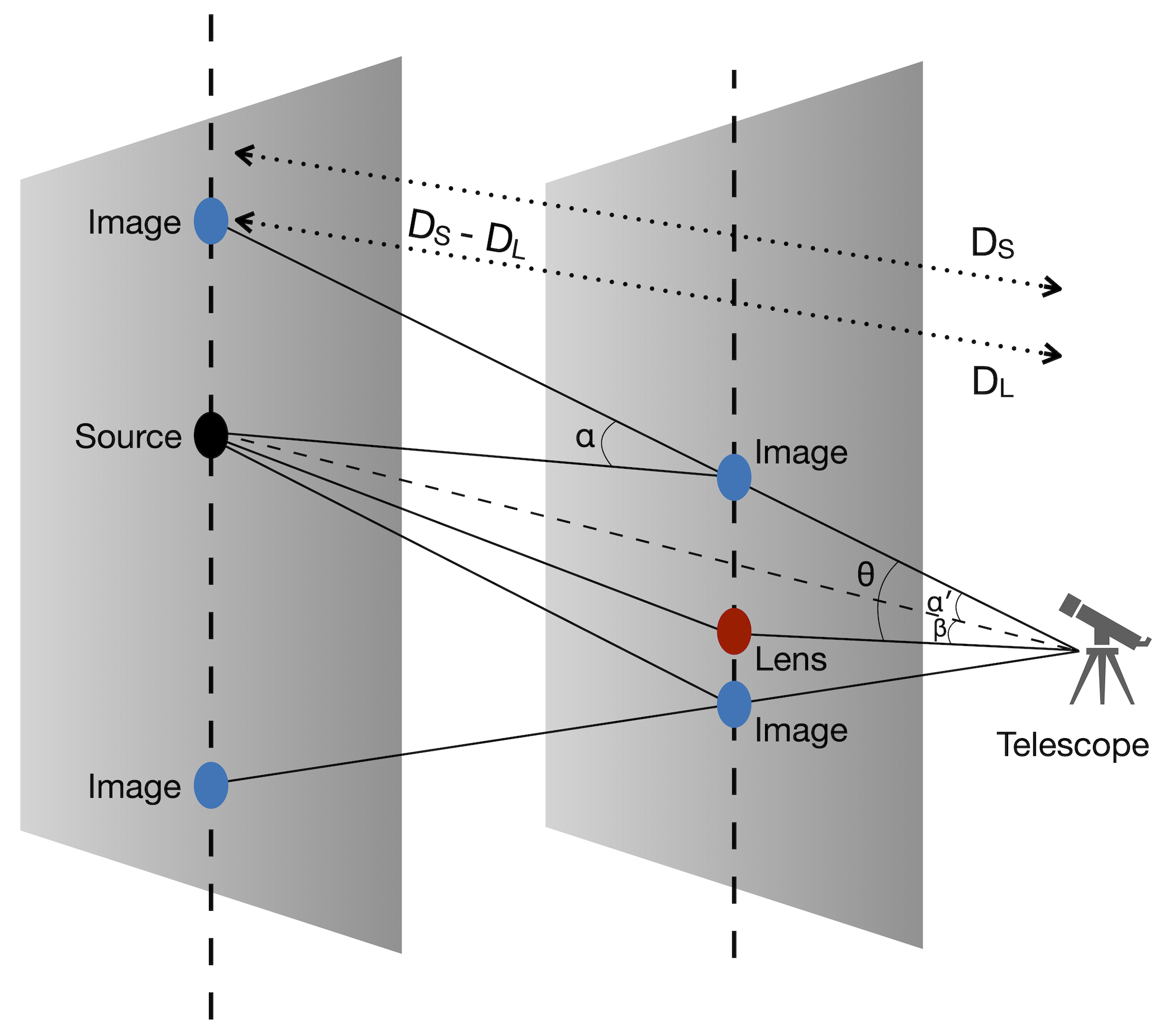}
    \caption[The geometry of a gravitational microlensing phenomenon]{The geometry of a gravitational microlensing phenomenon. The gray parallelograms represent the planes of the sky corresponding to the source star (black oval) and the lens star (crimson oval). The lens star is located at a distance $D_{L}$, whereas the source star is at a distance $D_{S}$ away from the Earth. The lens star produces two images of the source star (blue ovals). The top image is an angle $\alpha^{\prime}$ away from the source star and an angle $\theta$ from the lens star. The $\beta$ represents the angular separation between the source and lens stars. In addition, $\alpha$ is the deflection of light due to the gravity of the lens star, which is equal to $4GM/bc^{2}$, where $G$ is the gravitational constant, $c$ is the speed of light, $M$ is the mass of the lens, and $b$ is the closest distance between the lens and the photon. \textit{[The image has been reproduced from \citet{johnson2015you}.]}}
    \label{figure1.11}
\end{figure}

Figure~\ref{figure1.11} demonstrates a lens star at a distance $D_{L}$ and a source star at a distance $D_{S}$ away from the observer. Two images of the source star have been produced by the lens star, one closer to the source star than the other. From the observer's point-of-view, the nearby image is an angle $\alpha^{\prime}$ away from the source star and an angle $\theta$ from the lens star. Furthermore, the angular separation between source and lens is $\beta$. Angle $\alpha$ is the deflection of the photon due to the gravitational effect of the lens star. According to Einstein's theory of gravity, the deflection angle is obtained by equation~\ref{eq1.22}, where $M$ is the mass of the lens, $b$ is the closest distance between the lens and the photon, $G$ is the gravitational constant, and $c$ is the speed of light.

\begin{equation}
\alpha=\frac{4GM}{bc^{2}}
\label{eq1.22}
\end{equation}

As both $D_{S}-D_{L}$ and $D_{S}$ are very large, we can assume that $\alpha/(D_{S}-D_{L})\approx\alpha/D_{S}$. Then, $\alpha$ and $\alpha^{\prime}$ can geometrically be related to each other by equation~\ref{eq1.23}. Using deflection angle $\alpha$ and replacing $b$ with $\theta D_{L}$, the angular separation between the lens star and the source star can be achieved by equation~\ref{eq1.24}. Now, if the lens and source stars fall along the observer's line-of-sight ($\beta=0$), the Einstein ring is produced. This ring has an angular radius of $\theta_{E}$, which is known as the Einstein's radius. As proposed by \citet{gaudi2012microlensing}, the Einstein's radius can be rewritten in terms of a new parameter $k=4G/(c^{2}\text{AU})=8.14\;\text{mas}/M_{\odot}$ and a relative parallax $\pi_{\text{rel}}=\text{AU}(D_{L}^{-1}-D_{S}^{-1})$, as expressed in equation~\ref{eq1.25}. AU stands for astronomical unit, and mas for milliarcsecond. By introducing two new variables, $u=\beta/\theta_{E}$ and $y=\theta/\theta_{E}$, equation~\ref{eq1.24} is simplified to equation~\ref{eq1.26}, whose roots indicate the locations of the top ($y_{+}$) and bottom ($y_{-}$) images (see equation~\ref{eq1.27}).

\begin{equation}
\alpha^{\prime}=\alpha\biggl(\frac{D_{S}-D_{L}}{D_{S}}\biggr)
\label{eq1.23}
\end{equation}

\begin{equation}
\beta=\theta-\alpha^{\prime}=\theta-\frac{4GM}{\theta c^{2}}\biggl(\frac{D_{S}-D_{L}}{D_{S}D_{L}}\biggr)
\label{eq1.24}
\end{equation}

\begin{equation}
\theta_{E}=\sqrt{\frac{4GM}{c^{2}}(D_{L}^{-1}-D_{S}^{-1})}=\sqrt{kM\pi_{\text{rel}}}
\label{eq1.25}
\end{equation}
\begin{gather}
\nonumber u=y-y^{-1}\\
y^{2}-uy-1=0
\label{eq1.26}
\end{gather}
\begin{gather}
\nonumber y_{+}=\frac{1}{2}\bigl(\sqrt{u^{2}+4}+u\bigr)\\
y_{-}=\frac{1}{2}\bigl(\sqrt{u^{2}+4}-u\bigr)
\label{eq1.27}
\end{gather}
As a celestial source radiates isotropically, the observer can only see a limited fraction of light that ends at his telescope. Yet, during a gravitational lensing event, he is able to observe some light that is bent by the lens onto new paths. Thus, the distorted images of the source star add more light to the observer's line-of-sight, making the source appear brighter, which is called magnification. The magnification event is quantitatively given by equation~\ref{eq1.28} (see figure~\ref{figure1.11}).

\begin{equation}
A_{\pm}=\biggl|\frac{y_{\pm}}{u}\frac{dy_{\pm}}{du}\biggr|
\label{eq1.28}
\end{equation}

\begin{figure}[b!]
\centering
\captionsetup{width=0.60\paperwidth}
	\includegraphics[width=0.60\paperwidth]{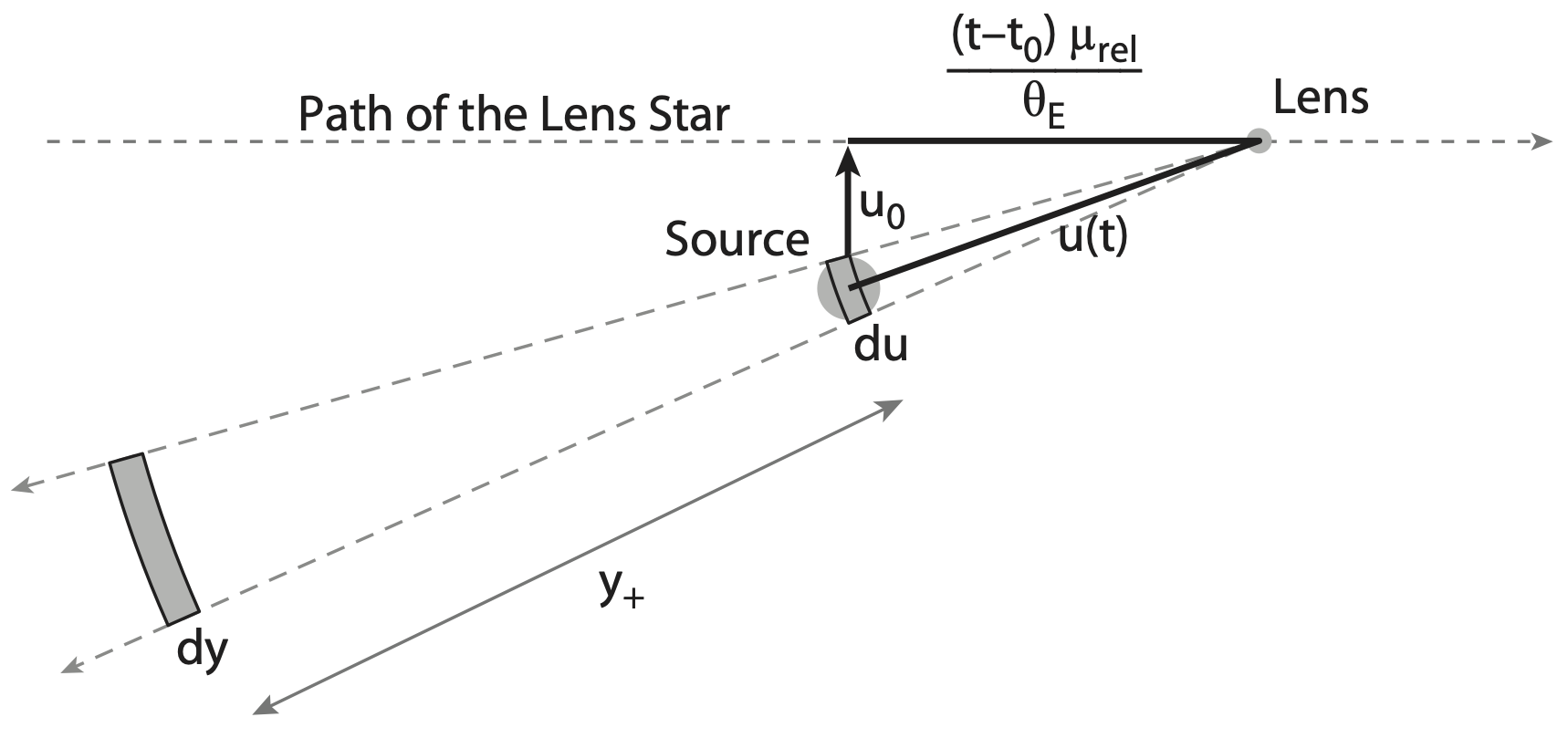}
    \caption[The source and lens stars moving through the sky]{The source and lens stars moving through the sky. The dashed horizontal line depicts the path of the lens star. The scaled distance between the lens and source stars is presented by $u(t)$, and $u_{0}$ is the minimum scaled impact parameter. The solid black lines depict the relationship between $u(t)$, $u_{0}$, and the path of the lens star. \textit{[Image source: \citet{johnson2015you}.]}}
    \label{figure1.12}
\end{figure}

Figure~\ref{figure1.12} depicts the source and lens stars moving through the sky, known as proper motion. The relative angular motion of stars $\mu_{\text{rel}}$, which is measured in units of $\text{mas}/\text{year}$ ($\text{mas}$ denotes $10^{-3}$ arcseconds), causes the projected separation between the stars, $u$, to be time-dependent. In figure~\ref{figure1.12}, the distance between the source star and one of the magnified lensed images is presented by $y_{+}$. The lensed image is compressed in the radial direction by $dy_{+}/du$ and extended tangentially by $y_{+}/u$. Thus, the magnification as a function of time is given by equation~\ref{eq1.29}. The first-order of the series expansion of $A(u)$, while $u\rightarrow0$, gives $A(u)=1/u$, which is useful to assess a microlensing light curve.

\begin{equation}
A(t)=A_{+}+A_{-}=\frac{u(t)^{2}+2}{u(t)\sqrt{u(t)^{2}+4}}
\label{eq1.29}
\end{equation}

The projected separation between the stars, $u$, is time-dependent. As the stars move closer, $u$ decreases toward a minimum value $u_{0}$ when the stars are closest together in the sky, and the magnification is at the maximum value. Then, the projected separation between the stars increases, and $A$ decreases symmetrically with the rise in magnification. $u$ can be derived using the geometry of figure~\ref{figure1.12}. If the source star is located in the origin and angles are measured as a fraction of the Einstein radius, the time-dependent projected separation is given by equation~\ref{eq1.30}, where $t_{0}$ is the time of closest approach. $\mu_{\text{rel}}(t-t_{0})$ is the perpendicular distance traveled by the lens star from $u_{0}$. By assuming an angular speed $\mu_{\text{rel}}$ for the source star, the time needed for it to move across the angle $\theta_{E}$ is given by $\theta_{E}/\mu_{\text{rel}}$. With this new variable, equation~\ref{eq1.30} becomes equation~\ref{eq1.31}. Equation~\ref{eq1.32} is a more quantitative form of $t_{E}$, which has been derived using equation~\ref{eq1.25}.

\begin{equation}
u(t)=\sqrt{u_{0}^2+(t-t_{0})^2\biggl(\frac{\theta_{E}}{\mu_{\text{rel}}}\biggl)^{-2}}
\label{eq1.30}
\end{equation}

\begin{equation}
u(t)=\sqrt{u_{0}^{2}+\biggl(\frac{t-t_{0}}{t_{E}}\biggl)^{2}}
\label{eq1.31}
\end{equation}

\begin{equation}
t_{E}\approx35\;\text{d}\;\biggl(\frac{M}{M_{\odot}}\biggl)^{1/2}\biggl(\frac{\pi_{\text{rel}}}{125\;\mu\text{as}}\biggl)^{1/2}\biggl(\frac{\mu_{\text{rel}}}{10.5\;\text{mas}\;\text{yr}^{-1}}\biggl)^{-1}
\label{eq1.32}
\end{equation}

\begin{figure}[b!]
\centering
\captionsetup{width=0.65\paperwidth}
	\includegraphics[width=0.65\paperwidth]{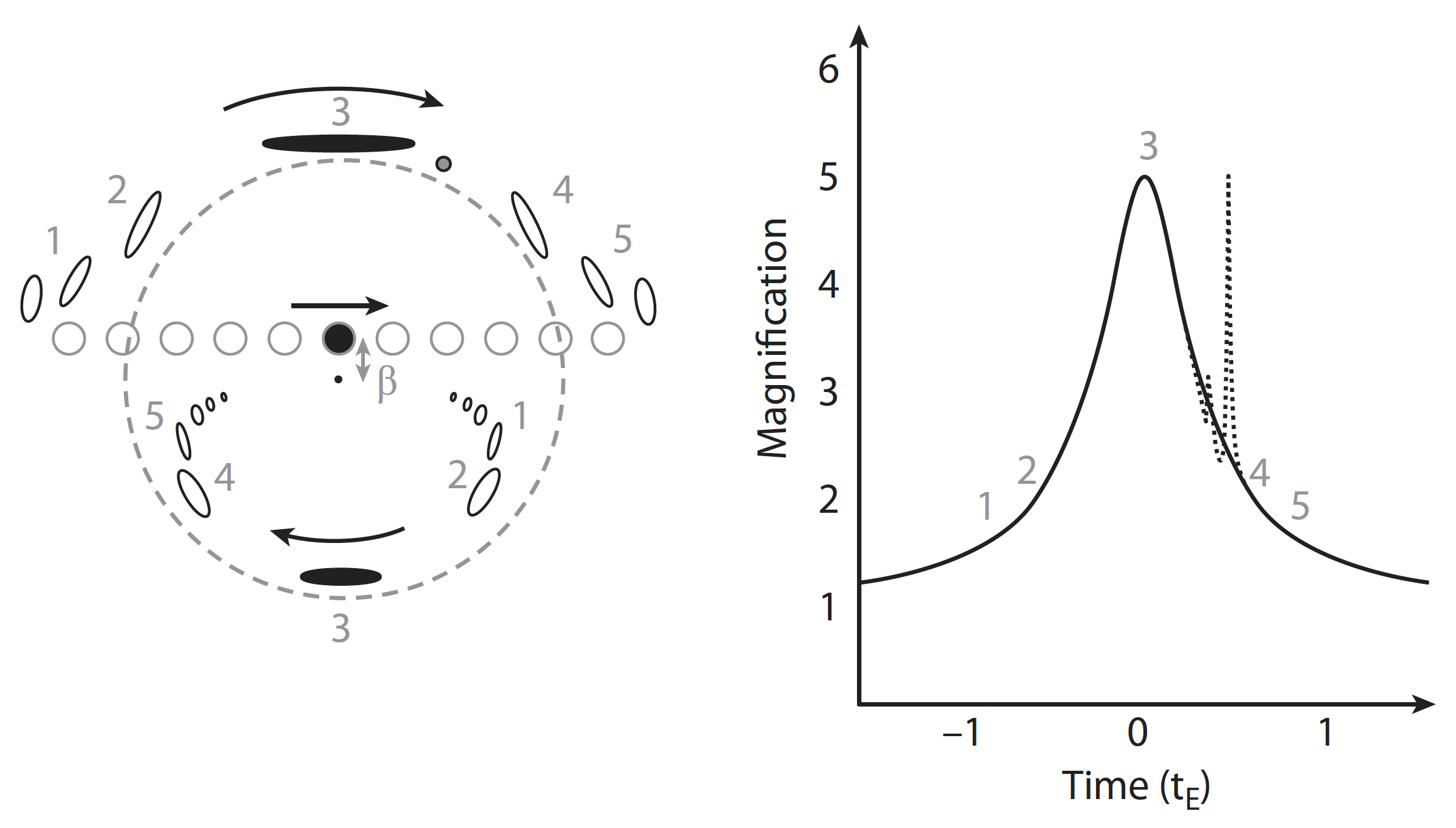}
    \caption[Exoplanet detection using gravitational microlensing]{Exoplanet detection using gravitational microlensing. Left: gray circles represent the source star, and the small dot at the center is the lens star with an impact parameter $\beta$. The gray dashed circle represents the Einstein ring. Two images are produced (black open ovals): the minor image interior to the ring and the major image exterior to the ring. These images move clockwise from 1 to 5 as the source star moves from left to right due to its relative proper motion to the lens star. The small gray dot between positions 3 and 4 is the lens star’s planet. Right: the microlensing light curve. It presents the total flux from the lens star, source star, and images, which produces a magnification relative to the source light. There is an additional magnification between positions 3 and 4 due to the planet’s gravitational field. \textit{[Image source: \citet{2010exop.book...79G}.]}}
    \label{figure1.13}
\end{figure}

Therefore, a microlensing light curve is parameterized by the time of the closest approach (or the time of the maximum magnification), $t_{0}$, the Einstein ring crossing time, $t_{E}$, and the scaled impact parameter, $u_{0}$. The width of the light curve and the position of the maximum light determine $t_{E}$ and $t_{0}$, respectively. $u_{0}$ can also be estimated based on the level of maximum magnification under the assumption that $A_{\text{max}}=1/u_{0}$, where $A_{\text{max}}$ is measured with respect to a nonlensed baseline level.

Figure~\ref{figure1.13} illustrates a microlensing time series as projected onto the sky. Two images are produced during a microlensing event: the minor image of the interior of the ring and the major image of the exterior of the ring. The shorter deviation in the microlensing light curve (between positions 3 and 4) is due to the planet’s gravitational field. One will be able to observe the microlensing signals of the planet along with the source star’s microlensing signals if the planet has a projected separation comparable to $\theta_{E}$. In this situation, the planet will fall on top of the Einstein ring, lensing the light in the ring and causing the ring to become slightly brighter.

The planet acts like an independent lens with its value of $t_{E}$ and $\theta_{E}$. Due to the fact that $t_{E}\sim M^{1/2}$, the duration of the planet’s microlensing signal is much less than the star’s. For instance, a Jupiter around a $0.5M_{\odot}$ star produces a signal with a duration of one day. An Earth-mass planet produces a signal with a duration of only $\sim1$ hour.

Since the exact distances to the lens and source stars are generally unknown, the mass cannot be measured exactly. However, the planet-to-star mass ratio ($q$) can be solved using equation~\ref{eq1.33} as the lens star and its planet are at the same $D_{L}$. $t_{E,s}$ and $t_{E,p}$ are the Einstein ring crossing times of the star and planet, with masses $M_{s}$ and $M_{p}$, respectively. Therefore, the light curve widths of the lens star and its planet can provide a measure of $q$.

\begin{equation}
q\equiv \sqrt{\frac{M_{p}}{M_{s}}}=\frac{t_{E,p}}{t_{E,s}}
\label{eq1.33}
\end{equation}

Moreover, the delay between the maxima of the lens star and planet magnification can provide the star-planet separation ($s$). Nonetheless, this time is measured relative to $t_{E,s}$ and only provides a scaled separation between the planet and the star, $s\equiv a_{p}/R_{E}$, where $a_{p}$ is the physical separation between the planet and the star projected along the sky plane. Therefore, like the other methods, one must know the properties of the host star to characterize the exoplanet properly.

\subsection{Direct imaging}
As the name would suggest, the direct imaging method consists of capturing images of exoplanets directly by searching for their faint light. It is the oldest detection method, dating back to the discovery of the solar system planets. So far, more than 200 exoplanets have been detected using this method. The direct imaging method has two primary difficulties: planets are located very close to their parent stars and are much fainter than them. These two problems are called the angular resolution and contrast. The angular resolution power is the smallest angle between close objects that can be seen clearly to be separate, which is limited by the wave nature of light. The star-planet separation is far less than the distance between the system and the observer. So, an instrument with high angular resolution power is needed to distinguish the planet from its parent star. The contrast is the ratio of star-planet flux, which is typically very large, making it difficult to distinguish their light \citep{2010exop.book...79G}.

\subsubsection{Angular resolution}
As shown in equation~\ref{eq1.34}, the limiting angular resolution of a telescope, which is also known as the diffraction limit, is related to the telescope’s diameter ($D$) and the wavelength at which the observation is made ($\lambda$). For example, for observations made by a 10-meter telescope at near-infrared wavelengths like 1.6 microns, the diffraction limit equals 0.04 arcsecond.

\begin{equation}
\theta_{\text{min}}=\frac{1.22\;\lambda}{D}
\label{eq1.34}
\end{equation}

The angular separation between a planet and its parent star, which is located at a distance $d$ from the observer, is given by $\theta=sin^{-1}(a/d)$, where $a$ is the planet’s semi-major axis, assumed to be viewed pole-on for simplicity. As $d\gg a$, the angular separation formula becomes $\theta\approx a/d$. For the planet Jupiter observed from a 100 pc distance, the angular separation is 0.05 arcsecond, which is slightly more than the diffraction limit of a 10-meter near-infrared telescope. However, in practice, some factors could prevent such a detection.

The most complicating factor is related to the Earth’s atmosphere. The atmosphere of Earth has pockets of warm and cool air that move with the wind. These temperature fluctuations can affect the observations made by ground-based telescopes. As air pockets have different indices of refraction due to their different temperatures, the light of astronomical objects is bent in different directions, known as the twinkling of stars. The atmosphere also causes the images of astronomical objects to be smeared out, especially over the long exposure times. This blurring effect in astronomical images is called “seeing.” An observation site with very stable air has a good seeing, probably 0.4 arcseconds at infrared wavelengths. Thus, the Earth’s atmosphere sets a limit that can be orders of magnitude worse than the diffraction limit.

\begin{figure}[b!]
\centering
\captionsetup{width=0.60\paperwidth}
	\includegraphics[width=0.60\paperwidth]{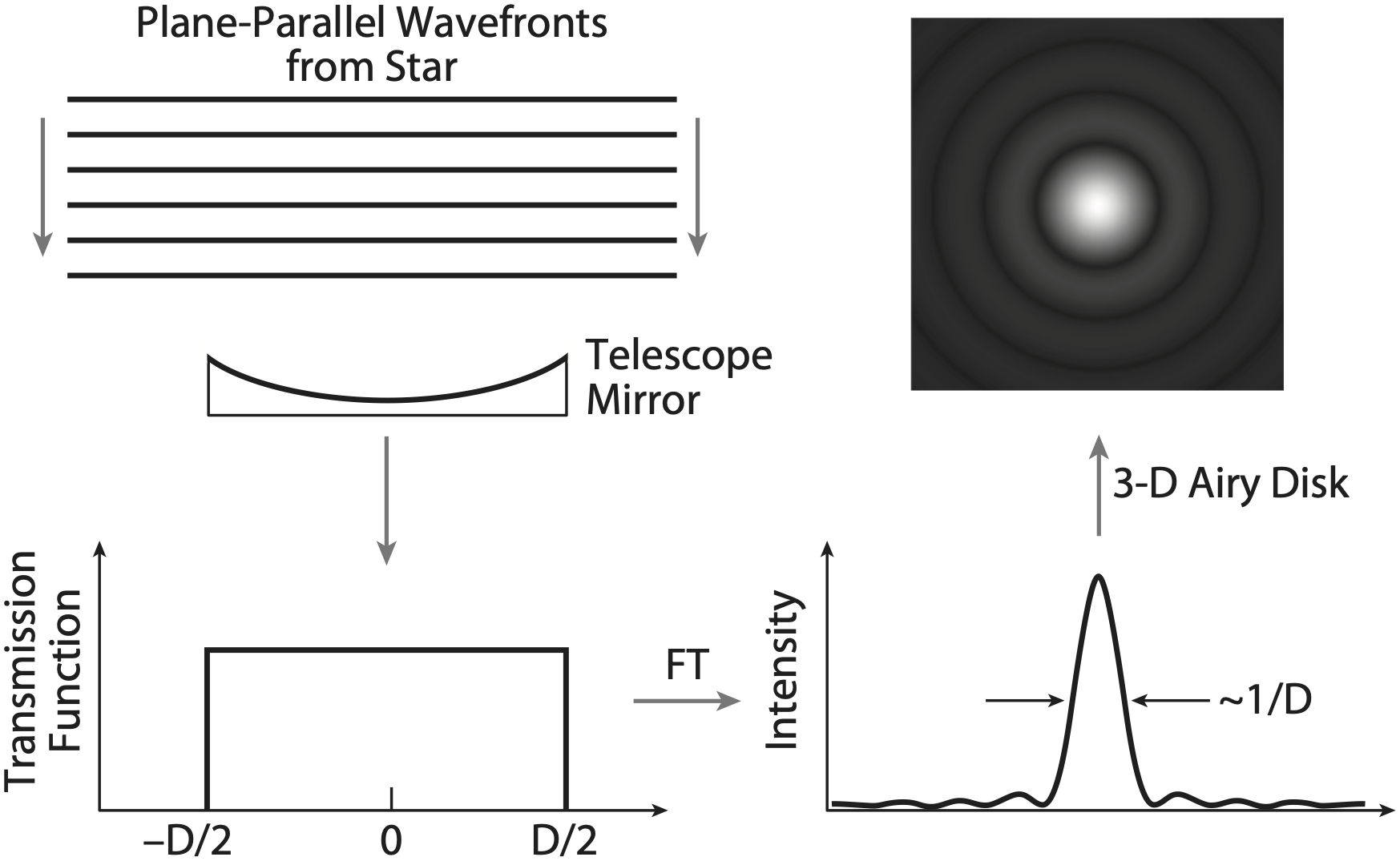}
    \caption[An illustration of the optical path of a telescope]{An illustration of the optical path of a telescope. Upper left-hand panel: the plane-parallel wavefronts from a star are incident upon a telescope. Lower left-hand panel: the resulting transmission function is a tophat function. Lower right-hand panel: the optical system performs a Fourier transform of the tophat function, resulting in a sinc function, whose width is inversely related to the diameter of the telescope’s mirror. Upper right-hand panel: the two-dimensional equivalent of the sinc function is the Airy function or the telescope’s point spread function (PSF). \textit{[Image source: \citet{2010exop.book...79G}.]}}
    \label{figure1.14}
\end{figure}

To solve this issue and obtain diffraction-limited images, observational astronomers use adaptive optics. In this technique, the distorted light of an astronomical object is compensated by deforming the telescope’s optics. As shown in figure~\ref{figure1.14}, in the upper left-hand panel, the plane-parallel wavefronts from a star are incident upon a telescope aperture with width $D$. The resulting transmission function is a tophat function, with nonzero transmission for points between $-D/2$ and $+D/2$ from the mirror’s axis (lower left-hand panel). The optical system performs a Fourier transform of the tophat function, resulting in a sinc function of the form $sin(x/D)/x$ (lower right-hand panel). So, the width of this sinc function is inversely related to the diameter of the telescope’s mirror. The upper right-hand panel of figure~\ref{figure1.14} shows the Airy function or the telescope’s point spread function (PSF), which is the two-dimensional equivalent of the sinc function. In this pattern, the angular distance to the first null is the diffraction limit of the telescope. The ideal PSF is when there is no atmosphere, and the telescope optics are well-designed. In practice, the PSF of a telescope is broader and less ideal.

\begin{figure}[b!]
\centering
\captionsetup{width=0.50\paperwidth}
	\includegraphics[width=0.50\paperwidth]{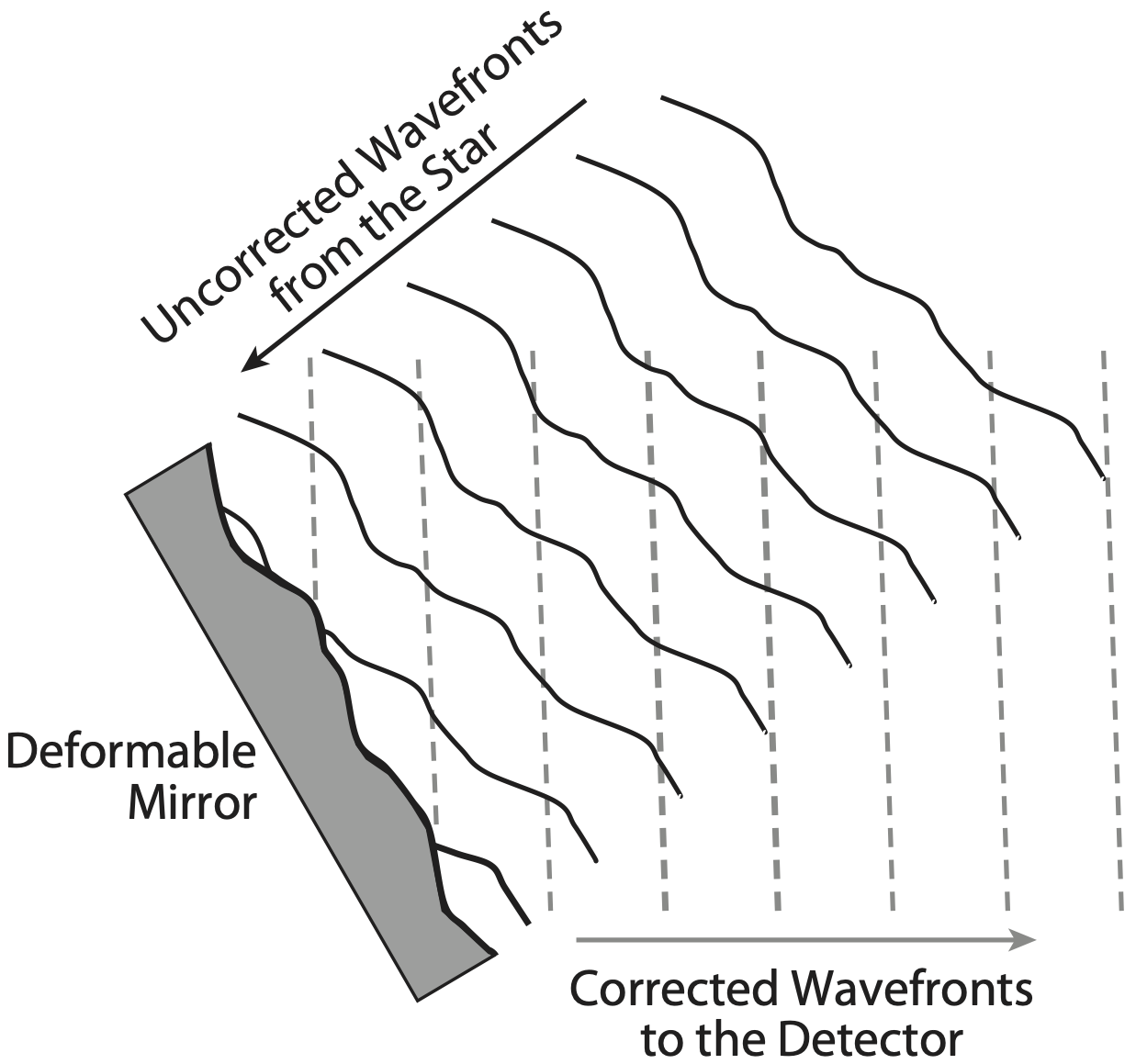}
    \caption[The action of an adaptive optics system’s deformable mirror]{The action of an adaptive optics system’s deformable mirror. The temperature fluctuations in the Earth’s atmosphere deform the stars’ wavefronts, which causes a blurry image. An adaptive optics system reshapes these uncorrected incoming wavefronts. \textit{[Image source: \citet{2010exop.book...79G}.]}}
    \label{figure1.15}
\end{figure}

An adaptive optics system helps observers remove these harmful effects and recover a diffraction-limited PSF. An adaptive optics system consists of a small deformable mirror, which is mounted after the primary mirror’s focus. The shapes of uncorrected incoming wavefronts are measured, and necessary corrections are relayed to the deformable mirror of the adaptive optics system, which results in corrected parallel wavefronts (see figure~\ref{figure1.15}). Thus, the blurry images of astronomical objects caused by temperature fluctuations in the Earth’s atmosphere can be enhanced to much sharper images by adaptive optics systems.

\subsubsection{Contrast}
The second problem in detecting exoplanets by this method is that planets are faint and stars are bright. A planet has two sources of light emission. The first one is the reflection of starlight by the planet’s surface. The second radiation source is due to the planet’s thermal emission. The newly formed planet’s slow gravitational collapse generates a large amount of thermal energy. After reaching the equilibrium radius, the planet still maintains some of its thermal energy along with the energy received from the parent star. Nonetheless, the planet’s radiation is far less than the star’s. For instance, the planet-star contrast between Jupiter and the Sun is $3\times 10^{-9}$.

There are two ways to surmount this difficulty. The first way is to make observations at wavelengths, where the planet emits most of its energy. While a Sun-like star emits most of its energy at visible wavelength, a Jupiter-sized planet will put out most of its energy at infrared wavelengths. Wien’s displacement law, $\lambda_{\text{max}}T=2.898\times 10^{-3}m.K$, which provides a relationship between the wavelength of maximum emission ($\lambda_{\text{max}}$) and the temperature of a blackbody ($T$), predicts that the Jupiter with $T=134 K$ will radiate most of its energy at around 22 microns. The Sun, however, radiates much less energy at such a long wavelength. Therefore, the planet-star contrast increases from $3\times 10^{-9}$, when all wavelengths are considered, to $2\times 10^{-5}$ at infrared wavelengths.

The second way is to target young stars that host correspondingly young planets. A planet has the highest thermal emission in its lifetime when it is young. The younger a planet is, the hotter it is. For example, Jupiter was 3.7 times hotter and 1.3 times larger at roughly ten million years after its formation because it was still experiencing gravitational contraction. As a consequence, such a planet emits most of its energy at 7 microns, and the planet-star contrast is near $2.5\times 10^{-4}$.

\begin{figure}[h!]
\centering
\captionsetup{width=0.50\paperwidth}
	\includegraphics[width=0.50\paperwidth]{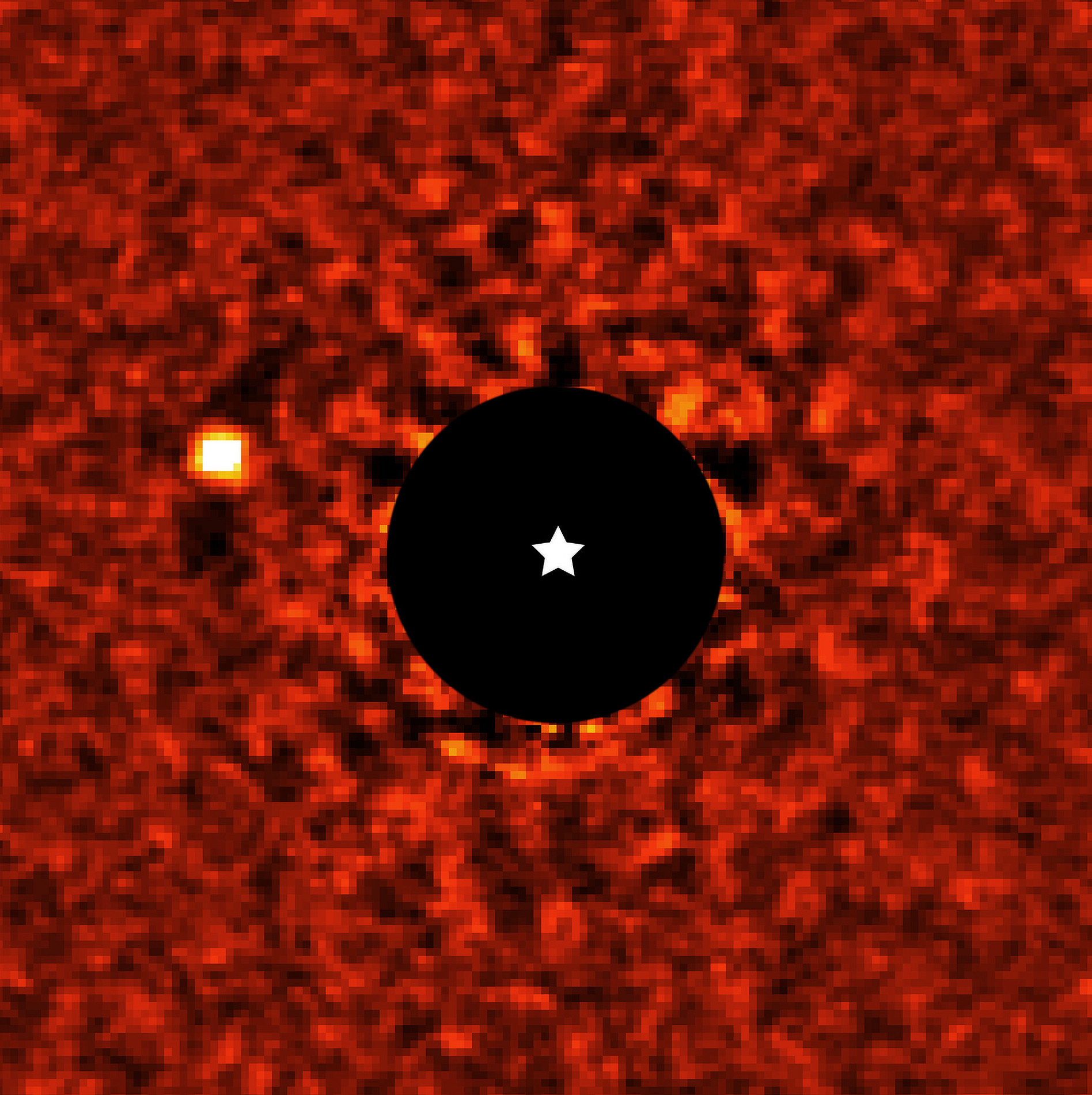}
    \caption[A directly imaged exoplanet named AF Lep b]{A directly imaged exoplanet named AF Lep b. This Jupiter-like planet has been imaged using the SPHERE instrument on ESO’s Very Large Telescope (VLT). The SPHERE instrument corrects the blurring effect caused by atmospheric turbulence using adaptive optics. Also, it blocks the light from the star with a coronagraph, revealing the planet next to it. The AF Leporis system is only 24 million years old, about 200 times younger than the Sun \citep{2023A&A...672A..93M,2023A&A...672A..94D}.}
    \label{figure1.16}
\end{figure}

There is also a technological way to solve the contrast problem, where astronomers use a tool called coronagraph. This instrument improves the planet-star contrast and allows the planet hunters to discover exoplanets. A coronagraph blocks the light from the star and allows the light from the angular region around the star to pass into the detector. Figure~\ref{figure1.16} shows a Jupiter-like planet named AF Lep b next to its parent star. This exoplanet has been directly imaged using the coronagraph and adaptive optics system of the SPHERE instrument on ESO’s Very Large Telescope (VLT) \citep{2023A&A...672A..93M,2023A&A...672A..94D}.

It should be noted that due to the limitations and imperfections in the optics of the imaging instruments, there are other obstacles to seeing the dim light from an exoplanet. Nonetheless, these optical issues can be addressed through a variety of observational techniques. Due to optical aberrations as well as background stars, not all the imaged tiny faint points near stars are planets. To verify an imaged planet near the target star, astronomers typically have to wait one or two years to see if the candidate planet moves with its parent star across the sky.

In principle, the planet’s mass and projected semi-major axis can be determined during a direct imaging process. The projected semi-major axis ($a_{\text{proj}}$) can be computed using the angular separation ($\theta$) and the star’s distance ($d$), given by $a_{\text{proj}}=d\theta$, when $d\gg a$. Determining the planet’s mass, however, is a difficult model-dependent process. Astronomers usually use the theoretical models of the interior structures and atmospheric properties of planets to obtain a planet’s physical parameters, such as its mass.

\subsection{Timing variation}
This method is limited to systems that show very regular periodic behavior. The transit timing variation method is an extension of the general transit method. It follows changes in the transit time of an observed exoplanet caused by the gravitational influence of other unseen, non-transiting exoplanets in the same system. For example, Kepler-19b shows transit timing variation with an amplitude of approximately five minutes, revealing the existence of Kepler-19c \citep{ballard2011kepler}. This method can also help discover bodies that are too faint for radial velocity.

In addition, if a celestial object like a planet orbits a pulsar, the pulsating radio source will not remain stationary and starts to circle the barycenter of the system. So, the pulsar periodically moves towards and away from the observer. As a result, it takes these highly regular signals a bit shorter to reach us when the pulsar approaches the observer and vice versa. By measuring the timing of the pulsar's radio signals, this radial movement can be detected. This effect helped detect two exoplanets around PSR 1527+12. PSR 1527+12 is a pulsar that emits radio waves every 6.2 milliseconds, and the orbiting planets change the pulsation period by 15 picoseconds \citep{wolszczan1992planetary}.

\subsection{Orbital brightness modulation}
When an exoplanet revolves around its star, the amount of light reflected from the star changes periodically. Therefore, an exoplanet can affect the total brightness of the system even if it does not transit its host star. The orbital brightness modulation method uses this effect to detect exoplanets. However, it can be very challenging to differentiate between modulations caused by variable stars and those caused by orbiting planets. Large, close-in planets with high albedos reflect more starlight, so they are the best candidates for this method. Kepler-70b and Kepler-70c are two exoplanets that have been discovered through this effect \citep{charpinet2011compact}.

Relativistic beaming is a phenomenon in which the apparent luminosity of an emitting source changes at high speeds close to the speed of light. A source that emits light in all directions when at rest will emit most of its light along its direction of motion when it moves at high speed with respect to a stationary inertial observer. According to this effect, the radial movement of a star can result in tiny luminosity variations, allowing for the detection of exoplanets. \citet{faigler2013beer} used this effect to discover the Kepler-76b. They also measured the amount of starlight reflected by the exoplanet and the brightness changes due to the star's ellipsoidal shape caused by tidal forces.

\subsection{Astrometry}
The activity of measuring the positions of astronomical objects relative to the background of relatively static stars is known as astrometry \citep{johnson2015you}. In an exoplanetary system, the position of the host star might change slightly in the sky plane as it wobbles around the barycenter. By measuring this type of star motion, one can identify exoplanets. But to detect such tiny changes in the position of the stars, very precise instruments are needed, and for this reason, only one exoplanet named DENIS-P J082303.1-491201b has been discovered by astrometry method \citep{2013A&A...556A.133S}.

\subsection{Polarimetry}
It is known that the light emitted by stars is unpolarized. When the starlight is reflected by the planetary atmosphere, it becomes polarized. Although the polarimetry method has not yet led to the identification of an exoplanet, this effect has been detected for the exoplanet HD 189733b. So, it could potentially reveal unknown exoplanets \citep{berdyugina2007first}.





\section{Titius-Bode law and prediction of planets}
In our solar system, there is a simple logarithmic spacing between planets, which has been known for over two centuries as the Titius-Bode (TB) law. Its classical relation is:
\begin{equation}
a_{n}=0.4+0.3\times2^{n},
\label{eqTB}
\end{equation}
where $a_{n}$ represents the orbital semi-major axis of the $n^{th}$ planet in AU. The planet Mercury corresponds to $n=-\infty$, Venus to $n=0$, Earth to $n=1$ and so on \citep{1974PhT....27e..54N}.

\begin{figure}
\centering
\captionsetup{width=0.60\paperwidth}
\includegraphics[width=0.60\paperwidth]{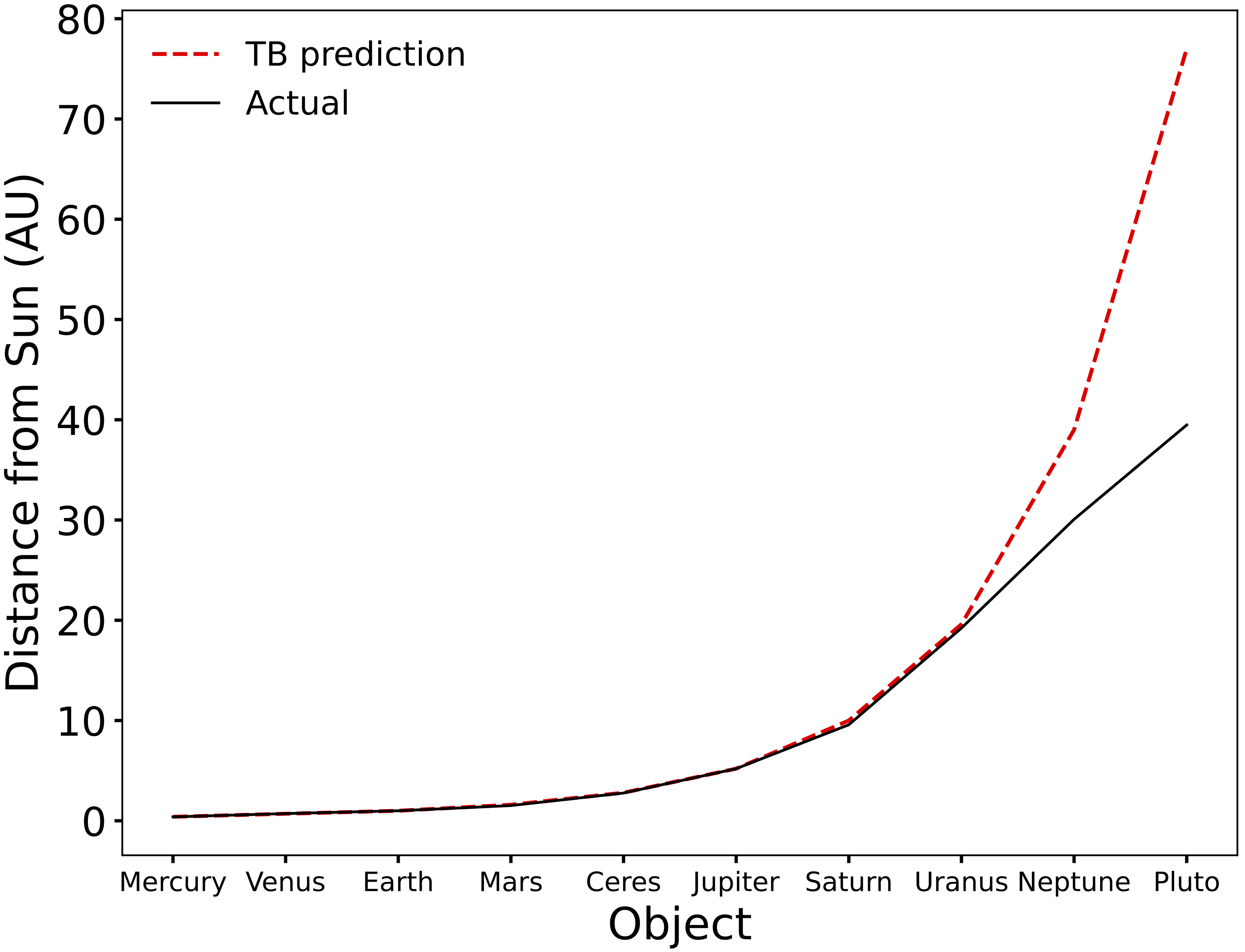}
\caption[Actual and TB predicted distances of the planets]{The comparison between actual (black solid line) and TB predicted (red dashed line) distances of the solar system planets, Ceres, and Pluto from the Sun.}
\label{figure1.17}
\end{figure}

After the discovery of the planet Uranus in 1781 by Frederick William Herschel, it was recognized that the TB law predicted this planet's semi-major axis \citep{1974PhT....27e..54N}. The discovery of the planet Uranus and the TB relation motivated many observation programs to investigate and detect the lost fifth planet, which eventually led to the discovery of the dwarf planet Ceres \citep{1948JRASC..42..241S}. The predictions made using the TB relation also played a key role in exploring the planet Neptune, but not as accurately as Uranus. Interestingly, the satellite systems of the giant planets also follow a TB relation \citep{1960VA......3...25L, 1970CeMec...3...67B}. Figure~\ref{figure1.17} compares the actual distance of planets in the solar system from the Sun (black solid line) and the predicted distances by TB law (red dashed line).

\par The TB relation was used effectively to predict lost undetected objects in our solar system. It was believed that this relation could help make similar predictions in detected multiple exoplanet systems, too. The five-planet 55 Cnc system was one of the first multiple- exoplanet systems where \citet{2008RMxAA..44..243P} applied TB to predict the undetected planets. They found that a simple exponential TB relation reproduces the semi-major axes of the five observed planets. They also predicted two additional planets at distances of 2 and 15 AU. Using the 55 Cnc system, \citet{2008JASS...25..239C} also checked whether the TB relation is enforceable on exoplanetary systems by statistically analyzing the distribution of the ratio of periods of two planets in the 55 Cnc system, by comparing it with that derived from the TB relation. \citet{2010JASS...27....1C} again repeated this calculation for 31 multiple exoplanet systems and concluded that the adherence of the solar system's planets to the TB relation might not be fortuitous; thus, we could not ignore the possibility of using the TB relationship in exoplanetary systems.

Moreover, \citet{2012AIPC.1479.2356L} showed that like 55 Cnc, ten other planetary systems (ups And, GJ 876, HD 160691, GJ 581, Kepler-223, HR 8799, Kepler-20, Kepler-33, HD 10180, and Kepler-11) host four or more planets that also obey a similar (but not identical) TB relation.

\citet{2013MNRAS.435.1126B} (hereafter, \citetalias{2013MNRAS.435.1126B}) used a sample of 68 multiple exoplanet systems with at least four planets, including samples of both confirmed and candidate systems. They identified a sample of exoplanet systems that are likely to be more complete and tested their adherence to the TB relation. They found that most of these exoplanetary systems adhere to the TB relation better than the solar system. Using a generalized TB relation, they predicted 141 additional exoplanets, including a planet with a low radius ($R<2.5R_{\oplus}$) and within the HZ of the Kepler-235.

Using the predictions made by \citetalias{2013MNRAS.435.1126B}, \citet{2014MNRAS.442..674H} analyzed Kepler's long-cadence data to search for 97 of those predicted planets and obtained a detection rate of $\sim5$ percent. Considering the possibility that the remaining predicted planets were not detected because of their small size or non-coplanarity, the detection rate was estimated to be less than the lower limit of the expected number of detections. They concluded that applying the TB relation to exoplanetary systems and using its predictive power in \citetalias{2013MNRAS.435.1126B} could be questionable.

\citet{2015MNRAS.448.3608B} (hereafter, \citetalias{2015MNRAS.448.3608B}) used the \citetalias{2013MNRAS.435.1126B} method to predict the existence of additional planets in 151 Kepler multiple exoplanet systems that contain at least three transiting planets. They found 228 undetected exoplanets and, on average, $2\pm1$ planets in the HZ of each star. They found that apart from the five planets detected by \citet{2014MNRAS.442..674H}, one additional planet in Kepler-271 had been detected around its predicted period. The completeness of observational data was taken into account as an important issue in their analysis. They declared a completeness factor due to the intrinsic noise of the host stars, planet radius, highly inclined orbits, and the detection techniques. Thus they did not expect all predicted planets to be detected. They estimated the geometric probability to transit for all 228 predicted planets and highlighted a list of 77 planets with high transit probability, resulting in an expected detection rate of $\sim15$ percent, which was about three times higher than the detection rate measured \citetalias{2013MNRAS.435.1126B}.

Recently, \citet{2020PASJ...72...24L} used data from 27 exoplanetary systems with at least five planets and applied their proposed method to find the reliability of the TB relation and its predictive capability to search for planets. They removed planets from the system one by one and used TB relation to recover them, where they were able to recover the missing one 78\% of the time. This number was much higher than when they tried this with random planetary systems, where 26\% of planets were recovered. Using statistical tests, they showed that the planetary orbital periods in exoplanetary systems were not consistent with a random distribution and concluded it to be an outcome of the interactions between true planets.
\section{The relationship between physical parameters in planetary systems}
\begin{figure}[b!]
\centering
\captionsetup{width=0.60\paperwidth}
\includegraphics[width=0.60\paperwidth]{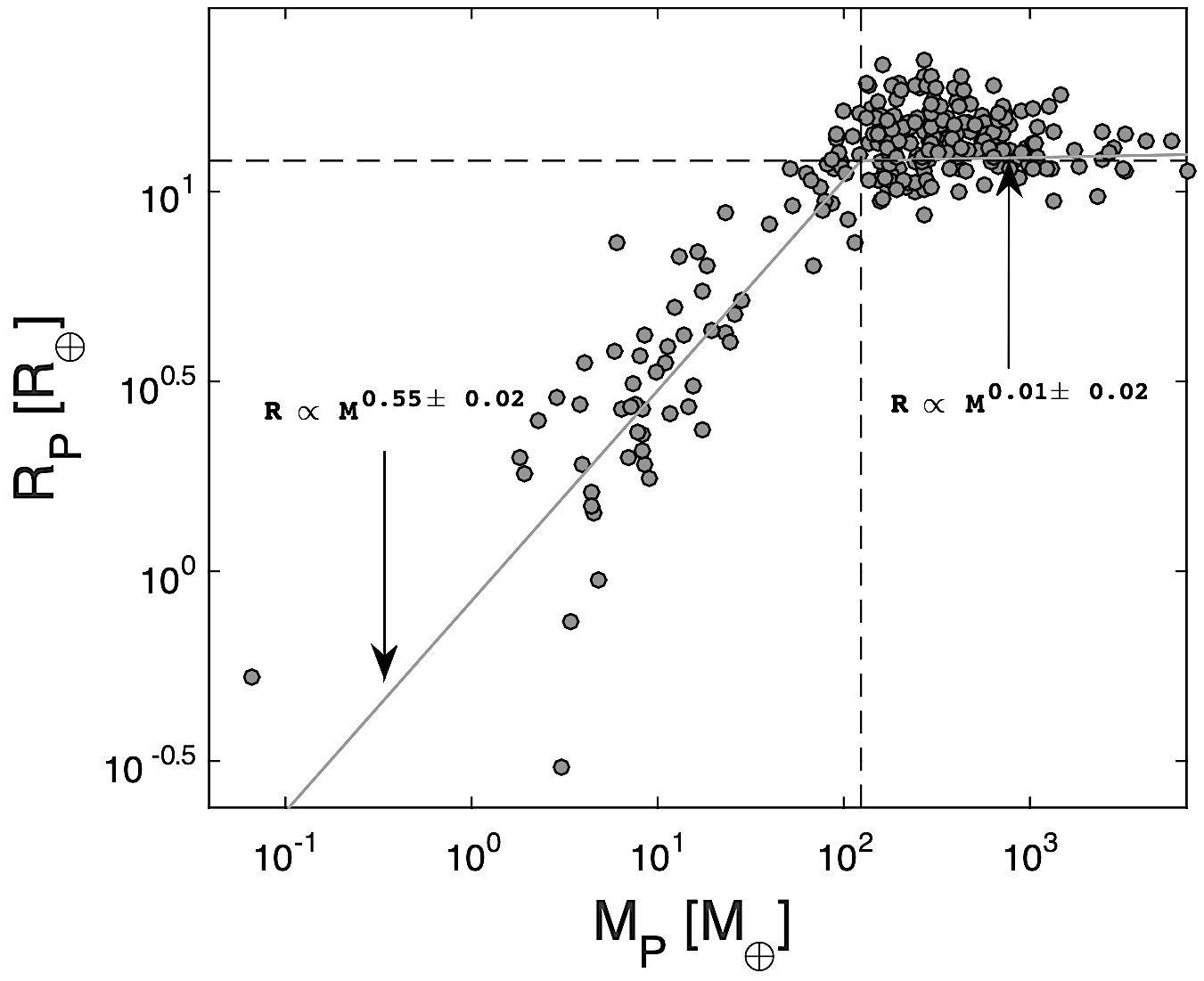}
\caption[The mass-radius relations obtained by \citet{2017A&A...604A..83B}]{The mass-radius relations and the derived best-fit curves obtained by \citet{2017A&A...604A..83B}.}
\label{figure1.18}
\end{figure}

There is a clear correlation between the radius and mass of a planet that can be described with a polytropic relation \citep{1993RvMP...65..301B,2000ARA&A..38..337C}. Many works have investigated the relationship between planetary parameters, particularly mass, and radius, and to deduce the composition and structure of exoplanets \citep{2007ApJ...669.1279S,2012ApJ...744...59S,2020A&A...634A..43O}. \citet{2013ApJ...768...14W} divided the planets into two groups, those with masses greater than and lower than 150$M_{\oplus}$, and presented a power law relation for the mass-radius distribution of each group. \citet{2017A&A...604A..83B} revised this mass breakpoint to $124\pm7M_{\oplus}$ and proposed $R_{p}\propto M_{p}^{0.55\pm0.02}$ for small planets and $R_{p}\propto M_{p}^{0.01\pm0.02}$ for large planets. Figure\ref{figure1.18} reveals the derived best-fit curves obtained by \citet{2017A&A...604A..83B}.

Assuming a power law description of the mass-radius relation, for the first time, a probabilistic model for planets with radii lower than 8$R_{\oplus}$ was presented by \citet{2016ApJ...825...19W}. \citet{2017ApJ...834...17C} implemented this idea to an extended dataset, forecasting the mass or radius of planets. Moreover, they calculated the forecasted mass for $\sim$7000 Kepler Objects of Interest \citep{2018MNRAS.473.2753C}.

Most previous works use a power law model to explore the mass-radius relation and have assumptions that are not pliable enough to consider principal attributes in such diagrams. Consequently, \cite{2018ApJ...869....5N} developed a non-parametric approach using a sequence of Bernstein polynomials and the sample of \citet{2016ApJ...825...19W}. The same method was used in follow-up work to analyze the mass-radius relation of exoplanets orbiting M dwarfs \citep{2019ApJ...882...38K}.

The correlation between physical parameters in planetary systems is not limited to the planet's mass and radius. It has been demonstrated that the radius of a giant planet is related to other parameters such as the orbital semi-major axis, the planetary equilibrium temperature, the tidal heating rate, and the stellar irradiation and metallicity \citep{2006A&A...453L..21G,2007ApJ...659.1661F,2012A&A...540A..99E}. \citet{2002ApJ...568L.113Z} reported a possible correlation between the mass and period of an exoplanet. They also showed that planets revolving around a binary host star might have an opposite correlation. \citet{2014ApJ...783L...6W} studied a restricted dataset containing 65 exoplanets smaller than 4$R_{\oplus}$ with orbital periods shorter than 100 days. They showed that planets smaller than 1.5$R_{\oplus}$ are consistent with a positive linear density–radius relation, but for planets larger than 1.5$R_{\oplus}$, density decreases with radius. \citet{2015ApJ...810L..25H} presented the mass-density relationship in a logarithmic space for objects ranging from planets ($M\approx0.01M_{J}$) to stars ($M>0.08M_{\odot}$). They divided the mass-density distribution into three regions based on changes in the slope of the relationship and introduced a new definition for giant planets.

\citet{2016arXiv160700322B} used a Random Forest regression model to evaluate the influence of different physical parameters on planet radii. Applying this model to different groups of giant planets, they found that the planet's mass and equilibrium temperature have the greatest effect on determining the radius of a hot-Saturn ($0.1<M_{p}<0.5M_{J}$). They also showed that the equilibrium temperature is more important for more massive planets. Moreover, \citet{2019A&A...630A.135U} (hereafter, \citetalias{2019A&A...630A.135U}) introduced Random Forest as a promising algorithm for obtaining exoplanet properties. They used Random Forest to predict the exoplanet radii based on several planetary and stellar parameters. Similar to previous results, an exoplanet's mass and equilibrium temperature were the fundamental parameters.


\chapter{Methodology} 

\label{chapter2} 



\newpage
\section{Sample I}\label{data}
We use fundamental physical parameters of exoplanets, including planetary orbital period ($P$), radius ($R_{p}$), and mass ($M_{p}$), and the stellar mass ($M_{s}$) and radius ($R_{s}$) from two exoplanet databases: the NASA Exoplanet Archive\footnote{\url{https://exoplanetarchive.ipac.caltech.edu/}} and the Extrasolar Planets Encyclopedia\footnote{\url{http://exoplanet.eu/}}. We also use conservative and optimistic limits of the HZ of available stars from the Habitable Zone Gallery\footnote{\url{http://hzgallery.org/}}.

\begin{figure}
\centering
\captionsetup{width=0.60\paperwidth}
\includegraphics[width=0.60\paperwidth]{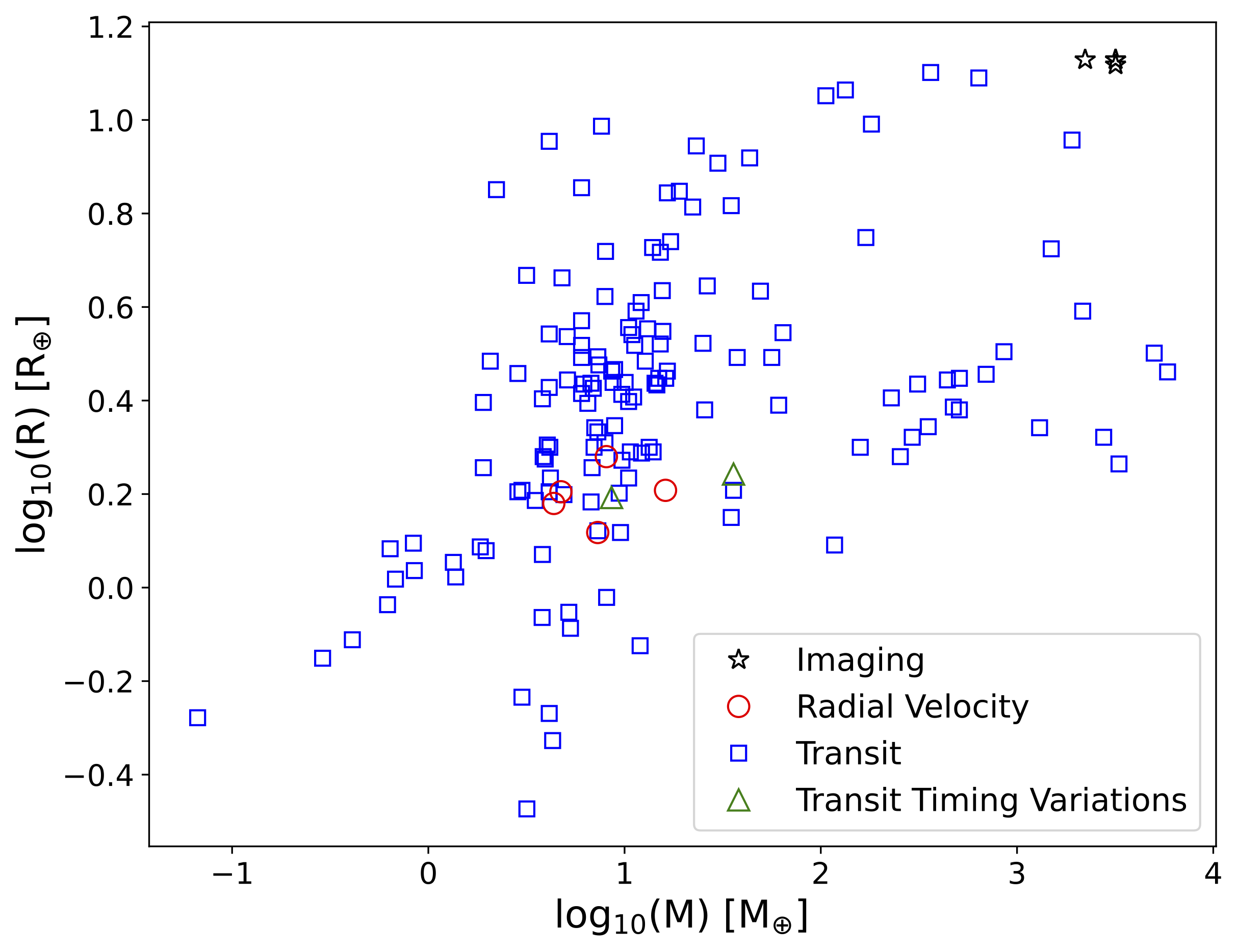}
\caption[The radius-mass distribution of sample I]{The radius-mass distribution of the exoplanets in our sample I, separated into four groups based on their detection methods: imaging (black stars), radial velocity (red circles), transit (blue squares), and transit timing variations (green triangles).}
\label{figure2.1}
\end{figure}

\begin{figure}
\centering
\captionsetup{width=0.60\paperwidth}
\includegraphics[width=0.60\paperwidth]{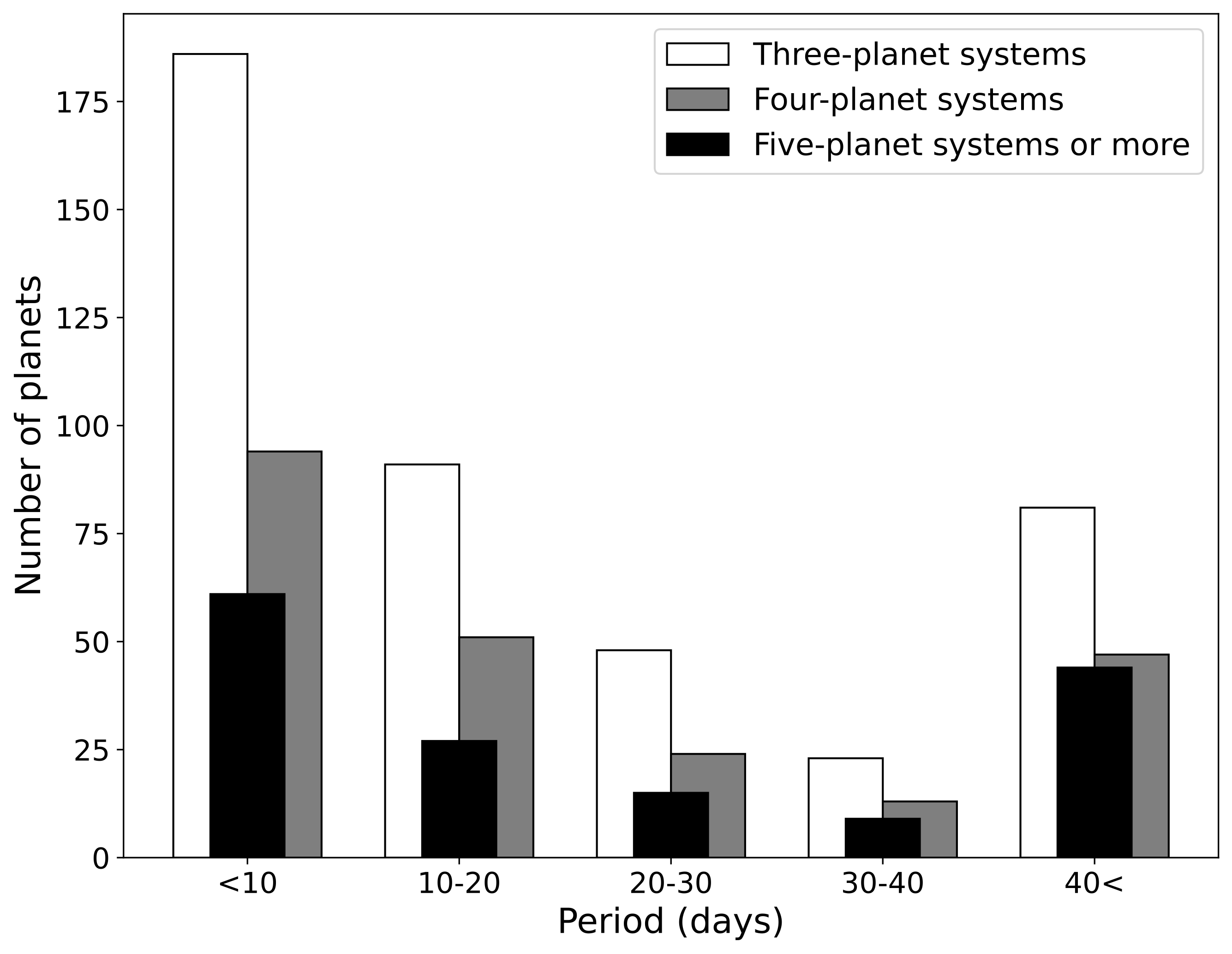}
\caption[The orbital period distribution of sample I]{The distribution of the orbital period of 813 member exoplanets in 229 multiple-exoplanet systems (sample I) hosting at least three planets (white bars), four planets (gray bars), and five (or more) planets (black bars).}
\label{figure2.2}
\end{figure}

The total number of multiple exoplanet systems with at least three confirmed planets available is 230 systems to date, hosting 818 planets. 81.5\% of these planets have been detected using the transit method and 16.4\% using the radial velocity method. The remaining 2.1\% of planets have been identified using transit timing variations, imaging, pulsar timing, and orbital brightness modulation methods. Figure~\ref{figure2.1} represents the mass-radius distribution of exoplanets for four groups of planets, separated according to their detection methods. Of the 230 systems, 142 systems host three planets, 59 systems host four planets, 20 systems include five planets, and seven systems host six planets. TRAPPIST-1 is the only system that contains seven planets, and KOI-351 the only one that contains eight. Figure~\ref{figure2.2} represents the distribution of the orbital period of exoplanets in various exoplanetary systems used in this study.

The exoplanetary systems 55 Cnc, GJ 676 A, HD 125612, K2-136, Kepler-132, Kepler-296, Kepler-47, Kepler-68, and ups And consist of binary stars. GJ 667 C and Kepler-444 are also triple star systems. The system Kepler-132 possesses four planets such that two of them (Kepler-132 b and Kepler-132 c) have roughly similar orbital periods of 6.1782 days and 6.4149 days. After more detailed studies, it was found that Kepler-132 b and Kepler-132 c cannot orbit the same star \citep{2014ApJ...784...44L}. Consequently, we exclude Kepler-132 from our analyzes; our final sample I of exoplanetary systems thus contains 229 systems.

\section{Prediction of additional exoplanets in multiple-exoplanet systems}
We use the TB relation to predict additional undetected planets for all multiple exoplanet systems in sample I, which have at least three confirmed planets.

The TB relation (see equation~\ref{eqTB}) can be written in terms of the orbital periods as follows:
\begin{equation}
P_{n}=P\alpha^{n}, n=0,1,2,...,N-1,
\label{equation2.1}
\end{equation}
where $P_{n}$ is the orbital period of the $n^{\text{th}}$ planet, and P and $\alpha$ are fitting parameters.

\begin{figure}[t!]
\centering
\captionsetup{width=0.60\paperwidth}
\includegraphics[width=0.60\paperwidth]{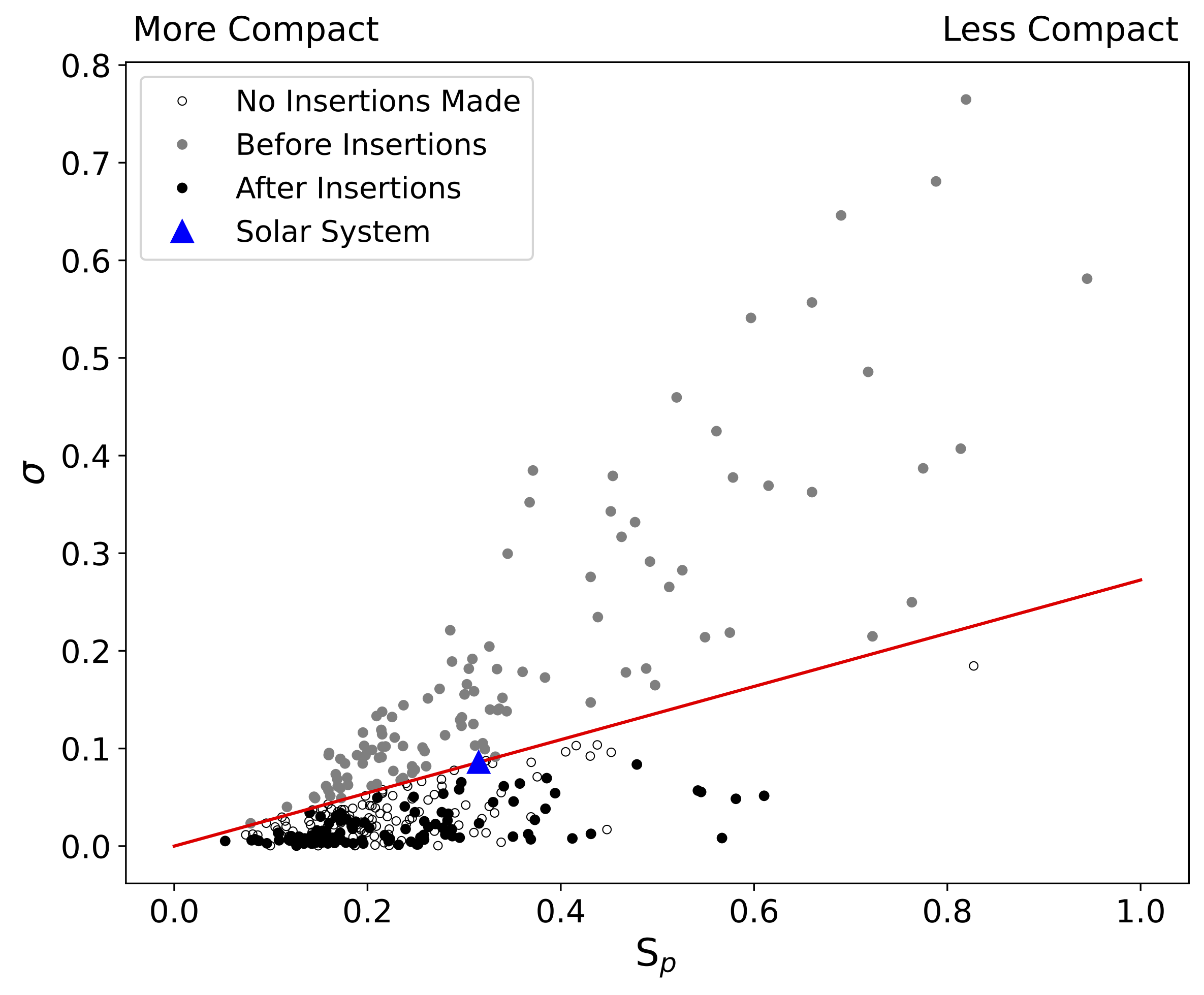}
\caption[$\sigma$ as a function of $S_p$]{$\sigma$ as a function of the average log period spacing between planets, $S_p$ (equation~\ref{equation2.4}), of exoplanet systems. The red line goes through two points: the origin (0,0) and ($S_{p}$,$\sigma$), where $S_{p}$ is the compactness of the solar system and $\sigma$ is the value required for the solar system to yield $\chi^2/dof=1$ in equation~\ref{equation2.3}. The blue triangle shows the solar system, and the empty circles show the exoplanet systems with no planet insertions (systems with $\chi^2/dof\leq1$). Gray circles indicate systems before planet insertions, and black circles indicate the ($S_{p}$,$\sigma$) of the systems after insertions have been made.}
\label{figure2.3}
\end{figure}

In analyzing the exoplanetary systems, the solar system is used to guide how well each system adheres to the TB relation. In logarithmic space, the TB relation is written as follows:
\begin{equation}
\log P_{n}=\log P + n\log\alpha=b+m\times n, n=0,1,2,...,N-1,
\label{equation2.2}
\end{equation}
where b=$\log P$ and m=$\log\alpha$ are intercept and slope of the relation, respectively. For a system with N planets, the $\chi^2/dof$ value can be calculated by:
\begin{equation}
\frac{\chi^2(b,m)}{N-2}=\frac{1}{N-2}\sum_{n=0}^{N-1}[\frac{(b+m\times n)-\log P_{n}}{\sigma}]^2,
\label{equation2.3}
\end{equation}
N-2 is the number of degrees of freedom (N planets - 2 fitted parameters (b and m)), and $\sigma$ represents the system's sparseness or compactness. The sparseness/compactness of a system is calculated from $\sigma$= 0.273 $S_p$, where $S_p$ represents the average log period spacing between planets as defined by:
\begin{equation}
S_{p}=\frac{\log P_{N-1}-\log P_{0}}{N},
\label{equation2.4}
\end{equation}
where N is the number of planets in the system and $P_{N-1}$, and $P_{0}$ are the largest and smallest orbital periods in the system, respectively. We plot the $\sigma$ values (see equation~\ref{equation2.3}) as a function of sparseness/compactness ( $S_{p}$; see equation~\ref{equation2.4}) in figure~\ref{figure2.3}. In this figure, the red line goes through two points: the origin and the specific $S_{p}$ and $\sigma$ values for the solar system ($\sigma$ is the value required for solar system to yield $\chi^2/dof=1$ in equation \ref{equation2.3}).

Using a proper value for $\sigma$, $\chi^2/dof=1$ is adjusted for the solar system case. If the detected planets in a system adhere to the TB relation better than the planets of the solar system ($\chi^2/dof\leq1$), we only predict an extrapolated planet beyond the outermost detected planet. If the detected planets adhere worse ($\chi^2/dof>1$), we begin the interpolation process. For the first step of interpolation, one new specific planet, from at-least 5,000 hypothetical planets that have a random period between the innermost and outermost detected planets, is inserted into the system. This new planet covers all possible locations between the two adjacent planets, and a new $\chi^2/dof$ value is calculated for each possibility. The new specific inserted planet is chosen from 5,000 cases when producing the minimum value of $\chi^2/dof$. Similarly, inserting up to 10 new specific planets for each system step by step (two planets for the second step, three planets for the third step, etc.) covers all possible locations and combinations between two adjacent planets. The period uncertainty of an inserted planet ($u_{ins}$) is calculated using the uncertainties of detected planets in the same system (e.g., $u_{1}$, $u_{2}$, $u_{3}$,...) as follows:
\begin{equation}
u_{\text{ins}}=\sqrt{u_{1}^{2}+u_{2}^{2}+u_{3}^{2}+...}.
\label{equation2.5}
\end{equation}
We adopt the highest value of the parameter $\gamma$, which is the improvement in the $\chi^2/dof$ per inserted planet, for identifying the best combination of detected and predicted planets. $\gamma$ is defined by:
\begin{equation}
\gamma=\frac{(\frac{\chi_{i}^2-\chi_{j}^2}{\chi_{j}^2})}{n_{\text{ins}}},
\label{equation2.6}
\end{equation}
where $\chi_{i}^2$ and $\chi_{j}^2$ are the $\chi^2$ values before and after inserting of $n_{\text{ins}}$ planets. We analyze each system's data separately and use the MCMC method to quantify the uncertainties of the best-fit parameters.

By applying the lowest signal-to-noise ratio (SNR) of the detected planets in the same system to the predicted planet's orbital period, the maximum mass or maximum radius of the predicted planets is calculated \citepalias{2013MNRAS.435.1126B,2015MNRAS.448.3608B}. For transiting detected planets, the maximum radius is calculated by:
\begin{equation}
R_{\text{max}}=R_{\text{minSNR}}(\frac{P_{\text{predicted}}}{P_{\text{minSNR}}})^{0.25},
\label{equation2.7}
\end{equation}
and for radial velocity detected planets, the maximum mass is calculated by:
\begin{equation}
M_{\text{max}}=M_{\text{minSNR}}(\frac{P_{\text{predicted}}}{P_{\text{minSNR}}})^{7/6},
\label{equation2.8}
\end{equation}
where $R_{\text{minSNR}}$, $M_{\text{minSNR}}$, and $P_{\text{minSNR}}$ are the radius, mass, and orbital period of the detected planet with the lowest SNR, respectively. After calculating the maximum radius or maximum mass of the predicted planets, using the mass-radius relationship established by \citet{2017A&A...604A..83B} (hereafter, \citetalias{2017A&A...604A..83B}) ($R_{p} \propto M_{p}^{0.55}$ and $R_{p} \propto M_{p}^{0.01}$ for the small and large planets, respectively), we convert the radius values to mass values, and vice versa.

\subsection{Dynamical stability}
We use the dynamical spacing criterion ($\Delta$) to analyze how our predicted objects could be stable in their positions in the exoplanetary system. The dynamical spacing $\Delta$ between two adjacent planets with masses $M_{1}$ and $M_{2}$ and orbital periods $P_{1}$ and $P_{2}$ orbiting a host star with a mass of $M_{s}$ was defined by \citet{1993Icar..106..247G} and \citet{1996Icar..119..261C} as follows:
\begin{equation}
\Delta=\frac{2M_{s}^{1/3}(P_{2}^{2/3}-P_{1}^{2/3})}{(M_{1}+M_{2})^{1/3}(P_{2}^{2/3}+P_{1}^{2/3})}.
\label{equation2.9}
\end{equation}
In a planetary system, two adjacent planets are less likely to be stable in their positions when the value of their dynamical spacing is small ($\Delta \leq 10$). However, if they are stable, they are more likely to be in orbital resonance with each other. Using this criterion, we analyze systems in our sample I, which have predicted planets within their HZs. We calculate the $\Delta$ value for all pairs in these systems and investigate whether they are in resonance with each other or not (for planets whose mass values have not been reported in catalogues, we use the mass-radius relation of \citetalias{2017A&A...604A..83B} to calculate their masses). To analyze the possibility of resonance, we calculate the period ratios of all adjacent planets considering an arbitrary threshold; the planet pairs that have $x\leq2\%$ are considered to be pairs in orbital resonance, where
\begin{equation}
x=\frac{\mid \frac{N_{j}}{N_{i}}-\frac{P_{n+1}}{P_{n}}\mid}{\frac{N_{j}}{N_{i}}},
\label{equation2.10}
\end{equation}
$N_{i}$ and $N_{j}$ are positive integers with $N_{i}<N_{j}\leq5$, and $P_{n}$ and $P_{n+1}$ are the orbital periods of two adjacent planets.

\subsection{Transit probability}
The transit phenomenon of a planet is only observable if the planetary orbit plane is close to the line-of-sight between the observer and the host star. In other words, the pole of the planetary orbit must be within the angle $d_{s}/a_{p}$, where $d_{s}$ is the stellar diameter, and $a_{p}$ is the planetary orbital radius \citep{2013PASP..125..933S}. Then, the geometric transit probability ($P_{\text{tr}}$) of a planet can be estimated using $d_{s}/2a_{p}$, where $P_{\text{tr}}=0.5\%$ for an Earth-size planet at 1 AU orbiting a solar-size star \citep{1984Icar...58..121B,1996chz..conf..229K}. According to the models of planetary systems, multi-planetary systems, like the solar system, are assumed to be formed out of common protoplanetary disks \citep{2015ARA&A..53..409W}. Therefore, the orbital planes should have small relative inclinations so that the Kepler multi-planetary systems are highly coplanar \citepalias{2015MNRAS.448.3608B}. We use this criterion to estimate the transit probability of predicted exoplanets in sample I and prioritize how soon the predicted planets can be detected.

\section{Sample II}\label{data2}
\begin{figure}[t!]
\centering
\captionsetup{width=0.60\paperwidth}
\includegraphics[width=0.60\paperwidth]{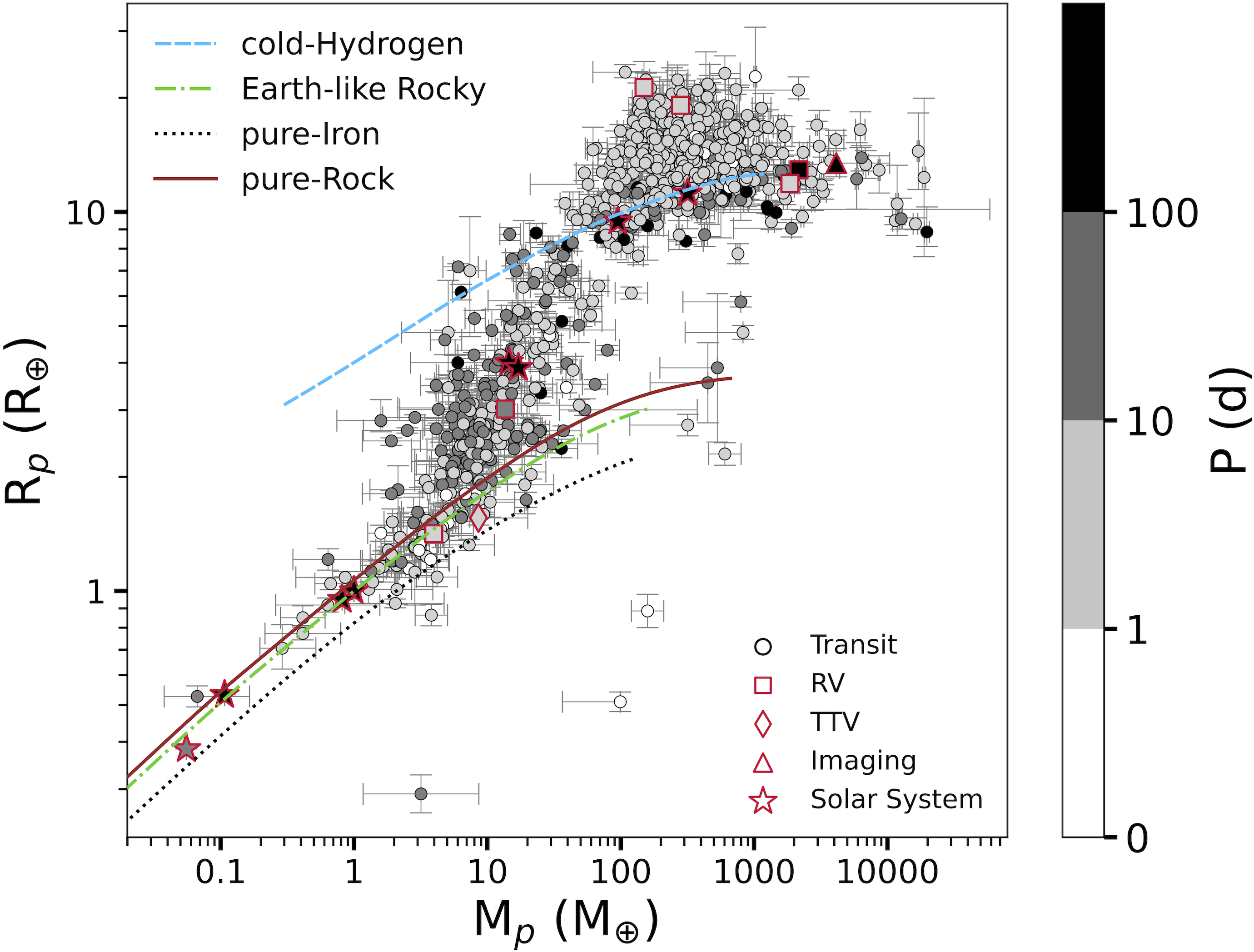}
\caption[The radius-mass distribution of sample II]{The radius-mass distribution of 770 planets (sample II) color coded by orbital period. This figure separates exoplanets into four groups based on detection methods: transit (black circles), radial velocity (red squares), transit timing variations (red diamond), and imaging (red triangle). Red stars show the solar system's planets. Four sample mass-radius relations are also shown: cold-hydrogen (blue dashed line), Earth-like rocky (green dash-dotted line), pure-iron (black dotted line), and pure rocky (solid crimson line) planets \citep{2010ApJ...712L..73M,2014ApJS..215...21B}.}
\label{figure2.4}
\end{figure}

We use the NASA Exoplanet Archive\footnote{\url{https://exoplanetarchive.ipac.caltech.edu/}} and the Extrasolar Planets Encyclopedia\footnote{\url{http://exoplanet.eu/}} to extract the data of sample II\footnote{The data was last extracted on March 22, 2022.}. These two catalogs are comprehensive, up-to-date, and available to the public, and also provide access to relevant publications \citep{2011epsc.conf....3S,2013PASP..125..989A}. There are 762 confirmed exoplanets with reported physical parameters, including the orbital period ($P$) and eccentricity ($e$), planetary mass ($M_{p}$) and radius ($R_{p}$), and the stellar mass ($M_{s}$), radius ($R_{s}$), metallicity (Fe/H), and effective temperature ($T_{\text{eff}}$). Note that exoplanets with only a minimum mass are not considered. Among these 762 exoplanets, six have been discovered by the radial velocity method, and each of the imaging and transit timing variation methods has identified only one exoplanet. In general, most exoplanets have been discovered by observing the slight decrease in brightness of the host star caused by the transit of a planet in front of it. Using NASA's Planetary Fact Sheet\footnote{\url{https://nssdc.gsfc.nasa.gov/planetary/factsheet/}}, we add the eight solar system planets to the sample II. Overall, the dataset contains 770 planets. Figure~\ref{figure2.4} shows the radius of planets plotted as a function of mass and color coded by orbital period. It is separated into four groups based on detection methods: transit, radial velocity, transit timing variation, and imaging. Distributions of cold-hydrogen, Earth-like rocky (32.5\% Fe+67.5\% MgSiO3), pure-iron (100\% Fe), and pure-rock (100\% MgSiO3) planets are also illustrated \citep{2010ApJ...712L..73M,2014ApJS..215...21B}.

We should note that, like other observational datasets, our sample II is also affected by detection biases. A dataset containing planets whose radius and mass have been measured suffers from the detection limits of both radial velocity and transit methods. Thus, it is impossible to draw a reliable conclusion about the occurrence of planets using our dataset.

\subsection{Data pre-processing}\label{pre}
In sample II, we aim to predict the planetary radius as the target variable, using other physical parameters as features. As the parameters have different ranges, we transfer them to a logarithmic space. In data analysis, the ultimate results might be affected by some unreliable measurements in the sample. In machine learning and statistics, there are diverse methods to detect these unusual observations in a dataset. We choose the Local Outlier Factor (LOF) method to identify and remove observations with abnormal distances from other values. The technique is often used with multidimensional datasets, like our eight-dimensional one, which has different densities and types of outliers. The LOF uses two hyper-parameters: neighborhood size ($k$), which defines the neighborhood for local density calculation, and contamination ($c$), which specifies the proportion of outliers in the dataset \citep{breunig2000lof,chandola2009anomaly}. By choosing $k=20$ and $c=0.05$, the LOF method is run twice: once for all parameters including $P$, $e$, $M_{p}$, $R_{p}$, $M_{s}$, $R_{s}$, Fe/H, and $T_{\text{eff}}$, then for planetary mass and radius, which are known to be highly correlated. Altogether, the LOF detects 76 data points as outliers. Appendix~\ref{appendixB.1} describes the process of identifying outliers in detail.

Choosing appropriate features plays a vital role in building an efficient ML model. Adding extra variables or those highly correlated with each other may reduce the overall predictive ability of the model and lead to wrong results. Feature selection (hereafter FS) methods rank features based on their usefulness and effectiveness in making predictions. The FS methods can be divided into three groups: filter, wrapper, and embedded methods \citep{guyon2008feature,chandrashekar2014survey,jovic2015review,brownlee2016machine}. In filter methods, features are filtered independently of any induction algorithm and based on some performance evaluation metrics calculated directly from the data \citep{sanchez2007filter,cherrington2019feature}.
In contrast, the selection process in wrapper methods is based on the performance of a specific ML algorithm operating with a subset of features \citep{ferri1994comparative,kohavi1997wrappers,hall1999feature}. Embedded methods combine the qualities of both filter and wrapper methods. They perform the FS in the training process and are usually specific to given learning machines \citep{lal2006embedded,bolon2013review}. We use five FS methods to identify the most important features in our dataset: Spearman's rank correlation test as a filter method, the Backward Elimination and Forward Selection as two wrapper methods, and the CART (classification and regression trees) and XGBoost (extreme gradient boosting) as two Embedded methods.

\section{Algorithms for exoplanet clustering}
Clustering as an unsupervised ML task involves grouping each data point with a specific type. In theory, data points belonging to a particular group should have similar properties \citep{xu2005survey,kaufman2009finding,2012arXiv1205.1117S}. Data clustering algorithms can be divided into hierarchical and partitional groups. Hierarchical algorithms find clusters using previously established clusters, while partitional algorithms find all clusters at once \citep{kononenko2007chapter,2012arXiv1205.1117S}.

We aim to group the planets of sample II into distinct, non-overlapping clusters, similar to exclusive clustering \citep{1988acd..book.....J}. We use ten ML clustering algorithms to include a wide range of clustering methods and examine their performance in exoplanet data. Since many diverse exoplanets have been discovered, implementing clustering algorithms can find potential exoplanet groups to investigate their characteristics. The algorithms are available in the Scikit-learn software ML library \citep{scikit-learn} and are as follows: Affinity Propagation, BIRCH (balanced iterative reducing and clustering using hierarchies), DBSCAN (density-based spatial clustering of applications with noise), Gaussian Mixture Model, Hierarchical Clustering, K-Means, Mean Shift, Mini-Batch K-Means, OPTICS (ordering points to identify the clustering structure), and Spectral Clustering \citep{davies1979cluster,ester1996density,zhang1996birch,ankerst1999optics,halkidi2001clustering,comaniciu2002mean,2007Sci...315..972F,von2007tutorial,schubert2017dbscan}.

BIRCH, Gaussian Mixture Model, Hierarchical Clustering, K-Means, Mini-Batch K-Means, and Spectral Clustering are algorithms that do not learn the number of clusters ($K$) from data. Therefore, we first perform the Elbow and Silhouette methods to find the optimal number of clusters. The Elbow method runs the K-Means clustering algorithm for $K$ values. Then, for each $K$, it computes the sum of squared distances (SSD) between data points and their assigned cluster centroids and uses them to propose an optimal number of clusters. The Silhouette method determines the degree of separation between clusters by choosing a range of $K$ values and calculating a coefficient for each $K$. The silhouette coefﬁcient for a particular data point is calculated by $(b^{i} - a^{i})/{max(a^i, b^i)}$. Here, $a^{i}$ represents the average distance from all data points in the same cluster, whereas $b^{i}$ is the average distance from data points that belong to the closest cluster. Provided that the sample is on or near the decision boundary between two neighboring clusters, the silhouette coefficient becomes 0. A coefficient close to +1 indicates that the sample is far from neighboring clusters. A negative coefficient value indicates that samples may have been assigned to the wrong cluster \citep{rousseeuw1987silhouettes,2012arXiv1205.1117S}.

\section{Models to predict a planet's radius}
In machine learning, different algorithms allow machines to learn information from a given dataset, uncover relationships, and make predictions \citep{brownlee2016machine,brownlee2016machine2}. In our case, we apply ML models to sample II in order to predict planet radii when other parameters are given. It is also possible to see how efficiently each parameter uses these models. The algorithms used to perform regression tasks are as follows: Decision Tree, K-Nearest Neighbors, Linear Regression, Multilayer Perceptron, M5P, and Support Vector Regression (SVR) \citep{hinton1990connectionist,quinlan1992learning,quinlan1993program,hastie2009elements,chang2011libsvm}. Bootstrap Aggregation and Random Forest are also used as ensemble algorithms that combine the predictions from multiple models \citep{breiman1996bagging,breiman2001random}. These algorithms are available in the Weka tool environment \citep{witten2005practical,hall2009weka}.

Linear Regression and M5P are two algorithms that can extract parametric equations. Linear Regression algorithm fits a linear model to the entire data. M5P performs a multiple linear regression model. This tree-based algorithm allocates linear regressions at the terminal nodes. It divides the entire dataset into several smaller subsets and fits a linear model to each subset \citep{quinlan1992learning}. They both, consequently, result in a basic linear equation like equation~\ref{equation2.11}, where $A$ and $C$ are fitting parameters, $Y$ is the dependent variable, $X$ is the independent variable, and $N$ represents the total number of independent variables. In our case, $Y$ is the planet’s radius, and $X$ represents other physical parameters. For these two algorithms, we use the MCMC to quantify the uncertainties of the best-fit parameters \citep{goodman10,emcee3}.

\begin{equation}
Y=C+\sum_{i=1}^{N}A_{i}X_{i}.
\label{equation2.11}
\end{equation}

To evaluate the quality of the predicted radius ($R_{pre}$) compared to the observed radius ($R_{obs}$) and to compare the efficiency of the models, root means square error (RMSE), mean absolute error (MAE), and coefficient of determination ($\rho^{2}$) are calculated. Equations~\ref{equation2.12} to~\ref{equation2.14} define RMSE, MAE, and $\rho^{2}$, respectively, where $R_{mean}$ is the mean of the $R_{obs}$ values and $n$ represents the total number of samples. Lower values of RMSE and MAE and higher values of $\rho^{2}$ indicate better accuracy of the models. It should be noted that hyper-parameters specific to each model are tuned to have the best performance. Also, a 10-fold cross-validation procedure, as a data resampling method, is used to evaluate the performance of models. Furthermore, each algorithm is executed for original and logarithmic datasets to understand the effect of data re-scaling (see appendix~\ref{appendixB.2}).

\begin{equation}
RMSE=\sqrt{\sum_{i=1}^{n}\frac{(R_{obs}-R_{pre})^2}{n}}.
\label{equation2.12}
\end{equation}

\begin{equation}
MAE=\frac{1}{n}\sum_{i=1}^{n}\mid R_{obs}-R_{pre}\mid.
\label{equation2.13}
\end{equation}

\begin{equation}
\rho^{2}=1-\frac{\sum_{i=1}^{n}(R_{obs}-R_{pre})^2}{\sum_{i=1}^{n}(R_{obs}-R_{mean})^2}.
\label{equation2.14}
\end{equation}


\chapter{Results and Discussion} 

\label{chapter3} 



\newpage
\section{Additional exoplanets in multi-planetary systems}
\subsection{Adherence to the TB relation and the predicted planets}
\begin{figure}
\centering
\captionsetup{width=\textwidth}
\includegraphics[width=\textwidth]{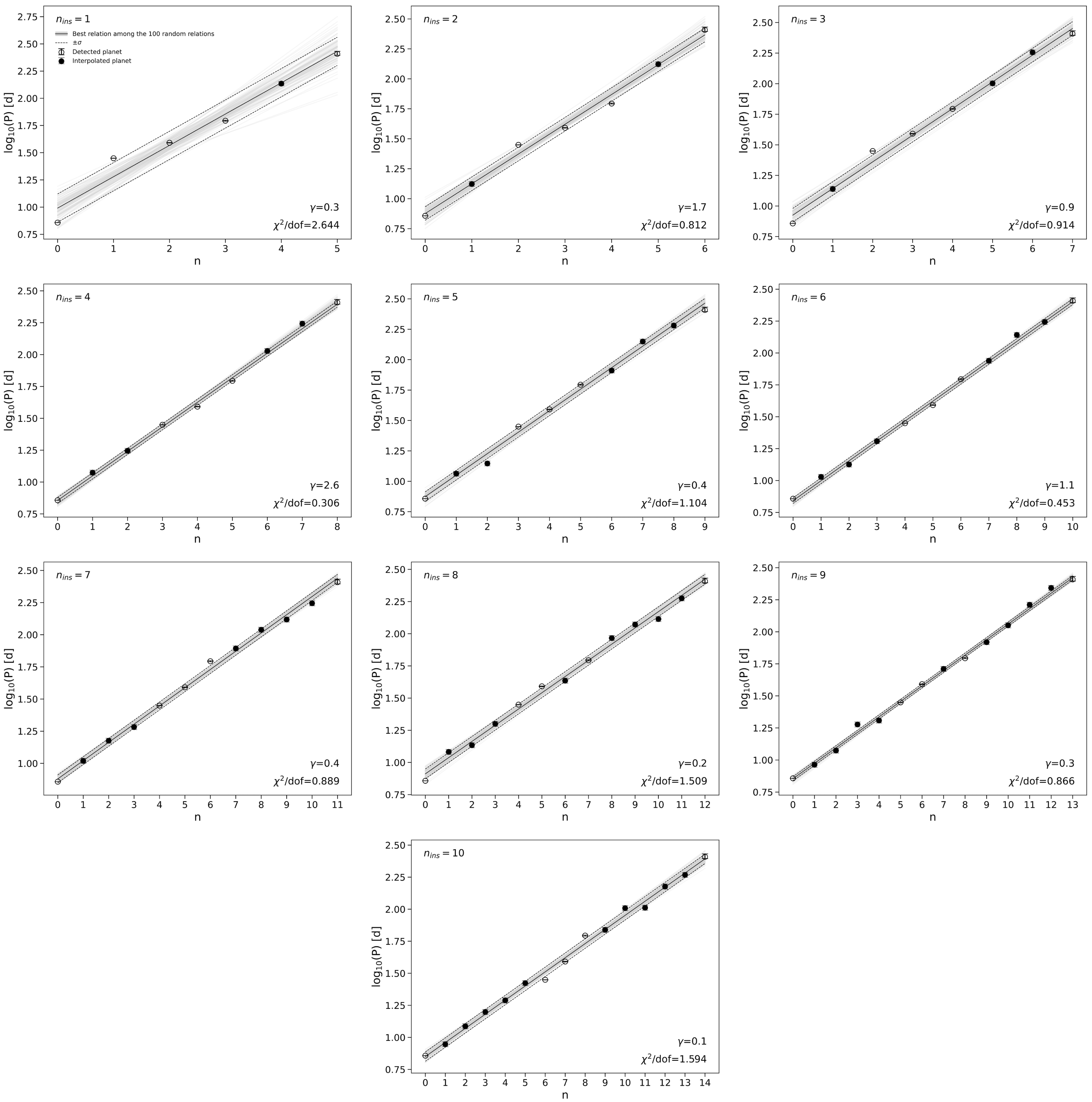}
\caption[Steps of linear regressions applied to the GJ 667 C system]{The TB relation and steps of linear regressions are applied to the data of system GJ 667 C. Detected and predicted planets are shown with empty and filled circles, respectively. For each step, the value of $n_{\text{ins}}$ represents the number of inserted planets. The values of $\gamma$ and $\chi^2/dof$ are also shown where the highest value of $\gamma$ is in the fourth step as the best combination of detected and predicted planets. The two black dashed lines show $\pm1\sigma$ uncertainties around the best scaling relation (black solid line). The gray lines are a set of 100 different realizations, drawn from the multivariate Gaussian distribution of the parameters (for the fourth step: m=0.1924, b=0.856 and ln<$\sigma$>=-3.73), and the scatter covariance matrix is estimated from the MCMC chain.}
\label{figure3.1}
\end{figure}

\begin{figure}[t!]
\centering
\captionsetup{width=0.60\paperwidth}
\includegraphics[width=0.60\paperwidth]{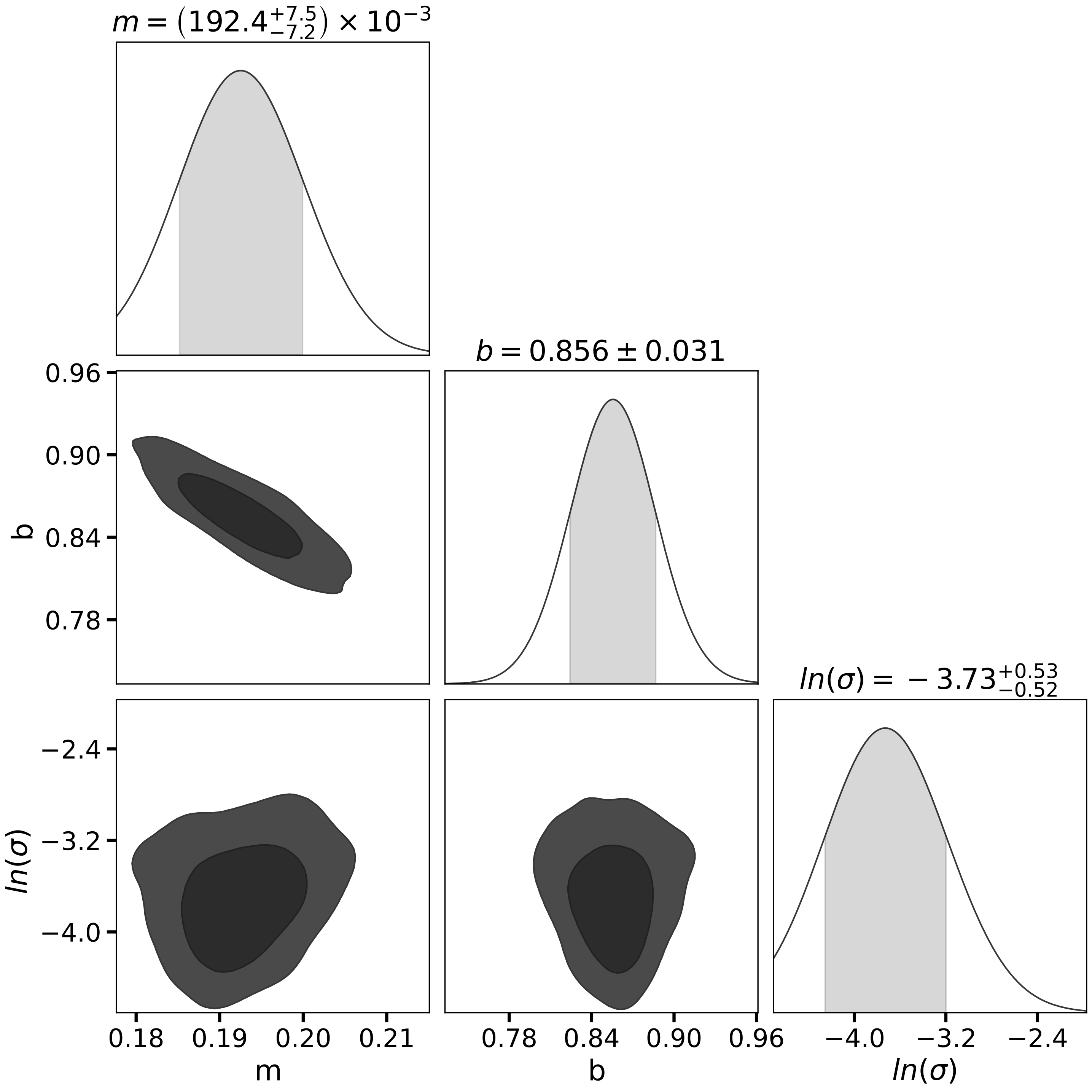}
\caption[Marginalized posterior distributions of the linear regression applied to the GJ 667 C system]{The one- and two-dimensional marginalized posterior distributions of the scaling relation parameters for the highest $\gamma$ value corresponding to the fourth step ($n_{\text{ins}}=4$) of the linear regression of the GJ 667 C system, as shown in figure \ref{figure3.1}.}
\label{figure3.2}
\end{figure}

To investigate the probability of the existence of additional planets in exoplanetary systems, we apply the TB relation and the MCMC method to analyze the data of a sample of 229 systems that contain at least three confirmed exoplanets (sample I). We find that 122 systems adhere to the TB relation better than the solar system without any need for interpolation. For those 107 systems that adhere to the TB relation worse than the solar system, we insert up to 10 new additional planets. For example, figure \ref{figure3.1} shows the best linear regression (TB relation) for up to 10 additional planet inserts in the GJ 667 C system. For GJ 667 C, the highest $\gamma$ value is in the fourth step, where the number of inserted planets ($n_{\text{\text{ins}}}$) is four. In this figure, the black line shows the best (mean) scaling relation, and the two dashed lines show the $\pm1\sigma$ uncertainties around this relation. The gray lines are a set of 100 different realizations, drawn from the multivariate Gaussian distribution of the parameters, where for the highest $\gamma$ value: m=0.1924, b=0.856, and ln<$\sigma$>=-3.73. The scatter covariance matrix is also estimated from the MCMC chain. Figure \ref{figure3.2} illustrates the one- and two-dimensional marginalized posterior distributions of the scaling relation parameters for the fourth step of the linear regression for the GJ 667  C system. Table~\ref{table3.1} represents the data corresponding to figure~\ref{figure3.1}.

After interpolating all 107 systems, we find that these systems adhere to the TB relation better than the solar system or approximately the same extent. Of these 107 interpolated systems, 50 systems need one, 33 systems need two, and the remaining 24 systems need more than two additional planets to be inserted. We also predict an extrapolated planet beyond the outermost detected planet for all systems in our sample I. We predict the existence of 426 possible additional exoplanets in these systems, of which 197 are predicted by interpolation. It should be noted that six of the predicted planets in Kepler-1388, Kepler-1542, Kepler-164, Kepler-374, Kepler-402, and Kepler-403 have been flagged as "Planetary Candidates" in the NASA Exoplanet Archive.

\twocolumn
\begin{table}
\centering
\captionsetup{width=0.5\textwidth}
\caption[Data corresponding to figure~\ref{figure3.1}]{Data corresponding to figure~\ref{figure3.1}.}
\label{table3.1}
\setlength{\tabcolsep}{8.5pt}
\begin{tabularx}{0.5\textwidth}{lcccc}
\hline
$n_{\text{ins}}^{a}$ & $\chi^2/dof$ & $\gamma^{b}$ & Period (d) & $\text{ON}^{c}$ \\
\hline
1     & 2.644 & 0.3   & 136.5 & 4 \\
2     & 0.812 & 1.7   & 13.3  & 1 \\
      &       &       & 132.6 & 5 \\
3     & 0.914 & 0.9   & 13.8  & 1 \\
      &       &       & 100.6 & 5 \\
      &       &       & 180.4 & 6 \\
4     & 0.306 & 2.6   & 11.9  & 1 \\
      &       &       & 17.6  & 2 \\
      &       &       & 106.7 & 6 \\
      &       &       & 174.8 & 7 \\
5     & 1.104 & 0.4   & 11.5  & 1 \\
      &       &       & 14.0    & 2 \\
      &       &       & 81.4  & 6 \\
      &       &       & 141.0   & 7 \\
      &       &       & 190.7 & 8 \\
6     & 0.453 & 1.1   & 10.7  & 1 \\
      &       &       & 13.4  & 2 \\
      &       &       & 20.3  & 3 \\
      &       &       & 86.9  & 7 \\
      &       &       & 138.6 & 8 \\
      &       &       & 175.1 & 9 \\
7     & 0.889 & 0.4   & 10.5  & 1 \\
      &       &       & 15.0    & 2 \\
      &       &       & 19.1  & 3 \\
      &       &       & 78.3  & 7 \\
      &       &       & 109.3 & 8 \\
      &       &       & 131.3 & 9 \\
      &       &       & 175.6 & 10 \\
8     & 1.509 & 0.2   & 12.1  & 1 \\
      &       &       & 13.6  & 2 \\
      &       &       & 19.9  & 3 \\
      &       &       & 43.2  & 6 \\
      &       &       & 92.5  & 8 \\
      &       &       & 117.9 & 9 \\
      &       &       & 130.0   & 10 \\
      &       &       & 188.5 & 11 \\
9     & 0.866 & 0.3   & 9.2   & 1 \\
      &       &       & 11.8  & 2 \\
      &       &       & 18.9  & 3 \\
      &       &       & 20.3  & 4 \\
      &       &       & 51.4  & 7 \\
      &       &       & 83.0    & 9 \\
      &       &       & 112.3 & 10 \\
      &       &       & 162.5 & 11 \\
      &       &       & 220.0   & 12 \\
\hline
\end{tabularx}
\end{table}

\begin{table}
\centering
\ContinuedFloat
\caption[]{continued}
\captionsetup{width=0.5\textwidth}
\setlength{\tabcolsep}{8.5pt}
\begin{tabularx}{0.5\textwidth}{lcccc}
\hline
$n_{\text{ins}}^{a}$ & $\chi^2/dof$ & $\gamma^{b}$ & Period (d) & $\text{ON}^{c}$ \\
\hline
10    & 1.594 & 0.1   & 8.8   & 1 \\
      &       &       & 12.2  & 2 \\
      &       &       & 15.8  & 3 \\
      &       &       & 19.4  & 4 \\
      &       &       & 26.5  & 5 \\
      &       &       & 69.0  & 9 \\
      &       &       & 101.9 & 10 \\
      &       &       & 102.6 & 11 \\
      &       &       & 150.3 & 12 \\
      &       &       & 185.2 & 13 \\
\hline
\end{tabularx}
\small
\begin{tablenotes}
\item$^{a}$Number of the inserted planet.\item$^{b}$$\gamma=(\chi_{i}^2-\chi_{j}^2)/(\chi_{j}^2\times n_{\text{ins}})$, where $\chi_{i}^2$ and $\chi_{j}^2$ are the $\chi^2$ values before and after inserting of $n_{\text{ins}}$ planets, respectively.\item$^{c}$The orbital number of the inserted planet.
\end{tablenotes}
\end{table}
\onecolumn

\begin{figure}
\centering
\includegraphics[width=0.60\paperwidth]{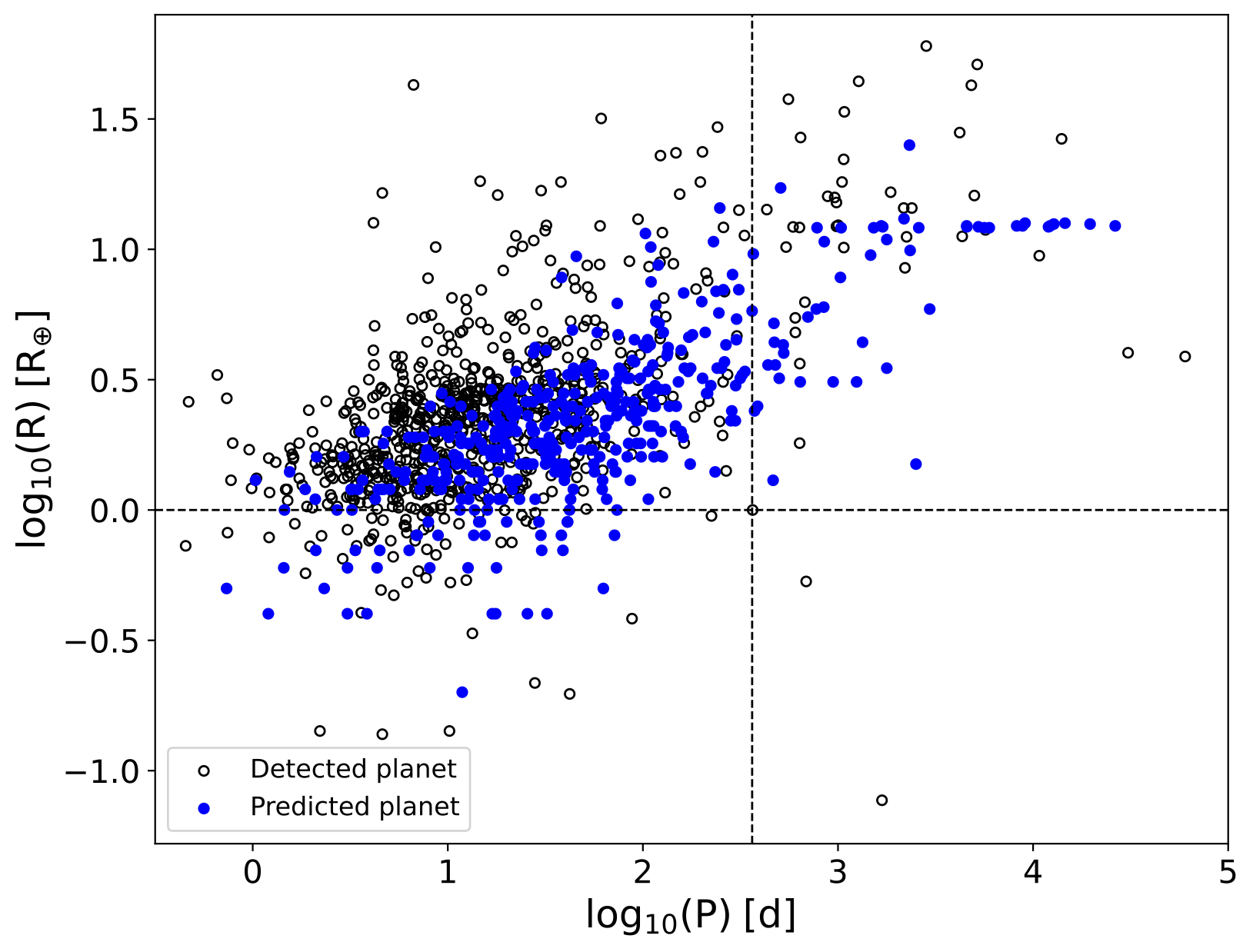}
\captionsetup{width=0.60\paperwidth}
\caption[The radius vs. period of detected and predicted exoplanets of sample I]{The radius vs. period of detected (empty circles) and predicted (filled circles) exoplanets of 229 multi-planetary systems used in this study (sample I). The dashed horizontal and vertical lines correspond to the radius and orbital period of the Earth, respectively.}
\label{figure3.3}
\end{figure}

Figure~\ref{figure3.3} illustrates the radius versus the orbital period of the detected (empty circles) and predicted (filled circles) exoplanets. We find similar trends for both detected and predicted exoplanets. As seen, the vast majority of exoplanets have larger radii and shorter orbital periods than Earth (the radius and orbital period of the Earth have been shown with vertical and horizontal dotted black lines) due to observational limitations.

Table~\ref{tableA.1} lists our predicted exoplanets by extrapolation and the best-fit TB relations including $\chi^2/dof$, m, and b. In this table, column 3 reports a flag that defines whether the system has already been analyzed by \citetalias{2015MNRAS.448.3608B} (or \citetalias{2013MNRAS.435.1126B}) (Y) or not (N). Column 8 reports whether the predicted period values in this paper and \citetalias{2015MNRAS.448.3608B} (or \citetalias{2013MNRAS.435.1126B}) are consistent within error (Y) or not (N). Table~\ref{tableA.2} lists the systems with interpolated and extrapolated planet predictions. In this table, we present the $\chi^2/dof$ before and after interpolation in the fourth and fifth columns. $\gamma$ and $\Delta\gamma$ (where $\Delta\gamma$=($\gamma_{1}$-$\gamma_{2}$)/$\gamma_{2}$; $\gamma_{1}$ and $\gamma_{2}$ are the highest and second-highest $\gamma$ values for system, respectively) are listed in columns 6 and 7, respectively. The definitions of other columns in this table are the same as in table~\ref{tableA.1}.

In figure~\ref{figureA.1}, we show the new systems containing the detected planets (empty circles) and the predicted planets (red circles) by extrapolation, while the predicted planets within the HZ are also shown as green circles. The sizes of the symbols are scaled based on the planet's radius. Similarly, figure~\ref{figureA.2} presents the detected exoplanets (empty circles) and the predicted exoplanets (red circles) by interpolation and extrapolation.

We note that the estimated masses (and radii) of GJ 676 A, Kepler-56, and WASP-47 are excluded because they are higher than the maximum possible limit of a typical planet. Furthermore, due to the lack of stellar parameters, the transit probabilities of HD 31527, HD 136352, and Kepler-402 are not calculated.

\subsection{Planets in the HZ}
To verify whether some of the predicted exoplanets are in the HZ of their parent stars, we use conservative and optimistic definitions of the HZ \citep{2012PASP..124..323K}. Among the predicted exoplanets, 47 exoplanets lie within the HZ of their host stars. 27 exoplanets out of 47 are located within the conservative HZ. Furthermore, 14 exoplanets in the HZ have been predicted by interpolation, and the remaining 33 exoplanets have been predicted by extrapolation. Table~\ref{table3.2} presents the 47 predicted exoplanets that lie within the HZ of the host star.

Kepler-167 is a four-planet system where we predict three interpolated additional exoplanets, in which two planets have an orbital period of 157.0 and 373.5 days, located within the conservative HZ. The host star Kepler-186 is also a five-planet system including two predicted exoplanets with orbital periods of 41.2 and 73.6 days; these two additional planets were found by interpolation and are located within the optimistic and conservative HZ, respectively.

\begin{table}
\setlength{\tabcolsep}{5.75pt}
\captionsetup{width=\textwidth}
\caption[Predicted exoplanets within the HZ of host stars in multi-planetary systems]{Predicted exoplanets within the HZ of host stars in multi-planetary systems. Columns 1 and 2 present the host star name and discovery method (Dis.). Columns 3, 4, and 5 present the orbital period in days, the distance from the parent star in AU, and the orbital number (ON). The estimated maximum radius ($R_{\text{Max}}$) and maximum mass ($M_{\text{Max}}$) in the Earth unit are presented in columns 6 and 7. Column 8 lists the transit probability ($P_{\text{tr}}$). The conservative ($HZ_{\text{Cons}}$) and optimistic ($HZ_{\text{Opt}}$) HZ limits in AU are presented in columns 9 and 10, respectively.}
\label{table3.2}
\small
\begin{tabularx}{\textwidth}{lccccccccc}
\hline
Host name & Dis.$^{a}$ & \makecell{Period\\(d)} & \makecell{a\\(AU)} & ON$^b$ & \makecell{$R_{\text{Max}}$\\($R_{\oplus}$)} & \makecell{$M_{\text{Max}}$\\($M_{\oplus}$)} & \makecell{$P_{\text{tr}}$\\(\%)} & \makecell{$HZ_{\text{Cons}}$\\(AU)} & \makecell{$HZ_{\text{Opt}}$\\(AU)} \\
\hline
GJ 163 & RV    & $72.9\substack{+2.1\\-2}$ & $0.25\substack{+0.01\\-0.01}$ & 2     & 1.4   & 2.5   & 2.05  & 0.14-0.28 & 0.11-0.30 \\
YZ Cet & RV    & $7.2\substack{+0.4\\-0.3}$ & $0.04\substack{+0.01\\-0.01}$ & 3 E   & 1.2   & 1.9   & 1.96  & 0.05-0.10 & 0.04-0.10 \\
Kepler-445 & Tr    & $13.4\substack{+0.7\\-0.6}$ & $0.06\substack{+0.01\\-0.01}$ & 3 E   & 1.4   & 2.5   & 1.61  & 0.07-0.13 & 0.05-0.14 \\
TOI-270 & Tr    & $20.4\substack{+8.1\\-5.8}$ & $0.11\substack{+0.02\\-0.02}$ & 3 E   & 2.0   & 4.5   & 1.59  & 0.14-0.26 & 0.11-0.28 \\
GJ 357 & RV    & $22.7\substack{+1.4\\-1.3}$ & $0.11\substack{+0.01\\-0.01}$ & 2     & 1.3   & 2.1   & 1.42  & 0.13-0.25 & 0.10-0.27 \\
Kepler-249 & Tr    & $33.1\substack{+0.2\\-0.2}$ & $0.16\substack{+0.01\\-0.01}$ & 3 E   & 1.9   & 4.3   & 1.38  & 0.19-0.37 & 0.15-0.39 \\
Kepler-186 & Tr    & $41.2\substack{+1.7\\-1.7}$ & $0.19\substack{+0.01\\-0.01}$ & 4     & 0.9   & 1.1   & 1.26  & 0.24-0.46 & 0.19-0.49 \\
       &       & $73.6\substack{+3.5\\-3.5}$ & $0.28\substack{+0.01\\-0.01}$ & 5     & 1.0   & 1.4   & 0.85  &       &  \\
K2-133 & Tr    & $40.7\substack{+1.7\\-1.6}$ & $0.18\substack{+0.01\\-0.01}$ & 6 E   & 1.9   & 4.4   & 1.18  & 0.19-0.36 & 0.15-0.38 \\
GJ 3138 & RV    & $41.0\substack{+11.4\\-9.4}$ & $0.20\substack{+0.01\\-0.03}$ & 2     & 1.0   & 1.2   & 1.12  & 0.22-0.42 & 0.17-0.44 \\
K2-72 & Tr    & $39.6\substack{+18\\-13.7}$ & $0.15\substack{+0.04\\-0.04}$ & 4 E   & 1.5   & 2.7   & 1.03  & 0.12-0.23 & 0.09-0.24 \\
Kepler-1388 & Tr    & $74.2\substack{+14.6\\-14.0}$ & $0.30\substack{+0.03\\-0.03}$ & 4 E,C & 2.9   & 9.3   & 0.97  & 0.31-0.59 & 0.25-0.63 \\
Kepler-26 & Tr    & $66.6\substack{+8.1\\-7.4}$ & $0.26\substack{+0.02\\-0.02}$ & 7 E   & 2.2   & 5.7   & 0.89  & 0.24-0.46 & 0.19-0.48 \\
Kepler-52 & Tr    & $77.8\substack{+12.9\\-10.4}$ & $0.29\substack{+0.03\\-0.03}$ & 3 E   & 2.4   & 6.4   & 0.89  & 0.31-0.58 & 0.25-0.61 \\
HD 40307 & RV    & $98.0\substack{+14.3\\-13.4}$ & $0.38\substack{+0.04\\-0.03}$ & 4     & 1.6   & 3.1   & 0.87  & 0.48-0.86 & 0.38-0.91 \\
HD 20781 & RV    & $200.4\substack{+79.7\\-55.4}$ & $0.59\substack{+0.15\\-0.11}$ & 4 E   & 6.3   & 38.0  & 0.85  & 0.69-1.23 & 0.54-1.30 \\
Kepler-331 & Tr    & $63.5\substack{+13.1\\-11.7}$ & $0.27\substack{+0.04\\-0.03}$ & 3 E   & 1.9   & 4.5   & 0.83  & 0.28-0.53 & 0.22-0.55 \\
Kepler-55 & Tr    & $105.6\substack{+58.3\\-37}$ & $0.37\substack{+0.13\\-0.09}$ & 5 E   & 2.8   & 8.6   & 0.77  & 0.38-0.71 & 0.30-0.74 \\
K2-3  & Tr    & $97.4\substack{+71.8\\-42.8}$ & $0.35\substack{+0.15\\-0.11}$ & 3 E   & 1.8   & 4.1   & 0.74  & 0.26-0.50 & 0.21-0.53 \\
K2-155 & Tr    & $98.4\substack{+75.2\\-42.1}$ & $0.36\substack{+0.17\\-0.11}$ & 3 E   & 2.4   & 6.5   & 0.74  & 0.32-0.60 & 0.25-0.63 \\
Kepler-169 & Tr    & $136.5\substack{+19.4\\-17.4}$ & $0.50\substack{+0.05\\-0.04}$ & 8 E   & 2.5   & 7.2   & 0.70  & 0.56-1.02 & 0.45-1.08 \\
Wolf 1061 & RV    & $61.9\substack{+2.5\\-2.5}$ & $0.20\substack{+0.01\\-0.01}$ & 2     & 1.2   & 1.8   & 0.70  & 0.10-0.21 & 0.08-0.22 \\
Kepler-235 & Tr    & $113.3\substack{+13.2\\-11}$ & $0.38\substack{+0.03\\-0.02}$ & 4 E   & 2.5   & 7.1   & 0.66  & 0.30-0.57 & 0.24-0.60 \\
Kepler-167 & Tr    & $157.0\substack{+36.3\\-29.2}$ & $0.52\substack{+0.08\\-0.07}$ & 4     & 2.0   & 4.6   & 0.64  & 0.52-0.94 & 0.41-0.99 \\
       &       & $373.5\substack{+106.6\\-83.3}$ & $0.93\substack{+0.17\\-0.14}$ & 5     & 2.4   & 6.8   & 0.36  &       &  \\
Kepler-166 & Tr    & $163.1\substack{+39.9\\-28}$ & $0.56\substack{+0.09\\-0.07}$ & 3 E   & 3.5   & 13.2  & 0.61  & 0.63-1.12 & 0.50-1.19 \\
Kepler-296 & Tr    & $113.6\substack{+8.2\\-7.2}$ & $0.36\substack{+0.02\\-0.01}$ & 5 E   & 2.1   & 5.0   & 0.61  & 0.21-0.40 & 0.16-0.42 \\
Kepler-351 & Tr    & $223.0\substack{+6.6\\-6.6}$ & $0.69\substack{+0.02\\-0.02}$ & 4 E   & 3.0   & 10.2  & 0.57  & 0.78-1.38 & 0.61-1.45 \\
HD 141399 & RV    & $477.4\substack{+45.5\\-41.5}$ & $1.22\substack{+0.08\\-0.07}$ & 2     & 3.6   & 13.5  & 0.55  & 1.12-1.98 & 0.88-2.09 \\
HD 181433 & RV    & $261.8\substack{+50.5\\-44.9}$ & $0.69\substack{+0.08\\-0.08}$ & 3     & 3.7   & 14.5  & 0.53  & 0.57-1.04 & 0.45-1.09 \\
Kepler-149 & Tr    & $285.5\substack{+41.1\\-33.5}$ & $0.82\substack{+0.08\\-0.06}$ & 4 E   & 2.4   & 6.8   & 0.53  & 0.80-1.43 & 0.63-1.51 \\
Kepler-251 & Tr    & $252.6\substack{+232.8\\-119.5}$ & $0.77\substack{+0.42\\-0.27}$ & 4 E   & 3.5   & 13.1  & 0.53  & 0.79-1.40 & 0.62-1.47 \\
HD 20794 & RV    & $322.7\substack{+155\\-98.2}$ & $0.82\substack{+0.24\\-0.18}$ & 4 E   & 3.3   & 11.9  & 0.52  & 0.91-1.63 & 0.72-1.72 \\
Kepler-56 & Tr    & $2324.8\substack{+325.9\\-305.1}$ & $3.77\substack{+0.34\\-0.34}$ & 7 E   & 25.1  &   -    & 0.52  & 2.97-5.41 & 2.34-5.70 \\
Kepler-150 & Tr    & $298.9\substack{+40.5\\-35.9}$ & $0.87\substack{+0.07\\-0.08}$ & 6     & 3.0   & 10.0  & 0.50  & 0.84-1.49 & 0.66-1.57 \\
Kepler-30 & Tr    & $310.2\substack{+108.4\\-81.9}$ & $0.89\substack{+0.2\\-0.16}$ & 3 E   & 7.0   & 46.6  & 0.49  & 0.83-1.48 & 0.66-1.56 \\
Kepler-65 & Tr    & $524.0\substack{+206.9\\-145.9}$ & $1.37\substack{+0.34\\-0.27}$ & 7 E   & 4.3   & 19.2  & 0.47  & 1.51-2.63 & 1.19-2.78 \\
61 Vir & RV    & $368.1\substack{+32.8\\-31.3}$ & $0.98\substack{+0.06\\-0.05}$ & 4 E   & 9.6   & 82.2  & 0.45  & 0.86-1.53 & 0.68-1.61 \\
Kepler-48 & Tr    & $333.4\substack{+124.5\\-88.6}$ & $0.90\substack{+0.21\\-0.17}$ & 4     & 3.4   & 12.5  & 0.45  & 0.71-1.27 & 0.56-1.34 \\
Kepler-298 & Tr    & $202.2\substack{+317.5\\-113.7}$ & $0.59\substack{+0.51\\-0.25}$ & 3 E   & 3.2   & 11.0  & 0.45  & 0.35-0.65 & 0.28-0.69 \\
Kepler-218 & Tr    & $437.8\substack{+160.6\\-114.6}$ & $1.14\substack{+0.26\\-0.21}$ & 4 E   & 3.6   & 14.1  & 0.43  & 0.93-1.65 & 0.73-1.74 \\
\hline
\end{tabularx}
\end{table}

\begin{table}
\setlength{\tabcolsep}{6.25pt}
\centering
\ContinuedFloat
\caption[]{continued}
\captionsetup{width=\textwidth}
\small
\begin{tabularx}{\textwidth}{lccccccccc}
\hline
Host name & Dis.$^{a}$ & \makecell{Period\\(d)} & \makecell{a\\(AU)} & ON$^b$ & \makecell{$R_{\text{Max}}$\\($R_{\oplus}$)} & \makecell{$M_{\text{Max}}$\\($M_{\oplus}$)} & \makecell{$P_{\text{tr}}$\\(\%)} & \makecell{$HZ_{\text{Cons}}$\\(AU)} & \makecell{$HZ_{\text{Opt}}$\\(AU)} \\
\hline
HD 219134 & RV    & $315.5\substack{+83.7\\-65.9}$ & $0.84\substack{+0.15\\-0.12}$ & 5     & 3.2   & 10.9  & 0.42  & 0.52-0.95 & 0.41-1.00 \\
KOI-351 & Tr    & $499.9\substack{+70.4\\-55.6}$ & $1.31\substack{+0.12\\-0.1}$ & 10 E  & 3.2   & 11.2  & 0.42  & 1.24-2.17 & 0.98-2.29 \\
Kepler-401 & Tr    & $640.4\substack{+299.4\\-188.2}$ & $1.54\substack{+0.46\\-0.32}$ & 3 E   & 3.1   & 10.6  & 0.40  & 1.39-2.43 & 1.10-2.56 \\
HIP 41378 & Tr    & $701.2\substack{+78.3\\-67.8}$ & $1.63\substack{+0.12\\-0.11}$ & 11 E  & 5.5   & 30.2  & 0.38  & 1.44-2.51 & 1.14-2.65 \\
Kepler-603 & Tr    & $527.1\substack{+1408.5\\-356.9}$ & $1.28\substack{+1.77\\-0.68}$ & 3 E   & 4.0   & 16.8  & 0.36  & 0.97-1.71 & 0.76-1.80 \\
HD 136352 & RV    & $302.2\substack{+462.8\\-188.2}$ & $0.82\substack{+0.71\\-0.39}$ & 3 E   & 5.4   & 28.6  & -     & 0.95-1.69 & 0.75-1.78 \\
\hline
\end{tabularx}
\begin{tablenotes}
\small
\item$^{a}$Discovery method of the system: `Tr' and `RV' represent transit and radial velocity, respectively.\item$^{b}$Orbital numbers (ON) followed by `E' indicate the extrapolated planets, and followed by `C' indicate that the corresponding orbital periods have been flagged as ``Planetary Candidate'' in the NASA Exoplanet Archive.
\end{tablenotes}
\end{table}

\citet{2011ApJ...736...19B} classified exoplanets into the following class sizes: Earth-size ($R_{p}<1.25R_{\oplus}$), super-Earth-size ($1.25R_{\oplus}\leq R_{p}<2R_{\oplus}$), Neptune-size ($2R_{\oplus}\leq R_{p}<6R_{\oplus}$) and Jupiter-size ($6R_{\oplus}\leq R_{p}<15R_{\oplus}$). Following this classification, five of our predicted exoplanets within HZ have maximum radii within the Earth-size range, 11 super-Earth-size, 27 Neptune-size, three Jupiter-size, and one with a maximum radius larger than twice that of Jupiter's. Using the proposed categories based on planet mass by \citet{2013PASP..125..933S}, our five predicted planets have maximum masses within the range of Earth ($0.1M_{\oplus}-2M_{\oplus}$). In addition, 22 and 19 predicted planets have maximum masses within the range of super-Earth ($2M_{\oplus}-10M_{\oplus}$) and Neptune ($10M_{\oplus}-100_{\oplus}$), respectively. As a result, among our 47 predicted exoplanets within HZ, there are only five exoplanets whose estimated maximum mass and radius are within the mass and radius range of Earth: the fourth and fifth planets of Kepler-186, the second planets of GJ 3138 and Wolf 1061, and the extrapolated planet of YZ Cet.

\subsubsection{Dynamically-stable planets}
We use the dynamical spacing criterion ($\Delta$) to investigate our predicted objects' stability at HZ. We calculate the $\Delta$ values for all adjacent planet pairs in these 45 systems, which host 47 predicted exoplanets within their HZ. We find that when inserting our predicted planets into systems, the average percentage of pairs with $\Delta\leq10$ increases from $\sim25\%$ to $\sim38\%$. Nevertheless, this alone cannot be a reason for instability. In our solar system, there are pairs of objects whose dynamical spacing values are small ($\Delta\leq10$); however, they are stable in their positions. Neptune and Pluto are one of those pairs; the dynamical spacing value for them is $\Delta\sim7.4$. We know that Neptune and Pluto are stable because they are in a 3:2 orbital resonance with each other. The same could be true for exoplanetary systems as found by \citet{2012ApJ...756L..11L}, \citet{2013AJ....145....1B}, and \citet{2014ApJ...790..146F}. We calculate the ratio of the orbital periods of planet pairs in these 45 systems and find that the number of orbital resonances increases when our predicted planets are considered.

\begin{figure}
\centering
\captionsetup{width=0.60\paperwidth}
\includegraphics[width=0.60\paperwidth]{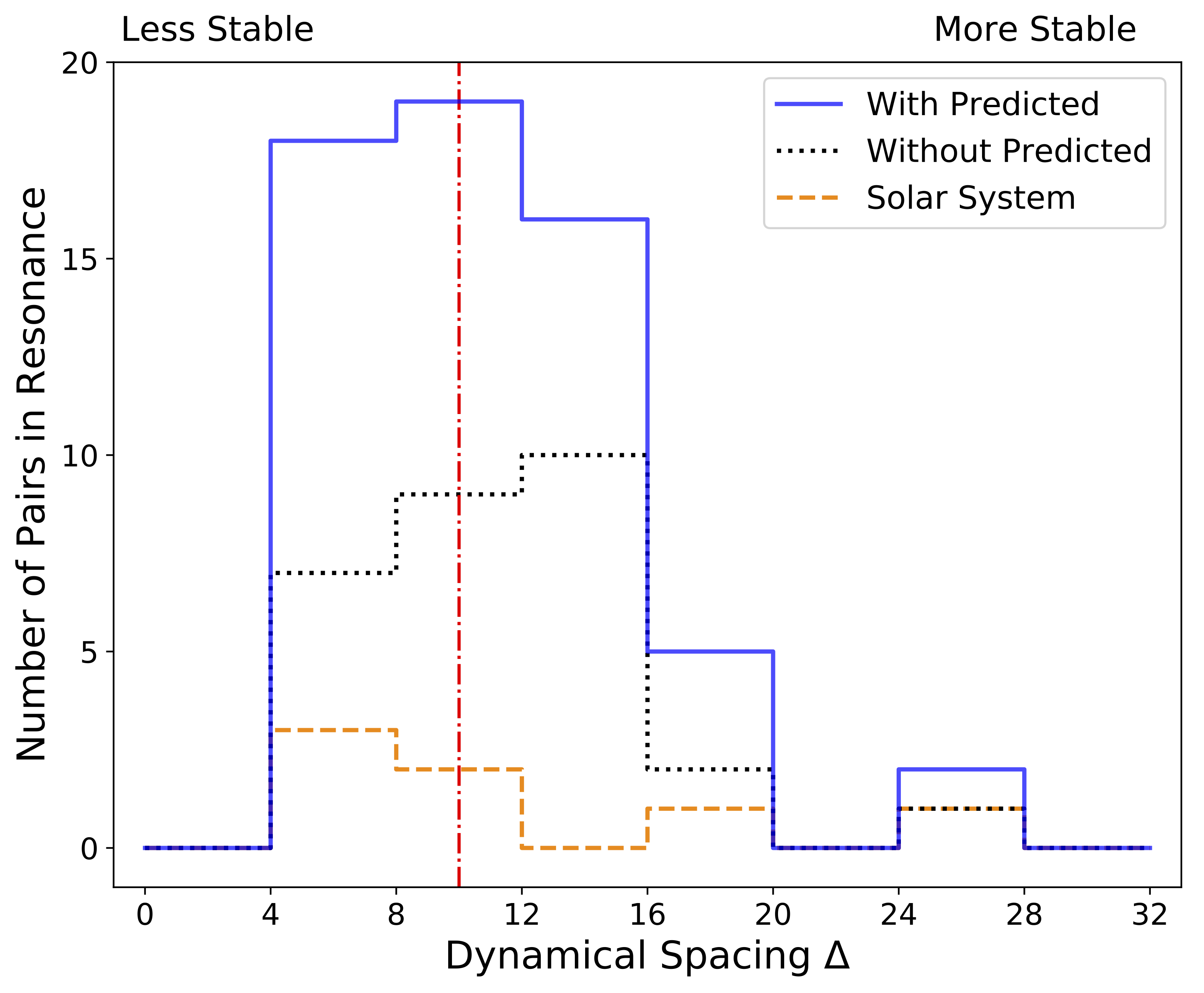}
\caption[$\Delta$ and the total number of adjacent exoplanet pairs that are in orbital resonance with each other]{Dynamical spacing $\Delta$ and the total number of adjacent exoplanet pairs that are in orbital resonance with each other. The solid blue line shows the number of resonance pairs considering our predicted planets, and the dotted black line shows the number before inserting any predicted planets into systems. The values for the solar system are also shown for reference via the orange dashed line. The vertical dash-dotted red line corresponds to $\Delta=10$ and separates the less and more stable adjacent planet pairs regimes.}
\label{figure3.4}
\end{figure}

Figure~\ref{figure3.4} shows the number of resonance pairs with and without considering our predicted planets. As shown in this figure, the $\Delta\leq10$ pairs are dominated by the pairs in resonance; hence, as a sample, the planets (both detected and predicted) of these 45 systems are more likely to be stable in their positions.

\begin{table}[!ph]
\centering
\setlength{\tabcolsep}{9.0pt}
\captionsetup{width=\textwidth}
\caption[Systems with detected planets since predictions made by \citetalias{2015MNRAS.448.3608B}]{Systems with detected planets since predictions made by \citetalias{2015MNRAS.448.3608B}.}
\begin{tabularx}{\textwidth}{lccccc}
\hline
Host name & \makecell{Detected\\period (d)} & \makecell{\citetalias{2015MNRAS.448.3608B} predicted\\period (d)} & \makecell{Our predicted\\period (d)} & \makecell{\citetalias{2015MNRAS.448.3608B}\\error (\%)} & \makecell{Our\\error (\%)} \\
\hline
Kepler-1388 & 75.73 & $73.0\pm8.0$ & $74.2\substack{+14.6\\-14.0}$ & 3.6   & 2.0 \\
Kepler-150 & 637.21 & N/A   & N/A   & N/A   & N/A \\
Kepler-1542 & 7.23  & $6.9\pm0.3$ & $6.9\substack{+0.2\\-0.2}$ & 4.6   & 4.6 \\
Kepler-20 & 34.94 & $39.1\pm5.4$ & $39.0\substack{+5.9\\-5.3}$ & 11.9  & 11.6 \\
Kepler-80 & 14.65 & $14.5\pm1.3$ & $14.6\substack{+0.8\\-0.7}$ & 1.0   & 0.3 \\
Kepler-82 & 75.73 & N/A   & N/A   & N/A   & N/A \\
KOI-351 & 14.45 & $15.4\pm1.7$ & $14.3\substack{+1.0\\-0.9}$ & 6.6   & 1.0 \\
\hline
\end{tabularx}
\label{table3.3}
\end{table}

\subsection{The reliability of planet predictions}
To examine the reliability of the predictions made by the TB relation, we look for some planetary systems in the literature that have had new exoplanets detected recently, notably, those systems that have been predicted to have additional planets by \citetalias{2015MNRAS.448.3608B}. We find the following seven systems with new planet detection: Kepler-1388, Kepler-150, Kepler-1542, Kepler-20, Kepler-80, Kepler-82, and KOI-351.

The detected planets in Kepler-150 and Kepler-82 were not predicted by \citetalias{2015MNRAS.448.3608B} \citep{2017AJ....153..180S,2019A&A...628A.108F}, while the rest of the five planets have been detected with orbital periods that agree with their predicted periods. For Kepler-1388 and Kepler-1542, two planetary candidates with orbital periods of 75.73 and 7.23 days have been detected, the detected periods of which are consistent with their predicted orbital periods, $73.0\pm8.0$ and $6.9\pm0.3$ days, by \citetalias{2015MNRAS.448.3608B} and also with our predictions $74.2\pm14.3$ and $6.9\pm0.2$ days.

\citetalias{2015MNRAS.448.3608B} could also predict an additional planet in Kepler-20 with an orbital period of $39.1\pm5.4$ days. This prediction was also confirmed by \citet{2016AJ....152..160B}. \citet{2018AJ....155...94S} used deep learning algorithms to classify potential planet signals, where they managed to validate two more new planets in Kepler-80 and KOI-351, which were previously predicted by \citetalias{2015MNRAS.448.3608B}.

Table~\ref{table3.3} summarizes these results and presents the predicted and detected periods, which are reasonably consistent, especially for Kepler-1542 and KOI-351 where \citetalias{2015MNRAS.448.3608B} highlighted them as predictions with a high geometric probability to transit. This demonstrates the potential and capability of such predictions to help search for new planets, using more precise observational data and new detection techniques such as ML. However, having seven systems is not statistically significant enough to establish either the reliability or unreliability of the TB relation and we require more follow-up observations. We must wait for ongoing exoplanet surveys such as TESS to be completed.

\subsection{Comparison between predictions by BL15 and this study}
This study uses the TB relation to predict the probability of finding additional planets in multiplanet systems with at least three member planets. This is similar to the method used by \citetalias{2015MNRAS.448.3608B}. In principle, we update the \citetalias{2015MNRAS.448.3608B} study and utilize the Markov Chain Monte Carlo (MCMC) simulation. The MCMC method is used to analyze systems and quantify the uncertainties of the best-fit parameters \citep{emcee3}. The main feature of our method is inserting thousands of hypothetical planets with random orbital periods into each of the systems and using the highest $\gamma$ value to achieve the most accurate predictions.

To compare our method and that of \citetalias{2015MNRAS.448.3608B}, we remove Ceres and the planet Uranus (two objects predicted by the TB law to exist) from the solar system and insert thousands of random planets into the solar system, covering all possible locations and combinations between the two adjacent planets, to achieve the minimum value of $\chi^2/dof$. This leads us to recover the actual solar system with the highest $\gamma$ value, where the predicted orbital periods' calculated errors are 20\% for Ceres and 23\% for Uranus. On the other hand, the \citetalias{2015MNRAS.448.3608B} method results in an error value of 23\% for both Ceres and Uranus. We remove the planets Earth and Mars, which are within the HZ range of the Sun, and apply the TB relation to the system to predict their orbital periods. This recovers the combination of the actual solar system where the calculated errors of orbital periods are 18\% for Earth and 41\% for Mars. We perform this process once again using the \citetalias{2015MNRAS.448.3608B} method, which gives us the error values of 18.5\% for Earth and 42\% for Mars. For a better comparison, we repeat this process for other planets in the solar system and exoplanetary systems, and with different combinations of planet removal.

\begin{figure}[t!]
\centering
 \includegraphics[width=0.60\paperwidth]{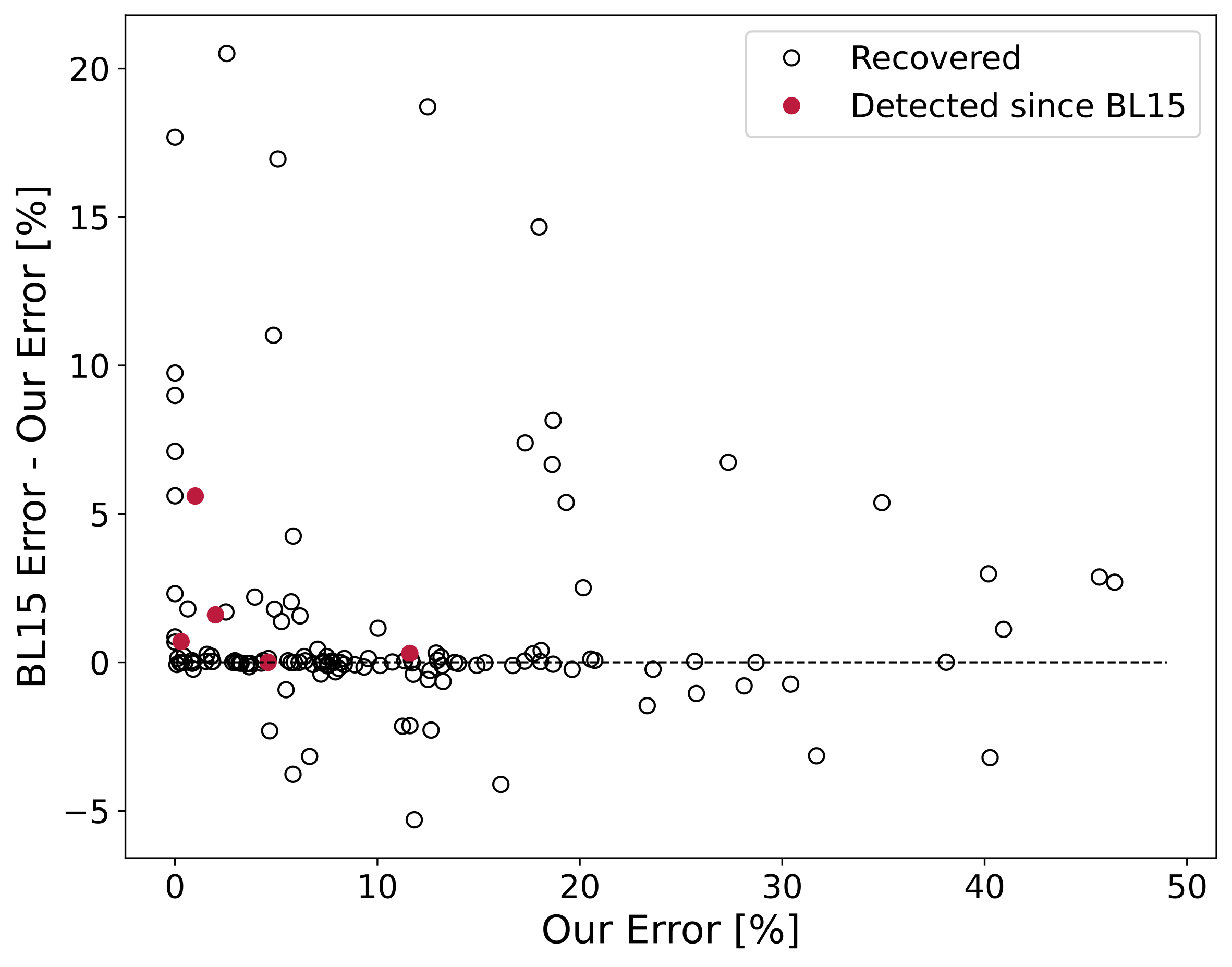}
 \captionsetup{width=0.60\paperwidth}
 \caption[Comparison of uncertainty on the predicted orbital periods calculated by \citetalias{2015MNRAS.448.3608B} and this study]{Comparison of uncertainty on the predicted orbital periods calculated by \citetalias{2015MNRAS.448.3608B} and this study. We remove planets from systems in various combinations and apply the TB relation to recover the orbital periods of removed planets. Each empty circle belongs to a specific combination of removed planets from systems, and five filled circles represent those detected planets after the predictions made by \citetalias{2015MNRAS.448.3608B} (see table~\ref{table3.3}).}
 \label{figure3.5}
\end{figure}

We also reapply our method to Kepler-1388, Kepler-1542, Kepler-20, Kepler-80, and KOI-351, regardless of the successfully prediction of planets by \citetalias{2015MNRAS.448.3608B}, to compare the validity of the methods used in predicting the additional planets. As shown in table~\ref{table3.3}, our predictions have fewer errors, and sometimes the predicted periods have smaller error bars. These fewer error bars can be interpreted as advantages of using the MCMC method in predicting exoplanets based on the TB relation. Using the paired samples t-test, the p-value is estimated to equal 0.025, which is less than 0.05 and statistically significant. Therefore, we conclude that, on average, there is evidence that applying the MCMC method does lead to more precise predictions than \citetalias{2015MNRAS.448.3608B}'s method. The difference between the calculated errors with our method and \citetalias{2015MNRAS.448.3608B}'s is illustrated in figure~\ref{figure3.5}, where each empty circle belongs to a planet recovery. Five filled circles represent those detected planets after the predictions made by \citetalias{2015MNRAS.448.3608B}.

\section{The revised insight into exoplanet populations}
\subsection{Feature importance}
Initially, we use the LOF algorithm to identify outliers in the dataset containing 770 planets (sample II). The LOF algorithm marks 76 data points as outliers, resulting in a study dataset of 694 planets. Appendix~\ref{appendixB.1} describes finding outliers and their effect on prediction accuracy. In general, we find that all regression models perform poorly considering the outliers in the dataset. Additionally, the effect of data re-scaling on planetary radius predictions is discussed in appendix~\ref{appendixB.2}. As a result, regression algorithms provide better results on a logarithmic scale.

It has been known that having inefficient and unnecessary features can cause declination in the performance of an ML model and lead to inaccurate results \citep{chandrashekar2014survey}. We apply five FS methods to find and eliminate the least important planetary, stellar, and orbital parameters in predicting planet radii. They are the Spearman’s rank correlation, Backward Elimination, Forward Selection, CART, and XGBoost. Features are orbital period ($P$), eccentricity ($e$), planetary mass ($M_{p}$), stellar mass ($M_{s}$), stellar radius ($R_{s}$), metallicity (Fe/H), and effective temperature ($T_{\text{eff}}$), and the target variable is planetary radius ($R_{p}$).

Spearman’s rank correlation ($r_{s}$) is a number between –1 and 1 that measures the monotonic correlation between two variables. This filter method reveals parameters that are strongly correlated with planetary radius: $M_{p}$ with a coefficient of 0.780 has the highest correlation, followed by $M_{s}$ with $r_{s}=0.590$. Furthermore, $T_{\text{eff}}$ with a coefficient of 0.568 and $R_{s}$ with a coefficient of 0.552 are in third and fourth place, respectively. Orbital period with $r_{s}=-0.389$ and eccentricity with $r_{s}=-0.151$ are two features that have a negative correlation with $R_{p}$. In addition, highly correlated stellar parameters ($R_{s}$, $M_{s}$, and $T_{\text{eff}}$) are also indicated by coefficients greater than 0.820. We should note that the estimated p-values are less than 0.001, which indicates strong certainty in the results. As an exception, the p-value corresponding to the coefficient between Fe/H and $R_{p}$ ($r_{s}=-0.037$) is greater than 0.1, which is statistically uncertain. To calculate uncertainties in the value of correlation coefficients, we apply the Monte Carlo error analysis, considering errors in each measurement \citep{2015ascl.soft04008C}.

\begin{figure}[b!]
\centering
\includegraphics[width=\textwidth]{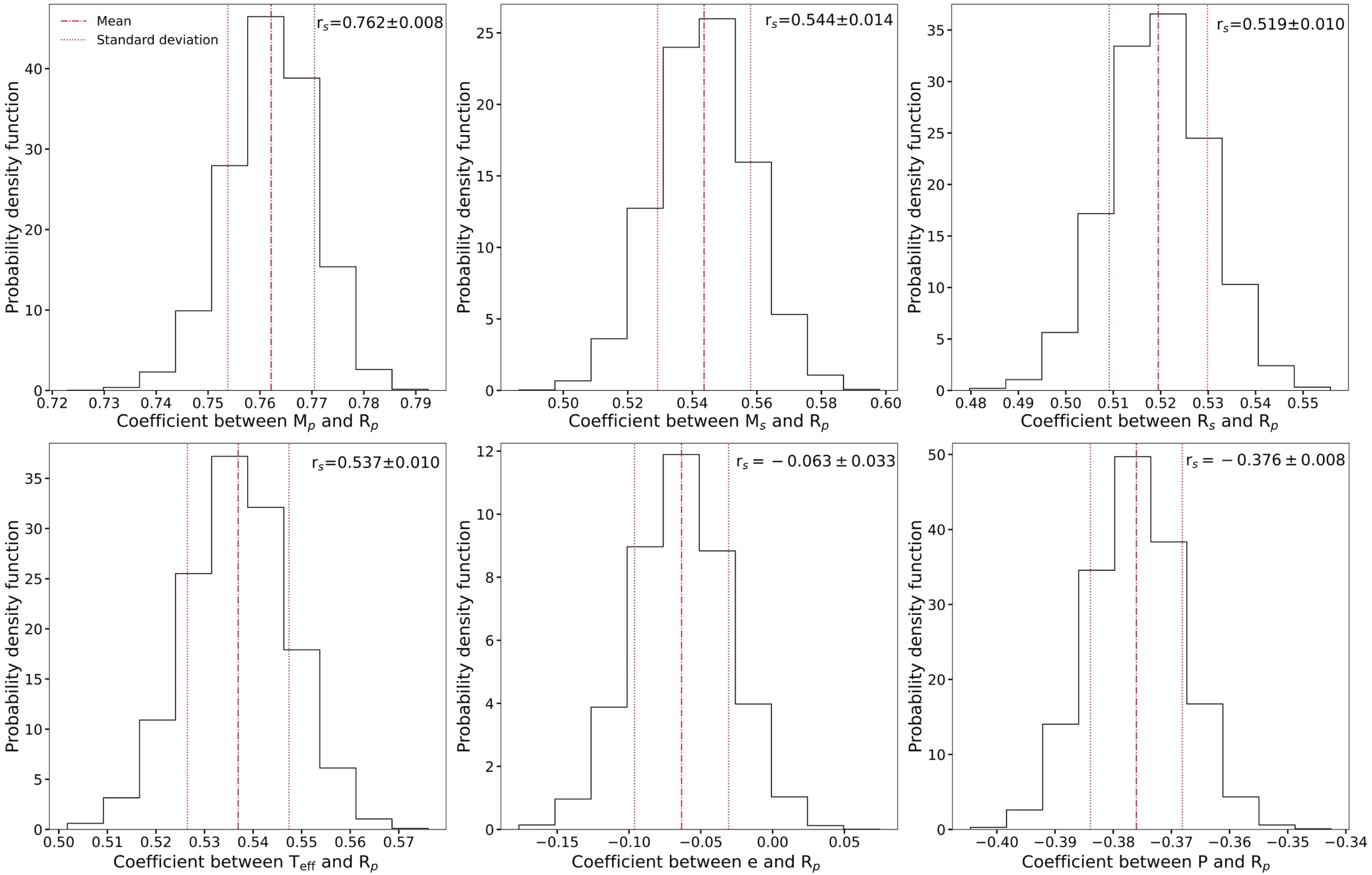}
\captionsetup{width=\textwidth}
    \caption[Distribution of correlation coefficient between planetary radius and other physical parameters]{Distribution of correlation coefficient ($r_{s}$) between planetary radius ($R_{p}$) and other physical parameters obtained by the Monte Carlo analysis. Parameters are planetary mass ($M_{p}$), stellar mass ($M_{s}$), radius ($R_{s}$), and effective temperature ($T_{\text{eff}}$), and orbital eccentricity ($e$) and period ($P$). Stellar metallicity (Fe/H), which displays a p-value greater than 0.1, has been excluded. The red dash-dotted line is the mean, and two red dotted lines represent uncertainties around it. The mean and uncertainty values are presented in the upper right corner of each panel. The absolute value of the coefficients determines the strength of the relationship. The larger the number, the stronger the relationship. It marks $M_{p}$ as the most and $e$ as the least relevant parameters to the planetary radius.}
    \label{figure3.6}
\end{figure}

Figure~\ref{figure3.6} illustrates the distribution of $r_{s}$ for each feature except Fe/H, which has a p-value greater than 0.1. By taking the standard deviation of distributions, we report the mean and uncertainty values in the upper right corner of each panel. As it is seen, the distributions of Spearman correlation coefficients for $M_{p}$, $M_{s}$, $R_{s}$, and $T_{\text{eff}}$ are constrained to positive values. In contrast, the distribution for $P$ is limited to negative values. In the case of $e$, even though the mean of the distribution is slightly negative, it is exceptionally wide, taking both negative and positive values. This more extensive distribution results from higher amounts of error in eccentricity measurement.

\begin{figure}[t!]
\centering
\includegraphics[width=0.60\paperwidth]{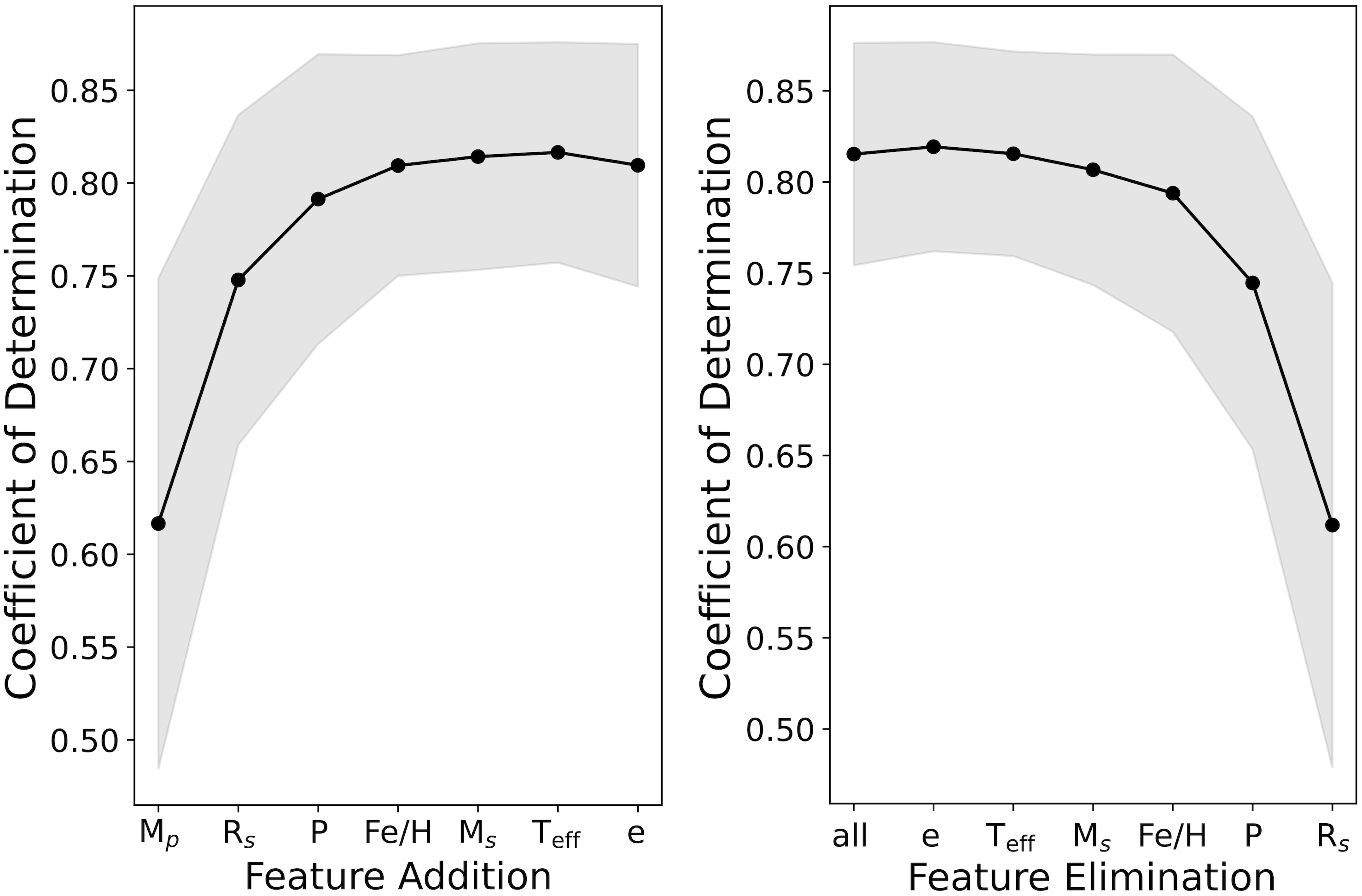}
\captionsetup{width=0.60\paperwidth}
    \caption[Wrapper FS methods]{Wrapper FS methods. Left: coefficients of determination ($\rho^{2}$) against feature sets for the Forward Selection technique. In the first step, it determines $M_{p}$ as the best feature, and in each following iteration, the best remaining feature is added to the set. Right: $\rho^{2}$ values against feature sets for the Backward Elimination technique. Unlike Forward Selection, it starts with all features and removes the worst one at each step. Features are planetary mass ($M_{p}$) and radius ($R_{p}$), orbital period ($P$) and eccentricity ($e$), and the stellar mass ($M_{s}$), radius ($R_{s}$), metallicity (Fe/H), and effective temperature ($T_{\text{eff}}$). Gray areas demonstrate standard errors. Both methods highlight $M_{p}$, $P$, and $R_{s}$ as three important parameters in predicting the planetary radius.}
    \label{figure3.7}
\end{figure}

Forward Selection and Backward Elimination are two wrapper FS methods. The procedure for Forward Selection starts with an empty set of features. Then, the best feature is determined and added to the set by applying a Random Forest regressor. In each subsequent iteration, the best remaining feature is determined and added until a complete set of features is reached. In contrast, Backward Elimination starts with a complete set of features and, at each step, eliminates the worst feature remaining in the set. Using a 10-fold cross-validation method, $\rho^{2}$ values, and corresponding standard errors are calculated for each step and shown in figure~\ref{figure3.7}, where the left-hand panel is Forward Selection, and the right-hand panel is Backward Elimination. Both methods highlight $M_{p}$, $P$, and $R_{s}$ as three important parameters.

Applied embedded methods include CART, which uses a decision tree regressor, and XGBoost, which implements a gradient boosting trees algorithm. These techniques score features based on their importance in computing the target variable. The ranking (and scores) obtained by the CART method are as follows: $M_{p}$ (0.905), $P$ (0.031), $R_{s}$ (0.026), $M_{s}$ (0.012), $T_{\text{eff}}$ (0.011), $e$ (0.008), and Fe/H (0.006). The XGBoost method ranks (and scores) features as follows: $M_{p}$ (0.851), $R_{s}$ (0.038), $P$ (0.029), $M_{s}$ (0.025), $T_{\text{eff}}$ (0.024), Fe/H (0.021), and $e$ (0.012). Similar importance is assigned to all features except planetary mass by CART and XGBoost.

Finally, we conclude that a set of features including $M_{p}$, $P$, and one of the stellar parameters ($M_{s}$, $R_{s}$, or $T_{\text{eff}}$) works well. Therefore, we select planetary mass, orbital period, and stellar mass as the main features.

\subsection{Clusters}
FS methods highlight planetary mass, stellar mass, and orbital period as vital features to estimate planet radii. We use ML clustering algorithms to investigate potential groups of exoplanets in sample II, which is a four-dimensional logarithmic space consisting of planetary mass and radius, stellar mass, and orbital period. The algorithms are Affinity Propagation, BIRCH, DBSCAN, Gaussian Mixture Model, Hierarchical Clustering, K-Means, Mean Shift, Mini-Batch K-Means, OPTICS, and Spectral Clustering.

\subsubsection{Number of clusters}\label{number}
When the number of clusters ($K$) is not known in advance, as in our case, hierarchical clustering is an appropriate technique to adopt \citep{landau2011cluster}. Using the Hierarchical Clustering algorithm as a distance-based method, one can have an assumption of $K$. Its agglomerative algorithm assigns each data point to an individual partition; then, at each iteration, the closest pair of partitions are merged until all data belong to a single partition.

\begin{figure}[t!]
\centering
	\includegraphics[width=0.60\paperwidth]{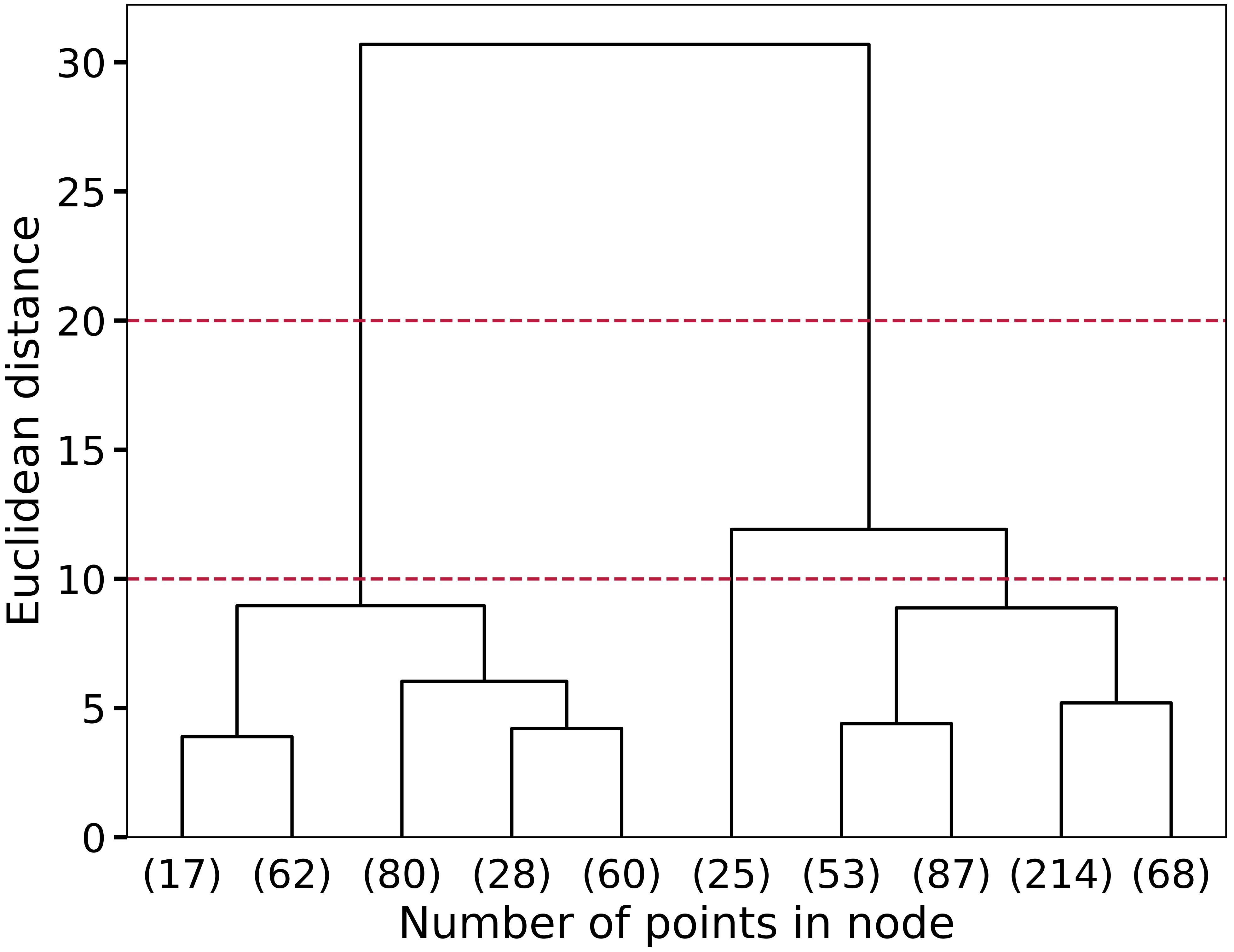}
 \captionsetup{width=0.60\paperwidth}
    \caption[Hierarchical Clustering dendrogram]{Hierarchical Clustering dendrogram. The greater the height of the vertical lines, the greater the distance between the clusters. Two red dashed lines are distance thresholds. The number of vertical lines intersecting the threshold line indicates the number of clusters. The larger threshold results in two clusters, while the smaller threshold results in three clusters. Due to illustration purposes, lower sequences have not been shown.}
    \label{figure3.8}
\end{figure}

Figure~\ref{figure3.8} shows the Hierarchical Clustering dendrogram, which records the sequences of merges. The greater the height of the vertical lines in the dendrogram, the greater the distance between the clusters. Clusters can be defined by trimming a dendrogram with a distance threshold. However, there is no universal method for setting thresholds. In general, a distance threshold on the dendrogram is set to intersect the longest vertical line. The number of vertical lines intersecting the threshold line indicates the $K$ value. Two distance thresholds are set to the dendrogram of our dataset (red dashed lines in figure~\ref{figure3.8}). The larger threshold results in two clusters with an Euclidean distance of 30.6, while the smaller threshold splits the larger cluster into two parts with a distance of 11.9 between them.

\begin{figure}
\centering
	\includegraphics[width=0.60\paperwidth]{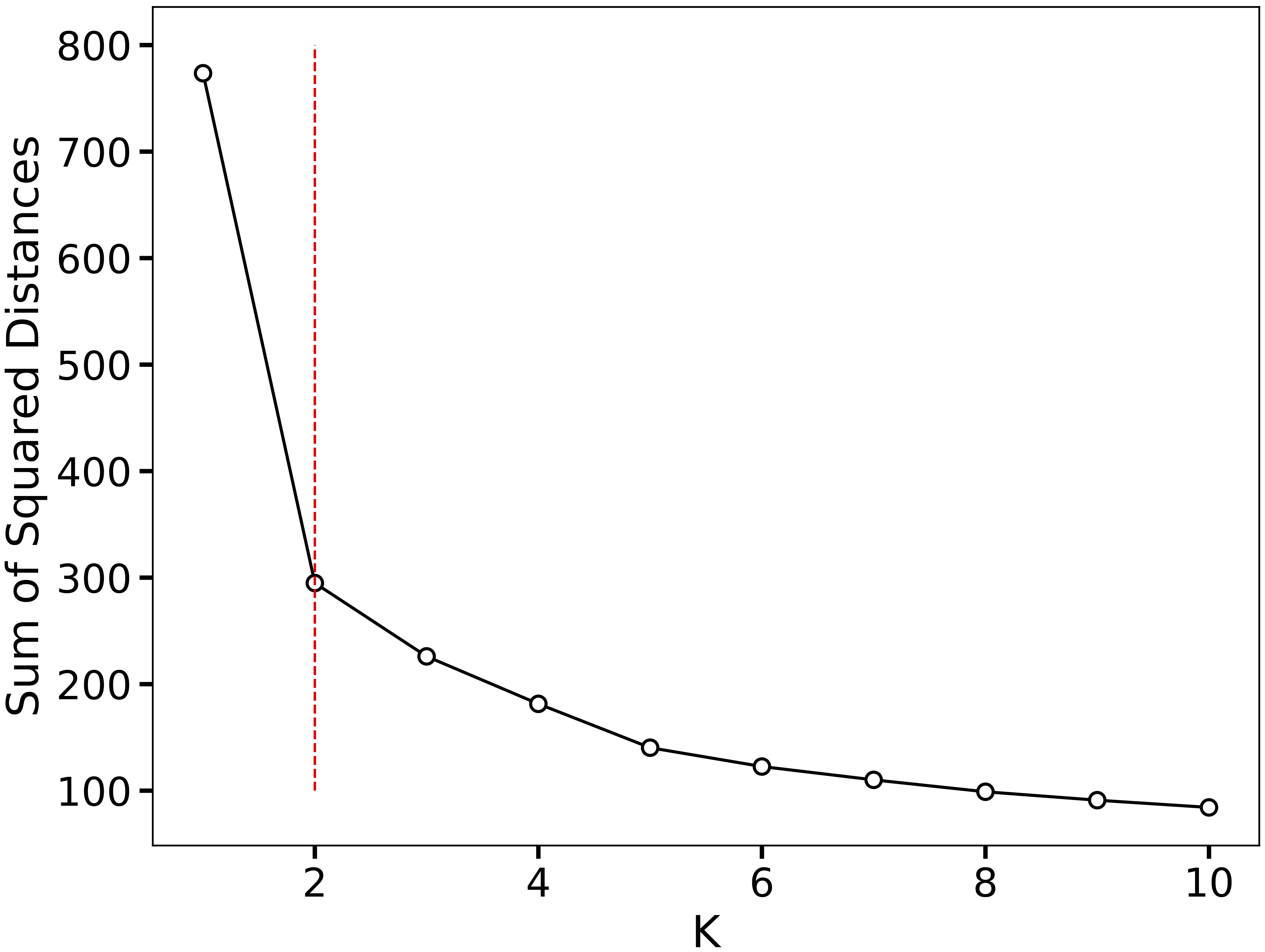}
 \captionsetup{width=0.60\paperwidth}
    \caption[SSD between data points and their assigned cluster centroids against $K$]{Sum of squared distances (SSD) between data points and their assigned cluster centroids against the number of clusters ($K$), calculated by the Elbow method. The red vertical dashed line corresponds to $K$=2, where the curve starts to flatten out. This point is chosen as the optimal number of clusters.}
    \label{figure3.9}
\end{figure}

For algorithms that do not learn the $K$ from data, we use Elbow and Silhouette methods to find the optimal value of $K$ and use it as an input parameter in clustering algorithms. The Elbow method performs the K-Means clustering algorithm for different values of $K$. Then, it calculates the sum of squared distances (SSD) between data points and their cluster centroids each time. Figure~\ref{figure3.9} demonstrates SSD values as a function of $K$. As shown, the curve starts to flatten out and form an elbow shape in $K$=2, which is chosen as the optimal number of clusters.

\begin{figure}[b!]
\centering
\captionsetup{width=0.60\paperwidth}
\includegraphics[width=0.60\paperwidth]{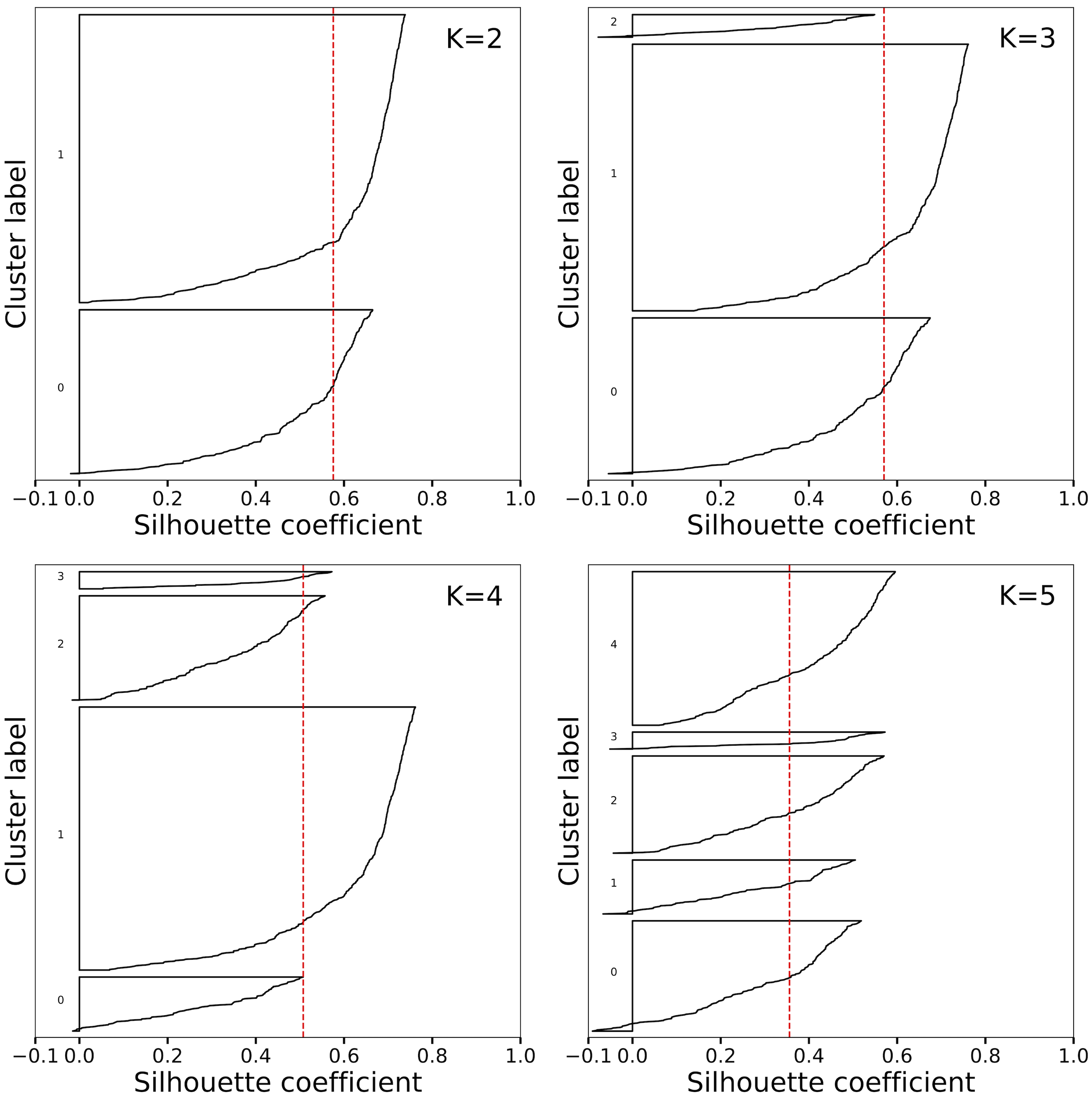}
    \caption[Silhouette plots for different numbers of clusters]{Silhouette plots for different numbers of clusters ($K$). The thickness of the plots indicates the cluster size and red vertical dashed lines represent the corresponding average Silhouette coefficients. If the sample is far from neighboring clusters, the coefficient becomes close to +1. The coefficient can be negative if the sample is assigned to the wrong cluster. Providing that the sample is on or near the decision boundary between two neighboring clusters, the coefficient becomes 0. Due to clusters with lower-than-average Silhouette scores and wide fluctuations in the size of plots, $K$=3, 4, and 5 are not appropriate. Scatter plots for each Silhouette plot are shown in figure\ref{figure3.11}.}
    \label{figure3.10}
\end{figure}

\begin{figure}
\centering
\captionsetup{width=\textwidth}
\includegraphics[width=\textwidth]{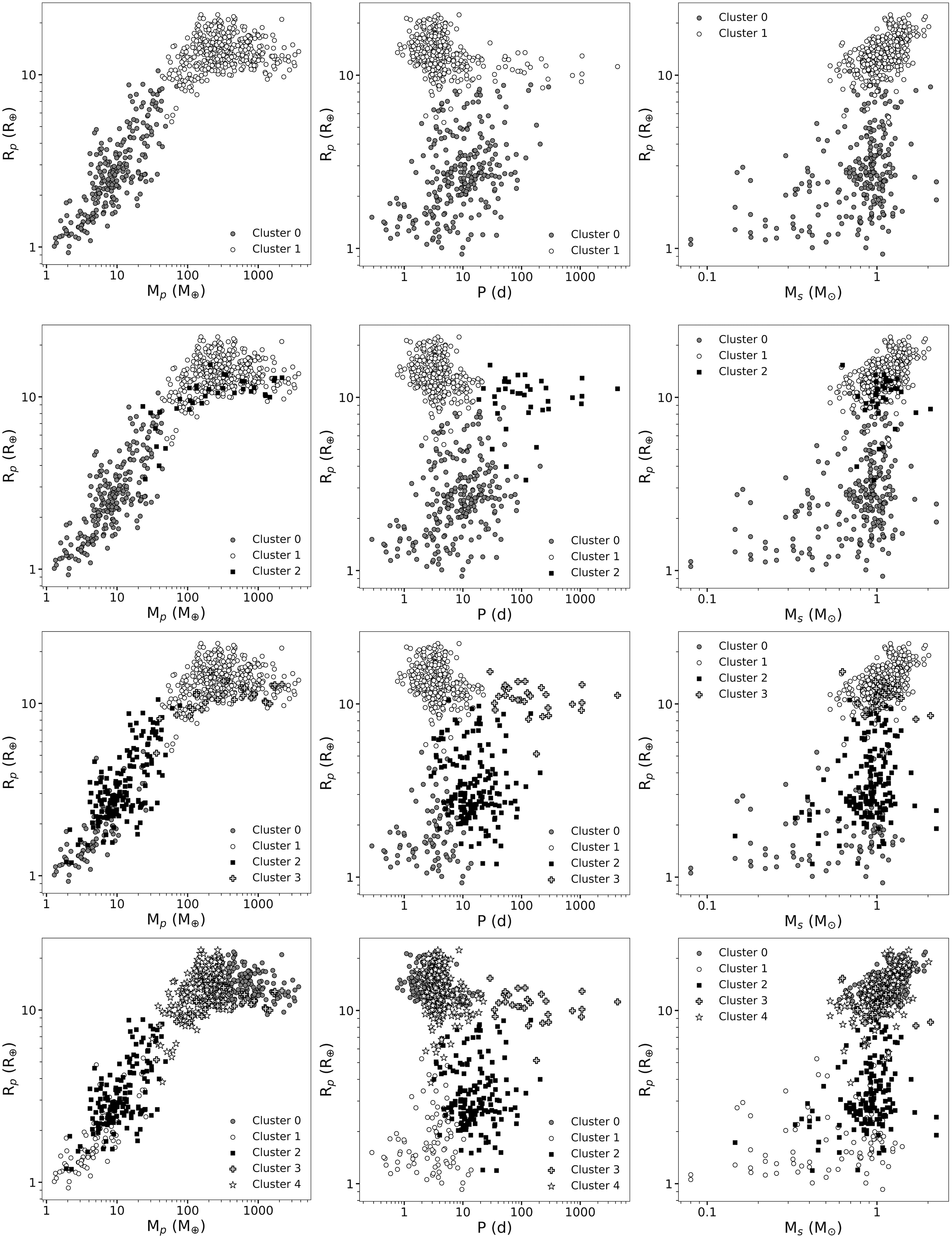}
    \caption[Matrix of scatter plots corresponding to figure~\ref{figure3.10}]{Matrix of scatter plots corresponding to figure~\ref{figure3.10}. Rows 1, 2, 3, and 4 represent the number of clusters ($K$) equal to two, three, four, and five, respectively. Columns 1, 2, and 3 illustrate the distribution of planetary radius ($R_{p}$) versus the planet’s mass ($M_{p}$), orbital period ($P$), and star’s mass ($M_{s}$), respectively. Gray-filled circles, white circles, black-filled squares, light gray-filled pluses, and white stars represent members of clusters 0, 1, 2, 3, and 4, respectively.}
    \label{figure3.11}
\end{figure}

Furthermore, the average Silhouette width can evaluate clustering reliability and may be used to estimate $K$ value \citep{rousseeuw1987silhouettes}. The Silhouette method computes a coefficient for different values of $K$ and uses it to determine the degree of separation between clusters. The coefficient becomes negative if the sample is assigned to the wrong cluster. Provided that the sample is far from neighboring clusters, the coefficient will be close to +1. If the sample is on or near the decision boundary between two neighboring clusters, the coefficient becomes 0. Figure~\ref{figure3.10} presents the Silhouette plots for $K$=2, 3, 4, and 5. In this figure, the thickness of the plots represents the cluster size, and each red vertical dashed line belongs to an average Silhouette score. $K$=3, 4, and 5 are not appropriate due to clusters with lower-than-average Silhouette scores and wide fluctuations in the size of plots. Like the Elbow method, this method proposes $K$=2 as the proper number of clusters.

Figure~\ref{figure3.11} illustrates a matrix of scatter plots for Silhouette plots, where each row has been assigned to a specific $K$, and columns 1, 2, and 3 show the distribution of planetary radius versus planet’s mass, orbital period, and star’s mass, respectively. Like exclusive clustering algorithms where each data point belongs exclusively to one cluster \citep{1988acd..book.....J}, we aim to group planets into distinct non-overlapping clusters. On the one hand, if $K=2$ (see the first row in figure~\ref{figure3.11}), two clusters are almost well-separated in all three spaces, with only a few data points overlapping. On the other hand, when planets are divided into more than two clusters (see the second, third, and fourth rows in figure~\ref{figure3.11}), the members of the clusters become less distinguishable from each other.

For $K=3$, planets with a longer orbital period are defined as a new cluster (black-filled squares). Although three clusters are separated in the $R_{p}$-$P$ space, in the $R_{p}$-$M_{p}$ and $R_{p}$-$M_{s}$ spaces, most members of cluster 2 are distributed over the other two clusters, especially cluster 1. Likewise, in the case of $K=4$, the cluster separation is better in the $R_{p}$-$P$ space than in the other two spaces, where clusters overlap.

For $K=5$, although clusters 0 and 4 are well distinguished in the planet’s mass-radius distribution, they overlap a lot in the $R_{p}$-$P$ and $R_{p}$-$M_{s}$ spaces. In the $R_{p}$-$P$ space, members of clusters 0, 1, 2, and 3 are almost separated; nevertheless, members of cluster 4 extremely overlap with those of cluster 0. In addition, cluster 3 members overlap with members of clusters 0 and 4 in the $R_{p}$-$M_{p}$ and $R_{p}$-$M_{s}$ spaces.

Consequently, we choose $K=2$ based on the results from the Hierarchical, Elbow, and Silhouette methods and the idea that planetary clusters are separate groups that do not overlap. Similarly, the Affinity Propagation and Mean Shift algorithms, which do not need to specify the number of clusters, give two clusters. Another point that should be taken into account is that this number of clusters chosen is consistent with published works (e.g., \citet{2013ApJ...768...14W} and \citet{2017A&A...604A..83B}), where two regimes were introduced in the planetary mass-radius relation by different techniques.
\newpage
As for DBSCAN and OPTICS algorithms, they give $K$=2 but cannot separate clusters well; thus, we exclude them from the analysis. The effectiveness of clustering methods depends on several factors, including the dataset's characteristics and underlying distribution. Different clustering algorithms make certain assumptions about the data structure and employ distinct approaches to identify clusters. Accordingly, their performance can vary based on how well these assumptions align with the dataset's properties. In cases where clustering algorithms fail to find appropriate clusters, there may be several potential explanations. One important factor is the distribution of the exoplanet data, which might not conform to the assumptions made by certain clustering algorithms. DBSCAN and OPTICS are two density-based methods that execute clustering by finding areas where data points are concentrated. They can discover arbitrarily shaped clusters, including non-spherical ones; they, however, might fail when the dataset is too sparse, and the density varies across the data, like in the case of exoplanet data \citep{moreira2005density,ahmad2015performance}.

\subsubsection{Planetary classes} 
The planets are divided into two groups by choosing $K=2$ for BIRCH, Gaussian Mixture Model, Hierarchical Clustering, K-Means, Mini-Batch K-Means, and Spectral Clustering. Moreover, the Affinity Propagation and Mean Shift algorithms, which learn the number of clusters from data, result in two clusters. To introduce a boundary between two groups in the planet’s mass and radius spaces, we construct a one-dimensional Gaussian kernel density estimation for each cluster and find the intersection point. Table~\ref{table3.4} lists the results of the clustering algorithms, which are almost similar (except for DBSCAN and OPTICS, as discussed in section~\ref{number}).

\begin{figure}[t!]
\centering
\captionsetup{width=0.60\paperwidth}
\includegraphics[width=0.60\paperwidth]{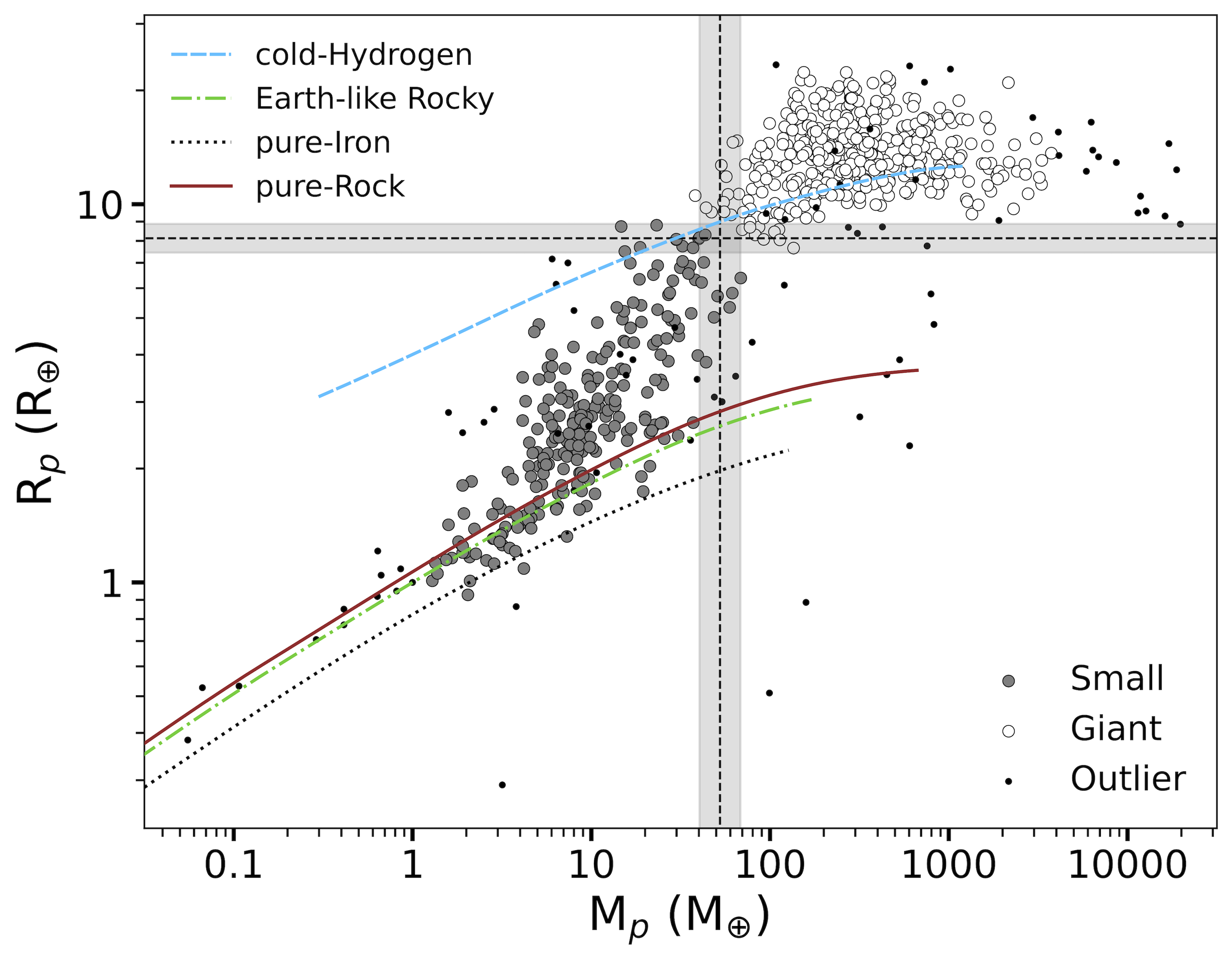}
\caption[The mass-radius distribution of clustered data]{The mass-radius distribution of clustered data. Data are separated into two classes using $R_{p}=8.13R_{\oplus}$ (horizontal dashed line) and $M_{p}=52.48M_{\oplus}$ (vertical dashed line). Gray areas demonstrate mean errors of mass and radius. Exoplanets with $R_{p}\leq8.13R_{\oplus}$ and $M_{p}\leq52.48M_{\oplus}$ are defined as small planets (gray circles), and those with $R_{p}>8.13R_{\oplus}$ and $M_{p}>52.48M_{\oplus}$ as giant planets (white circles). There are 254 small planets and 440 giant planets. Black dots are outlier planets found by the LOF method (see appendix~\ref{appendixB.1}). Four iso-density curves are also drawn: cold-hydrogen (blue dashed line), Earth-like rocky (green dash-dotted line), pure-iron (black dotted line), and pure rocky (solid crimson line) planets \citep{2010ApJ...712L..73M,2014ApJS..215...21B}.}
\label{figure3.12}
\end{figure}

Ultimately, we separate data into two classes using an average value of $\log R_{p}=0.91$ ($R_{p}=8.13R_{\oplus}$) for radius breakpoint ($B_{\text{Radius}}$) and $\log M_{p}=1.72$ ($M_{p}=52.48M_{\oplus}$) for mass breakpoint ($B_{\text{Mass}}$). Exoplanets with $R_{p}\leq8.13R_{\oplus}$ and $M_{p}\leq52.48M_{\oplus}$ are defined as small planets, and those with $R_{p}>8.13R_{\oplus}$ and $M_{p}>52.48M_{\oplus}$ as giant planets. Figure~\ref{figure3.12} shows the mass-radius distribution of clustered and outlier data. For several planets that lie outside the boundaries ($B_{\text{Radius}}$ and $B_{\text{Mass}}$), we use the criterion of their closeness to the boundaries to assign them to either of the classes.

\begin{table}
\centering
\setlength{\tabcolsep}{7.5pt}
\captionsetup{width=\textwidth}
\caption[Results obtained by different clustering algorithms]{Clustering algorithms, breakpoints of radius ($B_{\text{Radius}}$) and mass ($B_{\text{Mass}}$) in logarithmic space, number of planets in the first ($N_{1}$) and second ($N_{2}$) clusters along with adjusted hyper-parameters introduced in the Scikit-learn library \citep{scikit-learn}. The one-dimensional Gaussian distribution is used to find the intersection point and introduce the relevant breakpoints. Default values are set for hyper-parameters that are not listed. The Elbow method and Silhouette score have been used to find the optimal number of clusters equal to 2. DBSCAN and OPTICS algorithms fail to provide appropriate clusters.}
\begin{tabularx}{\textwidth}{lccccl}
\hline
Algorithm & $B_{\text{Radius}}$ & $B_{\text{Mass}}$ & $N_{1}$ & $N_{2}$ & Adjusted parameter \\
\hline
Affinity Propagation & 0.93     & 1.73     & 254     & 440     & damping=0.9, preference=-60 \\
BIRCH & 0.90   & 1.72  & 247   & 447   & n\_clusters=2, threshold=0.01 \\
DBSCAN & --     & --     & --     & --     & eps=0.2, min\_samples=25 \\
Gaussian Mixture Model & 0.95  & 1.80   & 271   & 423   & n\_components=2 \\
Hierarchical Clustering & 0.90   & 1.72  & 247   & 447   & n\_clusters=2 \\
K-Means & 0.92  & 1.72  & 252   & 442   & n\_clusters=2 \\
Mean Shift & 0.93    & 1.73     & 257     & 437     & bandwidth=0.9 \\
Mini-Batch K-Means & 0.92  & 1.72  & 252   & 442   & n\_clusters=2 \\
OPTICS & --     & --     & --     & --     & min\_samples=40 \\
Spectral Clustering & 0.89  & 1.64  & 239   & 455   & n\_clusters=2 \\
\hline
\end{tabularx}
\label{table3.4}
\end{table}

According to a traditional definition, small exoplanets are planets with radii smaller than $4R_{\oplus}$ and masses lower than $\sim30M_{\oplus}$ \citep{2010Sci...330..653H,2014PNAS..11112655M,2014ApJ...783L...6W}; however, this customary definition of small and large planets does not exactly match previous studies that have investigated a transition point in the mass-radius distribution of exoplanets. Table~\ref{table3.5} compares our breakpoints with those found by others in the literature and shows a considerable difference between mass breakpoints. Besides an increasing number of exoplanets and the evolution of their mass-radius distribution, this difference could result from applying different methods to find a cut-off point in planetary masses. Performed methods vary from a simple visual investigation of mass-radius and mass-density distributions \citep{2013ApJ...768...14W} to using different slope criteria in the planetary parametric relations \citep{2015ApJ...810L..25H,2017ApJ...834...17C,2017A&A...604A..83B}. As a result, the mass and radius breakpoints identified in this work ($52.48M_{\oplus}$ and $8.13R_{\oplus}$, respectively) are closer to traditional breakpoints than those found in previous studies.

\begin{table}
\centering
\setlength{\tabcolsep}{24pt}
\captionsetup{width=\textwidth}
\caption{Breakpoints of mass ($B_{\text{Mass}}$) and radius ($B_{\text{Radius}}$) derived by previous studies and in this work.}
\label{table3.5}
\begin{tabularx}{\textwidth}{lcc}
\hline
Study & $B_{\text{Mass}}$ (M$_{\oplus}$) & $B_{\text{Radius}}$ (R$_{\oplus}$) \\
\hline
\citet{2013ApJ...768...14W} & 150   & -- \\
\citet{2015ApJ...810L..25H} & 95    & -- \\
\citet{2017ApJ...834...17C} & $130\pm{22}$ & -- \\
\citet{2017A&A...604A..83B} & $124\pm{7}$ & $12.1\pm{0.5}$ \\
This work & 52.48 & 8.13 \\
\hline
\end{tabularx}
\end{table}

\begin{figure}[b!]
\centering
\captionsetup{width=\textwidth}
\includegraphics[width=\textwidth]{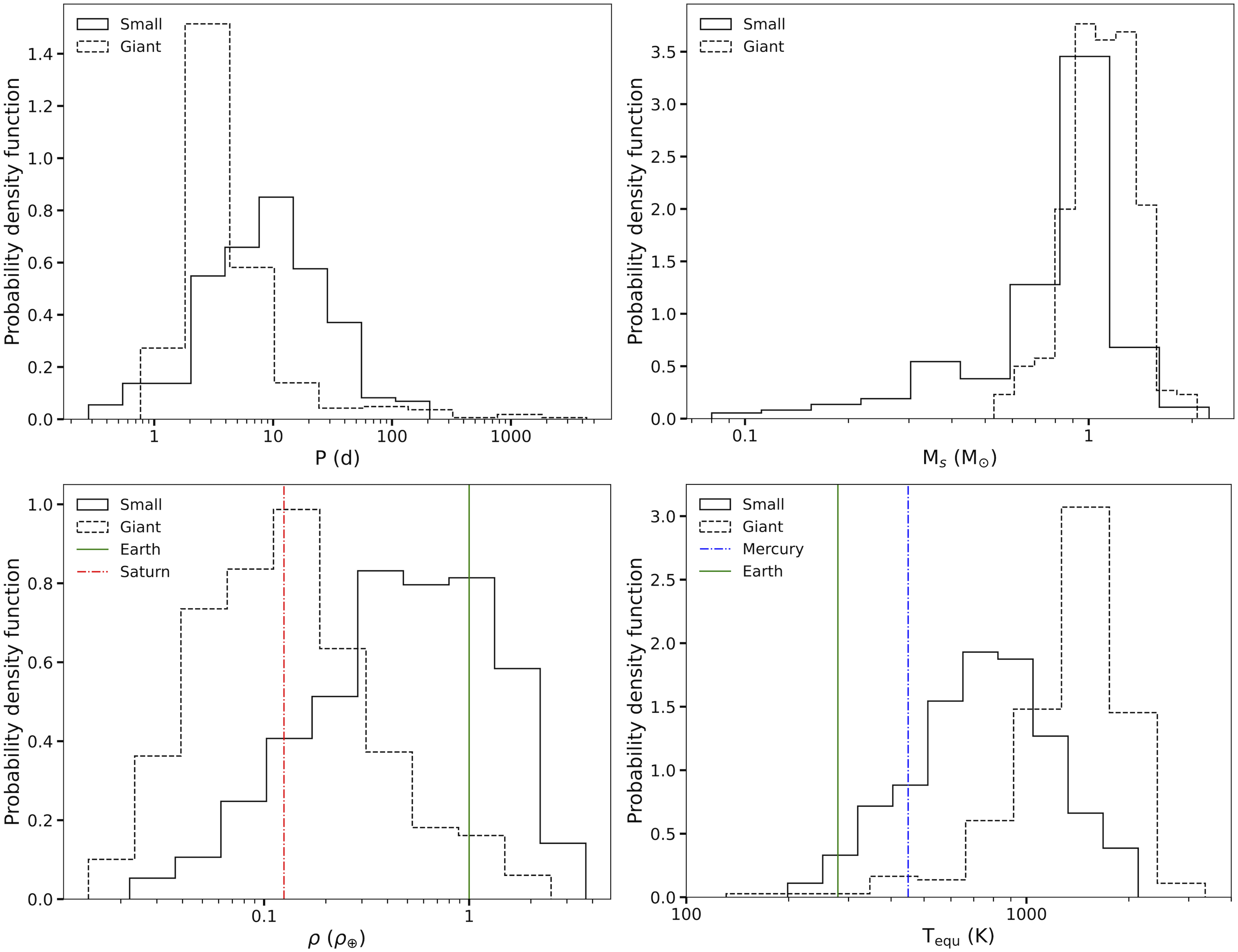}
\caption[Histograms of the orbital period, stellar mass, average density, and equilibrium temperature]{Histograms of the orbital period (upper-left panel), stellar mass (upper-right panel), average density (lower-left panel), and equilibrium temperature (lower-right panel) for small (solid-border bars) and giant (dashed-border bars) planets. The red dash-dotted line in the lower-left panel represents Saturn as the least dense planet, and the green solid line is Earth as the densest planet in the solar system. In the lower-right panel, the solid green line represents Earth's equilibrium temperature, and the blue dash-dotted line is Mercury, which has the highest equilibrium temperature in the solar system. The orbital period distribution shows that giant planets are closer to their host star than small planets. The stellar mass distribution demonstrates that, for most planets, the host star's mass is around the Sun's mass. Comparing planets' average density and calculated equilibrium temperature demonstrates that giant planets are hotter and less dense than small planets.}
\label{figure3.13}
\end{figure}

There are 254 small planets and 440 giant planets. The distributions of the orbital period, stellar mass, average density, and calculated equilibrium temperature for small and giant planets are demonstrated in figure~\ref{figure3.13}. The upper-left panel demonstrates that most giant planets are closer to their host star than small planets are. They, nonetheless, have a $P$ varying from 0.77 to 4331.01 days, while for small planets, it is between 0.28 and 207.62 days. The upper-right panel shows that the host star mass is frequently in a range around the Solar mass for most planets. It results from a selection effect: exoplanet-search programs often concentrate on Sun-like stars. In addition to this, lower-mass stars are not as likely to host exoplanets with sufficient mass to be identified by the radial-velocity technique \citep{2005A&A...443L..15B,2008PASP..120..531C}. Small planets revolve around stars with $M_{s}$ between 0.08 and 2.24 $M_{\odot}$, whilst, for host stars of giant planets, they vary from 0.53 to 2.07 $M_{\odot}$.

The lower-left panel compares small and giant planets' average density distribution. As expected, giant planets are generally less dense than small planets. The density distribution of giant planets peaks at about Saturn’s density. Hence, a significant fraction of giant planets has an average density similar to Saturn, the least dense planet in the solar system. On the contrary, small planets cover a higher density range that includes Earth, the densest planet in the solar system. This, in turn, suggests that giant planets are composed mainly of hydrogen and helium envelopes, whereas heavier elements dominate small planets.

Comparing the calculated equilibrium temperature of planets (lower-right panel) demonstrates that giant planets are hotter than small planets. This temperature difference between the two planet classes is expected because giant planets are closer to their host star (as shown in the upper-left panel); in addition to this, the host stars are almost of the same spectral type (equivalently, stellar mass, as shown in the upper-right panel).

It is important to note that our sample of giant planets is dominated by gas giant exoplanets with orbital periods of less than 10 days, commonly referred to as hot Jupiters. It is believed that these planets are more likely to be found around metal-rich stars than around stars with low stellar metallicity \citep{2018A&A...612A..93M,2020MNRAS.491.4481O,2023ApJ...949L..21Y}.

\subsection{Prediction of the planetary radius}
To find the radius of a planet based on physical parameters, ML predictive algorithms are applied to our dataset of 694 planets (sample II). The physical parameters are the planet's mass ($M_{p}$), orbital period ($P$), and host star's mass ($M_{s}$), selected by FS methods. Bootstrap Aggregation, Decision Tree, K-Nearest Neighbors, Linear Regression, Multilayer Perceptron, M5P, Random Forest, and Support Vector Regression (SVR) are implemented algorithms. These algorithms are applied separately to entire, small, and giant planets. The hyper-parameters are tuned to have the best performance of algorithms. Furthermore, a 10-fold cross-validation procedure is used to assess the performance of models. The root means square error (RMSE), mean absolute error (MAE), and coefficient of determination ($\rho^{2}$) are calculated as validation metrics (see equations~\ref{equation2.12},~\ref{equation2.13}, and~\ref{equation2.14}).

\begin{figure}[!htb]
\centering
\captionsetup{width=0.60\paperwidth}
\includegraphics[width=0.60\paperwidth]{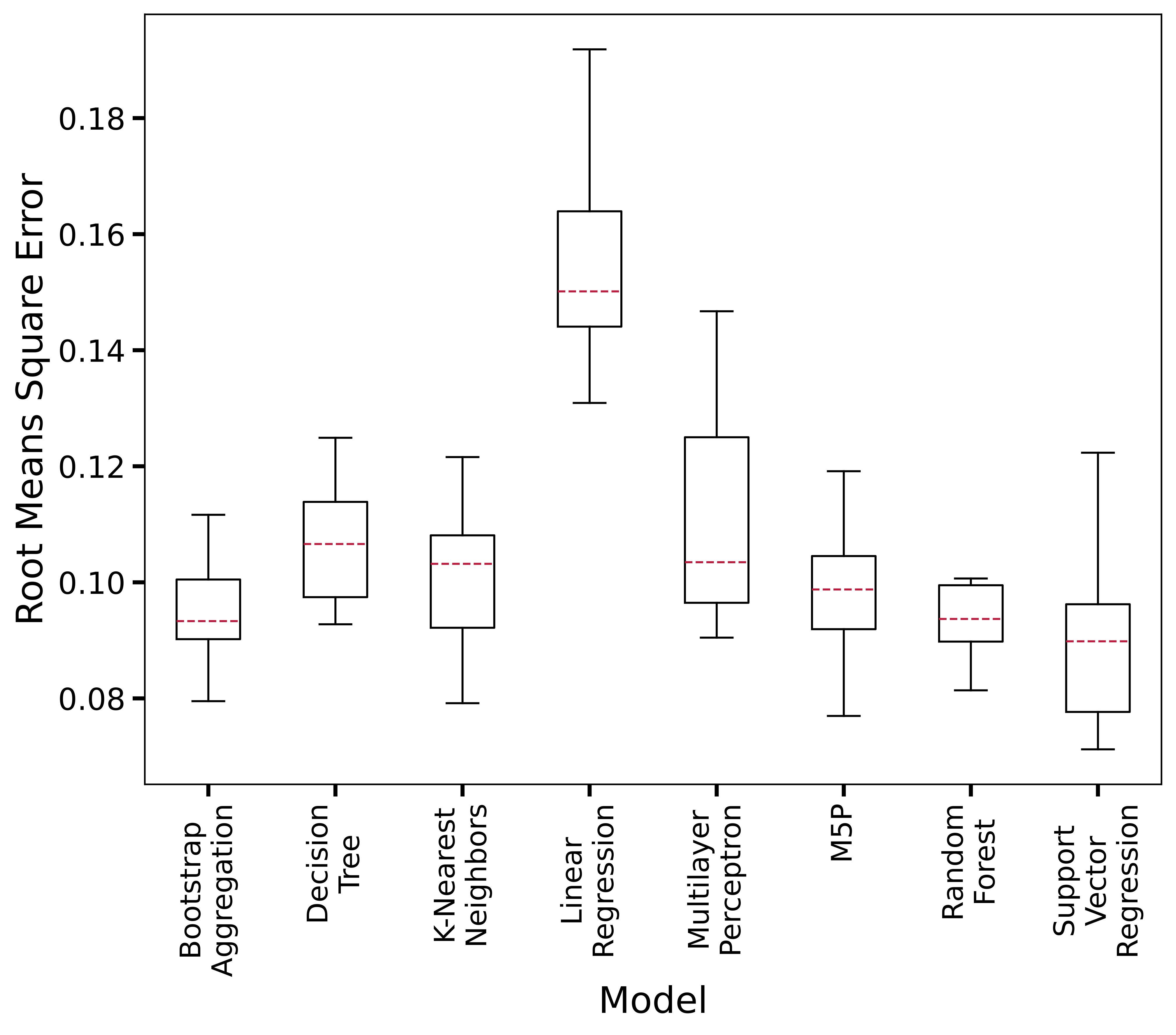}
\caption[The spread of accuracy in the 10-fold cross-validation for predictive algorithms]{Box plots showing the spread of accuracy in the 10-fold cross-validation for predictive algorithms. On each box, the red dashed mark is the median, and the edges of the box are the $25^{th}$ and $75^{th}$ percentiles. Support Vector Regression with an RMSE of 0.093 is the best-performing model for the entire dataset, followed by Bootstrap Aggregation, Random Forest, and M5P with RMSEs of 0.096, 0.097, and 0.098, respectively.}
\label{figure3.14}
\end{figure}

\begin{figure}[!htb]
\centering
\captionsetup{width=0.60\paperwidth}
\includegraphics[width=0.60\paperwidth]{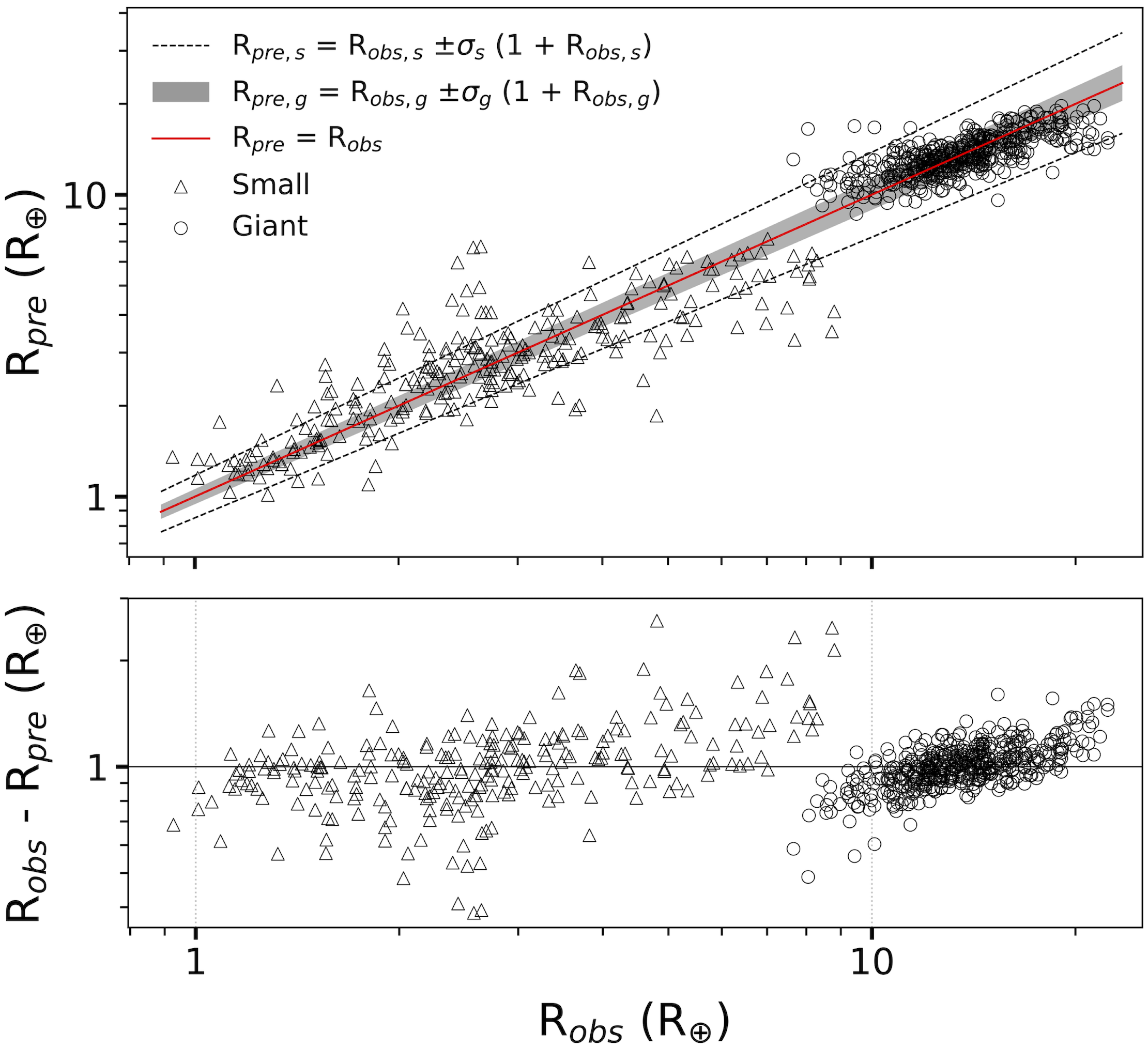}
\caption[Comparison between observed and predicted planetary radius along with residual values obtained by the SVR model]{Comparison between observed ($R_{\text{obs}}$) and predicted ($R_{\text{pre}}$) planetary radius (upper panel) along with residual values (lower panel) obtained by the SVR model, which has been applied separately to small (triangles) and giant (circles) planets. The red line indicates $R_{\text{pre}}=R_{\text{obs}}$ along with two dashed lines and a gray area that illustrate the normalized median absolute deviation (NMAD) for small and giant planets, respectively. Considering \citet{1983ured.book.....H}, NMAD is calculated using the equation~\ref{equation3.1}.}
\label{figure3.15}
\end{figure}

RMSE, MAE, and $\rho^{2}$ of entire, small, and giant planets, along with the tuned hyper-parameters, are listed in table~\ref{table3.6}. Figure~\ref{figure3.14} shows the box plots of accuracy in the 10-fold cross-validation for models. SVR with an RMSE of 0.093 is the best-performing model for the entire dataset, followed by Bootstrap Aggregation, Random Forest, and M5P with RMSEs of 0.096, 0.097, and 0.098, respectively. They also have lower MAE and higher $\rho^{2}$ values than other models. The SVR performs better than other algorithms for both subsets of small and giant planets. Figure~\ref{figure3.15} compares the observed ($R_{\text{obs}}$) and predicted ($R_{\text{pre}}$) radius (upper panel) along with residual values (lower panel) obtained by SVR as the best-performing model. In this figure, the model has been applied separately to small and giant planets, resulting in a gap and a relatively higher dispersion around $\sim8R_{\oplus}$. $\rho^{2}$ value is 0.710 for small planets and 0.510 for giant planets. The normalized median absolute deviation (NMAD) has been calculated for small and giant planets predictions, which is obtained by equation~\ref{equation3.1} \citep{1983ured.book.....H}.
\vspace*{-15pt}
\begin{gather}
\nonumber R_{\text{pre}}=R_{\text{obs}}\pm\sigma(1+R_{\text{obs}}),\\
\sigma=1.48\times median[\mid R_{\text{pre}}-R_{\text{obs}} \mid/(1+R_{\text{obs}})].
\label{equation3.1}
\end{gather}

As observed in the lower panel of figure~\ref{figure3.15}, a distinct linear trend with a slope of 0.459 is noticeable in the residuals between predicted and observed radii of giant planets. This trend could be attributed to factors such as systematic observation errors, the determination of physical parameters, and calculation issues, including hyper-parameter tuning, algorithm characteristics and limitations, and selected features. We adjust the learning algorithms to address this challenge by re-configuring their hyper-parameters. Interestingly, a similar pattern emerges in the residual values across all eight predictive algorithms, suggesting that the constraints of SVR alone cannot explain it. Moreover, we discover that excluding or including the parameter $M_{p}$ significantly affects the slope of the linear trend ($\pm0.08$) compared to other parameters. However, modifying the subset of features does not eliminate this trend. The most gradual trend appears when utilizing the feature subset of $M_{p}$, $P$, and $M_{s}$.

It is important to note that this trend minimally impacts radius prediction. Additionally, the residuals between predicted and observed radii of giant planets fall within the range seen for small planets. It is plausible that this trend is linked to systematic issues in exoplanet observations or the determination of observed parameters (e.g., mass, orbital period, and radius). As the current sample of exoplanets detected through the transit method is substantial, further investigation of this effect could be undertaken when a statistically significant sample of exoplanets detected through other detection methods becomes available.

\begin{landscape}
\begin{table}
\centering
\setlength{\tabcolsep}{5.75pt}
\captionsetup{width=1.55\textwidth}
\caption[Results obtained by different regression algorithms]{Results obtained by different regression algorithms. These algorithms have been separately applied to entire, small, and giant planets. Column 1 presents the name of the algorithm. Columns 2, 3, and 4 present the root mean square error (RMSE), mean absolute error (MAE), and coefficient of determination ($\rho^{2}$) for the whole dataset. Columns 5 to 10 present RMSE, MAE, and $\rho^{2}$ for small and giant planets, respectively. The last column lists adjusted hyper-parameters introduced in the Weka environment \citep{witten2005practical,hall2009weka}.}
\label{table3.6}
\begin{tabularx}{1.55\textwidth}{lcccccccccl}
\hline
Algorithm & RMSE  & MAE   & $\rho^{2}$ & RMSE$_{1}$ & MAE$_{1}$ & $\rho^{2}_{1}$ & RMSE$_{2}$ & MAE$_{2}$ & $\rho^{2}_{2}$ & \multicolumn{1}{l}{Adjusted parameter} \\
\hline
Bootstrap Aggregation & 0.096 & 0.068 & 0.933 & 0.126 & 0.096 & 0.692 & 0.064 & 0.047 & 0.489 & numIterations=100, classifier=REPTree \\
Decision Tree & 0.108 & 0.077 & 0.916 & 0.142 & 0.109 & 0.616 & 0.071 & 0.051 & 0.392 & maxDepth=-1, minNum=2 \\
K-Nearest Neighbors & 0.104 & 0.075 & 0.921 & 0.140 & 0.106 & 0.627 & 0.072 & 0.053 & 0.390 & KNN=3, distanceFunction=Euclidean \\
Linear Regression & 0.155 & 0.127 & 0.822 & 0.128 & 0.099 & 0.683 & 0.069 & 0.051 & 0.402 & eliminateColinearAttributes=True \\
& & & & & & & & & & attributeSelectionMethod=M5 \\
Multilayer Perceptron & 0.112 & 0.084 & 0.909 & 0.131 & 0.103 & 0.670 & 0.064 & 0.048 & 0.491 & learningRate=0.1, decay=False \\
& & & & & & & & & & momentum=0.1, hiddenLayers=a \\
M5P   & 0.098 & 0.071 & 0.930 & 0.130 & 0.101 & 0.688 & 0.073 & 0.053 & 0.391 & unpruned=False \\
Random Forest & 0.097 & 0.068 & 0.932 & 0.128 & 0.096 & 0.682 & 0.064 & 0.047 & 0.485 & numIterations=100, maxDepth=0 \\
& & & & & & & & & & numFeatures=2 \\
SVR   & 0.093 & 0.065 & 0.937 & 0.123 & 0.088 & 0.710 & 0.063 & 0.046 & 0.510 & c=1, kernel=Puk \\
\hline
\end{tabularx}
\end{table}
\end{landscape}

\begin{figure}[h]
\centering
\captionsetup{width=0.60\paperwidth}
\includegraphics[width=0.60\paperwidth]{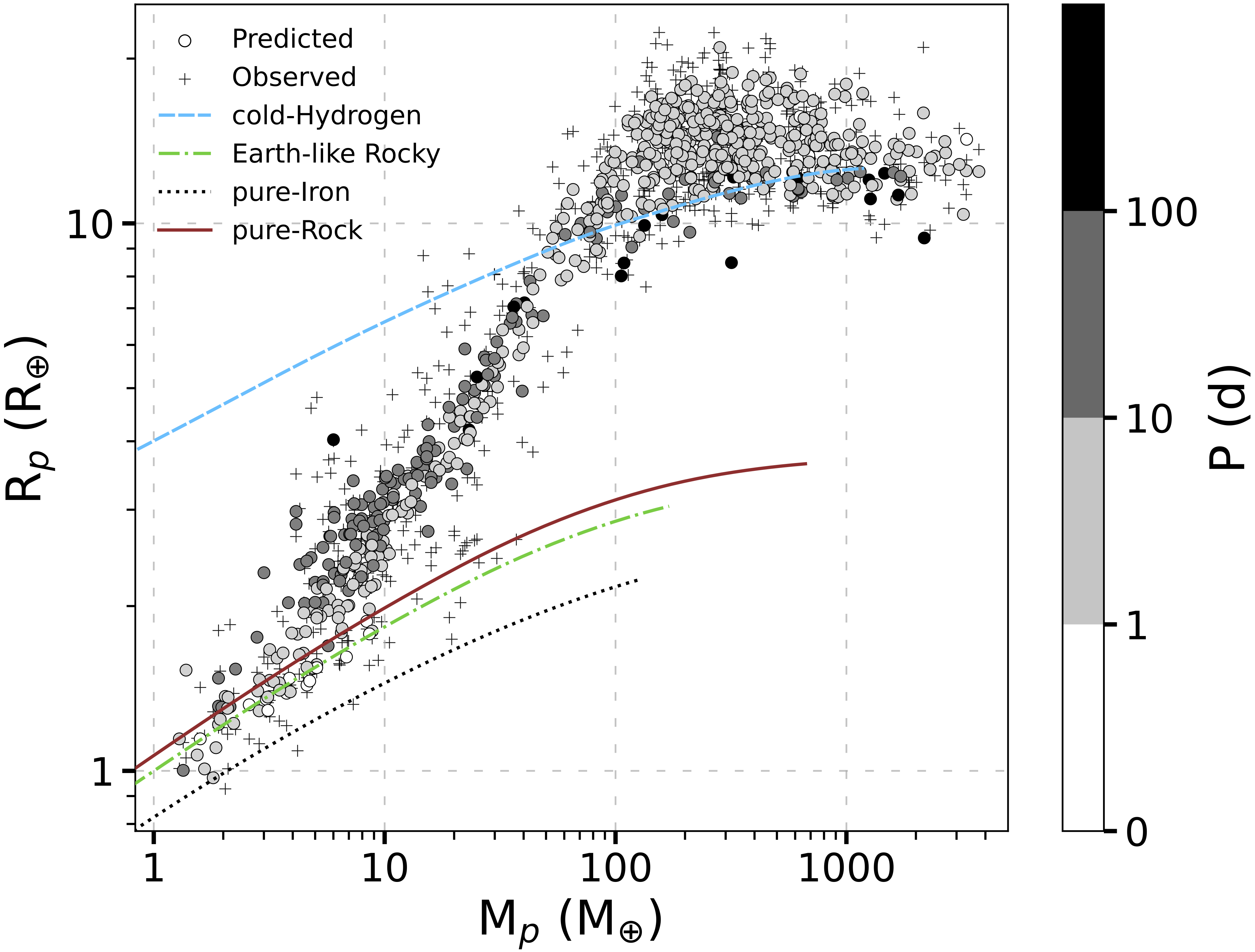}
\caption[Predicted and observed radius as a function of mass and orbital period obtained by the SVR model]{Predicted (circles) and observed (pluses) radius as a function of mass and orbital period obtained by the SVR model for all planets in the sample II. The distribution of cold-hydrogen (blue dashed line), Earth-like rocky (green dash-dotted line), pure-iron (black dotted line), and pure rocky (solid crimson line) planets are also illustrated \citep{2010ApJ...712L..73M,2014ApJS..215...21B}.}
\label{figure3.16}
\end{figure}

Figure~\ref{figure3.16} shows the predicted and observed radii as a function of a mass and orbital period obtained by the SVR model. The model has been applied to the entire sample in this figure, resulting in an $\rho^{2}$ of 0.937. SVR can efficiently reproduce the spread in radius.

Linear Regression and M5P can derive parametric equations between physical parameters by fitting linear models to the exoplanet data. A linear model is fitted to the real data in the Linear Regression algorithm. At the same time, M5P splits the entire dataset into several subsets and fits a multivariate linear function to each subset. Equation~\ref{equation3.2} presents a linear fit between the planetary radius, planetary mass, orbital period, and stellar mass derived by Linear Regression and M5P, where $A_{M_{p}}$, $A_{P}$, $A_{M_{s}}$, and $C$ are fitting parameters.
\begin{equation}
\log \biggl(\frac{R_{p}}{R_{\oplus}}\biggr)=A_{M_{p}}\log \biggl(\frac{M_{p}}{M_{\oplus}}\biggr)+A_{P}\log \biggl(\frac{P}{d}\biggr)+A_{M_{s}}\log \biggl(\frac{M_{s}}{M_{\odot}}\biggr)+C.
\label{equation3.2}
\end{equation}

Best-fit parameters obtained by Linear Regression and M5P are listed in table~\ref{table3.7}. Row 1 presents the linear fit of all planets provided by the Linear Regression algorithm. Running this algorithm independently for clusters produces individual linear fit for each cluster (rows 2 and 3). For small planets $A_{M_{s}}=0$, and for giant planets $A_{M_{p}}=0$, implying no dependence between the planetary radius and stellar mass of small planets, as well as between the radius and mass of giant planets.

M5P divides the planets into two groups using a mass breakpoint of $\log M_{p}=1.717$ ($M_{p}=52.12M_{\oplus}$). Interestingly, clustering algorithms also find this breakpoint (see table~\ref{table3.4}). In table~\ref{table3.7}, rows 4 and 5 present multivariate linear fits of small and giant planets produced by the M5P algorithm. Splitting the data provides M5P with much better results than Linear Regression (see table~\ref{table3.6}).

To estimate the uncertainty values of the M5P's best-fit parameters, we use the MCMC method. The likelihood function and initial values implemented in the MCMC analysis are the same as those acquired by the M5P. In table~\ref{table3.7}, rows 6 and 7 present the linear fits of small and giant planets, respectively, together with the uncertainty values obtained by the MCMC method. In addition, the corresponding regression surfaces are depicted beside the distribution of the scaling relation parameters in appendix~\ref{appendixC}. Figure~\ref{figure3.17} presents the distributions of the predicted and observed radii as a function of mass and orbital period, obtained by the M5P model for small (lower panel) and giant (upper panel) planets. The value of $\rho^{2}$ for the whole sample is 0.930. The predicted radii reproduce the spread in radius, especially for giant planets.

\begin{table}[b!]
\centering
\setlength{\tabcolsep}{7.75pt}
\captionsetup{width=\textwidth}
\caption[Parametric equations obtained by different regression algorithms]{Parametric equations obtained by different regression algorithms. Rows 1 to 7 present a linear fit, which equates the logarithm of planetary radius ($R_{p}$) to logarithms of planetary mass ($M_{p}$), orbital period ($P$), and stellar mass ($M_{s}$) plus a constant term ($C$) (see equation~\ref{equation3.2}). Row 8 presents a linear fit between planetary mass and radius as $\log(R_{p}/R_{\oplus})=A_{M_{p}}\log(M_{p}/M_{\oplus})+C$. The linear fit between the planetary radius and stellar mass is also presented in row 9 as $\log(R_{p}/R_{\oplus})=A_{M_{s}}\log(M_{s}/M_{\odot})+C$. The first column shows the row number. Column 2 presents the used dataset: the entire dataset (sample II), small planets, or giant planets. Best-fit parameters are listed in columns 3 to 6. The last column presents the applied algorithm: Linear Regression, M5P, or MCMC.}
\label{table3.7}
\begin{tabularx}{\textwidth}{lcccccl}
\hline
\# & Dataset & A$_{M_{p}}$ & A$_{P}$ & A$_{M_{s}}$ & C & Algorithm \\
\hline
1     & Entire & 0.367 & -0.030 & 0.280 & 0.191 & Linear Regression \\
2     & Small & 0.467 & 0.090 & 0     & -0.103 & Linear Regression \\
3     & Giant & 0     & -0.069 & 0.480 & 1.157 & Linear Regression \\
4     & Small & 0.481 & 0.076 & 0.016 & -0.095 & M5P \\
5     & Giant & 0.012 & -0.067 & 0.489 & 1.123 & M5P \\
6     & Small & $0.482\substack{+0.025\\-0.024}$ & $0.078\substack{+0.017\\-0.016}$ & $0.031\substack{+0.041\\-0.042}$ & -$0.099\substack{+0.030\\-0.030}$ & M5P and MCMC \\
7     & Giant & $0.013\substack{+0.010\\-0.009}$ & -$0.070\substack{+0.007\\-0.007}$ & $0.492\substack{+0.036\\-0.036}$ & $1.121\substack{+0.024\\-0.024}$ & M5P and MCMC \\
8     & Small & $0.497\substack{+0.023\\-0.022}$ & --     & --     & -$0.050\substack{+0.024\\-0.024}$ & MCMC \\
9     & Giant & --     & --     & $0.480\substack{+0.036\\-0.037}$ & $1.109\substack{+0.004\\-0.004}$ & MCMC \\
\hline
\end{tabularx}
\end{table}

\begin{figure}[p]
\centering
\captionsetup{width=0.60\paperwidth}\centering
	\includegraphics[width=0.60\paperwidth]{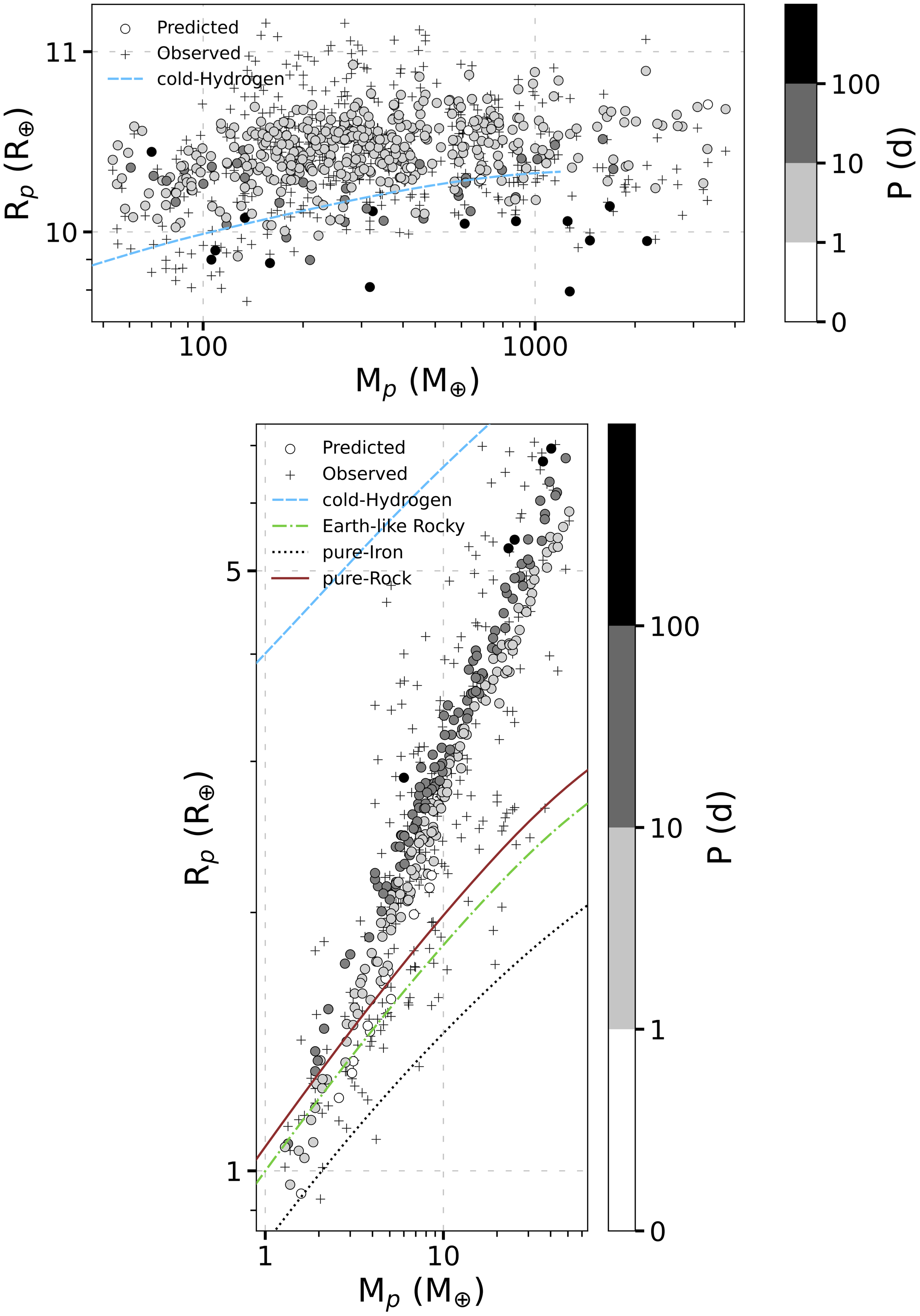}
    \caption[Predicted and observed radius as a function of mass and orbital period obtained by the M5P model]{Predicted (circles) and observed (pluses) radius as a function of mass and orbital period obtained by the M5P model for small (lower panel) and giant (upper panel) planets. Four iso-density curves are also drawn: cold-hydrogen (blue dashed line), Earth-like rocky (green dash-dotted line), pure-iron (black dotted line), and pure rocky (solid crimson line) planets \citep{2010ApJ...712L..73M,2014ApJS..215...21B}.}
\label{figure3.17}
\end{figure}

\newpage
\subsection{Dependence of planetary radius on host star's mass}
There are inconsistent assertions in the literature about the dependence of planetary parameters on the host star's mass. \citet{2018ApJ...856L..28P} claimed that the mass of the most common exoplanets depends on their host star mass. They investigated G, K, and M stars and suggested that planets around relatively low-mass stars (with a mass lower than $1M_{\odot}$) are lower in mass and smaller in radius. In contrast, \citet{2018ApJ...858...58N} showed that the mass-radius relation of small planets has no strong dependence on stellar mass. \citet{2019ApJ...874...91W} discussed a linear relationship between exoplanet mass and host star mass and the lack of correlation between the planetary radius and stellar metallicity. In addition, \citet{2021A&A...652A.110L} studied exoplanets with radii up to $8R_{\oplus}$ and masses up to $20M_{\oplus}$ surrounding G and K stars. They confirmed that exoplanets revolving around more massive stars tend to be larger and more massive.

\begin{figure}[!ph]
\centering
\captionsetup{width=0.50\paperwidth}\centering
	\includegraphics[width=0.50\paperwidth]{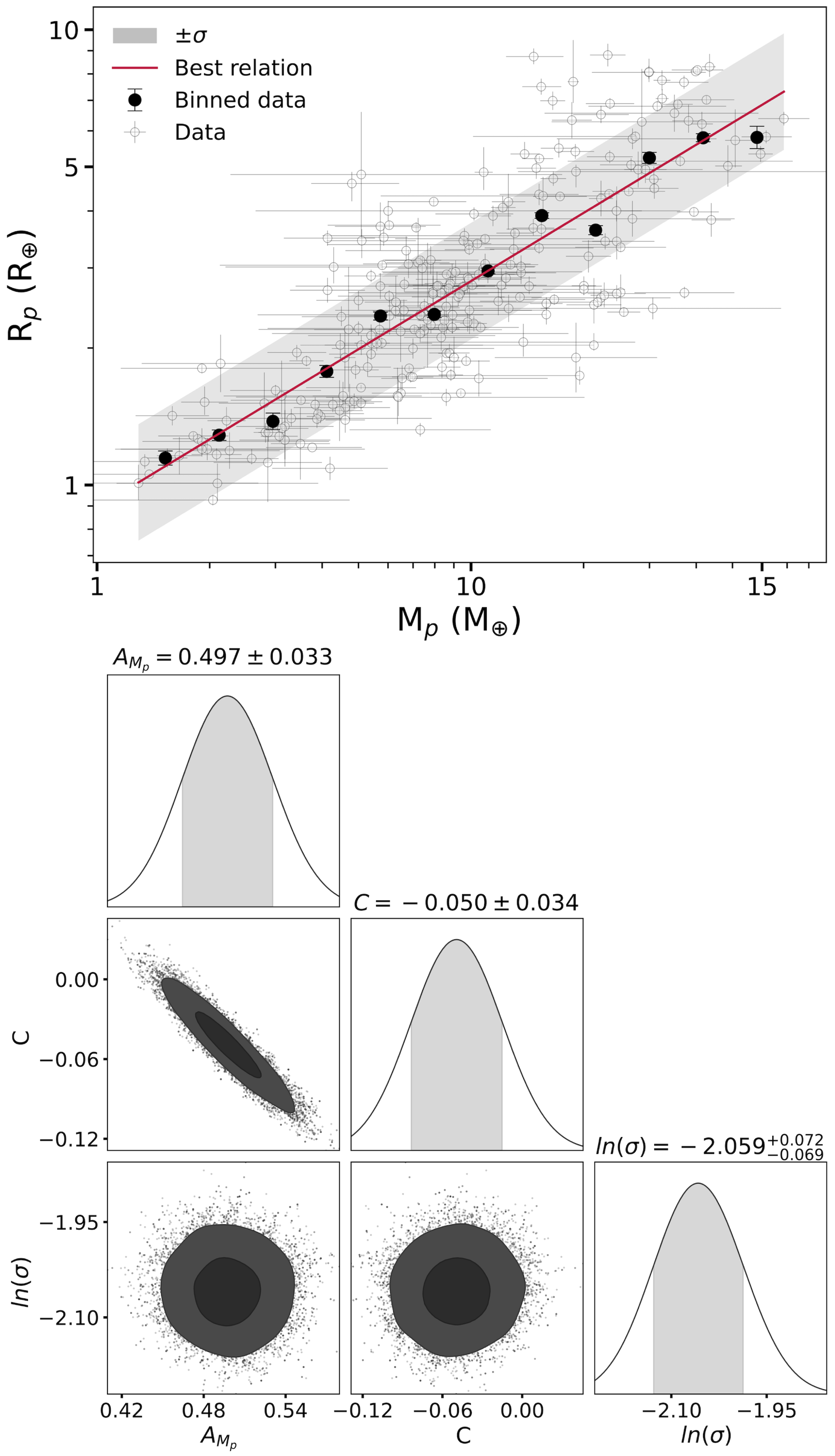}
    \caption[The relation between mass and radius of small planets obtained by the MCMC method]{Upper panel: the relation between mass and radius of small planets obtained by the MCMC method. The red line is the best scaling relation plotted using $\log(R_{p}/R_{\oplus})=0.497\log(M_{p}/M_{\oplus})-0.050$ (see table~\ref{table3.7}, row 8). The gray area is $\pm1\sigma$ uncertainties around the best scaling relation. For better illustration, the data points have been binned with a width of 0.05 $\log M_{p}$. The binned and actual data are shown in black and gray circles. Lower panel: the one- and two-dimensional marginalized posterior distributions of the scaling relation parameters obtained by the MCMC method. $A_{M_{p}}$ and $C$ are the slope and intercept, respectively. The uncertainty around the best scaling relation is shown by $\sigma$.}
    \label{figure3.18}
\end{figure}

\begin{figure}[!ph]
\centering
\captionsetup{width=0.50\paperwidth}\centering
	\includegraphics[width=0.50\paperwidth]{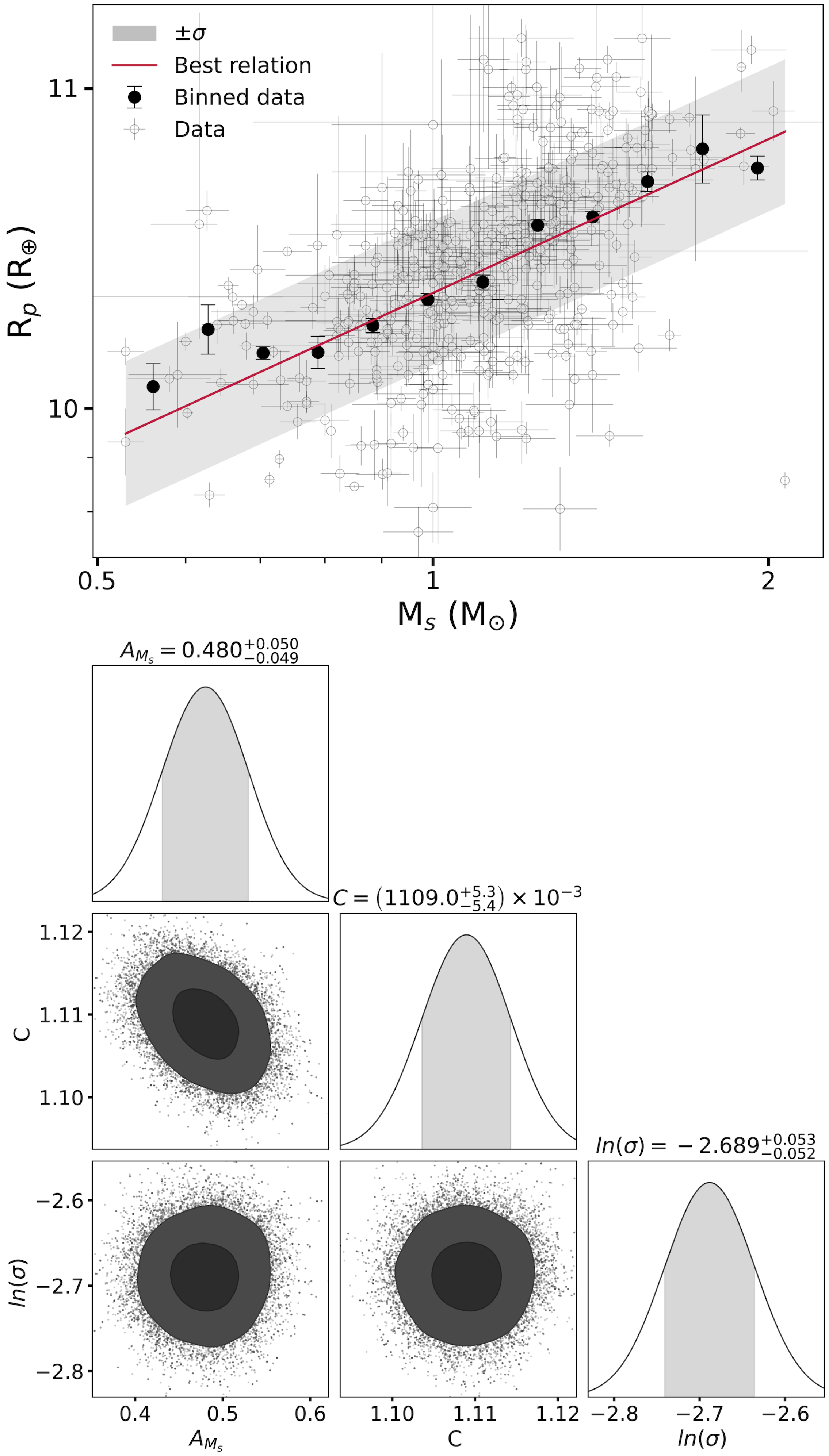}
    \caption[The relation between the radius of giant planets and the host star's mass obtained by the MCMC method]{Upper panel: the relation between the radius of giant planets and the host star's mass obtained by the MCMC method. The red line is the best scaling relation plotted using $\log(R_{p}/R_{\oplus})=0.480\log(M_{s}/M_{\odot})+1.109$ (see table~\ref{table3.7}, row 9). The gray area is $\pm1\sigma$ uncertainties around the best scaling relation. For better illustration, the data points have been binned with a width of 0.05 $\log M_{s}$. The binned and actual data are shown in black and gray circles. Lower panel: the one- and two-dimensional marginalized posterior distributions of the scaling relation parameters obtained by the MCMC method. $A_{M_{s}}$ and $C$ are the slope and intercept, respectively. The uncertainty around the best scaling relation is shown by $\sigma$.}
\label{figure3.19}
\end{figure}

As can be seen in table~\ref{table3.7}, the radius of a small planet shows a strong dependency on its mass. Furthermore, there is no strong correlation between stellar mass and planetary radius for small planets. In comparison, the radius of a giant planet depends weakly on its mass because above $\sim8R_{\oplus}$ the electron degeneracy pressure dominates \citep{1969ApJ...158..809Z,2007ApJ...669.1279S,2012ApJ...744...59S}. In addition to this, the planetary radius and stellar mass of giant planets have a strong linear correlation. We apply the MCMC as a supportive method to find the best scaling relations between the radius and mass of small planets and between the radius and stellar mass of giant planets while considering the reported errors of physical values. Row 8 of table~\ref{table3.7} presents the linear fit between the radius and mass of small planets. Additionally, the linear fit between the radius of giant planets and the mass of their host stars is presented in row 9. The related diagrams are depicted in figure~\ref{figure3.18} and \ref{figure3.19}.

Giant planets are less dense than small planets (see the lower-left panel of figure~\ref{figure3.13}) and are mostly composed of volatile elements (hydrogen and helium envelopes). On the other hand, giant planets orbit stars more massive than $\sim1M_{\odot}$, whereas the hosts of small planets include low-mass stars; that is, they have a mass greater than $0.08M_{\odot}$ (see the upper-right panel of figure.~\ref{figure3.13}). Concentrating on a limited sample of exoplanets, \citet{2021A&A...652A.110L} concluded that planets forming around massive stars accrete more H-He atmospheres than those that form around low-mass stars. Our extensive sample suggests a similar scenario for giant planets. Hence, the dependence of the radius of giant planets on the host star's mass may result from their different planetary composition. It should be noted that, in addition to naturally correlated parameters, this trend with stellar mass might be a consequence of observational biases. The larger transiting exoplanets are more detectable around luminous stars with larger masses \citep{2016arXiv160700322B}.

\subsection{The effect of equilibrium temperature, semi-major axis, and luminosity}
\citet{2016arXiv160700322B} used a Random Forest model to assess the effect of different physical parameters on predicting a planet's radius. They concluded that the planet's mass and equilibrium temperature have the greatest effect. In a similar work, \citetalias{2019A&A...630A.135U} presented planetary mass, equilibrium temperature, semi-major axis, stellar radius, mass, luminosity, and effective temperature as important parameters, and stellar metallicity, orbital period, and eccentricity as the three least important parameters.

\begin{figure}[b!]
\centering
\captionsetup{width=0.60\paperwidth}\centering
\includegraphics[width=0.60\paperwidth]{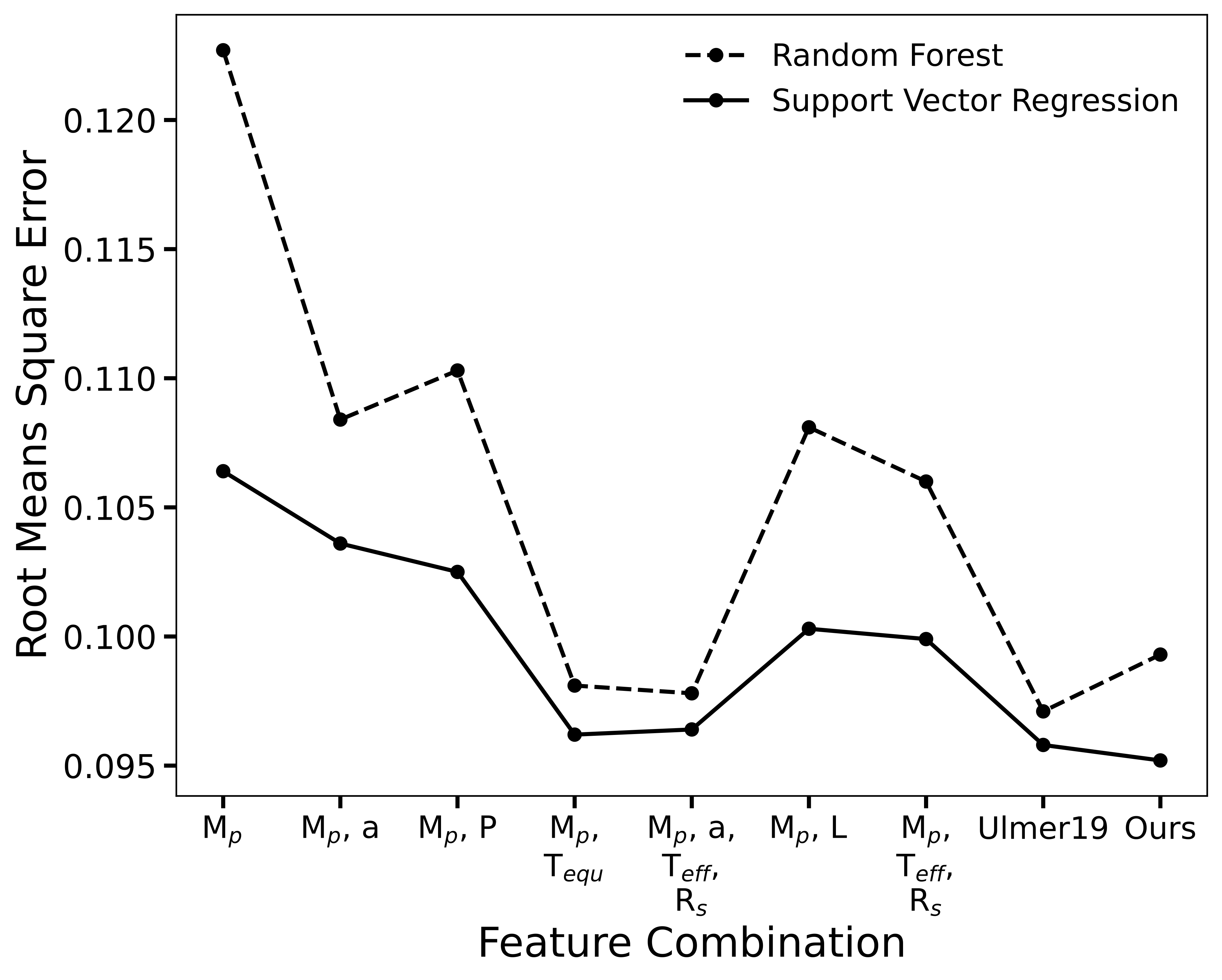}
\caption[Comparison of the performance of Random Forest and SVR models for different feature combinations]{Comparison of the performance of Random Forest and Support Vector Regression (SVR) models for different feature combinations. The dataset consists of 506 planets collected by \citetalias{2019A&A...630A.135U} and transferred to a logarithmic space. The dashed and solid lines represent root means square errors obtained by Random Forest and SVR, respectively. Features are as follows: orbital period ($P$), planetary mass ($M_{p}$), semi-major axis ($a$), equilibrium temperature ($T_{\text{equ}}$), luminosity ($L$), stellar mass ($M_{s}$), stellar radius ($R_{s}$), and effective temperature ($T_{\text{eff}}$). The eighth set consists of $M_{p}$, $T_{\text{equ}}$, $a$, $R_{s}$, $M_{s}$, $L$, and $T_{\text{eff}}$, selected by \citetalias{2019A&A...630A.135U}. The last set is our feature combination, which contains $M_{p}$, $P$, and $M_{s}$. The SVR model performs better than the Random Forest model for all combinations.}
    \label{figure3.20}
\end{figure}

We add orbital periods to the dataset collected by \citetalias{2019A&A...630A.135U} and transfer it to a logarithmic space. Hence, a new dataset consisting of 506 planets is provided. The features of this dataset are as follows: orbital period ($P$), planetary mass ($M_{p}$), semi-major axis ($a$), equilibrium temperature ($T_{\text{equ}}$), luminosity ($L$), stellar mass ($M_{s}$), stellar radius ($R_{s}$), and effective temperature ($T_{\text{eff}}$). To evaluate the effect of equilibrium temperature, semi-major axis, and luminosity in predicting planetary radius and to compare the performance of Random Forest and SVR models in \citetalias{2019A&A...630A.135U}'s dataset, we implement models on the nine feature combinations. As the most important parameter, planetary mass is added to all combinations. Figure~\ref{figure3.20} presents the RMSE values obtained by Random Forest and SVR versus different feature combinations. The SVR model performs better than the Random Forest for all combinations.

According to Kepler's third law, the orbital period and semi-major axis are correlated to each other \citep{cox2015allen}. So, as expected, considering the SVR model, the RMSE values of the second (0.104) and third (0.103) combinations are not significantly different. The second combination consists of $M_{p}$ and $a$, while the third set includes $M_{p}$ and $P$. Moreover, the equilibrium temperature of a planet can be calculated using equation~\ref{equation3.3} without considering the effect of albedo and eccentricity \citep{2015trge.book..673L}.
\begin{equation}
T_{\text{equ}}=\sqrt{\frac{R_{s}}{2a}}\times T_{\text{eff}}.
\label{equation3.3}
\end{equation}
The fourth combination includes planetary mass and equilibrium temperature, and the fifth combination includes planetary mass and constituent parameters of equilibrium temperature ($a$, $R_{s}$, and $T_{\text{eff}}$). The SVR model's RMSE values corresponding to the fourth and fifth sets are almost identical (0.096).

The luminosity of a star is correlated to its radius and effective temperature ($L\propto R_{s}^{2}\times T_{\text{eff}}^{4}$). The sixth combination is planetary mass and luminosity, whose RMSE value (0.100) is almost the same as the seventh combination, which includes planetary mass and constituent parameters of luminosity ($T_{\text{eff}}$ and $R_{s}$). Additionally, the eighth set corresponds to features selected by \citetalias{2019A&A...630A.135U}, including $M_{p}$, $T_{\text{equ}}$, $a$, $R_{s}$, $M_{s}$, $L$, and $T_{\text{eff}}$, and the last set consists of $M_{p}$, $P$, and $M_{s}$, which we have selected. Using the SVR model, it is clear that there is no remarkable difference between the results obtained by these two feature sets. RMSE of our feature set equals 0.095 while 0.096 for that used by \citetalias{2019A&A...630A.135U}.

Although \citetalias{2019A&A...630A.135U} considered the orbital period as an inconsequential parameter, here we show that the orbital period (or semi-major axis), along with the planet's mass and one of the stellar parameters, have remarkable effects on predictions. Moreover, contrary to the results acquired by \citet{2016arXiv160700322B} and \citetalias{2019A&A...630A.135U}, it seems that considering stellar luminosity and planetary equilibrium temperature as features does not improve the accuracy of planetary radius predictions. Luminosity and equilibrium temperature are two physically dependent parameters. Thus, similar results can be achieved only by considering their constituent parameters.

\section{Conclusions and future work}
In the first part of this work, we apply the TB relation to all available exoplanetary systems with three or more confirmed exoplanets, a total of 229 systems (sample I), to examine their adherence to the TB relation in comparison to the solar system and to predict the existence of possible additional planets. For those systems that adhere to the TB relation better than the solar system, we extrapolate one additional planet. For each of the remaining systems that adhere worse, we interpolate up to 10 specific planets between the detected planets and identify new possible additional planets in the systems. We present a list of 229 analyzed exoplanetary systems (of which 123 of them have not been previously analyzed by either \citetalias{2013MNRAS.435.1126B} or \citetalias{2015MNRAS.448.3608B}) containing their unique TB relations and a total of 426 additional predicted exoplanets, of which 47 are located within the HZ of their parent stars. We also estimate that five of the predicted planets in HZ have maximum mass and radius limits within the Earth's mass and radius range.

As an important result, the planets of $\sim53\%$ of our sample I systems adhere to a logarithmic spacing relation better than the planets of the solar system. Therefore, there is a need to work with more comprehensive data on multi-planetary systems to reveal the probable dynamical or gravitational aspects of the TB relation. Using new planet detections for seven exoplanetary systems after the predictions made by \citetalias{2015MNRAS.448.3608B} and comparing the detected and predicted orbital periods, we find that both the detected and predicted orbital periods agree very well within errors.

Our predictions also agree roughly better than those made by \citetalias{2015MNRAS.448.3608B}, indicating that using a precise modeling method and measurements could improve our predictions and uncertainties as well. However, to claim the (un)reliability of the TB relation in predicting the presence of additional planets in exoplanetary systems, we require much more data and further follow-up observations of the exoplanetary systems. Thus, we must wait for upcoming new exoplanet surveys or ongoing surveys such as TESS.

In the second part of this work, we conduct a comprehensive analysis of a sample comprising 762 exoplanets and eight solar system planets (sample II). Our main objective is to investigate the characteristics of these exoplanets and explore the correlations between various features. The dataset includes essential parameters such as orbital period ($P$) and eccentricity ($e$), planetary mass ($M_{p}$), and radius ($R_{p}$), and the stellar mass ($M_{s}$), radius ($R_{s}$), metallicity (Fe/H), and effective temperature ($T_{\text{eff}}$). To ensure the reliability of our analysis, we employ the Local Outlier Factor (LOF) algorithm, which allows us to identify and filter out data points that deviate significantly from the overall dataset. This process leads us to a refined dataset consisting of 76 anomalous objects, which can be considered robust and reliable measurements for our subsequent analysis.

By utilizing Feature Selection (FS) methods, we determine the most influential factors in predicting the radius of exoplanets. Our findings highlight that planetary mass ($M_{p}$) plays a pivotal role in this regard, whereas eccentricity ($e$) and metallicity (Fe/H) demonstrate relatively lesser significance in the prediction process. To further understand the underlying structure of the dataset, we employ various clustering algorithms and evaluation techniques such as the Elbow, Silhouette, and Hierarchical methods. Based on the outcomes of these analyses and in alignment with the conclusions drawn by \citet{2017A&A...604A..83B}, we opt to divide the dataset into two distinct clusters: small and giant planets. Notably, we observe distinct breakpoints in the mass-radius space at $M_{p}=52.48M_{\oplus}$ and $R_{p}=8.13R_{\oplus}$ for these clusters. Our analysis uncovers significant disparities between small and giant planets. Giant planets tend to exhibit higher masses, larger radii, and lower densities, suggesting a prevalence of volatile-rich exoplanets in this category.

Moreover, these giant planets tend to orbit their host stars at closer distances and possess higher equilibrium temperatures. On the other hand, small planets predominantly consist of elements heavier than hydrogen and helium, exhibiting lower equilibrium temperatures.

To predict the planetary radius, we employ various Machine Learning (ML) regression models. Among these models, the Support Vector Regression (SVR) demonstrates superior performance, yielding a Root Mean Squared Error (RMSE) of $0.093$. A discernible linear trend appears in the residuals between predicted and observed radii of giant planets, which is not attributed to the restrictions of the predictive models or calculation issues. This evident trend has no significant impact on the radius predictions and is possibly related to the systematic issues in the exoplanet observations or the determination of physical parameters.

Additionally, utilizing Linear Regression, M5P, and Markov Chain Monte Carlo (MCMC) methods, we establish a positive linear mass-radius relationship for small planets. In contrast, the radius of giant planets exhibits a positive correlation with the mass of their host stars, consistent with the findings presented by \cite{2021A&A...652A.110L}, which suggest a connection between volatile-rich planets and more massive host stars. Nonetheless, as most of our sample consists of transiting exoplanets, besides naturally correlated parameters, the observational bias in the detection method can explain this result.

Furthermore, our analysis reveals that a carefully selected subset of features, encompassing planetary mass, orbital period, and one of the stellar parameters (stellar mass, radius, or effective temperature), is sufficient for accurate radius prediction. The inclusion of additional features such as semi-major axis, equilibrium temperature, and luminosity does not yield substantial improvements in the predictive capability.

Looking ahead, a comprehensive understanding of exoplanet composition and structure, as well as the testing of theories related to planetary formation and evolution, necessitates further follow-up observations.  The James Webb Space Telescope (JWST) and future missions such as the Extremely Large Telescope (ELT), the Atmospheric Remote-sensing Infrared Exoplanet Large-survey (ARIEL), and the Planetary Transits and Oscillations of Stars (PLATO) mission will undoubtedly contribute invaluable insights into the atmospheric characteristics of exoplanets and the determination of stellar ages, thereby facilitating a more detailed exploration of exoplanetary systems.



\appendix 



\chapter{Predicted Exoplanets} 



\begin{landscape}
\begin{table}
\setlength{\tabcolsep}{13.5pt}
\centering
\captionsetup{width=1.55\textwidth}
\caption[Systems with only extrapolated planet predictions]{Systems with only extrapolated planet predictions. Columns 1 and 2 present the host star name and discovery method (Dis.). Column 3 reports a flag that defines whether the system has already been analyzed by \citetalias{2015MNRAS.448.3608B} (or \citetalias{2013MNRAS.435.1126B}) (Y) or not (N). Columns 4 to 7 present $\chi^2/dof$, slope (m), intercept (b), and predicted orbital period, respectively. Column 8 reports whether the predicted period values in this paper and \citetalias{2015MNRAS.448.3608B} (or \citetalias{2013MNRAS.435.1126B}) are consistent within error (Y) or not (N). Columns 9 to 12, respectively, list the orbital number (ON), estimated maximum radius ($R_{\text{Max}}$), and maximum mass ($M_{\text{Max}}$) in the Earth radius and mass unit, and the transit probability ($P_{\text{tr}}$). The columns have been sorted in descending order based on the transit probability.}
\label{tableA.1}
\begin{tabularx}{1.55\textwidth}{lccccccccccc}
\hline
Host name & Dis.$^{a}$ & $F_{1}$ & $\frac{\chi^2}{dof}$ & m & b     & \makecell{Period\\(d)} & $F_{2}$ & ON$^{b}$    & \makecell{$R_{\text{Max}}$\\($R_{\oplus}$)} & \makecell{$M_{\text{Max}}$\\($M_{\oplus}$)} & \makecell{$P_{\text{tr}}$\\(\%)} \\
\hline
Solar System & -     & Y     & 1.000 & $0.358\substack{+0.011\\-0.012}$ & $1.885\substack{+0.053\\-0.056}$ & -     & -     & -     & -     & -     & - \\
Kepler-207 & Tr    & Y     & 0.001 & $0.281\substack{+0.001\\-0.001}$ & $0.207\substack{+0.001\\-0.001}$ & $11.2\substack{+0.1\\-0.1}$ & Y     & 3     & 2.1   & 5.2   & 7.31 \\
Kepler-217 & Tr    & Y     & 0.676 & $0.174\substack{+0.036\\-0.027}$ & $0.579\substack{+0.039\\-0.045}$ & $12.6\substack{+5.1\\-3.2}$ & Y     & 3     & 1.8   & 3.9   & 6.90 \\
Kepler-374 & Tr    & Y     & 0.320 & $0.213\substack{+0.033\\-0.021}$ & $0.284\substack{+0.030\\-0.040}$ & $8.4\substack{+2.9\\-1.8}$ & N     & 3 C   & 1.4   & 2.5   & 5.23 \\
Kepler-60 & Tr    & Y     & 0.327 & $0.111\substack{+0.017\\-0.014}$ & $0.849\substack{+0.020\\-0.018}$ & $15.2\substack{+2.7\\-2}$ & Y     & 3     & 2.1   & 5.2   & 4.83 \\
Kepler-23 & Tr    & Y     & 0.135 & $0.166\substack{+0.012\\-0.012}$ & $0.857\substack{+0.017\\-0.013}$ & $22.7\substack{+2.9\\-2.5}$ & Y     & 3     & 2.4   & 6.6   & 4.46 \\
Kepler-223 & Tr    & Y     & 0.318 & $0.146\substack{+0.008\\-0.009}$ & $0.864\substack{+0.017\\-0.016}$ & $28\substack{+3.5\\-3.2}$ & Y     & 4     & 4.2   & 18.3  & 4.23 \\
Kepler-431 & Tr    & Y     & 0.306 & $0.121\substack{+0.013\\-0.013}$ & $0.828\substack{+0.019\\-0.018}$ & $15.5\substack{+2.2\\-1.9}$ & Y     & 3     & 0.8   & 0.9   & 4.18 \\
HR 858 & Tr    & N     & 0.234 & $0.250\substack{+0.026\\-0.020}$ & $0.544\substack{+0.025\\-0.035}$ & $19.7\substack{+5.4\\-3.9}$ & N     & 3     & 2.5   & 7.1   & 4.02 \\
Kepler-256 & Tr    & Y     & 0.236 & $0.269\substack{+0.016\\-0.016}$ & $0.233\substack{+0.032\\-0.029}$ & $20.3\substack{+5.1\\-4}$ & Y     & 4     & 2.8   & 8.7   & 3.99 \\
Kepler-107 & Tr    & Y     & 0.410 & $0.220\substack{+0.016\\-0.016}$ & $0.488\substack{+0.026\\-0.031}$ & $23.3\substack{+5.3\\-4.6}$ & Y     & 4     & 1.1   & 1.6   & 3.92 \\
K2-219 & Tr    & N     & 0.010 & $0.228\substack{+0.005\\-0.003}$ & $0.592\substack{+0.004\\-0.006}$ & $18.9\substack{+0.9\\-0.7}$ & N     & 3     & 1.9   & 4.3   & 3.91 \\
Kepler-226 & Tr    & Y     & 0.474 & $0.157\substack{+0.022\\-0.023}$ & $0.586\substack{+0.030\\-0.030}$ & $11.5\substack{+2.9\\-2.3}$ & Y     & 3     & 1.3   & 2.2   & 3.68 \\
Kepler-271 & Tr    & Y     & 0.001 & $0.149\substack{+0.001\\-0.001}$ & $0.720\substack{+0.001\\-0.001}$ & $14.7\substack{+0.1\\-0.1}$ & N     & 3     & 0.9   & 1.1   & 3.64 \\
Kepler-444 & Tr    & Y     & 0.227 & $0.109\substack{+0.005\\-0.004}$ & $0.558\substack{+0.011\\-0.011}$ & $12.7\substack{+1.1\\-0.9}$ & Y     & 5     & 0.6   & 0.5   & 3.56 \\
Kepler-208 & Tr    & Y     & 0.599 & $0.193\substack{+0.018\\-0.018}$ & $0.650\substack{+0.032\\-0.034}$ & $26.4\substack{+7.3\\-5.7}$ & Y     & 4     & 1.5   & 2.8   & 3.54 \\
Kepler-203 & Tr    & Y     & 0.591 & $0.280\substack{+0.048\\-0.039}$ & $0.483\substack{+0.048\\-0.064}$ & $21\substack{+11.7\\-7.1}$ & Y     & 3     & 1.7   & 3.5   & 3.40 \\
Kepler-758 & Tr    & Y     & 0.215 & $0.208\substack{+0.013\\-0.012}$ & $0.684\substack{+0.021\\-0.027}$ & $32.8\substack{+6\\-5.2}$ & Y     & 4     & 2.2   & 5.6   & 3.11 \\
Kepler-339 & Tr    & Y     & 0.193 & $0.163\substack{+0.015\\-0.014}$ & $0.691\substack{+0.018\\-0.020}$ & $15.2\substack{+2.4\\-2}$ & Y     & 3     & 1.3   & 2.2   & 3.07 \\
Kepler-272 & Tr    & Y     & 0.177 & $0.285\substack{+0.030\\-0.021}$ & $0.480\substack{+0.026\\-0.031}$ & $21.6\substack{+6.5\\-4.2}$ & Y     & 3     & 2.4   & 6.6   & 3.06 \\
Kepler-350 & Tr    & Y     & 0.207 & $0.184\substack{+0.018\\-0.017}$ & $1.054\substack{+0.023\\-0.027}$ & $40.4\substack{+7.9\\-6.7}$ & Y     & 3     & 2.5   & 7.1   & 2.93 \\
K2-148 & Tr    & N     & 0.411 & $0.173\substack{+0.023\\-0.024}$ & $0.652\substack{+0.034\\-0.029}$ & $14.8\substack{+4\\-3.1}$ & N     & 3     & 1.8   & 3.9   & 2.90 \\
Kepler-1254 & Tr    & Y     & 0.171 & $0.224\substack{+0.022\\-0.017}$ & $0.548\substack{+0.024\\-0.027}$ & $16.6\substack{+3.8\\-2.7}$ & Y     & 3     & 1.7   & 3.5   & 2.88 \\
K2-138 & Tr    & N     & 0.007 & $0.183\substack{+0.001\\-0.001}$ & $0.369\substack{+0.003\\-0.003}$ & $19.3\substack{+0.4\\-0.4}$ & N     & 5     & 2.7   & 8.2   & 2.83 \\
Kepler-24 & Tr    & Y     & 0.737 & $0.215\substack{+0.021\\-0.020}$ & $0.655\substack{+0.038\\-0.044}$ & $32.7\substack{+10.6\\-8.1}$ & Y     & 4     & 2.8   & 8.7   & 2.83 \\
\hline
\end{tabularx}
\end{table}

\begin{table}
\centering
\setlength{\tabcolsep}{13.75pt}
\ContinuedFloat
\captionsetup{width=1.55\textwidth}
\caption[]{continued}
\begin{tabularx}{1.55\textwidth}{lccccccccccc}
\hline
Host name & Dis.$^{a}$ & $F_{1}$ & $\frac{\chi^2}{dof}$ & m & b     & \makecell{Period\\(d)} & $F_{2}$ & ON$^{b}$    & \makecell{$R_{\text{Max}}$\\($R_{\oplus}$)} & \makecell{$M_{\text{Max}}$\\($M_{\oplus}$)} & \makecell{$P_{\text{tr}}$\\(\%)} \\
\hline
Kepler-18 & Tr    & Y     & 0.129 & $0.315\substack{+0.026\\-0.024}$ & $0.552\substack{+0.034\\-0.033}$ & $31.5\substack{+9.3\\-6.8}$ & Y     & 3     & 2.7   & 8.2   & 2.69 \\
Kepler-114 & Tr    & Y     & 0.097 & $0.178\substack{+0.012\\-0.011}$ & $0.718\substack{+0.015\\-0.018}$ & $17.9\substack{+2.3\\-1.9}$ & Y     & 3     & 1.7   & 3.5   & 2.57 \\
Kepler-305 & Tr    & Y     & 0.610 & $0.233\substack{+0.023\\-0.020}$ & $0.499\substack{+0.037\\-0.045}$ & $27.0\substack{+9.2\\-6.8}$ & Y     & 5     & 3.0   & 9.9   & 2.50 \\
Kepler-304 & Tr    & Y     & 0.568 & $0.264\substack{+0.024\\-0.023}$ & $0.205\substack{+0.043\\-0.047}$ & $18.3\substack{+7\\-5}$ & Y     & 4     & 2.2   & 5.6   & 2.44 \\
Kepler-206 & Tr    & Y     & 0.052 & $0.240\substack{+0.016\\-0.013}$ & $0.887\substack{+0.017\\-0.019}$ & $40.5\substack{+6.6\\-5}$ & Y     & 3     & 1.4   & 2.5   & 2.38 \\
Kepler-197 & Tr    & Y     & 0.390 & $0.214\substack{+0.016\\-0.017}$ & $0.769\substack{+0.031\\-0.031}$ & $42.1\substack{+10.2\\-8.5}$ & Y     & 4     & 1.0   & 1.3   & 2.34 \\
Kepler-398 & Tr    & Y     & 0.001 & $0.223\substack{+0.001\\-0.001}$ & $0.611\substack{+0.001\\-0.001}$ & $19.1\substack{+0.1\\-0.1}$ & Y     & 3     & 1.1   & 1.6   & 2.34 \\
Kepler-338 & Tr    & Y     & 0.580 & $0.228\substack{+0.020\\-0.019}$ & $0.942\substack{+0.036\\-0.034}$ & $71.6\substack{+21.6\\-16.3}$ & Y     & 4     & 2.6   & 7.6   & 2.29 \\
Kepler-450 & Tr    & Y     & 0.126 & $0.290\substack{+0.023\\-0.020}$ & $0.883\substack{+0.028\\-0.028}$ & $56.5\substack{+14.1\\-10.3}$ & Y     & 3     & 1.4   & 2.5   & 2.27 \\
Kepler-446 & Tr    & Y     & 0.257 & $0.259\substack{+0.029\\-0.026}$ & $0.204\substack{+0.037\\-0.036}$ & $9.6\substack{+3.2\\-2.2}$ & Y     & 3     & 1.5   & 2.8   & 2.21 \\
Kepler-301 & Tr    & Y     & 0.185 & $0.371\substack{+0.037\\-0.029}$ & $0.384\substack{+0.037\\-0.055}$ & $31.5\substack{+12.7\\-8.8}$ & Y     & 3     & 2.1   & 5.2   & 2.18 \\
K2-239 & Tr    & N     & 0.818 & $0.143\substack{+0.026\\-0.026}$ & $0.727\substack{+0.036\\-0.041}$ & $14.3\substack{+4.3\\-3.5}$ & N     & 3     & 1.2   & 1.9   & 2.07 \\
L 98-59 & Tr    & N     & 0.624 & $0.258\substack{+0.042\\-0.042}$ & $0.341\substack{+0.055\\-0.045}$ & $13\substack{+6.7\\-4.2}$ & N     & 3     & 1.2   & 1.9   & 2.04 \\
Kepler-191 & Tr    & Y     & 0.074 & $0.238\substack{+0.013\\-0.013}$ & $0.768\substack{+0.015\\-0.017}$ & $30.3\substack{+4\\-3.6}$ & Y     & 3     & 1.8   & 3.9   & 2.02 \\
YZ Cet & RV    & N     & 0.012 & $0.187\substack{+0.005\\-0.004}$ & $0.296\substack{+0.006\\-0.006}$ & $7.2\substack{+0.4\\-0.3}$ & N     & 3 H   & 1.2   & 1.9   & 1.96 \\
Kepler-85 & Tr    & Y     & 0.106 & $0.160\substack{+0.005\\-0.005}$ & $0.927\substack{+0.010\\-0.011}$ & $37\substack{+2.6\\-2.6}$ & Y     & 4     & 1.4   & 2.5   & 1.95 \\
Kepler-292 & Tr    & Y     & 0.396 & $0.232\substack{+0.012\\-0.012}$ & $0.381\substack{+0.027\\-0.030}$ & $34.7\substack{+7.8\\-6.4}$ & Y     & 5     & 2.5   & 7.1   & 1.91 \\
Kepler-221 & Tr    & Y     & 0.135 & $0.269\substack{+0.012\\-0.013}$ & $0.463\substack{+0.023\\-0.021}$ & $34.7\substack{+6.3\\-5.4}$ & Y     & 4     & 3.1   & 10.5  & 1.89 \\
K2-198 & Tr    & N     & 0.007 & $0.352\substack{+0.005\\-0.007}$ & $0.524\substack{+0.008\\-0.007}$ & $38.2\substack{+2\\-2.3}$ & N     & 3     & 2.6   & 7.6   & 1.75 \\
Kepler-334 & Tr    & Y     & 0.253 & $0.330\substack{+0.035\\-0.036}$ & $0.752\substack{+0.048\\-0.044}$ & $55.1\substack{+23.3\\-16.3}$ & Y     & 3     & 1.7   & 3.5   & 1.70 \\
Kepler-102 & Tr    & Y     & 0.898 & $0.179\substack{+0.014\\-0.013}$ & $0.688\substack{+0.033\\-0.035}$ & $38.2\substack{+10.4\\-7.9}$ & Y     & 5     & 0.8   & 0.9   & 1.66 \\
Kepler-92 & Tr    & Y     & 0.032 & $0.278\substack{+0.010\\-0.010}$ & $1.141\substack{+0.012\\-0.014}$ & $94.3\substack{+10\\-8.9}$ & Y     & 3     & 2.4   & 6.6   & 1.64 \\
Kepler-445 & Tr    & N     & 0.013 & $0.219\substack{+0.005\\-0.004}$ & $0.473\substack{+0.006\\-0.007}$ & $13.4\substack{+0.7\\-0.6}$ & N     & 3 H   & 1.4   & 2.5   & 1.61 \\
Kepler-127 & Tr    & N     & 0.604 & $0.266\substack{+0.053\\-0.038}$ & $1.169\substack{+0.054\\-0.067}$ & $92.8\substack{+58.5\\-31.6}$ & N     & 3     & 2.2   & 5.6   & 1.60 \\
TOI-270 & Tr    & N     & 0.427 & $0.265\substack{+0.033\\-0.034}$ & $0.515\substack{+0.047\\-0.045}$ & $20.4\substack{+8.1\\-5.8}$ & N     & 3 H   & 2.0   & 4.7   & 1.59 \\
Kepler-54 & Tr    & Y     & 0.451 & $0.211\substack{+0.032\\-0.024}$ & $0.892\substack{+0.031\\-0.043}$ & $33.4\substack{+11.3\\-7.8}$ & Y     & 3     & 1.6   & 3.2   & 1.58 \\
Kepler-164 & Tr    & Y     & 0.257 & $0.380\substack{+0.039\\-0.042}$ & $0.689\substack{+0.051\\-0.053}$ & $67.4\substack{+31.5\\-22.8}$ & N     & 3 C   & 2.7   & 8.2   & 1.56 \\
Kepler-224 & Tr    & Y     & 0.215 & $0.261\substack{+0.015\\-0.016}$ & $0.508\substack{+0.029\\-0.028}$ & $35.6\substack{+8.1\\-6.8}$ & Y     & 4     & 2.3   & 6.1   & 1.56 \\
\hline
\end{tabularx}
\end{table}

\begin{table}
\centering
\setlength{\tabcolsep}{13.5pt}
\ContinuedFloat
\captionsetup{width=1.55\textwidth}
\caption[]{continued}
\begin{tabularx}{1.55\textwidth}{lccccccccccc}
\hline
Host name & Dis.$^{a}$ & $F_{1}$ & $\frac{\chi^2}{dof}$ & m & b     & \makecell{Period\\(d)} & $F_{2}$ & ON$^{b}$    & \makecell{$R_{\text{Max}}$\\($R_{\oplus}$)} & \makecell{$M_{\text{Max}}$\\($M_{\oplus}$)} & \makecell{$P_{\text{tr}}$\\(\%)} \\
\hline
Kepler-257 & Tr    & Y     & 0.353 & $0.506\substack{+0.052\\-0.060}$ & $0.352\substack{+0.074\\-0.075}$ & $73.9\substack{+52\\-33}$ & Y     & 3     & 6.2   & 37.1  & 1.54 \\
Kepler-244 & Tr    & Y     & 0.083 & $0.334\substack{+0.021\\-0.020}$ & $0.642\substack{+0.026\\-0.031}$ & $44\substack{+9.9\\-8.4}$ & Y     & 3     & 1.2   & 1.9   & 1.53 \\
Kepler-295 & Tr    & Y     & 0.131 & $0.213\substack{+0.013\\-0.018}$ & $1.110\substack{+0.023\\-0.018}$ & $56\substack{+8.8\\-8.4}$ & Y     & 3     & 1.5   & 2.8   & 1.53 \\
Kepler-104 & Tr    & Y     & 0.026 & $0.328\substack{+0.011\\-0.009}$ & $1.054\substack{+0.013\\-0.013}$ & $109.1\substack{+12.6\\-9.9}$ & Y     & 3     & 4.3   & 19.1  & 1.49 \\
Kepler-172 & Tr    & Y     & 0.044 & $0.359\substack{+0.009\\-0.008}$ & $0.458\substack{+0.016\\-0.015}$ & $78.1\substack{+9.6\\-8.3}$ & Y     & 4     & 3.4   & 12.4  & 1.46 \\
Kepler-84 & Tr    & Y     & 0.548 & $0.253\substack{+0.015\\-0.015}$ & $0.648\substack{+0.036\\-0.035}$ & $81.9\substack{+24.3\\-18.5}$ & Y     & 5     & 2.6   & 7.6   & 1.45 \\
Kepler-247 & Tr    & Y     & 0.435 & $0.392\substack{+0.057\\-0.053}$ & $0.544\substack{+0.075\\-0.074}$ & $52.4\substack{+39.9\\-21.8}$ & Y     & 3     & 3.3   & 11.8  & 1.42 \\
Kepler-327 & Tr    & Y     & 0.520 & $0.373\substack{+0.068\\-0.047}$ & $0.385\substack{+0.065\\-0.078}$ & $31.9\substack{+27.4\\-12.7}$ & Y     & 3     & 1.6   & 3.2   & 1.41 \\
Kepler-770 & Tr    & N     & 0.726 & $0.553\substack{+0.109\\-0.100}$ & $0.140\substack{+0.128\\-0.152}$ & $62.8\substack{+116.8\\-40.7}$ & N     & 3     & 2.8   & 8.7   & 1.41 \\
Kepler-238 & Tr    & Y     & 0.822 & $0.336\substack{+0.025\\-0.025}$ & $0.386\substack{+0.061\\-0.062}$ & $116.5\substack{+62.3\\-40.6}$ & Y     & 5     & 5.3   & 27.9  & 1.40 \\
Kepler-249 & Tr    & Y     & 0.001 & $0.334\substack{+0.001\\-0.001}$ & $0.519\substack{+0.001\\-0.001}$ & $33.1\substack{+0.2\\-0.2}$ & Y     & 3 H   & 1.9   & 4.3   & 1.38 \\
Kepler-81 & Tr    & Y     & 0.313 & $0.271\substack{+0.029\\-0.030}$ & $0.786\substack{+0.039\\-0.040}$ & $39.7\substack{+13.3\\-10.2}$ & Y     & 3     & 1.4   & 2.5   & 1.36 \\
Kepler-83 & Tr    & Y     & 0.079 & $0.296\substack{+0.023\\-0.014}$ & $0.706\substack{+0.020\\-0.031}$ & $39.3\substack{+9.1\\-6.2}$ & Y     & 3     & 2.8   & 8.7   & 1.36 \\
K2-16 & Tr    & N     & 0.069 & $0.423\substack{+0.021\\-0.020}$ & $0.444\substack{+0.025\\-0.027}$ & $51.7\substack{+11.5\\-9.5}$ & N     & 3     & 2.2   & 5.6   & 1.28 \\
K2-58 & Tr    & N     & 0.105 & $0.474\substack{+0.028\\-0.026}$ & $0.396\substack{+0.048\\-0.040}$ & $65.5\substack{+22.9\\-15.7}$ & N     & 3     & 2.2   & 5.6   & 1.28 \\
Kepler-106 & Tr    & Y     & 0.331 & $0.280\substack{+0.016\\-0.020}$ & $0.817\substack{+0.034\\-0.028}$ & $86.2\substack{+21.4\\-18.9}$ & Y     & 4     & 1.3   & 2.2   & 1.26 \\
Kepler-122 & Tr    & Y     & 0.907 & $0.246\substack{+0.021\\-0.021}$ & $0.813\substack{+0.047\\-0.050}$ & $110.3\substack{+47\\-32.8}$ & Y     & 5     & 2.1   & 5.2   & 1.25 \\
Kepler-299 & Tr    & Y     & 0.066 & $0.368\substack{+0.010\\-0.011}$ & $0.465\substack{+0.022\\-0.019}$ & $86.7\substack{+13.2\\-11.5}$ & Y     & 4     & 2.3   & 6.1   & 1.25 \\
KOI-94 & Tr    & N     & 0.185 & $0.384\substack{+0.023\\-0.020}$ & $0.594\substack{+0.035\\-0.041}$ & $134.8\substack{+45.5\\-32.6}$ & N     & 4     & 4.2   & 18.3  & 1.25 \\
K2-32 & Tr    & N     & 0.960 & $0.297\substack{+0.035\\-0.032}$ & $0.658\substack{+0.061\\-0.062}$ & $69.8\substack{+40.7\\-24.6}$ & N     & 4     & 1.5   & 2.8   & 1.23 \\
Kepler-222 & Tr    & Y     & 0.037 & $0.428\substack{+0.023\\-0.014}$ & $0.588\substack{+0.019\\-0.030}$ & $74.6\substack{+16.9\\-11.6}$ & Y     & 3     & 4.7   & 22.4  & 1.21 \\
Kepler-282 & Tr    & Y     & 0.501 & $0.230\substack{+0.019\\-0.021}$ & $0.940\substack{+0.042\\-0.037}$ & $72.7\substack{+22.3\\-17.6}$ & Y     & 4     & 1.7   & 3.5   & 1.18 \\
Kepler-53 & Tr    & Y     & 0.065 & $0.298\substack{+0.017\\-0.015}$ & $0.983\substack{+0.019\\-0.022}$ & $75.5\substack{+13.1\\-11}$ & Y     & 3     & 3.5   & 13.1  & 1.17 \\
Kepler-332 & Tr    & Y     & 0.003 & $0.326\substack{+0.004\\-0.004}$ & $0.881\substack{+0.005\\-0.005}$ & $72.2\substack{+2.7\\-2.6}$ & Y     & 3     & 1.4   & 2.5   & 1.07 \\
K2-233 & Tr    & N     & 0.141 & $0.502\substack{+0.050\\-0.031}$ & $0.373\substack{+0.045\\-0.057}$ & $75.8\substack{+42.9\\-22}$ & N     & 3     & 2.4   & 6.6   & 1.06 \\
Kepler-215 & Tr    & Y     & 0.893 & $0.290\substack{+0.036\\-0.035}$ & $0.933\substack{+0.062\\-0.070}$ & $123.9\substack{+75.2\\-47.5}$ & Y     & 4     & 2.0   & 4.7   & 1.05 \\
Kepler-245 & Tr    & Y     & 0.073 & $0.354\substack{+0.011\\-0.009}$ & $0.515\substack{+0.019\\-0.022}$ & $85.3\substack{+13.5\\-11}$ & Y     & 4     & 3.2   & 11.1  & 1.05 \\
Kepler-325 & Tr    & Y     & 0.027 & $0.467\substack{+0.019\\-0.014}$ & $0.650\substack{+0.019\\-0.024}$ & $112.4\substack{+21.6\\-16.2}$ & Y     & 3     & 3.6   & 13.8  & 1.05 \\
\hline
\end{tabularx}
\end{table}

\begin{table}
\centering
\setlength{\tabcolsep}{12.75pt}
\ContinuedFloat
\captionsetup{width=1.55\textwidth}
\caption[]{continued}
\begin{tabularx}{1.55\textwidth}{lccccccccccc}
\hline
Host name & Dis.$^{a}$ & $F_{1}$ & $\frac{\chi^2}{dof}$ & m & b     & \makecell{Period\\(d)} & $F_{2}$ & ON$^{b}$    & \makecell{$R_{\text{Max}}$\\($R_{\oplus}$)} & \makecell{$M_{\text{Max}}$\\($M_{\oplus}$)} & \makecell{$P_{\text{tr}}$\\(\%)} \\
\hline
K2-72 & Tr    & N     & 0.980 & $0.219\substack{+0.027\\-0.032}$ & $0.722\substack{+0.057\\-0.056}$ & $39.6\substack{+18\\-13.7}$ & N     & 4 H   & 1.5   & 2.8   & 1.03 \\
Kepler-79 & Tr    & Y     & 0.631 & $0.260\substack{+0.024\\-0.027}$ & $1.155\substack{+0.050\\-0.042}$ & $157.1\substack{+62.5\\-45.6}$ & Y     & 4     & 4.1   & 17.5  & 1.00 \\
Kepler-1388 & Tr    & Y     & 0.477 & $0.276\substack{+0.013\\-0.017}$ & $0.767\substack{+0.025\\-0.024}$ & $74.2\substack{+14.6\\-14.0}$ & Y     & 4 C,H & 2.9   & 9.3   & 0.97 \\
Kepler-154 & Tr    & Y     & 0.968 & $0.290\substack{+0.024\\-0.024}$ & $0.664\substack{+0.060\\-0.059}$ & $130.6\substack{+68\\-44.2}$ & N     & 5     & 2.9   & 9.3   & 0.96 \\
Kepler-20 & Tr    & Y     & 0.421 & $0.261\substack{+0.011\\-0.011}$ & $0.532\substack{+0.034\\-0.033}$ & $125.8\substack{+33.1\\-25.5}$ & Y     & 6     & 1.6   & 3.2   & 0.92 \\
Kepler-52 & Tr    & Y     & 0.039 & $0.333\substack{+0.016\\-0.014}$ & $0.891\substack{+0.018\\-0.019}$ & $77.8\substack{+12.9\\-10.4}$ & Y     & 3 H   & 2.4   & 6.6   & 0.89 \\
Kepler-176 & Tr    & Y     & 0.170 & $0.324\substack{+0.018\\-0.015}$ & $0.754\substack{+0.029\\-0.032}$ & $112\substack{+28.9\\-21.8}$ & Y     & 4     & 1.8   & 3.9   & 0.87 \\
Kepler-171 & Tr    & Y     & 0.210 & $0.487\substack{+0.044\\-0.046}$ & $0.602\substack{+0.058\\-0.053}$ & $115.7\substack{+63.2\\-41}$ & Y     & 3     & 2.5   & 7.1   & 0.86 \\
HD 20781 & RV    & N     & 0.261 & $0.393\substack{+0.025\\-0.024}$ & $0.728\substack{+0.047\\-0.045}$ & $200.4\substack{+79.7\\-55.4}$ & N     & 4 H   & 6.3   & 38.0  & 0.85 \\
Kepler-31 & Tr    & Y     & 0.001 & $0.312\substack{+0.001\\-0.001}$ & $1.319\substack{+0.002\\-0.001}$ & $179.5\substack{+1.9\\-1.8}$ & Y     & 3     & 4.7   & 22.4  & 0.85 \\
Kepler-229 & Tr    & Y     & 0.001 & $0.409\substack{+0.001\\-0.001}$ & $0.796\substack{+0.001\\-0.001}$ & $105.8\substack{+0.4\\-0.5}$ & Y     & 3     & 4.5   & 20.7  & 0.84 \\
Kepler-331 & Tr    & Y     & 0.101 & $0.289\substack{+0.019\\-0.020}$ & $0.935\substack{+0.024\\-0.029}$ & $63.5\substack{+13.1\\-11.7}$ & Y     & 3 H   & 1.9   & 4.3   & 0.83 \\
Kepler-357 & Tr    & Y     & 0.072 & $0.440\substack{+0.021\\-0.023}$ & $0.803\substack{+0.030\\-0.028}$ & $132.9\substack{+32\\-26.9}$ & Y     & 3     & 3.9   & 16.0  & 0.83 \\
Kepler-288 & Tr    & Y     & 0.024 & $0.485\substack{+0.018\\-0.014}$ & $0.790\substack{+0.018\\-0.022}$ & $175.7\substack{+31.3\\-23.6}$ & Y     & 3     & 3.5   & 13.1  & 0.80 \\
Kepler-55 & Tr    & Y     & 0.900 & $0.335\substack{+0.025\\-0.025}$ & $0.349\substack{+0.066\\-0.063}$ & $105.6\substack{+58.3\\-37}$ & Y     & 5 H   & 2.8   & 8.7   & 0.77 \\
K2-3  & Tr    & N     & 0.850 & $0.322\substack{+0.056\\-0.058}$ & $1.023\substack{+0.071\\-0.076}$ & $97.4\substack{+71.8\\-42.8}$ & N     & 3 H   & 1.8   & 3.9   & 0.74 \\
K2-155 & Tr    & N     & 0.516 & $0.404\substack{+0.058\\-0.057}$ & $0.782\substack{+0.073\\-0.071}$ & $98.4\substack{+75.2\\-42.1}$ & N     & 3 H   & 2.4   & 6.6   & 0.74 \\
Kepler-399 & Tr    & Y     & 0.276 & $0.304\substack{+0.035\\-0.031}$ & $1.146\substack{+0.041\\-0.043}$ & $114\substack{+44.9\\-30.4}$ & Y     & 3     & 1.6   & 3.2   & 0.70 \\
TRAPPIST-1 & Tr    & N     & 0.574 & $0.181\substack{+0.007\\-0.006}$ & $0.213\substack{+0.024\\-0.023}$ & $30\substack{+5.2\\-4.3}$ & N     & 7     & 0.8   & 0.9   & 0.68 \\
Kepler-19 & Tr    & N     & 0.659 & $0.415\substack{+0.061\\-0.067}$ & $0.991\substack{+0.083\\-0.081}$ & $172\substack{+145.6\\-81.9}$ & N     & 3     & 4.6   & 21.5  & 0.67 \\
Kepler-235 & Tr    & Y     & 0.032 & $0.383\substack{+0.008\\-0.007}$ & $0.522\substack{+0.016\\-0.016}$ & $113.3\substack{+13.2\\-11}$ & Y     & 4 H   & 2.5   & 7.1   & 0.66 \\
Kepler-130 & Tr    & Y     & 0.002 & $0.507\substack{+0.004\\-0.005}$ & $0.929\substack{+0.006\\-0.006}$ & $282.4\substack{+12.5\\-12.9}$ & Y     & 3     & 2.2   & 5.6   & 0.62 \\
Kepler-166 & Tr    & Y     & 0.019 & $0.672\substack{+0.023\\-0.018}$ & $0.197\substack{+0.026\\-0.029}$ & $163.1\substack{+39.9\\-28}$ & Y     & 3 H   & 3.5   & 13.1  & 0.61 \\
Kepler-296 & Tr    & Y     & 0.033 & $0.256\substack{+0.004\\-0.004}$ & $0.773\substack{+0.008\\-0.010}$ & $113.6\substack{+8.2\\-7.2}$ & Y     & 5 H   & 2.1   & 5.2   & 0.61 \\
Kepler-289 & Tr    & N     & 0.001 & $0.281\substack{+0.001\\-0.001}$ & $1.539\substack{+0.001\\-0.001}$ & $240.4\substack{+1.9\\-1.7}$ & N     & 3     & 3.5   & 13.1  & 0.59 \\
Kepler-51 & Tr    & Y     & 0.825 & $0.231\substack{+0.046\\-0.040}$ & $1.668\substack{+0.050\\-0.058}$ & $230.1\substack{+124.4\\-77.4}$ & Y     & 3     & 10.7  & 100.0 & 0.58 \\
Kepler-251 & Tr    & Y     & 0.891 & $0.423\substack{+0.046\\-0.046}$ & $0.712\substack{+0.100\\-0.093}$ & $252.6\substack{+232.8\\-119.5}$ & Y     & 4 H   & 3.5   & 13.1  & 0.53 \\
HD 20794 & RV    & N     & 0.702 & $0.305\substack{+0.029\\-0.027}$ & $1.288\substack{+0.053\\-0.049}$ & $322.7\substack{+155\\-98.2}$ & N     & 4 H   & 3.3   & 11.9  & 0.52 \\
\hline
\end{tabularx}
\end{table}

\begin{table}
\centering
\setlength{\tabcolsep}{12.5pt}
\ContinuedFloat
\captionsetup{width=1.55\textwidth}
\caption[]{continued}
\begin{tabularx}{1.57\textwidth}{lccccccccccc}
\hline
Host name & Dis.$^{a}$ & $F_{1}$ & $\frac{\chi^2}{dof}$ & m & b     & \makecell{Period\\(d)} & $F_{2}$ & ON$^{b}$    & \makecell{$R_{\text{Max}}$\\($R_{\oplus}$)} & \makecell{$M_{\text{Max}}$\\($M_{\oplus}$)} & \makecell{$P_{\text{tr}}$\\(\%)} \\
\hline
Kepler-30 & Tr    & Y     & 0.169 & $0.346\substack{+0.031\\-0.030}$ & $1.455\substack{+0.037\\-0.043}$ & $310.2\substack{+108.4\\-81.9}$ & Y     & 3 H   & 7.0   & 46.2  & 0.49 \\
Kepler-298 & Tr    & Y     & 0.966 & $0.442\substack{+0.102\\-0.075}$ & $0.981\substack{+0.103\\-0.134}$ & $202.2\substack{+317.5\\-113.7}$ & Y     & 3 H   & 3.2   & 11.1  & 0.45 \\
Kepler-401 & Tr    & Y     & 0.088 & $0.554\substack{+0.042\\-0.032}$ & $1.145\substack{+0.041\\-0.054}$ & $640.4\substack{+299.4\\-188.2}$ & Y     & 3 H   & 3.1   & 10.5  & 0.40 \\
Kepler-603 & Tr    & Y     & 0.754 & $0.656\substack{+0.137\\-0.111}$ & $0.753\substack{+0.155\\-0.157}$ & $527.1\substack{+1408.5\\-356.9}$ & Y     & 3 H   & 4.0   & 16.7  & 0.36 \\
GJ 3293 & RV    & N     & 0.871 & $0.311\substack{+0.033\\-0.036}$ & $1.130\substack{+0.061\\-0.072}$ & $237.1\substack{+132\\-92.8}$ & N     & 4     & 6.9   & 45.5  & 0.33 \\
HD 69830 & RV    & N     & 0.607 & $0.679\substack{+0.110\\-0.110}$ & $0.892\substack{+0.110\\-0.133}$ & $850.5\substack{+1496\\-557.7}$ & N     & 3     & 10.7  & 99.8  & 0.25 \\
Kepler-174 & Tr    & Y     & 0.824 & $0.625\substack{+0.096\\-0.114}$ & $1.101\substack{+0.136\\-0.129}$ & $944.4\substack{+1567.8\\-625.4}$ & Y     & 3     & 3.1   & 10.5  & 0.18 \\
tau Cet & RV    & N     & 0.481 & $0.499\substack{+0.040\\-0.041}$ & $1.253\substack{+0.078\\-0.071}$ & $1778.9\substack{+1300.9\\-739.7}$ & N     & 4     & 3.5   & 13.1  & 0.14 \\
HD 10180 & RV    & Y     & 0.617 & $0.513\substack{+0.026\\-0.024}$ & $0.699\substack{+0.080\\-0.078}$ & $5965.2\substack{+4366\\-2405.9}$ & Y     & 6     & 12.1  & 209.7 & 0.08 \\
GJ 676 A & RV    & N     & 0.669 & $1.136\substack{+0.100\\-0.110}$ & $0.543\substack{+0.186\\-0.194}$ & $121691.5\substack{+347768.7\\-93338.3}$ & N     & 4     & -     & -     & 0.01 \\
HR 8799 & Im    & Y     & 0.114 & $0.317\substack{+0.092\\-0.086}$ & $4.273\substack{+0.165\\-0.157}$ & $348901.9\substack{+841068.7\\-238461.9}$ & Y     & 4     & -     & -     & 0.01 \\
HD 31527 & RV    & N     & 0.764 & $0.609\substack{+0.109\\-0.116}$ & $1.186\substack{+0.147\\-0.136}$ & $1029.3\substack{+2046.7\\-692.1}$ & N     & 3     & 7.8   & 55.9  & - \\
HD 136352 & RV    & N     & 0.989 & $0.484\substack{+0.094\\-0.099}$ & $1.028\substack{+0.121\\-0.127}$ & $302.2\substack{+462.8\\-188.2}$ & N     & 3 H   & 5.4   & 28.6  & - \\
Kepler-402 & Tr    & Y     & 0.952 & $0.150\substack{+0.017\\-0.017}$ & $0.623\substack{+0.032\\-0.030}$ & $16.7\substack{+4.3\\-3.3}$ & N     & 4 C   & 1.6   & 3.2   & - \\
\hline
\end{tabularx}
\small
\begin{tablenotes}
\item$^{a}$Discovery method of the system: `Tr', `RV', and `Im' represent transit, radial velocity, and imaging, respectively.\item$^b$Orbital numbers (ON) followed by `H' indicate the predicted planets within the HZ, and ON followed by `C' indicate that the corresponding orbital periods have been flagged as ``Planetary Candidate'' in NASA Exoplanet Archive.
\end{tablenotes}
\end{table}

\begin{table}
\centering
\setlength{\tabcolsep}{7.5pt}
\captionsetup{width=1.55\textwidth}
\caption[Systems with interpolated and extrapolated planet predictions]{Systems with interpolated and extrapolated planet predictions. Columns 1 and 2 present the host star name and discovery method (Dis.). Column 3 reports a flag that defines whether the system has already been analyzed by \citetalias{2015MNRAS.448.3608B} (or \citetalias{2013MNRAS.435.1126B}) (Y) or not (N). Columns 4 and 5 present $\chi^2/dof$ before and after interpolation. Columns 6 to 10, respectively, list the $\gamma$, $\Delta\gamma$, slope (m), intercept (b), and predicted orbital period. Column 11 reports whether the predicted period values in this paper and \citetalias{2015MNRAS.448.3608B} (or \citetalias{2013MNRAS.435.1126B}) are consistent within error (Y) or not (N). Columns 12 to 15, respectively, list the orbital number (ON), estimated maximum radius ($R_{\text{Max}}$), and maximum mass ($M_{\text{Max}}$) in the Earth radius and mass unit, and the transit probability ($P_{\text{tr}}$). The columns have been sorted in descending order based on the transit probability.}
\begin{tabularx}{1.55\textwidth}{lcccccccccccccc}
\hline
Host name & Dis.$^{a}$  & $F_{1}$ & $(\frac{\chi^2}{dof})_{i}$ & $(\frac{\chi^2}{dof})_{j}$ & $\gamma$ & $\Delta\gamma$$^{b}$ & m & b & \makecell{Period\\(d)} & $F_{2}$ & ON$^{c}$    & \makecell{$R_{\text{Max}}$\\($R_{\oplus}$)} & \makecell{$M_{\text{Max}}$\\($M_{\oplus}$)} & \makecell{$P_{\text{tr}}$\\(\%)} \\
\hline
K2-183 & Tr    & N     & 7.718 & 0.024 & 107.1 & 3.1   & \multicolumn{1}{l}{$0.338\substack{+0.003\\-0.003}$} & \multicolumn{1}{l}{$-0.330\substack{+0.010\\-0.010}$} & $1\substack{+0.1\\-0.1}$ & N     & 1     & 1.3   & 2.2   & 22.31 \\
      &       &       &       &       &       &       &       &       & $2.1\substack{+0.1\\-0.1}$ & N     & 2     & 1.6   & 3.2   & 14.87 \\
      &       &       &       &       &       &       &       &       & $4.9\substack{+0.2\\-0.2}$ & N     & 3     & 2.0   & 4.7   & 8.11 \\
      &       &       &       &       &       &       &       &       & $49.8\substack{+3.6\\-3.2}$ & N     & 6 E   & 3.5   & 13.1  & 1.72 \\
Kepler-342 & Tr    & Y     & 10.136 & 0.203 & 16.3  & 0.7   & \multicolumn{1}{l}{$0.236\substack{+0.005\\-0.005}$} & \multicolumn{1}{l}{$0.212\substack{+0.019\\-0.017}$} & $2.7\substack{+0.2\\-0.1}$ & Y     & 1     & 1.0   & 1.3   & 16.91 \\
      &       &       &       &       &       &       &       &       & $4.7\substack{+0.3\\-0.3}$ & Y     & 2     & 1.2   & 1.9   & 11.27 \\
      &       &       &       &       &       &       &       &       & $8.5\substack{+0.7\\-0.6}$ & N     & 3     & 1.3   & 2.2   & 0.80 \\
\hline
\end{tabularx}
\label{tableA.2}
\end{table}

\begin{table}
\centering
\setlength{\tabcolsep}{6.25pt}
\captionsetup{width=1.55\textwidth}
\ContinuedFloat
\caption[]{continued}
\begin{tabularx}{1.55\textwidth}{lcccccccccccccc}
\hline
Host name & Dis.$^{a}$  & $F_{1}$ & $(\frac{\chi^2}{dof})_{i}$ & $(\frac{\chi^2}{dof})_{j}$ & $\gamma$ & $\Delta\gamma$$^{b}$ & m & b & \makecell{Period\\(d)} & $F_{2}$ & ON$^{c}$    & \makecell{$R_{\text{Max}}$\\($R_{\oplus}$)} & \makecell{$M_{\text{Max}}$\\($M_{\oplus}$)} & \makecell{$P_{\text{tr}}$\\(\%)} \\
\hline
Kepler-363 & Tr    & N     & 1.090 & 0.060 & 5.7   & 0.1   & \multicolumn{1}{l}{$0.104\substack{+0.001\\-0.002}$} & \multicolumn{1}{l}{$0.561\substack{+0.005\\-0.004}$} & $4.6\substack{+0.1\\-0.1}$ & N     & 1     & 1.2   & 1.9   & 11.42 \\
      &       &       &       &       &       &       &       &       & $6\substack{+0.1\\-0.1}$ & N     & 2     & 1.3   & 2.2   & 9.79 \\
      &       &       &       &       &       &       &       &       & $9.6\substack{+0.2\\-0.2}$ & N     & 4     & 1.5   & 2.8   & 7.62 \\
      &       &       &       &       &       &       &       &       & $15.4\substack{+0.5\\-0.5}$ & N     & 6 E   & 1.7   & 3.5   & 5.27 \\
KIC 10001893 & OBM   & N     & 3.259 & 0.076 & 41.9  & 3.9   & \multicolumn{1}{l}{$0.190\substack{+0.006\\-0.006}$} & \multicolumn{1}{l}{$-0.665\substack{+0.011\\-0.011}$} & $0.5\substack{+0.1\\-0.1}$ & N     & 2     & -     & -     & 11.27 \\
      &       &       &       &       &       &       &       &       & $1.2\substack{+0.1\\-0.1}$ & N     & 4 E   & -     & -     & 6.27 \\
55 Cnc & RV    & Y     & 1.441 & 0.139 & 4.7   & 4.0   & \multicolumn{1}{l}{$0.634\substack{+0.013\\-0.013}$} & \multicolumn{1}{l}{$-0.147\substack{+0.042\\-0.047}$} & $3.1\substack{+0.4\\-0.4}$ & Y     & 1     & 0.4   & 0.2   & 10.81 \\
\hline
\end{tabularx}
\end{table}

\begin{table}
\centering
\setlength{\tabcolsep}{7.75pt}
\captionsetup{width=1.55\textwidth}
\ContinuedFloat
\caption[]{continued}
\begin{tabularx}{1.55\textwidth}{lcccccccccccccc}
\hline
Host name & Dis.$^{a}$  & $F_{1}$ & $(\frac{\chi^2}{dof})_{i}$ & $(\frac{\chi^2}{dof})_{j}$ & $\gamma$ & $\Delta\gamma$$^{b}$ & m & b & \makecell{Period\\(d)} & $F_{2}$ & ON$^{c}$    & \makecell{$R_{\text{Max}}$\\($R_{\oplus}$)} & \makecell{$M_{\text{Max}}$\\($M_{\oplus}$)} & \makecell{$P_{\text{tr}}$\\(\%)} \\
\hline
Kepler-326 & Tr    & N     & 1.730 & 0.030 & 55.8  & 13.2  & \multicolumn{1}{l}{$0.158\substack{+0.003\\-0.004}$} & \multicolumn{1}{l}{$0.350\substack{+0.007\\-0.007}$} & $3.2\substack{+0.1\\-0.1}$ & N     & 1     & 1.0   & 1.3   & 8.98 \\
      &       &       &       &       &       &       &       &       & $9.6\substack{+0.5\\-0.5}$ & N     & 4 E   & 1.3   & 2.2   & 4.28 \\
Kepler-9 & Tr    & N     & 6.300 & 0.060 & 51.6  & 0.8   & \multicolumn{1}{l}{$0.349\substack{+0.007\\-0.007}$} & \multicolumn{1}{l}{$0.210\substack{+0.017\\-0.016}$} & $3.6\substack{+0.2\\-0.2}$ & N     & 1     & 2.0   & 4.7   & 8.83 \\
      &       &       &       &       &       &       &       &       & $8.1\substack{+0.6\\-0.5}$ & N     & 2     & 2.5   & 7.1   & 5.52 \\
      &       &       &       &       &       &       &       &       & $90.5\substack{+11.2\\-10}$ & N     & 5 E   & 4.5   & 20.7  & 1.10 \\
K2-19 & Tr    & N     & 4.627 & 0.012 & 188.6 & 14.9  & \multicolumn{1}{l}{$0.169\substack{+0.002\\-0.002}$} & \multicolumn{1}{l}{$0.397\substack{+0.004\\-0.004}$} & $3.7\substack{+0.1\\-0.1}$ & N     & 1     & 1.3   & 2.2   & 8.79 \\
      &       &       &       &       &       &       &       &       & $5.4\substack{+0.1\\-0.1}$ & N     & 2     & 1.4   & 2.5   & 6.59 \\
\hline
\end{tabularx}
\end{table}

\begin{table}
\centering
\setlength{\tabcolsep}{8.25pt}
\captionsetup{width=1.55\textwidth}
\ContinuedFloat
\caption[]{continued}
\begin{tabularx}{1.55\textwidth}{lcccccccccccccc}
\hline
Host name & Dis.$^{a}$  & $F_{1}$ & $(\frac{\chi^2}{dof})_{i}$ & $(\frac{\chi^2}{dof})_{j}$ & $\gamma$ & $\Delta\gamma$$^{b}$ & m & b & \makecell{Period\\(d)} & $F_{2}$ & ON$^{c}$    & \makecell{$R_{\text{Max}}$\\($R_{\oplus}$)} & \makecell{$M_{\text{Max}}$\\($M_{\oplus}$)} & \makecell{$P_{\text{tr}}$\\(\%)} \\
\hline
Kepler-228 & Tr    & N     & 2.456 & 0.004 & 607.6 & 115.8 & \multicolumn{1}{l}{$0.212\substack{+0.002\\-0.002}$} & \multicolumn{1}{l}{$0.407\substack{+0.003\\-0.003}$} & $6.8\substack{+0.1\\-0.1}$ & N     & 2     & 1.9   & 4.3   & 6.64 \\
      &       &       &       &       &       &       &       &       & $18\substack{+0.4\\-0.4}$ & N     & 4 E   & 2.5   & 7.1   & 3.32 \\
Kepler-142 & Tr    & N     & 3.852 & 0.075 & 25.2  & 0.2   & \multicolumn{1}{l}{$0.325\substack{+0.007\\-0.008}$} & \multicolumn{1}{l}{$0.326\substack{+0.018\\-0.018}$} & $9.5\substack{+0.8\\-0.7}$ & N     & 2     & 1.5   & 2.8   & 6.49 \\
      &       &       &       &       &       &       &       &       & $19.8\substack{+2\\-1.8}$ & N     & 3     & 1.8   & 3.9   & 4.17 \\
      &       &       &       &       &       &       &       &       & $89.7\substack{+12.2\\-11}$ & N     & 5 E   & 2.6   & 7.6   & 1.50 \\
Kepler-49 & Tr    & Y     & 2.424 & 0.085 & 27.5  & 13.1  & \multicolumn{1}{l}{$0.213\substack{+0.005\\-0.005}$} & \multicolumn{1}{l}{$0.415\substack{+0.013\\-0.012}$} & $4.2\substack{+0.2\\-0.2}$ & Y     & 1     & 1.1   & 1.6   & 6.13 \\
      &       &       &       &       &       &       &       &       & $30\substack{+2.9\\-2.5}$ & Y     & 5 E   & 1.8   & 3.9   & 1.72 \\
\hline
\end{tabularx}
\end{table}

\begin{table}
\centering
\setlength{\tabcolsep}{8.0pt}
\captionsetup{width=1.55\textwidth}
\ContinuedFloat
\caption[]{continued}
\begin{tabularx}{1.55\textwidth}{lcccccccccccccc}
\hline
Host name & Dis.$^{a}$  & $F_{1}$ & $(\frac{\chi^2}{dof})_{i}$ & $(\frac{\chi^2}{dof})_{j}$ & $\gamma$ & $\Delta\gamma$$^{b}$ & m & b & \makecell{Period\\(d)} & $F_{2}$ & ON$^{c}$    & \makecell{$R_{\text{Max}}$\\($R_{\oplus}$)} & \makecell{$M_{\text{Max}}$\\($M_{\oplus}$)} & \makecell{$P_{\text{tr}}$\\(\%)} \\
\hline
K2-80 & Tr    & N     & 4.976 & 0.004 & 610.9 & 61.8  & \multicolumn{1}{l}{$0.178\substack{+0.001\\-0.001}$} & \multicolumn{1}{l}{$0.750\substack{+0.002\\-0.002}$} & $8.5\substack{+0.1\\-0.1}$ & N     & 1     & 1.6   & 3.2   & 5.00 \\
      &       &       &       &       &       &       &       &       & $12.8\substack{+0.1\\-0.1}$ & N     & 2     & 1.8   & 3.9   & 4.00 \\
      &       &       &       &       &       &       &       &       & $43.5\substack{+0.7\\-0.7}$ & N     & 5 E   & 2.5   & 7.1   & 1.74 \\
Kepler-26 & Tr    & Y     & 2.209 & 0.288 & 2.2   & 0.5   & \multicolumn{1}{l}{$0.185\substack{+0.005\\-0.005}$} & \multicolumn{1}{l}{$0.528\substack{+0.017\\-0.018}$} & $5.1\substack{+0.3\\-0.3}$ & Y     & 1     & 1.2   & 1.9   & 4.99 \\
      &       &       &       &       &       &       &       &       & $7.8\substack{+0.5\\-0.5}$ & Y     & 2     & 1.3   & 2.2   & 3.72 \\
      &       &       &       &       &       &       &       &       & $27.7\substack{+2.8\\-2.6}$ & Y     & 5     & 1.8   & 3.9   & 1.56 \\
      &       &       &       &       &       &       &       &       & $66.6\substack{+8.1\\-7.4}$ & Y     & 7 E,H & 2.2   & 5.6   & 0.89 \\
\hline
\end{tabularx}
\end{table}

\begin{table}
\centering
\setlength{\tabcolsep}{7.75pt}
\captionsetup{width=1.55\textwidth}
\ContinuedFloat
\caption[]{continued}
\begin{tabularx}{1.55\textwidth}{lcccccccccccccc}
\hline
Host name & Dis.$^{a}$  & $F_{1}$ & $(\frac{\chi^2}{dof})_{i}$ & $(\frac{\chi^2}{dof})_{j}$ & $\gamma$ & $\Delta\gamma$$^{b}$ & m & b & \makecell{Period\\(d)} & $F_{2}$ & ON$^{c}$    & \makecell{$R_{\text{Max}}$\\($R_{\oplus}$)} & \makecell{$M_{\text{Max}}$\\($M_{\oplus}$)} & \makecell{$P_{\text{tr}}$\\(\%)} \\
\hline
HD 215152 & RV    & N     & 4.762 & 0.075 & 15.7  & 1.1   & \multicolumn{1}{l}{$0.091\substack{+0.001\\-0.001}$} & \multicolumn{1}{l}{$0.766\substack{+0.004\\-0.004}$} & $8.9\substack{+0.1\\-0.1}$ & N     & 2     & 0.8   & 0.9   & 4.36 \\
      &       &       &       &       &       &       &       &       & $13.7\substack{+0.2\\-0.2}$ & N     & 4     & 1.0   & 1.4   & 3.36 \\
      &       &       &       &       &       &       &       &       & $16.3\substack{+0.3\\-0.3}$ & N     & 5     & 1.1   & 1.7   & 2.92 \\
      &       &       &       &       &       &       &       &       & $20.4\substack{+0.5\\-0.4}$ & N     & 6     & 1.3   & 2.3   & 2.51 \\
      &       &       &       &       &       &       &       &       & $31\substack{+0.9\\-0.8}$ & N     & 8 E   & 1.7   & 3.7   & 1.90 \\
Kepler-250 & Tr    & N     & 1.237 & 0.131 & 8.5   & 0.7   & \multicolumn{1}{l}{$0.207\substack{+0.009\\-0.009}$} & \multicolumn{1}{l}{$0.629\substack{+0.016\\-0.018}$} & $11.1\substack{+0.9\\-0.9}$ & N     & 2     & 1.5   & 2.8   & 4.23 \\
      &       &       &       &       &       &       &       &       & $28.8\substack{+3.7\\-3.4}$ & N     & 4 E   & 1.8   & 3.9   & 2.19 \\
\hline
\end{tabularx}
\end{table}

\begin{table}
\centering
\setlength{\tabcolsep}{6.0pt}
\captionsetup{width=1.55\textwidth}
\ContinuedFloat
\caption[]{continued}
\begin{tabularx}{1.55\textwidth}{lcccccccccccccc}
\hline
Host name & Dis.$^{a}$  & $F_{1}$ & $(\frac{\chi^2}{dof})_{i}$ & $(\frac{\chi^2}{dof})_{j}$ & $\gamma$ & $\Delta\gamma$$^{b}$ & m & b & \makecell{Period\\(d)} & $F_{2}$ & ON$^{c}$    & \makecell{$R_{\text{Max}}$\\($R_{\oplus}$)} & \makecell{$M_{\text{Max}}$\\($M_{\oplus}$)} & \makecell{$P_{\text{tr}}$\\(\%)} \\
\hline
HIP 41378 & Tr    & N     & 2.078 & 0.804 & 0.3   & 0.1   & \multicolumn{1}{l}{$0.150\substack{+0.003\\-0.003}$} & \multicolumn{1}{l}{$-0.723\substack{+0.009\\-0.007}$} & $22.1\substack{+0.6\\-0.5}$ & N     & 1     & 2.3   & 6.1   & 3.80 \\
      &       &       &       &       &       &       &       &       & $44.8\substack{+2\\-1.7}$ & N     & 3     & 2.8   & 8.7   & 2.37 \\
      &       &       &       &       &       &       &       &       & $82.8\substack{+5.4\\-4.8}$ & N     & 5     & 3.2   & 11.1  & 1.57 \\
      &       &       &       &       &       &       &       &       & $266.5\substack{+21.4\\-18.8}$ & N     & 8     & 4.3   & 19.1  & 0.72 \\
      &       &       &       &       &       &       &       &       & $301.4\substack{+33.3\\-29}$ & N     & 9     & 4.5   & 20.7  & 0.67 \\
      &       &       &       &       &       &       &       &       & $701.2\substack{+78.3\\-67.8}$ & N     & 11 E,H & 5.5   & 29.8  & 0.38 \\
Kepler-46 & Tr    & N     & 5.195 & 0.003 & 861.2 & 93.4  & \multicolumn{1}{l}{$0.231\substack{+0.001\\-0.001}$} & \multicolumn{1}{l}{$0.831\substack{+0.003\\-0.002}$} & $11.6\substack{+0.1\\-0.1}$ & N     & 1     & 1.9   & 4.3   & 3.75 \\
\hline
\end{tabularx}
\end{table}

\begin{table}
\centering
\setlength{\tabcolsep}{6.75pt}
\captionsetup{width=1.55\textwidth}
\ContinuedFloat
\caption[]{continued}
\begin{tabularx}{1.55\textwidth}{lcccccccccccccc}
\hline
Host name & Dis.$^{a}$  & $F_{1}$ & $(\frac{\chi^2}{dof})_{i}$ & $(\frac{\chi^2}{dof})_{j}$ & $\gamma$ & $\Delta\gamma$$^{b}$ & m & b & \makecell{Period\\(d)} & $F_{2}$ & ON$^{c}$    & \makecell{$R_{\text{Max}}$\\($R_{\oplus}$)} & \makecell{$M_{\text{Max}}$\\($M_{\oplus}$)} & \makecell{$P_{\text{tr}}$\\(\%)} \\
\hline
Kepler-58 & Tr    & Y     & 2.970 & 0.044 & 66.2  & 10.7  & \multicolumn{1}{l}{$0.199\substack{+0.005\\-0.005}$} & \multicolumn{1}{l}{$1.003\substack{+0.008\\-0.008}$} & $25.2\substack{+1\\-1}$ & Y     & 2     & 2.6   & 7.6   & 3.25 \\
      &       &       &       &       &       &       &       &       & $62.8\substack{+4\\-3.9}$ & Y     & 4 E   & 3.3   & 11.8  & 1.73 \\
KOI-351 & Tr    & Y     & 5.448 & 0.448 & 5.6   & 1.1   & \multicolumn{1}{l}{$0.190\substack{+0.004\\-0.003}$} & \multicolumn{1}{l}{$0.801\substack{+0.020\\-0.018}$} & $23.6\substack{+1.7\\-1.5}$ & Y     & 3     & 1.5   & 2.8   & 3.25 \\
      &       &       &       &       &       &       &       &       & $35.9\substack{+3\\-2.5}$ & Y     & 4     & 1.7   & 3.5   & 2.40 \\
      &       &       &       &       &       &       &       &       & $499.9\substack{+70.4\\-55.6}$ & Y     & 10 E,H & 3.2   & 11.1  & 0.42 \\
HD 125612 & RV    & N     & 5.092 & 0.003 & 891.4 & 95.5  & \multicolumn{1}{l}{$0.710\substack{+0.003\\-0.003}$} & \multicolumn{1}{l}{$0.615\substack{+0.008\\-0.008}$} & $20.6\substack{+0.5\\-0.5}$ & N     & 1     & 2.6   & 7.4   & 3.22 \\
      &       &       &       &       &       &       &       &       & $110.1\substack{+3.4\\-3.5}$ & N     & 2     & 7.5   & 52.3  & 1.05 \\
\hline
\end{tabularx}
\end{table}

\begin{table}
\centering
\setlength{\tabcolsep}{6.5pt}
\captionsetup{width=1.55\textwidth}
\ContinuedFloat
\caption[]{continued}
\begin{tabularx}{1.55\textwidth}{lcccccccccccccc}
\hline
Host name & Dis.$^{a}$  & $F_{1}$ & $(\frac{\chi^2}{dof})_{i}$ & $(\frac{\chi^2}{dof})_{j}$ & $\gamma$ & $\Delta\gamma$$^{b}$ & m & b & \makecell{Period\\(d)} & $F_{2}$ & ON$^{c}$    & \makecell{$R_{\text{Max}}$\\($R_{\oplus}$)} & \makecell{$M_{\text{Max}}$\\($M_{\oplus}$)} & \makecell{$P_{\text{tr}}$\\(\%)} \\
\hline
V1298 Tau & Tr    & N     & 3.004 & 0.518 & 4.8   & 7.0   & \multicolumn{1}{l}{$0.229\substack{+0.034\\-0.046}$} & \multicolumn{1}{l}{$0.897\substack{+0.052\\-0.043}$} & $38.3\substack{+16.1\\-13.1}$ & N     & 3     & 7.8   & 56.3  & 2.69 \\
      &       &       &       &       &       &       &       &       & $109.7\substack{+72.2\\-51.2}$ & N     & 5 E   & 10.2  & 21.8  & 1.33 \\
PSR B1257+12 & PT    & N     & 3.673 & 0.151 & 23.3  & 0.3   & \multicolumn{1}{l}{$0.199\substack{+0.009\\-0.010}$} & \multicolumn{1}{l}{$1.407\substack{+0.017\\-0.019}$} & $40.5\substack{+2.5\\-2.6}$ & N     & 1     & -     & -     & 2.50 \\
      &       &       &       &       &       &       &       &       & $159.4\substack{+21.1\\-19.8}$ & N     & 4 E   & -     & -     & 1.00 \\
GJ 581 & RV    & N     & 1.221 & 0.139 & 7.8   & 0.1   & \multicolumn{1}{l}{$0.202\substack{+0.010\\-0.011}$} & \multicolumn{1}{l}{$0.509\substack{+0.020\\-0.017}$} & $8.2\substack{+0.8\\-0.7}$ & N     & 2     & 1.6   & 3.2   & 2.47 \\
      &       &       &       &       &       &       &       &       & $20.8\substack{+3.1\\-2.7}$ & N     & 4 E   & 2.9   & 9.6   & 1.33 \\
Kepler-169 & Tr    & Y     & 8.061 & 0.316 & 8.2   & 0.6   & \multicolumn{1}{l}{$0.200\substack{+0.005\\-0.005}$} & \multicolumn{1}{l}{$0.538\substack{+0.020\\-0.019}$} & $20.6\substack{+2\\-1.9}$ & Y     & 4     & 1.6   & 3.2   & 2.44 \\
\hline
\end{tabularx}
\end{table}

\begin{table}
\centering
\setlength{\tabcolsep}{7.25pt}
\captionsetup{width=1.55\textwidth}
\ContinuedFloat
\caption[]{continued}
\begin{tabularx}{1.55\textwidth}{lcccccccccccccc}
\hline
Host name & Dis.$^{a}$  & $F_{1}$ & $(\frac{\chi^2}{dof})_{i}$ & $(\frac{\chi^2}{dof})_{j}$ & $\gamma$ & $\Delta\gamma$$^{b}$ & m & b & \makecell{Period\\(d)} & $F_{2}$ & ON$^{c}$    & \makecell{$R_{\text{Max}}$\\($R_{\oplus}$)} & \makecell{$M_{\text{Max}}$\\($M_{\oplus}$)} & \makecell{$P_{\text{tr}}$\\(\%)} \\
\hline
Kepler-178 & Tr    & N     & 2.330 & 0.001 & 3627.8 & 1221.9 & \multicolumn{1}{l}{$0.335\substack{+0.001\\-0.001}$} & \multicolumn{1}{l}{$0.980\substack{+0.002\\-0.002}$} & $44.6\substack{+0.4\\-0.4}$ & N     & 2     & 3.3   & 11.8  & 2.05 \\
      &       &       &       &       &       &       &       &       & $208.7\substack{+2.9\\-3.1}$ & N     & 4 E   & 4.8   & 23.3  & 0.75 \\
GJ 163 & RV    & N     & 4.847 & 0.005 & 510.2 & 47.4  & \multicolumn{1}{l}{$0.462\substack{+0.003\\-0.003}$} & \multicolumn{1}{l}{$0.941\substack{+0.007\\-0.006}$} & $72.9\substack{+2.1\\-2}$ & N     & 2 H   & 1.4   & 2.5   & 2.05 \\
      &       &       &       &       &       &       &       &       & $214.6\substack{+7.4\\-7.3}$ & N     & 3     & 2.8   & 8.8   & 0.99 \\
      &       &       &       &       &       &       &       &       & $1778.7\substack{+86.2\\-85.9}$ & N     & 5 E   & 10.9  & 103.7 & 0.24 \\
Kepler-138 & Tr    & N     & 1.592 & 0.049 & 31.6  & 8.4   & \multicolumn{1}{l}{$0.116\substack{+0.003\\-0.003}$} & \multicolumn{1}{l}{$1.018\substack{+0.006\\-0.005}$} & $17.8\substack{+0.5\\-0.4}$ & N     & 2     & 0.6   & 0.5   & 1.84 \\
      &       &       &       &       &       &       &       &       & $30.3\substack{+1.2\\-1.2}$ & N     & 4 E   & 0.7   & 0.7   & 1.35 \\
\hline
\end{tabularx}
\end{table}

\begin{table}
\centering
\setlength{\tabcolsep}{7.25pt}
\captionsetup{width=1.55\textwidth}
\ContinuedFloat
\caption[]{continued}
\begin{tabularx}{1.55\textwidth}{lcccccccccccccc}
\hline
Host name & Dis.$^{a}$  & $F_{1}$ & $(\frac{\chi^2}{dof})_{i}$ & $(\frac{\chi^2}{dof})_{j}$ & $\gamma$ & $\Delta\gamma$$^{b}$ & m & b & \makecell{Period\\(d)} & $F_{2}$ & ON$^{c}$    & \makecell{$R_{\text{Max}}$\\($R_{\oplus}$)} & \makecell{$M_{\text{Max}}$\\($M_{\oplus}$)} & \makecell{$P_{\text{tr}}$\\(\%)} \\
\hline
HIP 57274 & RV    & N     & 1.946 & 0.011 & 169.9 & 44.9  & \multicolumn{1}{l}{$0.576\substack{+0.008\\-0.009}$} & \multicolumn{1}{l}{$0.918\substack{+0.015\\-0.015}$} & $116.5\substack{+9\\-8.8}$ & N     & 2     & 6.1   & 36.3  & 1.16 \\
      &       &       &       &       &       &       &       &       & $1665.6\substack{+198.4\\-190.8}$ & N     & 4 E   & 12.3  & 809.1 & 0.20 \\
Kepler-149 & Tr    & N     & 1.249 & 0.124 & 9.0   & 0.7   & \multicolumn{1}{l}{$0.244\substack{+0.010\\-0.009}$} & \multicolumn{1}{l}{$1.478\substack{+0.018\\-0.018}$} & $92.6\substack{+8.6\\-7.4}$ & N     & 2     & 1.8   & 3.9   & 1.13 \\
      &       &       &       &       &       &       &       &       & $285.5\substack{+41.1\\-33.5}$ & N     & 4 E,H & 2.4   & 6.6   & 0.53 \\
GJ 3138 & RV    & N     & 3.354 & 0.093 & 34.9  & 2.4   & \multicolumn{1}{l}{$0.778\substack{+0.027\\-0.032}$} & \multicolumn{1}{l}{$0.055\substack{+0.052\\-0.049}$} & $41\substack{+11.4\\-9.4}$ & N     & 2 H   & 0.9   & 1.2   & 1.12 \\
      &       &       &       &       &       &       &       &       & $1470.4\substack{+664.8\\-495.1}$ & N     & 4 E   & 9.5   & 80.1  & 0.10 \\
Kepler-351 & Tr    & N     & 2.535 & 0.008 & 331.6 & 42.1  & \multicolumn{1}{l}{$0.196\substack{+0.002\\-0.002}$} & \multicolumn{1}{l}{$1.566\substack{+0.004\\-0.004}$} & $90.5\substack{+1.8\\-1.7}$ & N     & 2     & 2.4   & 6.6   & 1.03 \\
\hline
\end{tabularx}
\end{table}

\begin{table}
\centering
\setlength{\tabcolsep}{7.75pt}
\captionsetup{width=1.55\textwidth}
\ContinuedFloat
\caption[]{continued}
\begin{tabularx}{1.57\textwidth}{lcccccccccccccc}
\hline
Host name & Dis.$^{a}$  & $F_{1}$ & $(\frac{\chi^2}{dof})_{i}$ & $(\frac{\chi^2}{dof})_{j}$ & $\gamma$ & $\Delta\gamma$$^{b}$ & m & b & \makecell{Period\\(d)} & $F_{2}$ & ON$^{c}$    & \makecell{$R_{\text{Max}}$\\($R_{\oplus}$)} & \makecell{$M_{\text{Max}}$\\($M_{\oplus}$)} & \makecell{$P_{\text{tr}}$\\(\%)} \\
\hline
HD 34445 & RV    & N     & 2.169 & 0.522 & 3.2   & 1.4   & \multicolumn{1}{l}{$0.339\substack{+0.025\\-0.019}$} & \multicolumn{1}{l}{$1.715\substack{+0.054\\-0.065}$} & $2594.4\substack{+1314.8\\-801.1}$ & N     & 5     & 12.1  & 181.4 & 0.16 \\
      &       &       &       &       &       &       &       &       & $12242.1\substack{+8542\\-4539.8}$ & N     & 7 E   & 12.3  & 1108.7 & 0.06 \\
47 Uma & RV    & N     & 14.799 & 0.500 & 28.6  & 18.5  & \multicolumn{1}{l}{$0.347\substack{+0.024\\-0.022}$} & \multicolumn{1}{l}{$3.033\substack{+0.001\\-0.001}$} & $5624.1\substack{+635.7\\-531.9}$ & N     & 2     & 12.1  & 179.9 & 0.09 \\
      &       &       &       &       &       &       &       &       & $26359.6\substack{+6565.8\\-4927.6}$ & N     & 4 E   & 12.3  & 1090.6 & 0.03 \\
\hline
\end{tabularx}
\small
\begin{tablenotes}
\item$^{a}$Discovery method of the system: `Tr', `RV', `OBM', and `PT' represent transit, radial velocity, orbital brightness modulation, and pulsar timing, respectively.\item$^b$$\Delta\gamma$=($\gamma_{1}$-$\gamma_{2}$)/$\gamma_{2}$, where $\gamma_{1}$ and $\gamma_{2}$ are the highest and second-highest $\gamma$ values for the system, respectively.\item$^c$Orbital numbers (ON) followed by `E' indicate the extrapolated planets, followed by `H' indicate the predicted planets within the HZ, and followed by `C' indicate that corresponding orbital periods have been flagged as ``Planetary Candidate'' in NASA Exoplanet Archive.
\end{tablenotes}
\end{table}
\end{landscape}

\begin{figure}
\centering
\captionsetup{width=0.55\paperwidth}
\includegraphics[width=0.55\paperwidth]{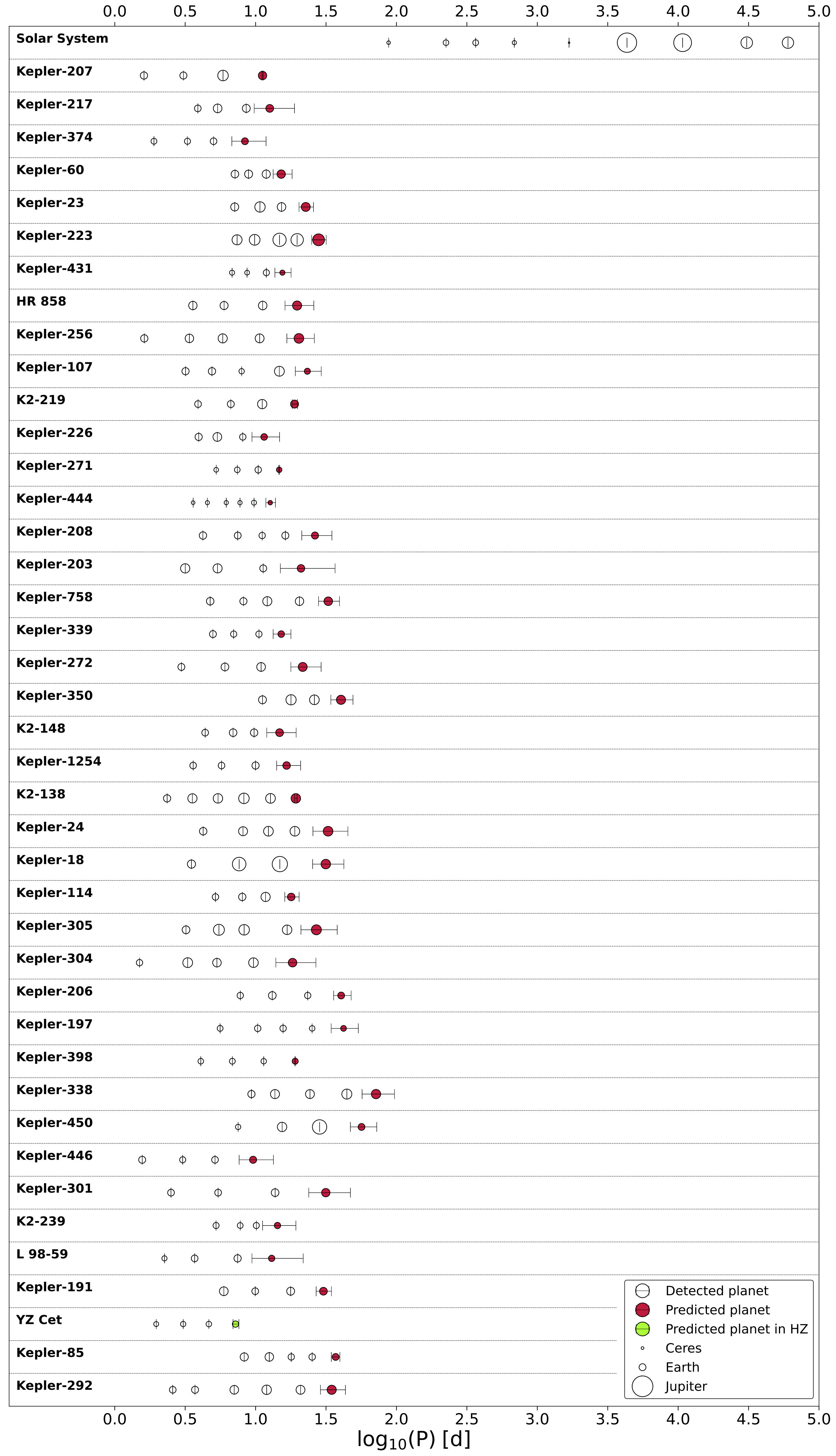}
\caption[Orbital periods and scaled radii of exoplanets in multiple-planet systems, including the predicted exoplanets from extrapolations]{Orbital periods and scaled radii of exoplanets in multiple-planet systems, including the predicted exoplanets from extrapolations. The empty and red circles indicate detected and predicted planets in systems, respectively. The green circles also indicate the predicted planets located within the HZ of their parent stars. The estimated radius of the predicted planet in GJ 676 A is higher than the maximum possible limit of a typical planet. Furthermore, due to the discovery method of HR 8799, the predicted planet's radius is not calculated. Therefore, GJ 676 A and HR 8799 are excluded.}
\label{figureA.1}
\end{figure}

\begin{figure}
\ContinuedFloat
\captionsetup{width=0.55\paperwidth}
\begin{center}
 \includegraphics[width=0.55\paperwidth]{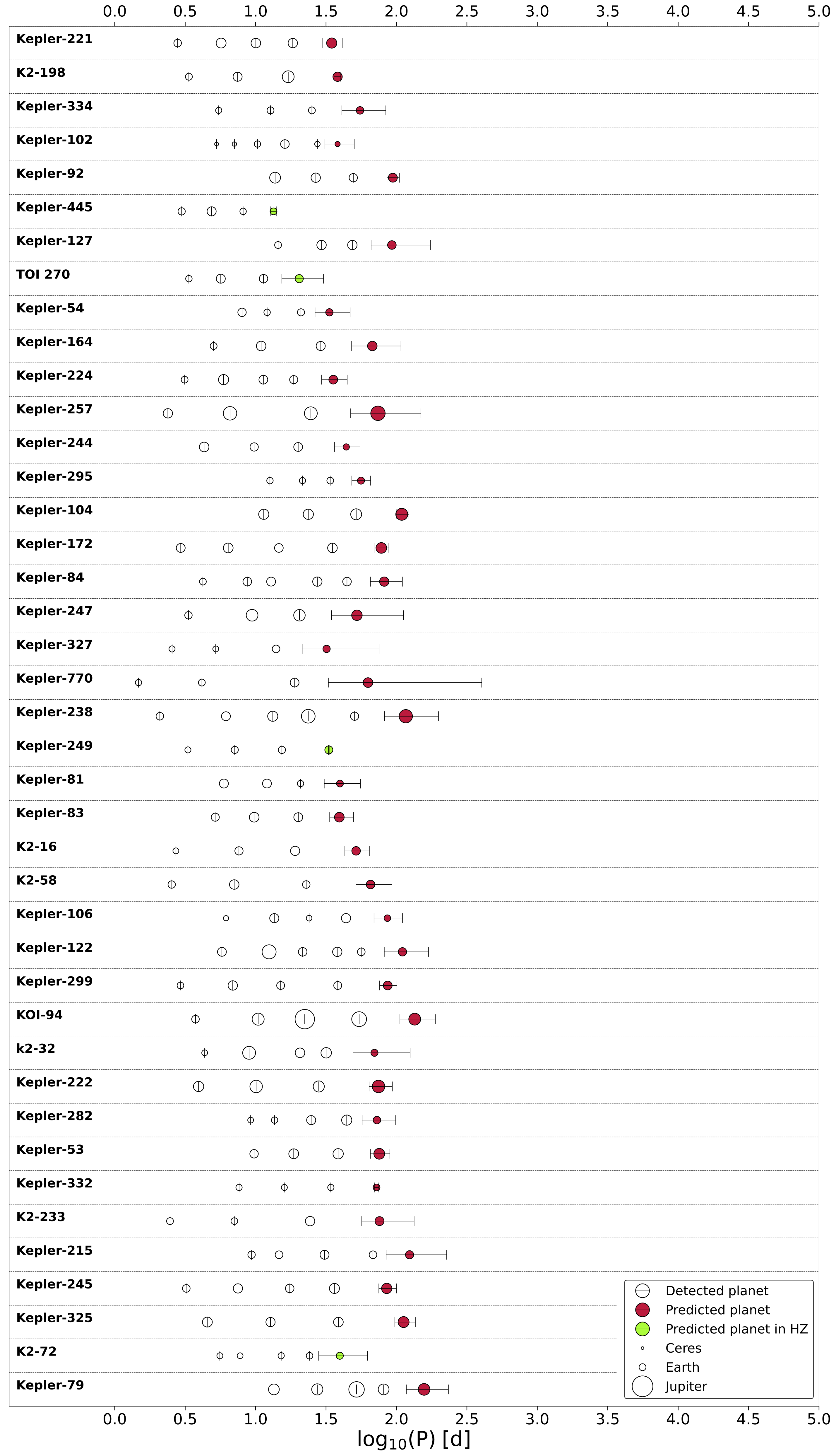}
 \end{center}
 \caption[]{continued}
\end{figure}

\begin{figure}
\ContinuedFloat
\captionsetup{width=0.55\paperwidth}
\begin{center}
 \includegraphics[width=0.55\paperwidth]{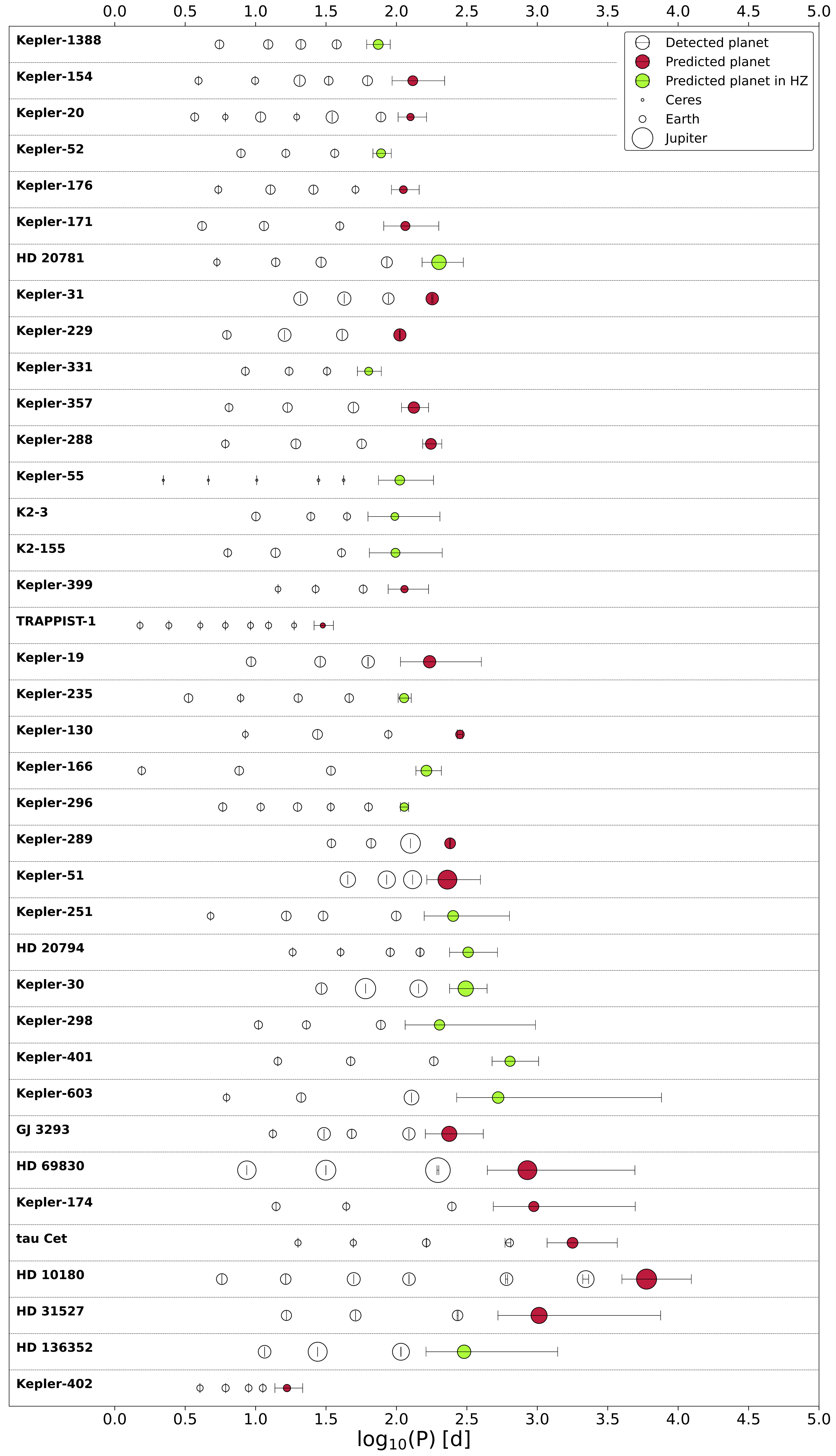}
 \end{center}
 \caption[]{continued}
\end{figure}

\begin{figure}
\centering
\captionsetup{width=0.55\paperwidth}
 \includegraphics[width=0.55\paperwidth]{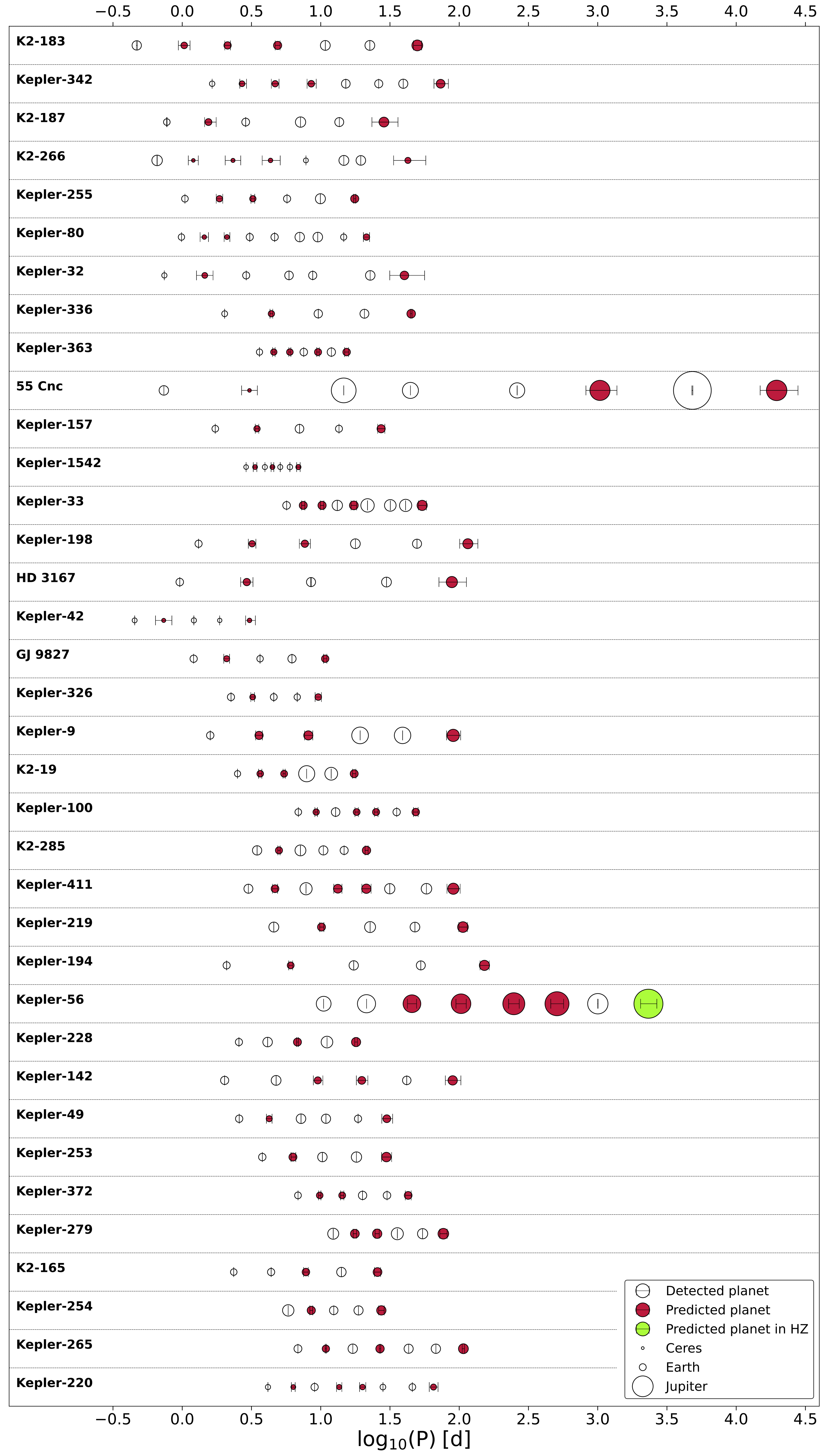}
 \caption[Orbital periods and scaled radii of exoplanets in multiple-planet systems with interpolated and extrapolated planet predictions]{Orbital periods and scaled radii of exoplanets in multiple-planet systems with interpolated and extrapolated planet predictions. The empty and red circles indicate detected and predicted planets in systems, respectively. The green circles also indicate the predicted planets within the HZ of their parent stars. Due to the discovery methods of KIC 10001893 and PSR B1257+12, the predicted planets' radii are not calculated. Therefore, KIC 10001893 and PSR B1257+12 are excluded.}
 \label{figureA.2}
\end{figure}

\begin{figure}
\ContinuedFloat
\captionsetup{width=0.55\paperwidth}
\begin{center}
 \includegraphics[width=0.55\paperwidth]{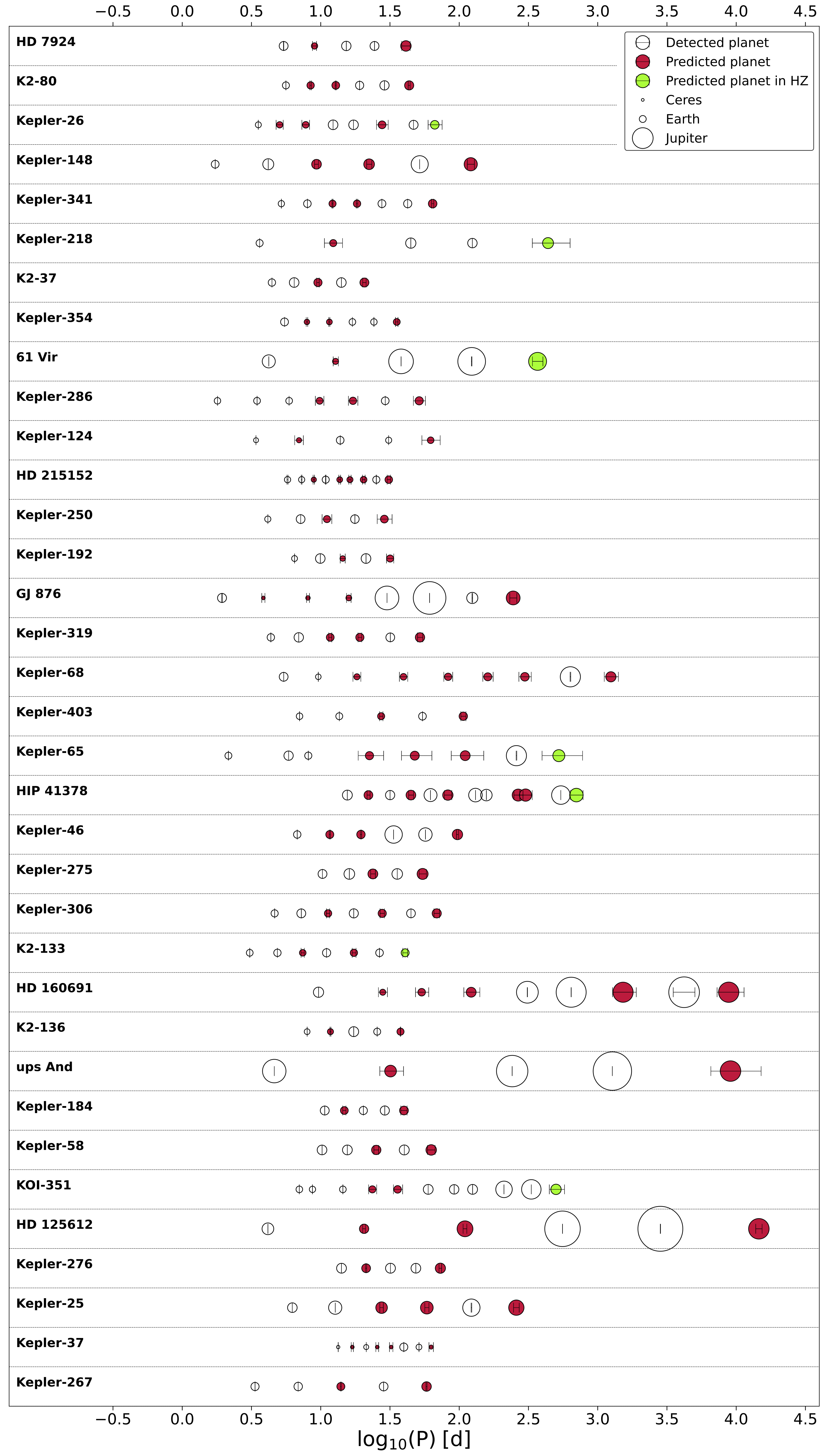}
 \end{center}
 \caption[]{continued}
\end{figure}

\begin{figure}
\ContinuedFloat
\captionsetup{width=0.55\paperwidth}
\begin{center}
 \includegraphics[width=0.55\paperwidth]{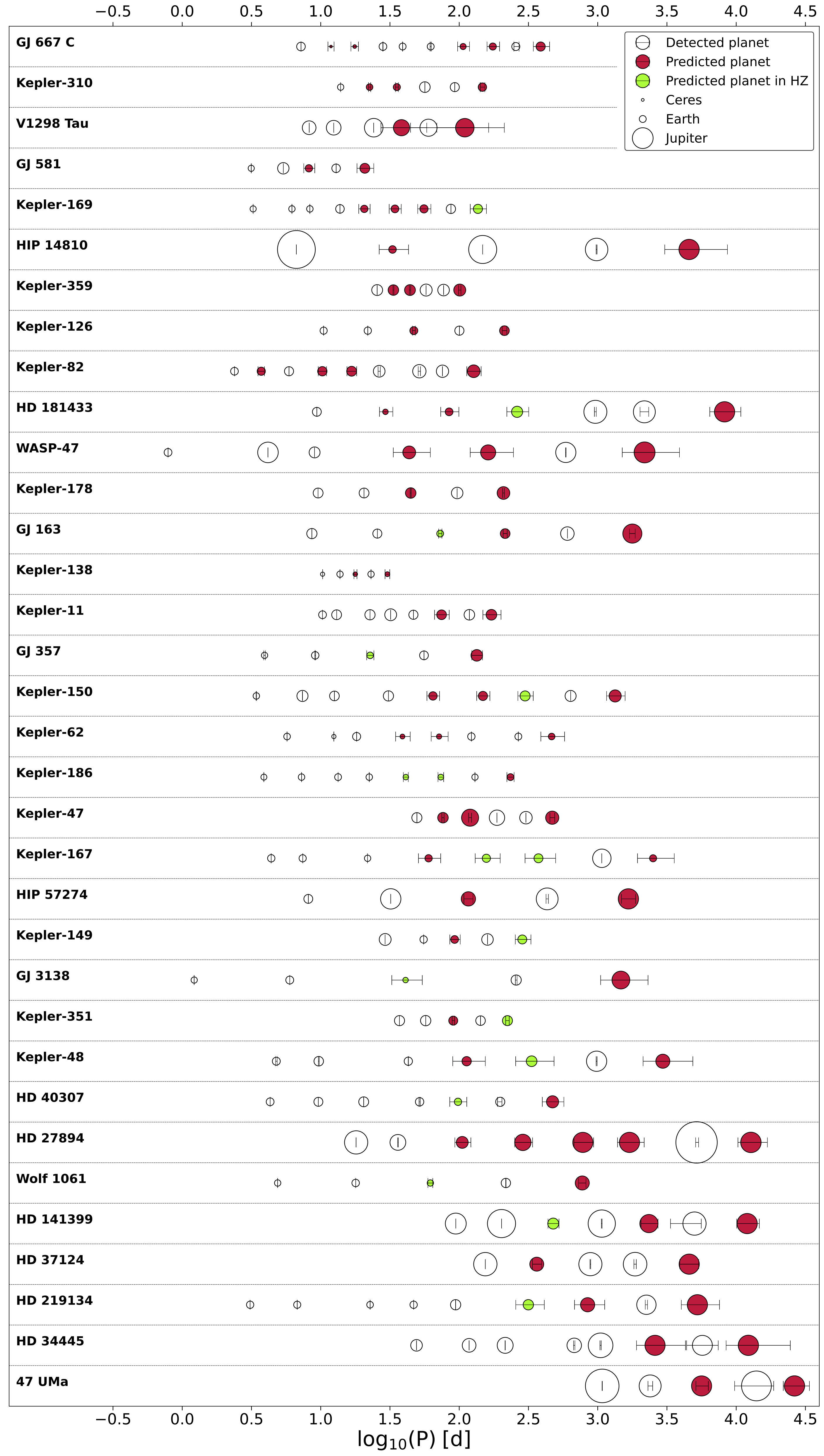}
 \end{center}
 \caption[]{continued}
\end{figure}

\chapter{Data Cleaning and Re-scaling and Their Impact on Prediction Accuracy}

\newpage
\section{Data-cleaning}\label{appendixB.1}
Our dataset contains 770 data points (sample 2). The Local Outlier Factor (LOF) method is chosen to identify outlier observations. It is first applied to all parameters including $P$, $e$, $M_{p}$, $R_{p}$, $M_{s}$, $R_{s}$, Fe/H, and $T_{\text{eff}}$, and then to $R_{p}$ and $M_{p}$. The first step determines 39 outliers with an average score of 1.951, where the higher the LOF score, the more abnormal the data point. In comparison, the average score of inliers is 1.113. The second step determines 37 outliers and assigns average scores of 2.113 and 1.061 to the outlier and inlier data points, respectively. In total, the LOF marks 76 data points as outliers. The outliers have an average Mahalanobis distance of 18.050, compared to 6.888 for the inliers. This indicates that the 76 outlier data points are farther away from the dataset's central point than the inliers. It is interesting to note that identified outliers have higher uncertainties than inliers. In a logarithmic scale, outlier data points have an average uncertainty of 0.19 for planetary mass and 0.05 for planetary radius. In comparison, these values for inlier data points are 0.11 and 0.04, respectively.

\begin{figure}[ht]
\centering
\captionsetup{width=0.60\paperwidth}
\includegraphics[width=0.60\paperwidth]{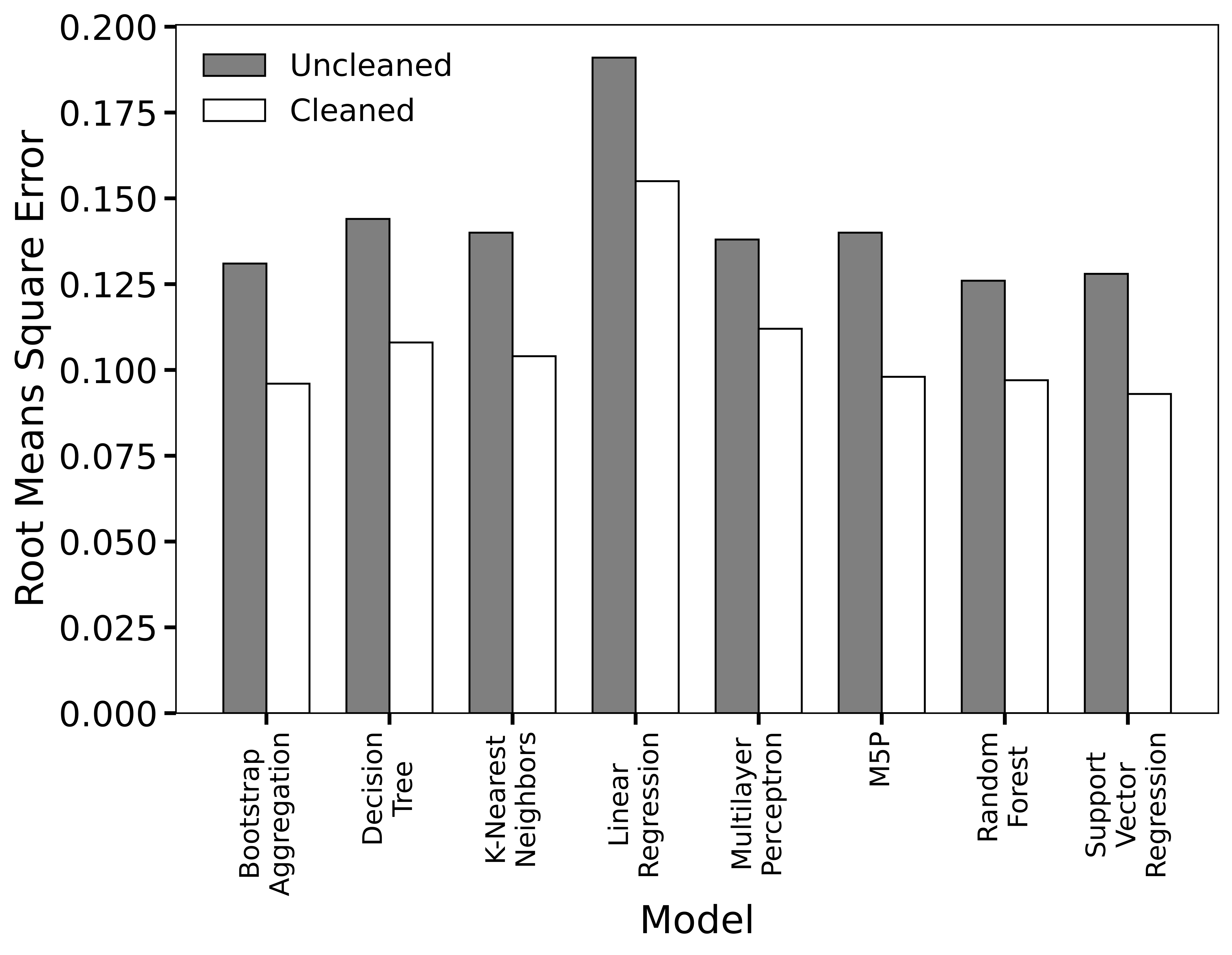}
\caption[RMSE values of different ML regression models implemented in uncleaned and cleaned datasets]{Root means square error (RMSE) values of different ML regression models implemented in uncleaned (gray bars) and cleaned (white bars) datasets. All models have higher RMSE values when outliers are included in the dataset.}
\label{figureB.1}
\end{figure}

To quantify the impact of outliers on prediction precisions, we run ML regression models on the dataset containing all 770 data points. Figure~\ref{figureB.1} compares the prediction accuracy obtained from the uncleaned (with outliers) and cleaned (without outliers) datasets. As can be seen, all models perform remarkably better when outliers are removed from the dataset, demonstrating the significance of the data-cleaning step in predicting the planetary radius.

\newpage
\section{Data re-scaling}\label{appendixB.2}
\begin{figure}[ht]
\centering
\captionsetup{width=0.60\paperwidth}
\includegraphics[width=0.60\paperwidth]{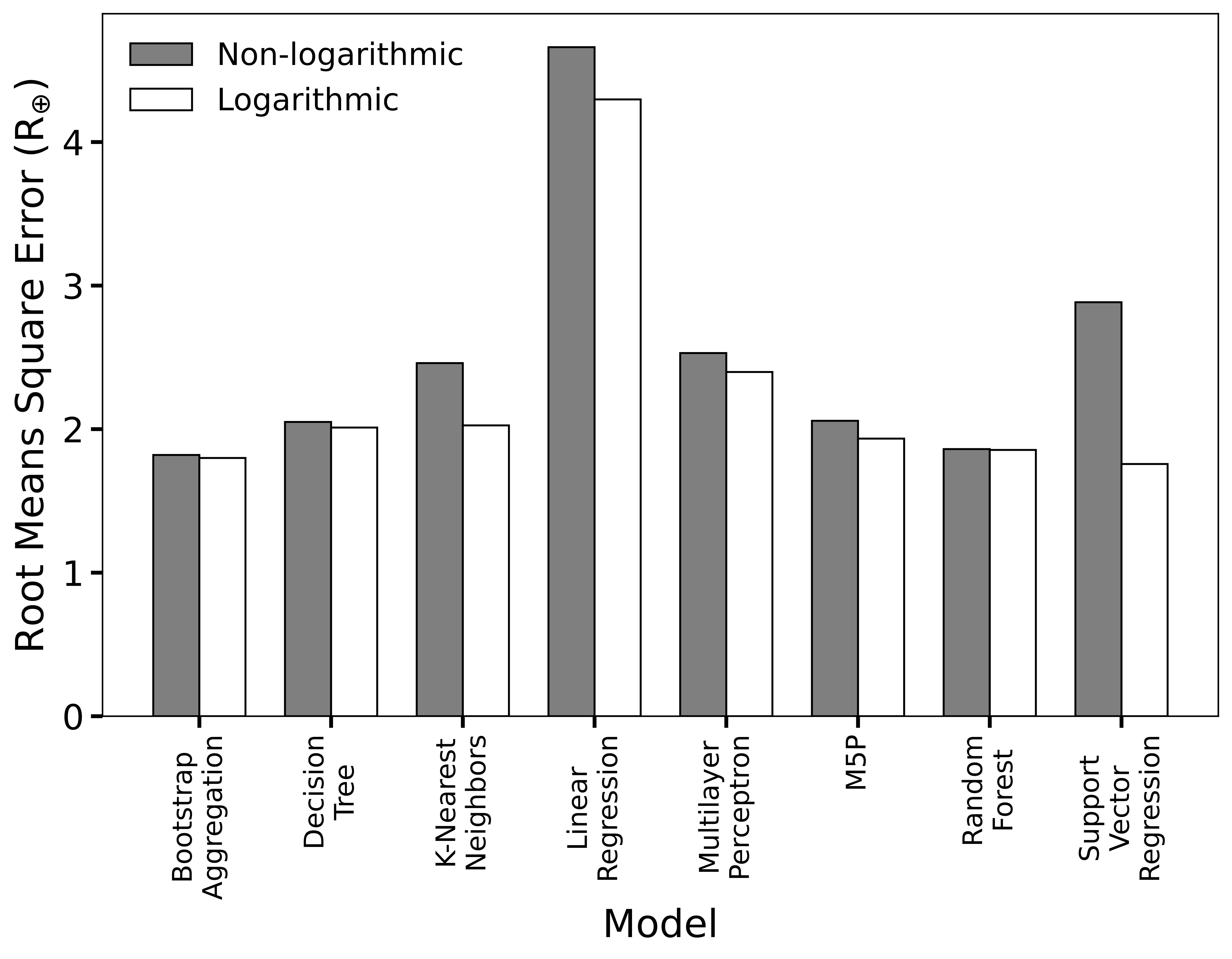}
\caption[RMSE values of different ML regression models implemented in logarithmic and non-logarithmic datasets]{Root means square error (RMSE) values of different ML regression models implemented in logarithmic (white bars) and non-logarithmic (gray bars) datasets. Bootstrap Aggregation and Random Forest models do not show a remarkable difference between logarithmic and non-logarithmic scaling. In contrast, other algorithms, particularly the Support Vector Regression, provide better results on a logarithmic scale.}
\label{figureB.2}
\end{figure}

We process both logarithmic and non-logarithmic datasets (sample 2) to understand the effect of data re-scaling on predicting planetary radius. Figure~\ref{figureB.2} compares the RMSE values corresponding to different models applied in non-logarithmic and logarithmic datasets. Bootstrap Aggregation and Random Forest slightly differ between logarithmic and non-logarithmic scaling. In contrast, other algorithms, particularly Support Vector Regression, provide better results on a logarithmic scale. Transforming the exoplanet data into a logarithmic space helps handle the wide range of values by compressing them, allowing the ML model better to capture the underlying patterns and relationships within the data. Moreover, logarithm transformation efficiently addresses data skewness and outliers. When data do not follow a normal distribution and contain extreme values, the logarithmic space helps mitigate the influence of outliers by compressing their impact and making the data more symmetrical.

\chapter{Extra Figures}
\label{appendixC} 

\newpage
\begin{figure}
\centering
\captionsetup{width=0.50\paperwidth}
\includegraphics[width=0.50\paperwidth]{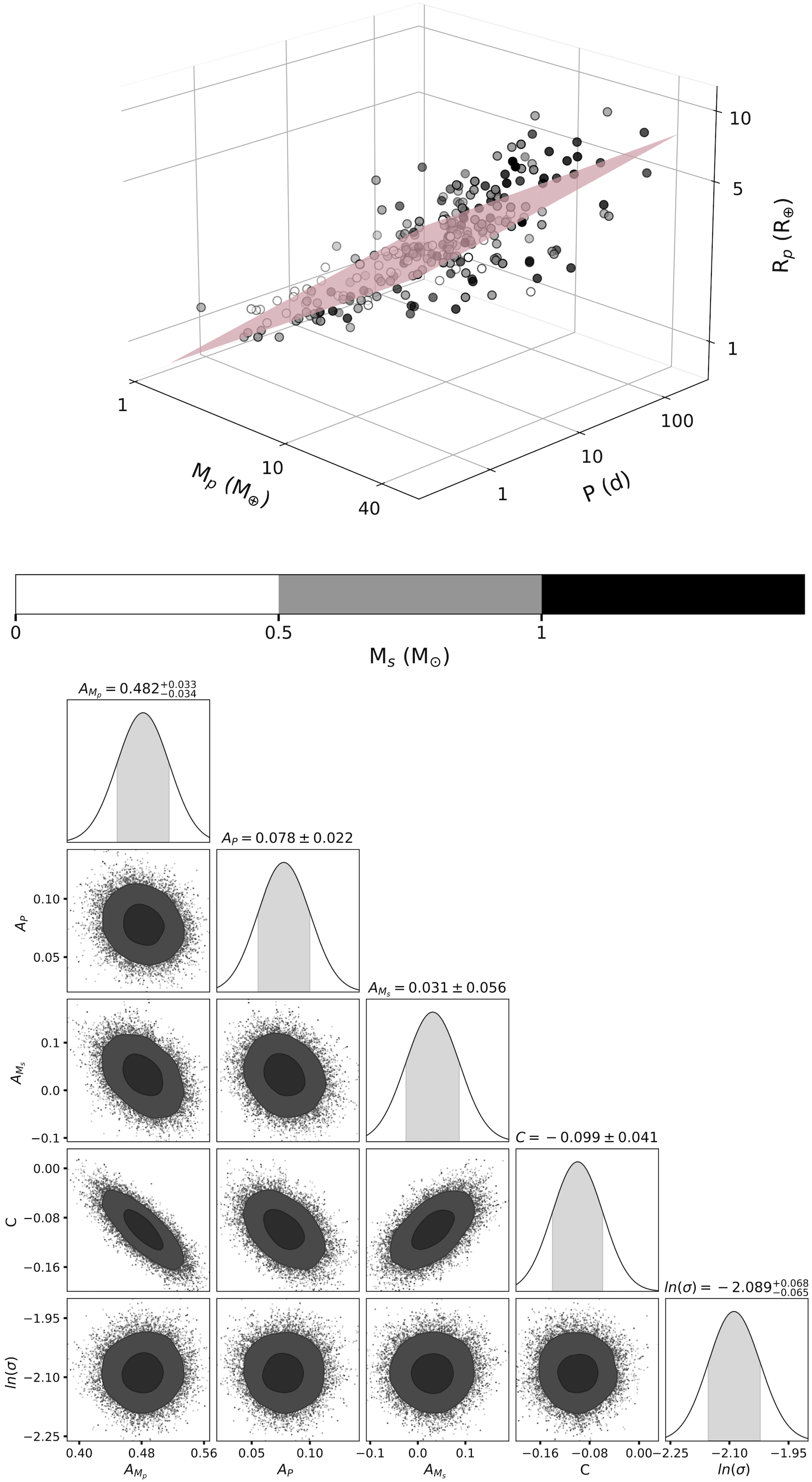}
\caption[The regression obtained by the M5P model for small planets]{Upper panel: the regression surface (red plane) obtained by the M5P model for small planets. This surface has been plotted using $\log(R_{p}/R_{\oplus})=0.482\log(M_{p}/M_{\oplus})+0.078\log(P/d)+0.031\log(M_{s}/M_{\odot})-0.099$ (see table~\ref{table3.7}, row 6), where the stellar mass that has the lowest coefficient has not been considered. The color of the circles indicates the $M_{s}$ value. Lower panel: the one- and two-dimensional marginalized posterior probability distributions of parameters obtained by the MCMC method. Coefficients of planetary mass, orbital period, and stellar mass along with constant term are shown by $A_{M_{p}}$, $A_{P}$, $A_{M_{s}}$, and $C$, respectively. The uncertainty around the best scaling relation is shown by $\sigma$.}
\label{figureC.1}
\end{figure}

\begin{figure}
\centering
\captionsetup{width=0.50\paperwidth}
\includegraphics[width=0.50\paperwidth]{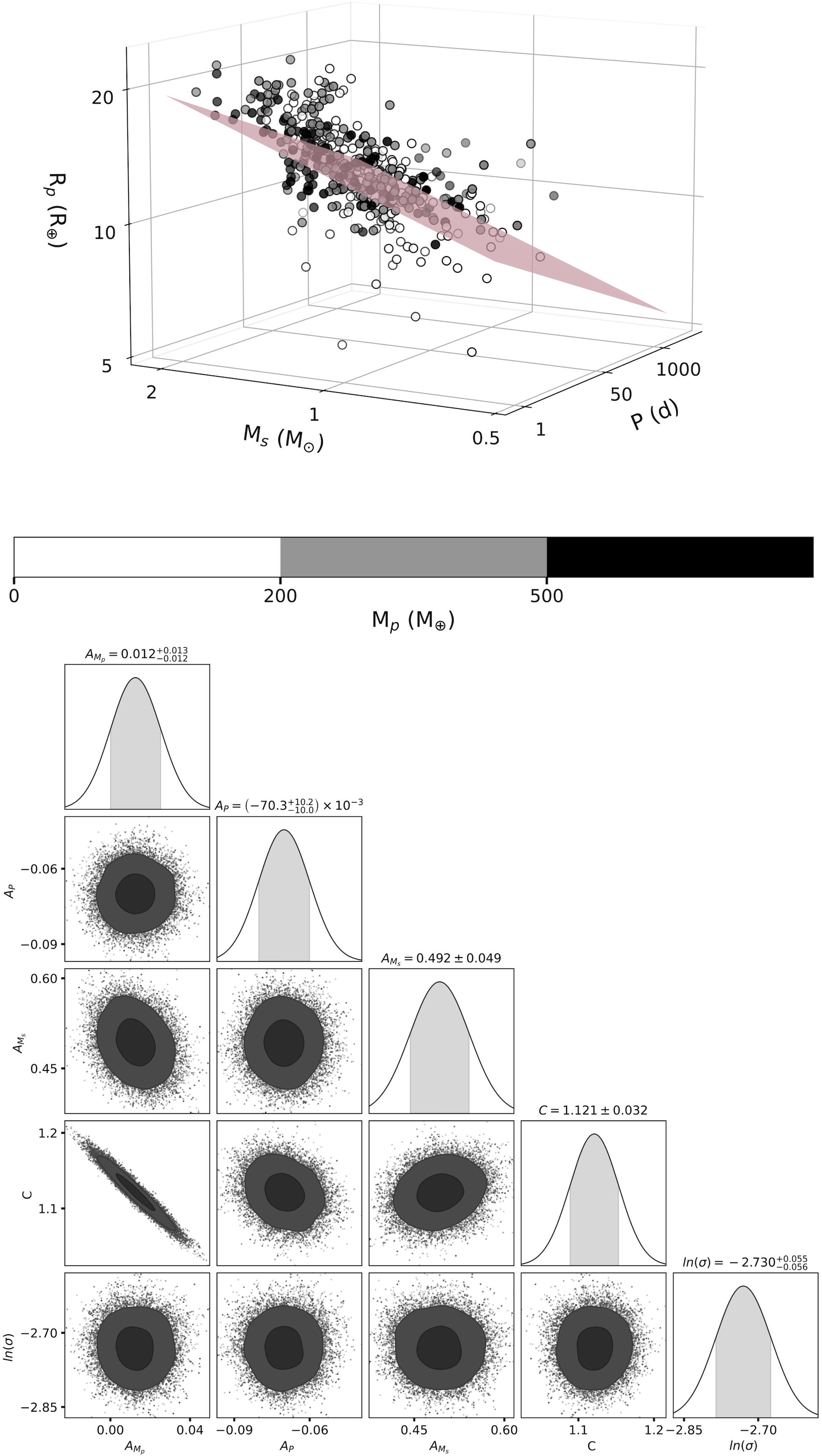}
\caption[The regression obtained by the M5P model for giant planets]{Upper panel: the regression surface (red plane) obtained by the M5P model for giant planets. This surface has been plotted using $\log(R_{p}/R_{\oplus})=0.013\log(M_{p}/M_{\oplus})-0.070\log(P/d)+0.492\log(M_{s}/M_{\odot})+1.121$ (see table~\ref{table3.7}, row 7), where the planetary mass that has the lowest coefficient has not been considered. The color of the circles indicates the $M_{p}$ value. Lower panel: the one- and two-dimensional marginalized posterior probability distributions of parameters obtained by the MCMC method. Coefficients of planetary mass, orbital period, and stellar mass along with constant term are shown by $A_{M_{p}}$, $A_{P}$, $A_{M_{s}}$, and $C$, respectively. The uncertainty around the best scaling relation is shown by $\sigma$.}
\label{figureC.2}
\end{figure}

\printbibliography[heading=bibintoc]

\newrefsegment
\newpage
\thispagestyle{empty}
\begin{center}
\includegraphics[scale=0.27]{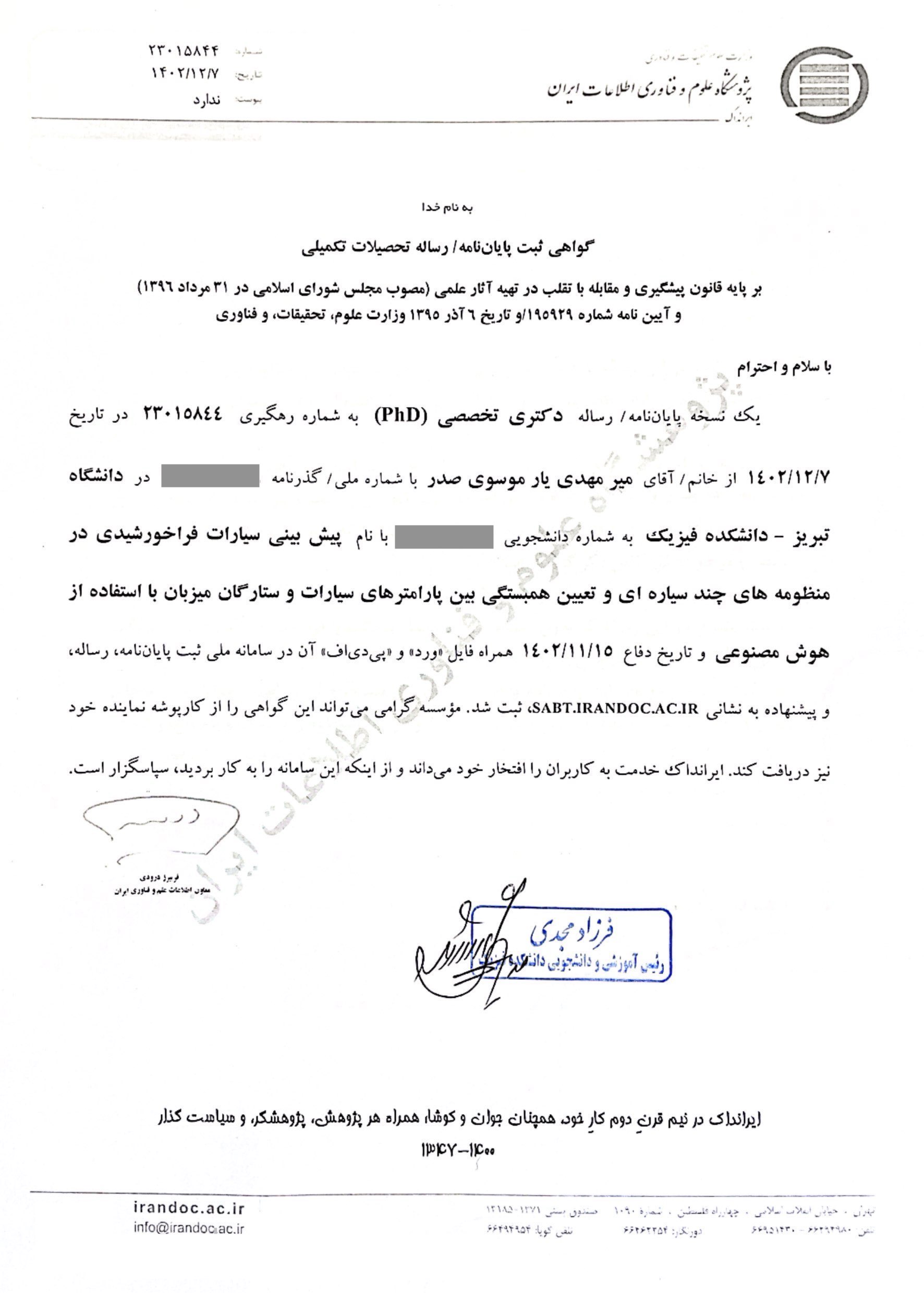}
\newpage
\thispagestyle{empty}
\includegraphics[scale=0.26]{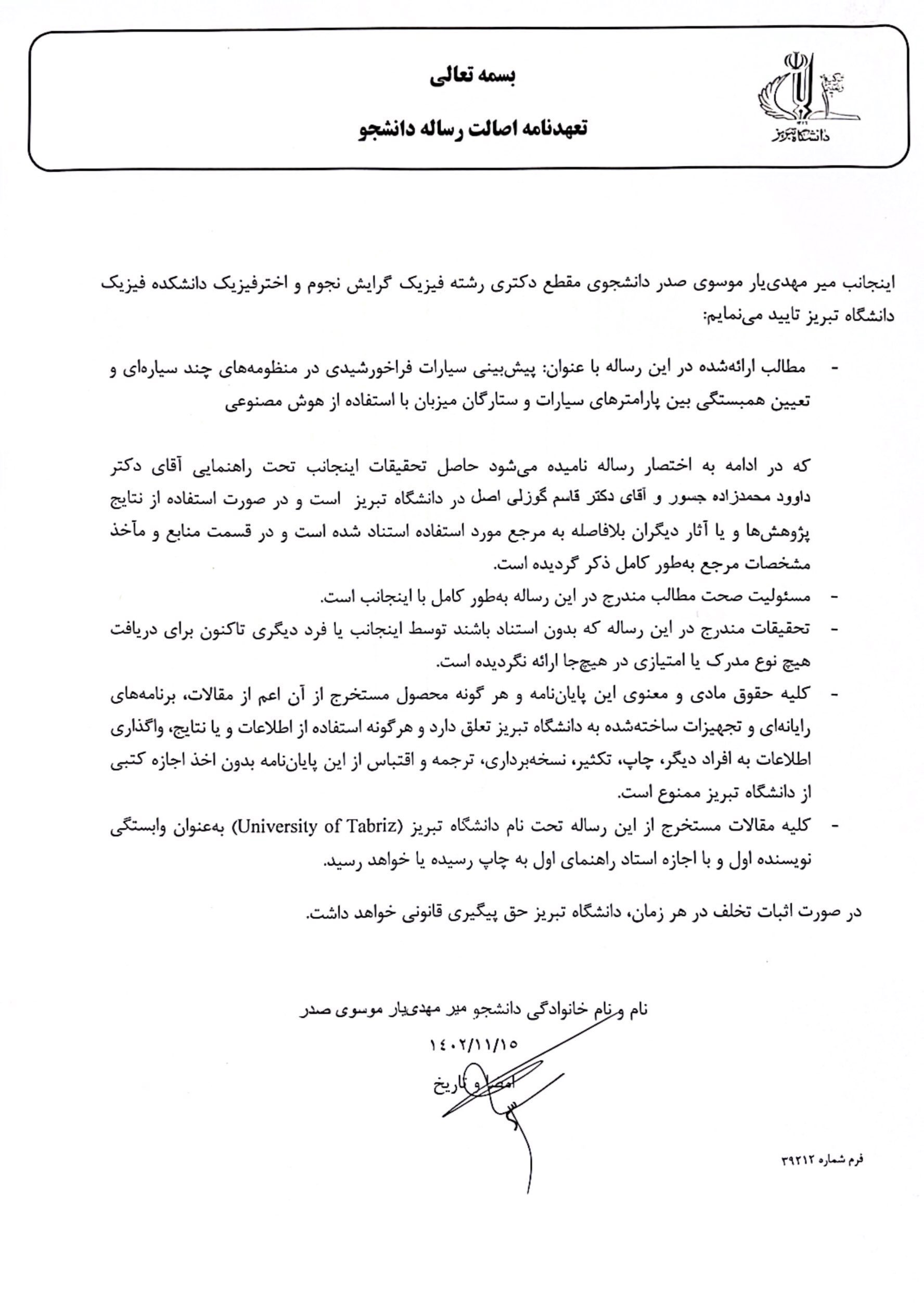}
\newpage
\thispagestyle{empty}
\includegraphics[scale=0.37]{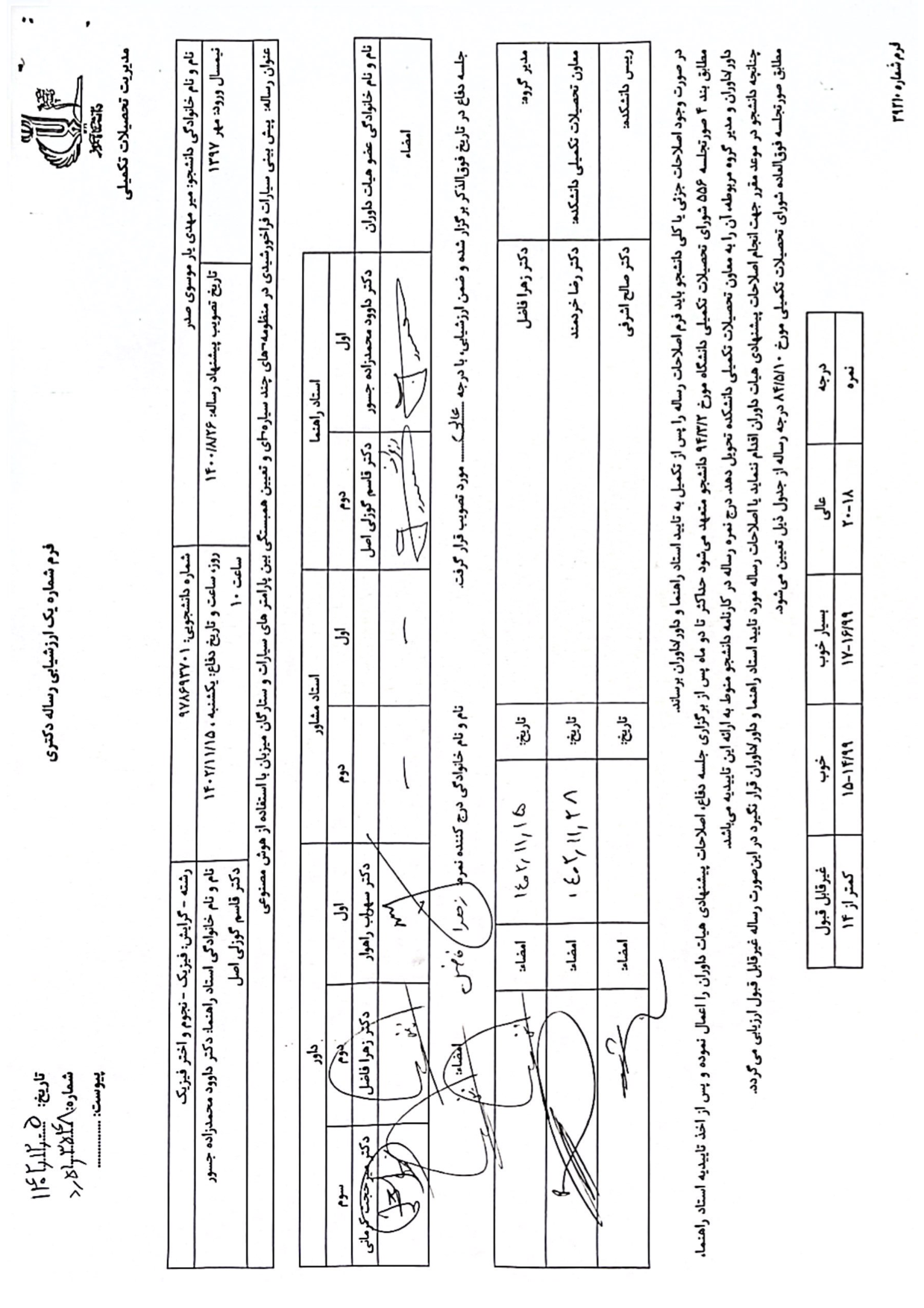}
\newpage
\thispagestyle{empty}
\includegraphics[scale=0.40]{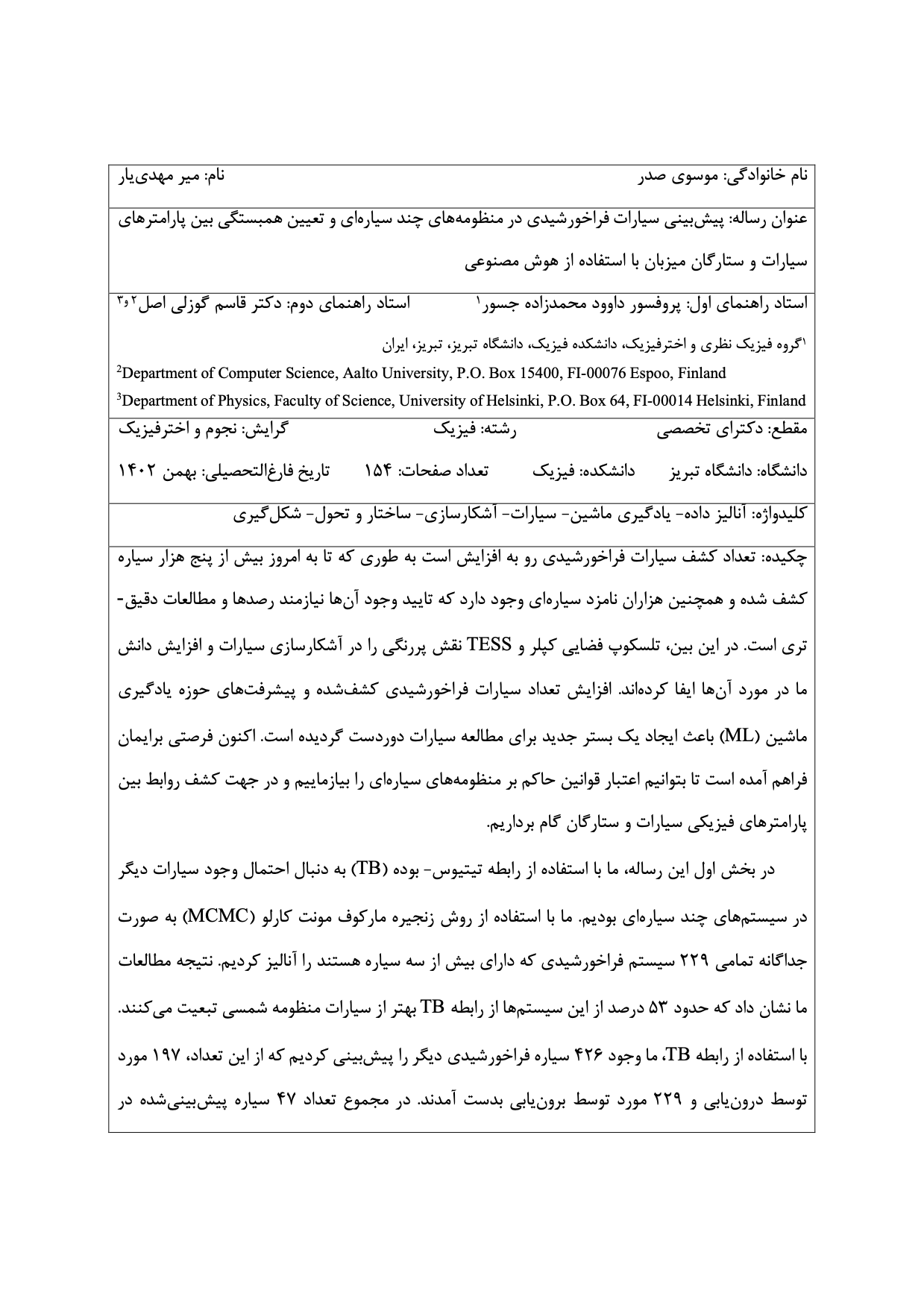}
\newpage
\thispagestyle{empty}
\includegraphics[scale=0.40]{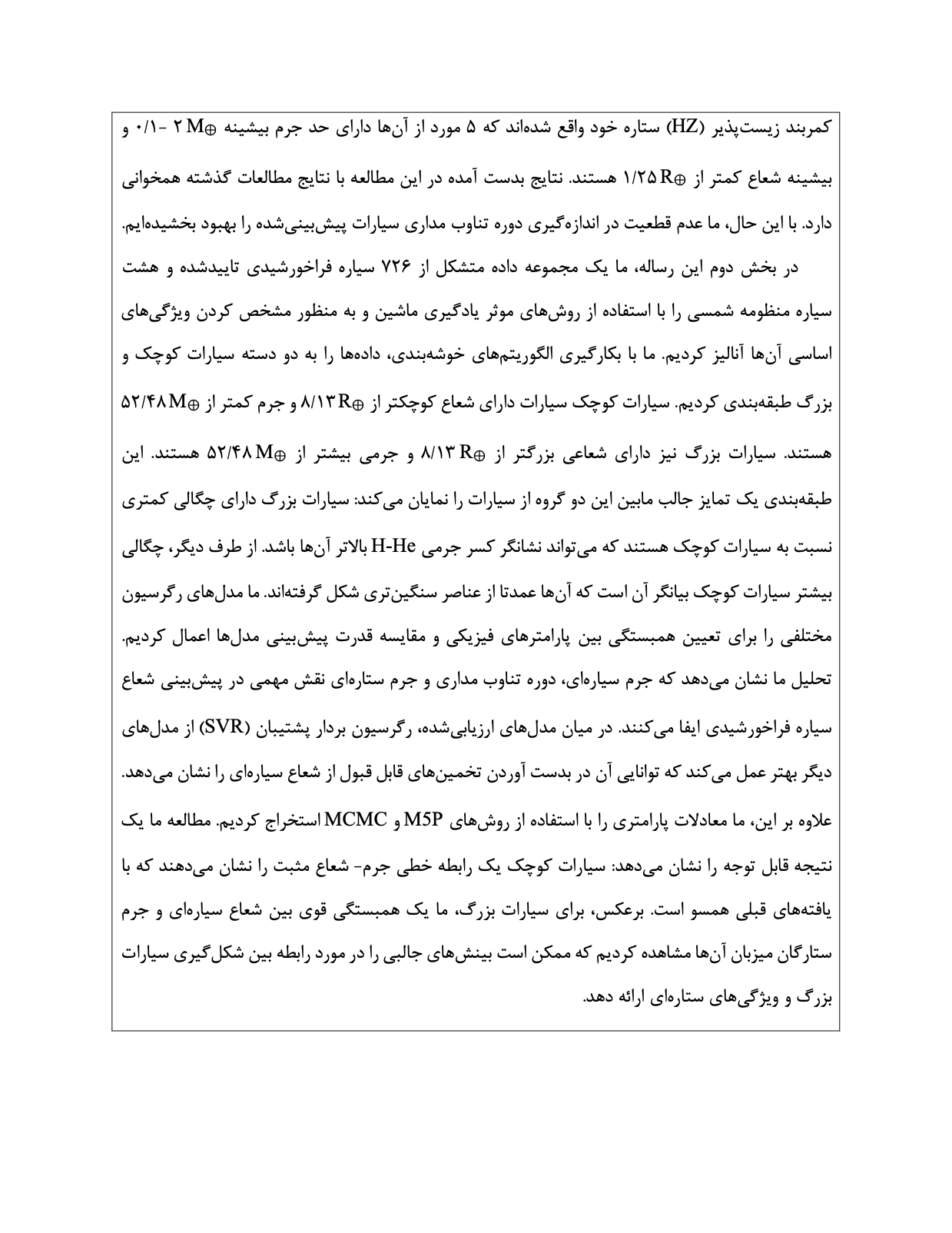}
\end{center}

\end{document}